\documentclass[a4paper,11pt]{article}
\pdfoutput=1 

\usepackage{jheppub} 
\usepackage[bottom]{footmisc}
\usepackage{amssymb}
\usepackage{amsmath}
\usepackage{amsthm}
\usepackage[usenames,dvipsnames]{xcolor}
\usepackage{epsfig}
\usepackage{dcolumn}
\usepackage{tikz}
\usetikzlibrary{shapes.geometric, arrows}
\usepackage{upgreek}
\usepackage{setspace}
\usepackage{subfig}
\usepackage{enumitem}
\usepackage{array,multirow,bigdelim,arydshln}
\usepackage{appendix}
\usepackage{xparse}
\usepackage{stmaryrd}
\usepackage[T1]{fontenc} 
\usepackage{mathtools}
\usepackage{physics} 
\usepackage[export]{adjustbox}
\usepackage{multirow}
\usepackage{graphicx} 
\usepackage{float} 
\graphicspath{{./images/}}
\usepackage[nottoc]{tocbibind}
\usepackage{hyperref}
\usepackage[utf8]{inputenc}
\usepackage{CJK}
\usetikzlibrary{decorations.text}
\usetikzlibrary{decorations.markings}
\usetikzlibrary{decorations.pathmorphing}
\usetikzlibrary{intersections}
\usetikzlibrary{calc}
\hypersetup{
	colorlinks,
	urlcolor=Maroon,
	linkcolor=Maroon,
	citecolor=Maroon
	}

\newcommand{\bub}{
\begin{tikzpicture}[scale=0.25]
    \draw (0,0) -- (0.5,0);
    \draw (1,0) circle (0.5);
    \draw (1.5,0) -- (2,0);
\end{tikzpicture}
}

\NewDocumentCommand{\binomial}{omm}
 {%
  \genfrac(){0pt}{}{#2}{#3}%
  \IfValueT{#1}{_{\!#1}}%
 }
\NewDocumentCommand{\eulerian}{omm}
 {%
  \genfrac<>{0pt}{}{#2}{#3}%
  \IfValueT{#1}{_{\!#1}}%
 }

\usepackage{latexsym}
\usepackage{tikz}

\theoremstyle{plain}

\theoremstyle{definition}

\def\bi{\begin{itemize}}
\def\ei{\end{itemize}}
\def\it{\textit}

\DeclareMathAlphabet{\mathbbold}{U}{bbold}{m}{n}

\def\bea#1\eea{\begin{eqnarray}#1\end{eqnarray}}
\def\be#1\ee{\begin{equation}#1\end{equation}}
\def\ba#1\ea{\begin{align}#1\end{align}}

\usepackage{amsmath}
\usepackage{multicol}
\usepackage{bbm}

\usepackage{amsthm}
\usepackage{mathrsfs}
\usepackage{upgreek}
\usepackage{amssymb}
\usepackage{bm}
\usepackage{setspace}
\usepackage{array,multirow,arydshln}
\usepackage{bigdelim}
\usepackage{scalerel}
\usepackage{diagbox}

\usepackage{tabularx}

\usepackage{comment}

\usepackage{tikz}

\usetikzlibrary{shapes.geometric,arrows,arrows.meta,decorations.pathmorphing,decorations.markings,patterns}

\def\<{\langle}
\def\>{\rangle}

\usepackage[compat=1.1.0]{tikz-feynman}
\tikzset{
    vector/.style={
        decoration={snake, aspect=0.75, mirror, segment length=2mm},
        decorate
    },
    photon/.style={decorate, decoration={snake, amplitude=1pt, segment length=4pt}}
}

\usepackage[percent]{overpic}
\usepackage{multirow} 
\usepackage{empheq}
\usepackage{slashed}

\usepackage{slashed}
\tikzset{>=latex}
\usetikzlibrary{patterns.meta} 

\title{How gluon leading singularities discover curves on surfaces}

\author[a]{Sérgio  Carrôlo,}
\author[b]{Carolina Figueiredo}

\affiliation[a]{Max-Planck-Institut für Physik, Werner-Heisenberg-Institut, Boltzmannstr, 8, 85748 Garching, Germany}
\affiliation[b]{Jadwin Hall, Princeton University, Princeton, NJ 08540, USA}

\emailAdd{scarrolo@mpp.mpg.de}
\emailAdd{cfigueiredo@princeton.edu}

\abstract{We study the leading singularities for pure gluon amplitudes obtained by on-shell gluing of three-particle amplitudes for an arbitrary graph in any number of dimensions. By encoding the polarization vector contractions in a graphical way, on-shell gluing ``discovers'' curves on surfaces, and we find that the leading singularity is determined by a simple combinatorial question: what are all ways of covering the graph with non-overlapping curves such that each edge is covered exactly once? This precisely matches the formula from the surfaceology formulation of gluons, where the leading singularities are given by maximal residues, with the combinatorial problem arising from the linearized form of the $u$ variables. At loop-level we describe how the novelties associated with spin sums (related with the need for ghosts when working off-shell using Lagrangians) can be easily encoded in this combinatorial picture. Matching the leading singularities also lets us settle an open question in the surface formulation of gluons, determining the exponents of the closed curves at any loop order.}

\begin{document}

\addtocontents{toc}{\protect\setcounter{tocdepth}{2}}
\maketitle

\section{Introduction}
Leading singularities \cite{Cachazo:2008vp,Arkani-Hamed:2009ljj,Bullimore:2009cb,Bern:2007ct,Britto:2004nc} (LS) are a fundamental feature of scattering amplitudes. At tree-level they encode the most singular part of the amplitude, and similarly at loop-level they show up as the rational coefficients of the terms in the amplitude with highest transcendental weight \cite{Henn:2021aco}. Given a graph, $\mathcal{G}^m_k$, at $m$-loops with $k$ loop-propagators, then if the spacetime dimension, $D$, is such that $D\times m\geq k$, we can reach the so-called maximal cut or leading singularity, which  is given as follows~\cite{Cutkosky:1960sp,Hannesdottir:2022bmo}: 
\begin{equation}
\text{Cut } \mathcal{I}_{\mathcal{G}^m_k} \propto \tilde{N} \times \int \prod_{a=1}^m \frac{d^D l_a}{i \pi ^{D/2}}  \prod_{k} \delta(q_k^2-m^2),  
\label{eq:UnitarityCut}
\end{equation}
where $\tilde{N}$ is a numerator function, evaluated on the locus where all propagators of the graph $\mathcal{G}^m_k$ are on-shell\footnote{Here we are considering that both the loop-propagators as well as the tree propagators are on-shell}. 

In this paper we focus on  studying this numerator function for the case of gluon amplitudes in non-supersymmetric Yang-Mills (YM) theory, which can be computed directly by on-shell gluing of the three-point gluon vertices -- from now on we will call this function $\tilde{N}$ the leading singularity. We assume the spacetime dimension to be sufficiently large so that we can always reach the maximal cut, and therefore provided we have $\tilde{N}$, we can reconstruct the full unitarity cut by multiplying it by the respective Lorentz-invariant phase space integral as given in \eqref{eq:UnitarityCut}.

In addition to their intrinsic importance, leading singularities have played a crucial role in discovering new ways of formulating amplitudes via geometry. For example, in the context of maximally supersymmetric YM, $\mathcal{N}=4$ SYM, the understanding of the 4-dimensional leading singularities as maximal residues of the positive Grassmannian \cite{Arkani-Hamed:2009ljj,Arkani-Hamed:2012zlh,postnikov2006totalpositivitygrassmanniansnetworks} was instrumental to understand the full connection between the amplitudes in this theory and the Amplituhedron \cite{Arkani-Hamed:2013jha}. 

More recently, it has been proposed that amplitudes of non-supersymmetric gluons in any number of dimensions can be extracted from the low-energy limit of a surface ``stringy'' integral \cite{Arkani-Hamed:2023swr,Gluons,Arkani-Hamed:2024tzl}. In this story, we start with the string integral giving amplitudes of Tr$(\phi^3)$ theory at low-energies in the surfaceology parametrization, and perform a kinematic shift -- the $\delta$-shift -- to obtain the scalar scaffolded gluon amplitude, where each gluon is produced from a pair of colored scalars. 

Let us briefly review the aspects of surfaceology which are important for this paper. Given an order in the topological expansion specified by a surface, $\mathcal{S}$, we associate the ``surface integral'' \cite{Arkani-Hamed:2019mrd,Arkani-Hamed:2019plo,Arkani-Hamed:2023lbd,Arkani-Hamed:2023mvg}
\begin{equation}     \mathcal{I}_{\mathcal{S}}= \int_0^\infty \prod_{P\in \mathcal{T}} \frac{dy_P}{y_P} \prod_{\mathcal{C}\in \mathcal{S}} u_{\mathcal{C}}[\{y_P\}]^{\alpha^\prime X_\mathcal{C}}\, \xrightarrow{\alpha^\prime X_\mathcal{C} \ll1 }  \sum_{\substack{\text{triang.} T \\
\text{of }\mathcal{S}}} \, \prod_{\mathcal{C}\in T} \frac{1}{X_\mathcal{C}} \equiv \mathcal{A}_\mathcal{S}^{\text{Tr}(\phi^3)}.
    \label{eq:trp3tree}
\end{equation}

This integral relies on a choice of a triangulation, $\mathcal{T}$ of $\mathcal{S}$ -- $i.e.$ a maximal collection of non-crossing curves that divide the surface into triangles. Given $\mathcal{T}$ we integrate over the positive coordinates, $y_P$, each associated to one of the curves $P$ in $\mathcal{T}$.  Specifying a triangulation is then equivalent to picking a representative cubic diagram/fatgraph, which we obtain by drawing a vertex inside each triangle (see figure \ref{fig:Lam}, for a triangulation of the $8$-point disk, given in red, as well as the dual fatgraph), so that the propagators in the fat-graph are dual to the curves $P\in \mathcal{T}$. 

In addition, in \eqref{eq:trp3tree},  we also have a product over all curves $\mathcal{C}$ in the surface (up to homotopy), of the so-called $u$-variables, $u_\mathcal{C}$, raised to $X_{\mathcal{C}}$,  a kinematic variable associated to curve $\mathcal{C}$. The $u$-variables, $u_\mathcal{C} \in [0,1]$, live in the support of the $u$-equations which read
\begin{equation}
    u_\mathcal{C}[\{y_P\}] + \prod_\mathcal{D} u_\mathcal{D}[\{y_P\}]^{\#(\mathcal{C},\mathcal{D})} = 1, 
    \label{eq:ueqns}
\end{equation}
where $\#(\mathcal{C},\mathcal{D})$ is the number of times the curves $\mathcal{C}$ and $\mathcal{D}$ intersect.
Simply the knowledge of these equations allows us to infer many remarkable features of string amplitudes and surface integrals~\cite{Arkani-Hamed:2023swr,Arkani-Hamed:2024fyd,Arkani-Hamed:2024nzc}, in particular that it correctly factorizes near poles. This is because if a given $X_{\mathcal{C}} \to 0$, the integral develops a singularity where $u_\mathcal{C} \to 0$, and from \eqref{eq:ueqns} all $u$'s associated with curves that \textit{cross} $\mathcal{C}$ must go to one, so that the product in \eqref{eq:trp3tree} appropriately factors.

Finally, if we extract the low energy limit from \eqref{eq:trp3tree}, corresponding to $\alpha^\prime X_{\mathcal{C}} \ll 1$, at leading order we get the field-theory Tr($\phi^3$) amplitude/integrand. The surprising observation made in \cite{Gluons,Arkani-Hamed:2023swr}, is that if we start with the integral for the scattering of $2n$ scalars, and perform the following kinematic shift
\begin{equation}
    \mathcal{I}_n^{\text{gluons}}(X_\mathcal{C}) = \mathcal{I}_{2n}(X_{\mathcal{C} }\equiv X_{i,j}\rightarrow X_{i,j} + \delta_{i,j}/\alpha^\prime), \quad \text{with } \delta_{i,j} = \begin{cases}
        +1,  \text{ if } (i,j) \text{ even }\\
         -1,  \text{ if } (i,j) \text{ odd } \\
         0, \text{ otherwise.}
    \end{cases}
    \label{eq:deltashift}
\end{equation}
where we label a curve $\mathcal{C}$ by the points where it starts and ends ($i$ and $j$); then at low energies we get the scalar-scaffolded gluon amplitude. 

We will momentarily review this formulation of gluon amplitudes in section \ref{sec:Dictionary}, and explain how to extract leading singularities for any graph from it by performing simple residue computations, as already described in \cite{Gluons}. This gives us a leading singularity as a polynomial in the $X_\mathcal{C}$ variables. We go further and recognize this computation of the leading singularities as being associated with a simple combinatorial question attached to the graph: 

\begin{center}
    \textit{Given the fatgraph for the LS under consideration, what are all possible collections of curves that fully cover the fatgraph  such that each edge is covered once and only once?}
\end{center}
That is, each monomial in $X_\mathcal{C}$ coming from the residue precisely specifies one such collection, so that the full LS has this very simple combinatorial interpretation. In this paper we will see how we can rediscover this picture directly, from the bottom up, by building the LS directly from gluing the YM $3$-point vertices. 

Quite remarkably, we find that simply by encoding the Lorentzian contractions coming from on-shell spin sums in a graphical way -- by drawing ``contraction'' curves on the fatgraph -- the two pictures precisely match each other. This is quite non-trivial since in both pictures, the same monomial arises in many ways in the course of the computation--for instance in on-shell gluing, different contractions generate the same monomial in $X_\mathcal{C}$, potentially with different signs such that they cancel when we add everything together. However, every single monomial generated from on-shell gluing (before dealing with possible cancellations) is also generated exactly with the same sign in the surface residue -- with the full combinatorial picture of curves ``tilling'' the fatgraph emerging purely from the pattern of Lorentz contractions!

In section \ref{sec:LSfromQFT}, we start at tree-level and explain how we can encode the standard polarization-sum tensor and different contractions by drawing curves on the fatgraph, as well as a simple way to go from contraction patterns to the monomials produced in terms of $X_{\mathcal{C}}$. As we explain, simply the process of translating from the contraction curve into the actual kinematic variables associated to curves $X_{\mathcal{C}}$, one automatically lands on the combinatorial formulation of LS. 

In sections \ref{sec:loop-level} and \ref{sec:HigherLoops}, we extend this picture to loop-level. As is well-known, at loop-level the gluing rule has to be done correctly -- spin indices can't simply be contracted with the metric tensor -- instead to correctly obtain a gauge-invariant LS one needs to include a correction, a fact famously reflected by the need for ghosts in the Lagrangian formulation of Yang-Mills theory. As it turns out, this correction to the naive tree-level gluing also has a simple graphical interpretation which precisely fits into the loop-level LS from the surface integral. In particular at loop-level there are a few new ingredients at the level of the surface integral, namely the presence of self-intersecting curves as well as closed curves around the puncture \cite{Gluons}, where the exponents of the latter encode the dependence on the spacetime dimension, $D$. In \cite{Gluons}, the exponents for the closed curves entering for planar one and two-loop diagrams were  given explicitly, but the general rule at any loop order remained an open question. As it turns out, all the new loop-level ingredients of the surface integral naturally arise simply by drawing the loop-level contractions as curves on surface, and by matching the two pictures we are able to determine the exponent for any closed curve contributing to the LS.  

As mentioned earlier, both at tree- and loop-level, different contraction patterns generate the same monomial in $X_\mathcal{C}$, with possibly different signs. In section \ref{sec:CancellationsVRule}, we describe how we can systematically account for these cancellations, and provide an explicit rule which tests whether a given monomial survives in the full answer.

Finally in section \ref{sec:FermLS} we take a first look into gluon LS with fermions running in the loop. This type of LS is of obvious interest in the context of QCD, and we  show that a similar graphical/combinatorial picture emerges in this setting. In addition to polarization contractions, we now also need to encode the different terms coming from the traces over the gamma matrices for the fermion loop. We present a simple way in which this can be done, leaving a systematic treatment for future work.

\section{Scalar-scaffolded gluon leading singularities and surfaces}
\label{sec:Dictionary}

In the gluon amplitude of equation~\eqref{eq:deltashift} the gluons are scalar-scaffolded, this is each gluon is being produced by a pair of scalars. In the limit where the gluon is on-shell, we can read off its polarization and momenta from the momenta of the scalars as follows
\begin{equation}
    \begin{gathered}
    \begin{tikzpicture}[line width=0.6,scale=1.2,baseline={([yshift=0.0ex]current bounding box.center)}]
        \begin{feynman}
    		\coordinate (a) at (0,0);
        		\coordinate (p1) at (-0.35,0.65);
        		\coordinate (p2) at (-0.35,-0.65);
        		\coordinate (q) at (1.2,0);
        		\coordinate (b) at (1,0);
        		\diagram*{
        			(a) -- [gluon] (b),
        			(a) -- (-35:-1);
        			(a) -- (35:-1);
       			};
                \node[scale=0.8] at (p1) {$p_2^\mu$};
       			\node[scale=0.8] at (p2) {$p_1^\mu$};
       			\node[scale=0.8] at (q) {$q_1^\mu$};
        		
    	\end{feynman}
    \end{tikzpicture}
\end{gathered} \propto \,g_{\text{YM}}(p_1-p_2)^\mu \quad \Rightarrow \quad \begin{cases}
q_1^\mu = p_1^\mu + p_2^\mu, \\ \epsilon_1^\mu = (p_1-p_2)^\mu +\alpha (p_1+p_2)^\mu, \end{cases}
\label{eq:GluonScalarMap}
\end{equation}
where the polarization vector is defined up to gauge transformations, $\epsilon^\mu \rightarrow \epsilon^\mu + \alpha q^\mu$, but $\alpha$ should drop out of the final answer. Thus, using \eqref{eq:GluonScalarMap}, for an n-gluon scattering process, we can effectively trade the kinematical data of the gluons with that of $2n$-scalars, and so the amplitude is given in terms of dot products of momenta for $2n$ scalars. By fixing a color-ordering for the $2n$-scalars, which we assume to be the standard ordering $[12\cdots 2n]$, we can draw the momentum of each scalar tip-to-toe according to this ordering, and due to momentum conservation this produces a closed polygon -- the so-called \textit{momentum-polygon}, where each edge labels a momentum, $p_i^\mu$, 
\begin{equation}
   \begin{gathered}
\begin{tikzpicture}[scale = 1]
  \foreach \i in {0,60,...,300} {
    \coordinate (P\i) at ({2*cos(\i)}, {2*sin(\i)});
  }
    \coordinate (X5) at ({2*cos(0)}, {2*sin(0)}) ; 
    \coordinate (X4) at ({2*cos(60)}, {2*sin(60)}) ;  
    \coordinate (X3) at ({2*cos(120)}, {2*sin(120)}) ;   
    \coordinate (X2) at ({2*cos(180)}, {2*sin(180)}) ;    
    \coordinate (X1) at ({2*cos(240)}, {2*sin(240)}) ;
    \coordinate (X6) at ({2*cos(300)}, {2*sin(300)}) ;
    
  \draw[thick] 
    (X1) -- (X2) -- (X3) -- (X4) -- (X5) -- (X6) -- cycle;

    \node at ({2*cos(0)+0.3}, {2*sin(0)}) {$x_5^\mu$}; 
    \node at ({2*cos(60)+0.2}, {2*sin(60)+0.2}) {$x_4^\mu$};  
    \node at ({2*cos(120)-0.2}, {2*sin(120)+0.2}) {$x_3^\mu$}; 
    \node at ({2*cos(180)-0.3}, {2*sin(180)}) {$x_2^\mu$};    
    \node at ({2*cos(240)-0.2}, {2*sin(240)-0.2}) {$x_1^\mu$};
    \node at ({2*cos(300)+0.2}, {2*sin(300)-0.2}) {$x_6^\mu$};
    
    \coordinate (p1) at ($(X1)!0.5!(X2)+(-0.2,-0.2)$);
    \node at ($(p1)+(-0.1,-0.1)$) {$p_1^\mu$};
    \node[very thick,rotate=120] at ($(X1)!0.5!(X2)$) {$>$};

    \coordinate (p2) at ($(X2)!0.5!(X3)+(-0.2,+0.2)$);
    \node at ($(p2) + (-0.1, 0.1)$) {$p_2^\mu$};
    \node[very thick,rotate=60] at ($(X2)!0.5!(X3)$) {$>$};

    \coordinate (q1) at ($(X1)!0.5!(X3)+(-0.3,0)$);
    \node at (q1) {$q_1^\mu$};
    \node[very thick,rotate=90] at ($(X1)!0.5!(X3)$) {$>$};

    \coordinate (p3) at ($(X3)!0.5!(X4)+(0,+0.3)$);
    \node at ($(p3)+(0,0.1)$) {$p_3^\mu$};
    \node[very thick,rotate=0] at ($(X3)!0.5!(X4)$) {$>$};

    \coordinate (p4) at ($(X4)!0.5!(X5)+(+0.2,+0.2)$);
    \node at ($(p4)+(0.1,0.1)$) {$p_4^\mu$};
    \node[very thick,rotate=-60] at ($(X4)!0.5!(X5)$) {$>$};

    \coordinate (q2) at ($(X3)!0.5!(X5)+(+0.2,0.2)$);
    \node at (q2) {$q_2^\mu$};
    \node[very thick,rotate=-30] at ($(X3)!0.5!(X5)$) {$>$};

    \coordinate (p5) at ($(X5)!0.5!(X6)+(+0.2,-0.2)$);
    \node at (p5) {$p_5^\mu$};
    \node[very thick,rotate=-120] at ($(X5)!0.5!(X6)$) {$>$};

    \coordinate (p6) at ($(X6)!0.5!(X1)+(0,-0.3)$);
    \node at ($(p6)+(0,-0.1)$) {$p_6^\mu$};
    \node[very thick,rotate=180] at ($(X6)!0.5!(X1)$) {$>$};

    \coordinate (q3) at ($(X5)!0.5!(X1)+(+0.2,-0.2)$);
    \node at (q3) {$q_3^\mu$};
    \node[very thick,rotate=-150] at ($(X5)!0.5!(X1)$) {$>$};
    \draw[thick] (X1) to (X3);
    \draw[thick] (X3) to (X5);
    \draw[thick] (X5) to (X1); 
\end{tikzpicture} 
\end{gathered}
\quad \Rightarrow \quad X_{i,j} = (x_{j} - x_{i})^2 = (p_i + \dots + p_{j-1})^2.
\label{eq:planarVar}
\end{equation}

Furthermore, labelling each vertex of the polygon with a dual coordinate -- $x_i^{\mu}$ -- allows us to write $ p_i^\mu = x_{i+1}^\mu - x_{i}^\mu$, and in general write the planar Mandelstam invariants, $X_{i,j}$, as (length)$^2$ of the chords of this polygon. The $X_{i,j}$'s are a particularly natural set of variables to describe color-ordered scalar amplitudes, as they correspond to the combinations of momenta that can appear as poles, and they define a basis of kinematic space, without breaking the underlying cyclic invariance of the  problem\footnote{For general $n$, the $X_{i,j}$'s are only independent variables, if we allow $D$ to be large enough, otherwise they live under the support of the Gram-determinant constraints. For the sake of this paper, we will assume $D$ is sufficiently large to allow for the $X_{i,j}$ to be independent, unless stated otherwise.}. In this language, asking for the gluons to be on-shell corresponds to having 
\begin{equation}
    X_{\text{scaff}}= \{X_{1,3},X_{3,5},\cdots,X_{1,2n-1}\} =0.
    \label{eq:scaffVars}
\end{equation}
So, if we consider the inscribed momentum polygon containing only \textit{odd} vertices, we obtain an $n$-gon which is precisely associated with the $n$-gluon momentum-polygon. 
To give a concrete example, let's look at the $3$-point gluon interaction, which in its standard representation in terms of gluon polarization and momentum is simply
\begin{equation}
   \begin{gathered}
    \begin{tikzpicture}[line width=0.6,scale=1.2,baseline={([yshift=0.0ex]current bounding box.center)}]
    	\begin{feynman}
    		\coordinate (a) at (0,0);
        		\coordinate (p1) at (-0.35,0.65);
        		\coordinate (p2) at (-0.35,-0.65);
        		\coordinate (q) at (1.2,0);
        		\coordinate (b) at (1,0);
        		\diagram*{
        			(a) -- [gluon] (b),
        			(a) --[gluon] (-35:-1);
        			(a) --[gluon] (35:-1);
       			};
                \node[scale=0.8] at (p1) {$q_2^\mu$};
       			\node[scale=0.8] at (p2) {$q_1^\mu$};
       			\node[scale=0.8] at (q) {$q_3^\mu$};
        		
    	\end{feynman}
    \end{tikzpicture}
\end{gathered} = [ \epsilon_1 \cdot \epsilon_2 ][\epsilon_3 \cdot (q_1-q_2)] + [\epsilon_2 \cdot \epsilon_3][ \epsilon_1 \cdot (q_2-q_3)]+ [\epsilon_1 \cdot \epsilon_3][ \epsilon_2 \cdot (q_3-q_1)].
\label{eq:3ptGluonPol}
\end{equation}
Applying the map \eqref{eq:GluonScalarMap} leads to the following expression in terms of the momentum of the six scalars:
\begin{equation}
   \mathcal{A}_3^{\text{YM}}(X_{i,j}) = X_{1,4}X_{2,6}+X_{3,6}X_{2,4}+X_{2,5}X_{4,6}-X_{2,5}X_{3,6}-X_{1,4}X_{3,6}-X_{1,4}X_{2,5}.
   \label{eq:3ptScaff}
\end{equation}

\subsection*{Scalar-scaffolded gluons from surfaces}

Let's now briefly review how the scalar-scaffolded gluon amplitude is encoded in \eqref{eq:deltashift}, with the goal of understanding how we can use it to efficiently compute gluon leading singularities. 

The first step in defining the shifted integral \eqref{eq:deltashift} is to define the surface $\mathcal{S}$ that the integral is associated with, as well as a triangulation of it -- that is equivalent to choosing a representative fatgraph whose propagators are dual to the chords/curves in the triangulation. For convenience, we always pick a graph containing the scaffolding chords $X_{\text{scaff}}$ -- corresponding to a \textit{scaffolding triangulation} -- in which case the integral reduces to \cite{Gluons} 
\begin{equation}
\mathcal{I}^{\text{gluons}}_{\mathcal{S}_n}[X_{i,j}] = \int \prod_{P \in \mathcal{T}_{2n}} \frac{dy_p}{y_p^2} \prod_{\mathcal{C} \in \mathcal{S}} u_{ij}(\{y_P\})^{ X_\mathcal{C}}\, ,
    \label{eq:SurfIntGluons2}
\end{equation}
where each $y_P$ is associated to one of the propagators $P$ entering the fatgraph.  Now every curve on the surface corresponds to a path on the fat-graph: given a curve going from marked points $i$ to $j$, we can draw it in the fat-graph as a curve entering the graph in edge $(i,i+1)$ and exiting via edge $(j,j+1)$ (see for example curve $(3,7)$ in fig. \ref{fig:Lam}). The kinematic invariant $X_\mathcal{C}$/momentum of a given curve can be read by \textit{homology}, which practically translates into the following simple rule~\cite{Arkani-Hamed:2023lbd}:
\begin{enumerate}
\itemsep 0em
    \item Assign momentum to each edge of the fatgraph, $i.e.$ the external edges between regions $(i,i+1)$ have ingoing momentum $p_i^\mu$, and momentum of the internal edges is determined by momentum conservation, other than those corresponding to loop-propagators, which carry loop momenta, $l_k^\mu$.
    \item Record the path of the $\mathcal{C}=(i,j)$ inside the fatgraph, $i.e.$ the set of edges it goes through as well as the set left/right turns it does at each vertex. 
    \item To determine its momentum, add the momentum of the edge in which it entered the graph, $p_i^{\mu}$, and then each time the curve makes a right turn, add the momentum of the edge coming into that vertex from the left. $X_\mathcal{C}$ is simply the square of this momentum. 
\end{enumerate}

At loop-level, this rule implies that (homotopically) different curves on the surface are assigned the same momentum, reflecting the fact that, at loop-level, different propagators carry the same momentum. However, at the level of the surface integral, we are free to assign each curve, $\mathcal{C}$, a given $X_\mathcal{C}$, and only in the end map these to physical momenta, using the map described above. This generalization of kinematics -- where there is a variable per homotopy class~\cite{Salvatori:2018aha,Salvatori:2019ehb} -- is called \textit{surface kinematics} and has been crucial in defining gluon loop-integrands which are gauge-invariant and factorize consistently on cuts \cite{Arkani-Hamed:2024tzl}. Finally, there are two other new features at loop-level worth highlighting: 
\vspace{2mm}

{\noindent\bfseries Infinitely many curves:} Under the presence of punctures, curves are allowed to loop around the punctures an arbitrary number of times, leading to infinitely many homotopically different curves. Therefore the product over $\mathcal{C}$ in \eqref{eq:SurfIntGluons2} is infinite. However, as explained in \cite{Gluons}, to correctly obtain the low-energy gluon amplitude, one can consistently truncate this product by keeping only curves have up to one self-intersection (around each puncture).
\vspace{2mm}

{\noindent\bfseries Closed curves:} In addition to open curves, going from marked points $i$ to $j$ or to the puncture $p_k$, we also have \textit{closed} curves around the punctures, encoding the dependence on the  spacetime dimension, $D$. To a closed curve enclosing a set of punctures, $J$, we associate the exponent  $X_{\mathcal{C}} \equiv \Delta_{J}$. In \cite{Gluons}, it was shown that at one-loop (planar) the exponent of the closed curve around the puncture must be $\Delta = 1- D$. However, at higher-loops there are different types of closed curves (enclosing different subsets of punctures), and the exponents of such curves remained an open question -- which we will adress later in this paper.  

\subsection{Leading singularities from linearized $u$'s}
\label{sec:LS_U}

Perhaps the most important ingredient entering the surface integrals \eqref{eq:SurfIntGluons2}  are the $u$-variables and their parametrization in terms of the positive $y_P$'s. In our case, the $y_P$'s are naturally divided into those associated with the scaffolding propagators, $y_s$ with $s=(2j-1,2j+1)$, and the remaining, $y_{P_i}$, which specify a particular triangulation, $P_i \in \mathcal{T}_{\text{int}}$, of the internal ``gluon'' surface bounded by the scaffolding curves, $\mathcal{S}^{\text{Scaff}}$.  

With simply the knowledge of the path of $\mathcal{C}$ in the fat-graph, we can derive $u_\mathcal{C}[\{y_p\}]$ (see \cite{Arkani-Hamed:2023lbd,Gluons}). However, to extract the leading singularities, all we need is the expansion of each $u_\mathcal{C}$'s up to linear order in the $y_P$'s. This is simply because residues of $\mathcal{A}_n^{\text{YM}}$ can be computed as residues of the $y$-space integrand in \eqref{eq:SurfIntGluons2}. In particular, already for the scaffolding residue \eqref{eq:scaffVars}, we can write it as a residue of the integrand as follows: 
\begin{equation}    
\begin{aligned}
\mathcal{A}_{n}^{\text{YM}}[X_{i,j}] &= \mathop{\mathrm{Res}}_{X_{\text{Scaff.}=0}} \left[  \mathcal{I}^{\text{gluons}}_{\mathcal{S}_n}[X_{i,j}] \right],\\
&=\int \prod_{P_i \in \mathcal{T}_{\text{int}}} \frac{d y_{P_i}}{y_{P_i}^2}\mathop{\mathrm{Res}}_{y_{\text{Scaff.}}=0} \left[ \prod_{s \in \text{Scaff.}} \frac{1}{y_s^2} \prod_{\mathcal{C}\in \mathcal{S}_{2n}} u_{\mathcal{C}}(\{y_{P_i},y_s\})^{ X_\mathcal{C}} \bigg \vert_{X_{\text{Scaff.}}=0} \right].
\end{aligned}
\label{eq:scaffRes}
\end{equation}

So to extract the LS corresponding to the gluon diagram with propagators $\tilde{P}_i$, we simply pick the triangulation of $\mathcal{S}^{\text{Scaff}}$,   $\mathcal{T}_{\text{int}}$, containing $\tilde{P}_i$, and extract the respective $y$-space residue: 
\begin{equation}
    \text{LS}_{\mathcal{T}_{\text{int}}}[\mathcal{A}_{n}^{\text{YM}} ] =  \mathop{\mathrm{Res}}_{\substack{y_{\tilde{P}}=0,\\y_{\text{Scaff.}}=0}} \left[ \prod_{\tilde{P}_i \in \mathcal{T}_{\text{int}}} \frac{1}{y_{P_i}^2}\prod_{s \in \text{Scaff.}} \frac{1}{y_s^2} \prod_{\mathcal{C}\in \mathcal{S}_{2n}} u_{\mathcal{C}}(\{y_{\tilde{P}_i},y_s\})^{ X_\mathcal{C}} \bigg \vert_{X_{\text{Scaff.}},X_{\tilde{P}}=0} \right].
    \label{eq:LeadSingRes}
\end{equation}

Therefore, after extracting the full LS, we kill all the integrations and are left with a polynomial in the $X_\mathcal{C}$s. The leading order part of this polynomial, with the smallest power in $X's$, gives us the pure YM LS, and the higher powers in $X$ correspond to LS with pure gluon vertices and $F^3$ vertices, all the way up to the pure $F^3$ LS. From \eqref{eq:LeadSingRes}, it is clear that to obtain this polynomial all we need to know is the dependence of each $u_\mathcal{C}$ in the respective $y$'s, up to linear order, so we can extract the piece of $\prod_{\mathcal{C}\in \mathcal{S}_{2n}} u_{\mathcal{C}}(\{y_{P_i},y_s\})^{ X_\mathcal{C}}$ which is \textit{linear} in all the $y$'s.
\begin{figure}[t]
    \centering
    \includegraphics[width=0.9\linewidth]{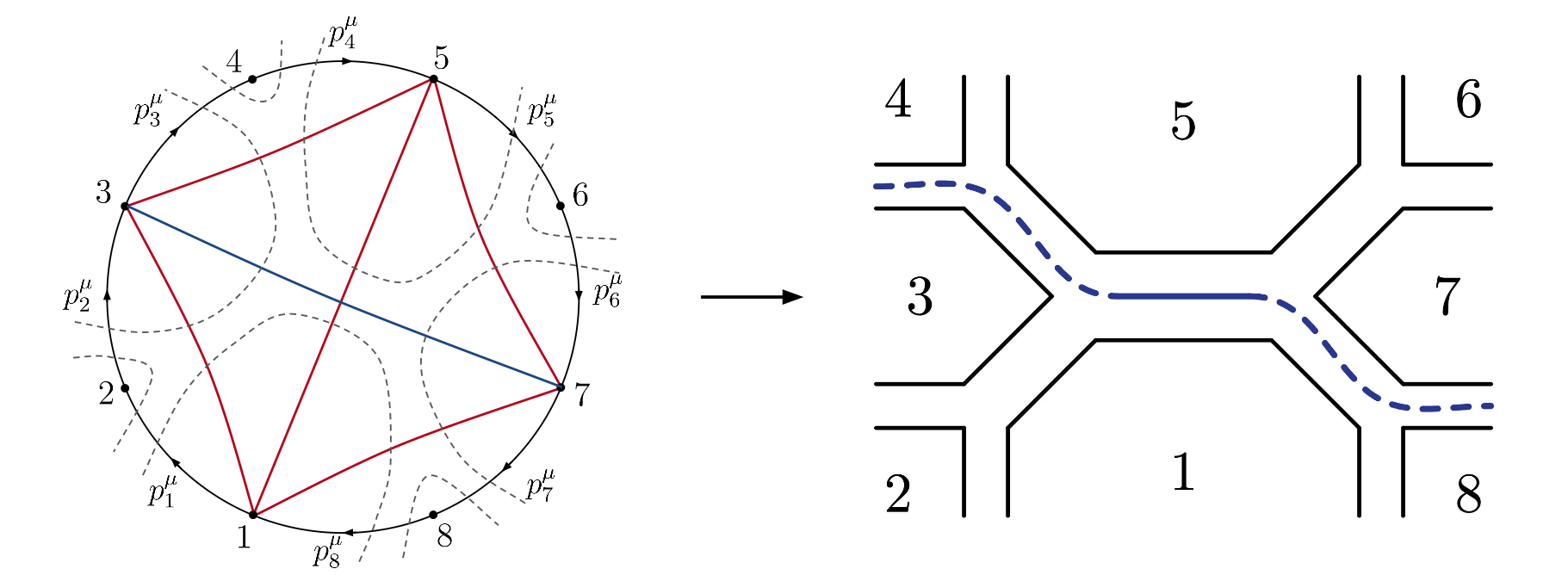}
    \caption{(Left) $s$-channel fatgraph and respective dual $\mathcal{T}_{\text{int}} = \{(1,5)\}$ for the $4$-point gluon amplitude. (Right) Correspondence between curves on the surface and paths on the fat-graph. The curve $(3,7)$ in blue LHS maps to the path in the RHS. The path on the fat-graph can be further truncated into a core piece (solid) plus its extensions (dashed).}
    \label{fig:Lam}
\end{figure}

As it turns out, the linear expansion of the $u$'s in $y$'s follows a very simple graphical rule described in \cite{Gluons}, but that we recast here in a slightly different way useful for our analysis. Let us look at a concrete example: consider the curve $(3,7)$, in the $4$-point tree scaffolded-gluon amplitude of figure \ref{fig:Lam}. As a curve on the disk, $(3,7)$ only intersects curve $(1,5)$, however, we when draw it on the fatgraph, it goes through edge $(1,5)$ as well as $(3,5)$ and $(1,7)$. This is simply because we defined that a curve starting at marked point $i$ enters the fatgraph in the edge $(i,i+1)$, which effectively means that, at the level of the disk, it is starting in the boundary between $i$ and $i+1$ -- this deformation of the ends of the curve is what is called a \textit{lamination}. So we have that the lamination of $(3,7)$ starts between $(3,4)$ and ends between $(7,8)$, which now does intersect all three curves, $(1,5)$, $(3,5)$ and $(1,7)$, just like the path drawn in the fatgraph. 

As a consequence, it is natural to divide the path in the fatgraph into the part corresponding to the \textit{core} of the curve, $C^{\mathcal{C}}$ -- the part which passes through the edges that the curve on the disk intersects -- and its \textit{extensions}, $E_L^{\mathcal{C}}, E_R^{\mathcal{C}}$ -- which are the rests of the path on the fatgraph coming from the lamination. In figure \ref{fig:Lam} (right), we represent in solid blue the core of curve $(3,7)$, and in dashed the two extensions, $E^{(3,7)}_L$ going through edge $(3,5)$ and $E^{(3,7)}_R$ through $(1,7)$. 

In general we can describe the core of a curve $(i,j)$, as the path inside the fatgraph (does not need to start in external edge) from region $i$ to region $j$ which goes through the minimal number of edges. In our working example, the minimal way to connect regions $3$ and $7$ is simply going through edge $(1,5)$ -- which is precisely the core of $(3,7)$. Starting from the core, we can easily produce the extensions, $E^{\mathcal{C}}_L$ and  $E^{\mathcal{C}}_R$, by going to each end of the core, and considering the path which turns right once and then left continuously. We can label each extension as an ordered list of the edges crossed since the end of the core and the point where the curve exits the fatgraph, so  $E^{\mathcal{C}}_L = \{p^\star_1,p^\star_2,\cdots,p^\star_{k_L}\}$, and similarly $E^{\mathcal{C}}_R = \{q^\star_1,q^\star_2,\cdots,q^\star_{k_R}\}$. Using this information, we can directly write the linearized expression for $u_{a,b}^{X_{a,b}}$ as
\begin{equation} \label{eq:linearu}
\begin{aligned}
	&u_{{a,b}}^{X_{a,b}} = 1 - X_{a,b} \prod_{P_i \in \mathcal{T}} y_P^{\text{Int}[\{P_i,(a,b)\}]} \left[1 - \left( y_{p^\star_1} + y_{p^\star_1} y_{p^\star_{2}} +\dots +y_{p^\star_1} y_{p^\star_{2}}\dots y_{p^\star_k}
	\right) \right.\\
	& \left.\quad \quad - \left( y_{q^\star_1} + y_{q^\star} y_{q^\star_{2}} +\dots +y_{q^\star_1} y_{q^\star_{2}}\dots y_{q^\star_l}
	\right) \right.\\
	& \left.\quad \quad + \left( y_{p^\star_1} + y_{p^\star_1} y_{p^\star_{2}} +\dots +y_{p^\star_1} y_{p^\star_{2}}\dots y_{p^\star_k}
	\right)\left( y_{q^\star_1} + y_{q^\star} y_{q^\star_{2}} +\dots +y_{q^\star_1} y_{q^\star_{2}}\dots y_{q^\star_l}
	\right) \right]\\
	& \quad \quad +\mathcal{O}(y^2) \, ,
\end{aligned}
\end{equation}
where $\text{Int}[\{P_i,(a,b)\}]$ is the number of times the core of $(a,b)$, $C^{(a,b)}$, crosses edge $P_i$ on the fatgraph. At tree-level, the intersection numbers are always zero or one, but they can be larger at higher-loops.

As for the closed curves on the surface, $\Delta_J$, we have the similar linearized form \cite{Gluons}
\begin{equation}
    u_{\Delta_J}^{\Delta_J} = 1- \Delta_J \prod_{P_i\in \mathcal{T}} y_{P_i}^{\text{Int}[\{P_i,\Delta_J\}]}  +\mathcal{O}(y^2),
    \label{eq:LinUDelta}
\end{equation}
where we simply get the product over all chords in the triangulation that intersect the closed curve $\Delta$\footnote{Note that these intersection numbers are defined as the \textit{minimal} number of times two curves intersect each other, where we can deform the curves to any homotopically equivalent configuration.}, raised to the respective intersection number.

Looking at \eqref{eq:linearu}-\eqref{eq:LinUDelta}, it follows that the leading singularity computed through the residue in \eqref{eq:LeadSingRes} is given by a sum of monomials in the $X_{i,j}$, where each monomial has a simple geometric interpretation:  it is associated to a collection of curves that cross all the chords in the triangulation \textit{once and only once}, or equivalently, a collection of curves covering all edges of the fatgraph \textit{once and only once}. Crucially, in defining this collection, it is important to keep in mind that due to \eqref{eq:linearu}, a given open curve can be thought of as simply its \textit{core} part, in which case it contributes with the $1$ inside brackets in \eqref{eq:linearu}, or it can be crossing certain edges via its \textit{extensions}, which correspond to the remaining terms in \eqref{eq:linearu}. As a result, there are different ways of producing the same monomial, which in general contribute to the LS with different signs, and might even cancel in the final answer. 

In the rest of the paper, we explain how we can systematically keep track of the different monomials and respective signs. Most interestingly, we do this by understanding how this picture for LS naturally emerges from the standard gluing of $3$-point vertices once we recast the Lorentz contractions in terms of ``paths'' on the fat graph. But, before that, let us first give some explicit examples of LS both at tree- and loop-level from the surface integral.

\subsection{Tree and loop examples}
\label{sec:examples_LS_SurfInt}

\paragraph{$3$-point interaction} As the simplest example at tree-level let us look at the 3-point gluon interaction given in \eqref{eq:3ptScaff}, but now extract it as a residue of the surface integral. In this case, the set of scaffolding chords automatically gives a triangulation of the surface (as shown in \eqref{eq:planarVar}), and so after seting $X_{\text{Scaff.}}=0$, we have only two different cyclic classes  $\{X_{2,4},X_{4,6},X_{2,6}\}$ and $\{X_{1,4},X_{2,5},X_{3,6}\}$, and so it's enough to write down the linearized form of $u_{2,4}^{X_{2,4}}$ and $u_{1,4}^{X_{1,4}}$, since all remaining are given by cyclic transformations. Using \eqref{eq:linearu}, we get
\begin{equation}
    u_{14}^{X_{1,4}} = 1- X_{1,4}(y_{3,5})(1-y_{1,3}), \quad 
    u_{24}^{X_{2,4}} = 1- X_{2,4}(y_{1,3}y_{3,5})(1),
\end{equation}
and replacing these in \eqref{eq:LeadSingRes}, we get
\begin{equation}
\begin{aligned}    &\text{LS}_3=\text{Res}_{y_{13},y_{35},y_{15}=0}\left( \frac{1}{y_{1,3}^2 y_{3,5}^2 y_{1,5}^2}u_{14}^{X_{14}}u_{24}^{X_{24}}u_{36}^{X_{36}}u_{46}^{X_{46}}u_{25}^{X_{25}}u_{26}^{X_{26}}\right) \\
    &=  \left[ \begin{gathered}
    \begin{tikzpicture}[line width=0.6,scale=0.2,line cap=round,every node/.style={font=\tiny}]
		\node (0) at (-2.75, 1.5) {};
		\node (1) at (-2, 2.25) {};
		\node (2) at (-1, -0.25) {};
		\node (3) at (-0.25, 0.5) {};
		\node (4) at (-1, -3.25) {};
		\node(5) at (0, -3.25) {};
		\node (6) at (0, -0.5) {};
		\node   (7) at (2.75, -0.5) {};
		\node   (8) at (2.75, 0.5) {};
		\node   (9) at (-4.25, 1.5) {};
		\node   (10) at (-2.5, 4.25) {};
		\node   (11) at (-4.25, 2.5) {};
		\node   (12) at (-3.25, 3.75) {};
		\node   (13) at (-3, 2.5) {};
		\node   (14) at (-2.25, -5) {};
		\node   (15) at (-1.25, -5) {};
		\node   (16) at (-0.5, -4) {};
		\node   (17) at (0.25, -5) {};
		\node   (18) at (1.25, -5) {};
		\node   (19) at (4.75, -2) {};
		\node   (20) at (4.75, -0.75) {};
		\node   (21) at (3.75, 0) {};
		\node   (22) at (4.75, 0.75) {};
		\node   (23) at (4.75, 2) {};
		\node   (24) at (-2.75, -0.75) {};
		\node   (25) at (-2.75, -0.75) {1};
		\node   (27) at (1, 2.25) {3};
		\node   (28) at (4.75, 0) {4};
		\node   (29) at (1.25, -2) {5};
		\node   (30) at (-0.5, -4.95) {6};
		\node   (31) at (-0.25, 0) {};
		\node   (32) at (3.25, 0) {};
		\node   (33) at (-2.5, 2) {};
		\node   (34) at (-0.5, 0) {};
		\node   (35) at (-0.5, -3.25) {};
		\node   (36) at (-3.75, 3) {};
		\node   (37) at (-3.75, 3) {2};
		\draw[thick] (0.center) to (2.center);
		\draw[thick] (2.center) to (4.center);
		\draw[thick] (6.center) to (5.center);
		\draw[thick] [in=135, out=-45] (1.center) to (3.center);
		\draw[thick] (3.center) to (8.center);
		\draw[thick]  (6.center) to (7.center);
		\draw[thick]  (9.center) to (0.center);
		\draw[thick]  (11.center) to (13.center);
		\draw[thick] (13.center) to (12.center);
		\draw[thick] (10.center) to (1.center);
		\draw[thick](14.center) to (4.center);
		\draw[thick](15.center) to (16.center);
		\draw[thick](16.center) to (17.center);
		\draw[thick](5.center) to (18.center);
	\draw[thick] (7.center) to (19.center);
		\draw[thick](8.center) to (23.center);
	\draw[thick](21.center) to (22.center);
		\draw[thick](21.center) to (20.center);
		\draw[very thick][style=Maroon] (31.center) to (32.center);
	\draw[very thick][style=Green] (33.center) to (34.center);
		\draw[very thick][style=Green] (34.center) to (35.center);
\end{tikzpicture}
\end{gathered}\right] X_{1,4}X_{2,6} - \left[ \begin{gathered}
    \begin{tikzpicture}[line width=0.6,scale=0.2,line cap=round,every node/.style={font=\tiny}]
		\node (0) at (-2.75, 1.5) {};
		\node (1) at (-2, 2.25) {};
		\node (2) at (-1, -0.25) {};
		\node (3) at (-0.25, 0.5) {};
		\node (4) at (-1, -3.25) {};
		\node(5) at (0, -3.25) {};
		\node (6) at (0, -0.5) {};
		\node   (7) at (2.75, -0.5) {};
		\node   (8) at (2.75, 0.5) {};
		\node   (9) at (-4.25, 1.5) {};
		\node   (10) at (-2.5, 4.25) {};
		\node   (11) at (-4.25, 2.5) {};
		\node   (12) at (-3.25, 3.75) {};
		\node   (13) at (-3, 2.5) {};
		\node   (14) at (-2.25, -5) {};
		\node   (15) at (-1.25, -5) {};
		\node   (16) at (-0.5, -4) {};
		\node   (17) at (0.25, -5) {};
		\node   (18) at (1.25, -5) {};
		\node   (19) at (4.75, -2) {};
		\node   (20) at (4.75, -0.75) {};
		\node   (21) at (3.75, 0) {};
		\node   (22) at (4.75, 0.75) {};
		\node   (23) at (4.75, 2) {};
		\node   (24) at (-2.75, -0.75) {};
		\node   (25) at (-2.75, -0.75) {1};
		\node   (27) at (1, 2.25) {3};
		\node   (28) at (4.75, 0) {4};
		\node   (29) at (1.25, -2) {5};
		\node   (30) at (-0.5, -4.95) {6};
		\node   (31) at (-0.25, 0) {};
		\node   (32) at (3.25, 0) {};
		\node   (33) at (-2.5, 2) {};
		\node   (34) at (-0.5, 0) {};
		\node   (35) at (-0.5, -3.25) {};
		\node   (36) at (-3.75, 3) {};
		\node   (37) at (-3.75, 3) {2};
		\draw[thick] (0.center) to (2.center);
		\draw[thick] (2.center) to (4.center);
		\draw[thick] (6.center) to (5.center);
		\draw[thick] [in=135, out=-45] (1.center) to (3.center);
		\draw[thick] (3.center) to (8.center);
		\draw[thick]  (6.center) to (7.center);
		\draw[thick]  (9.center) to (0.center);
		\draw[thick]  (11.center) to (13.center);
		\draw[thick] (13.center) to (12.center);
		\draw[thick] (10.center) to (1.center);
		\draw[thick](14.center) to (4.center);
		\draw[thick](15.center) to (16.center);
		\draw[thick](16.center) to (17.center);
		\draw[thick](5.center) to (18.center);
	\draw[thick] (7.center) to (19.center);
		\draw[thick](8.center) to (23.center);
	\draw[thick](21.center) to (22.center);
		\draw[thick](21.center) to (20.center);
		\draw[very thick][style=Maroon] (31.center) to (32.center);
	\draw[very thick][style=Green] (33.center) to (34.center);
		\draw[very thick][style=Green,dashed] (34.center) to (35.center);
\end{tikzpicture}
\end{gathered}\right] X_{1,4}X_{2,5} + \frac{1}{3} \left[ \begin{gathered}
    \begin{tikzpicture}[line width=0.6,scale=0.2,line cap=round,every node/.style={font=\tiny}]
		\node (0) at (-2.75, 1.5) {};
		\node (1) at (-2, 2.25) {};
		\node (2) at (-1, -0.25) {};
		\node (3) at (-0.25, 0.5) {};
		\node (4) at (-1, -3.25) {};
		\node(5) at (0, -3.25) {};
		\node (6) at (0, -0.5) {};
		\node   (7) at (2.75, -0.5) {};
		\node   (8) at (2.75, 0.5) {};
		\node   (9) at (-4.25, 1.5) {};
		\node   (10) at (-2.5, 4.25) {};
		\node   (11) at (-4.25, 2.5) {};
		\node   (12) at (-3.25, 3.75) {};
		\node   (13) at (-3, 2.5) {};
		\node   (14) at (-2.25, -5) {};
		\node   (15) at (-1.25, -5) {};
		\node   (16) at (-0.5, -4) {};
		\node   (17) at (0.25, -5) {};
		\node   (18) at (1.25, -5) {};
		\node   (19) at (4.75, -2) {};
		\node   (20) at (4.75, -0.75) {};
		\node   (21) at (3.75, 0) {};
		\node   (22) at (4.75, 0.75) {};
		\node   (23) at (4.75, 2) {};
		\node   (24) at (-2.75, -0.75) {};
		\node   (25) at (-2.75, -0.75) {1};
		\node   (27) at (1, 2.25) {3};
		\node   (28) at (4.75, 0) {4};
		\node   (29) at (1.25, -2) {5};
		\node   (30) at (-0.5, -4.95) {6};
		\node   (31) at (-0.25, 0) {};
		\node   (32) at (3.25, 0) {};
		\node   (33) at (-2.5, 2) {};
		\node   (34) at (-0.5, 0) {};
            \node   (34p) at (-0.5, -0.3) {};
		\node   (35) at (-0.5, -3.25) {};
		\node   (36) at (-3.75, 3) {};
		\node   (37) at (-3.75, 3) {2};
		\draw[thick] (0.center) to (2.center);
		\draw[thick] (2.center) to (4.center);
		\draw[thick] (6.center) to (5.center);
		\draw[thick] [in=135, out=-45] (1.center) to (3.center);
		\draw[thick] (3.center) to (8.center);
		\draw[thick]  (6.center) to (7.center);
		\draw[thick]  (9.center) to (0.center);
		\draw[thick]  (11.center) to (13.center);
		\draw[thick] (13.center) to (12.center);
		\draw[thick] (10.center) to (1.center);
		\draw[thick](14.center) to (4.center);
		\draw[thick](15.center) to (16.center);
		\draw[thick](16.center) to (17.center);
		\draw[thick](5.center) to (18.center);
	\draw[thick] (7.center) to (19.center);
		\draw[thick](8.center) to (23.center);
	\draw[thick](21.center) to (22.center);
		\draw[thick](21.center) to (20.center);
		\draw[very thick][style=Maroon] (31.center) to (32.center);
	\draw[very thick][style=Green] (33.center) to (34.center);
		\draw[very thick][style=Blue] (34p.center) to (35.center);
\end{tikzpicture}
\end{gathered}\right]X_{1,4}X_{2,5}X_{3,6}\\
    & \quad + (\text{cyclic by two}) ,
\end{aligned}
\label{eq:3ptLS_Surf}
\end{equation}
where at lowest mass-dimension, $X^2$, we find precisely the 3-point YM vertex as given in \eqref{eq:3ptScaff}, while the piece with mass dimension $X^3$ corresponds to scalar-scaffolded 3-point $\Tr(F^3)$ vertex.

Already in this simple example, we can see that indeed each monomial is nicely associated with a collection of curves (cores/extensions) covering all edges of the fatgraph \textit{once and only once}. Note that, in order generate all monomials it is important to consider the possible extensions of a curve (represented in dashed). 

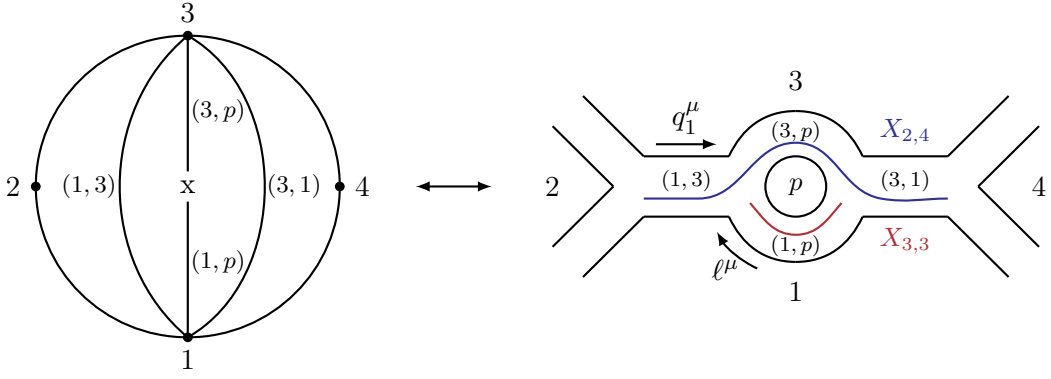
\begin{figure}
\begin{center}
	\begin{tikzpicture}[scale=1, thick]

\coordinate (S) at (0,-2); 
\coordinate (W) at (-2,0); 
\coordinate (N) at (0,2);  
\coordinate (E) at (2,0);  
\coordinate (X) at (0,0);  

\draw[thick](0,0) circle (2);

\draw (S) to[out=30,in=-30] (N);
\draw (S) to[out=140,in=-140] (N);
\draw (X) node{x};
\draw (0,0.2) to (N);
\draw (0,-0.2) to (S);

\node at (0.4,1) {\footnotesize{$(3,p)$}};
\node at (0.4,-1) {\footnotesize{$(1,p)$}};
\node at (-1.3,0) {\footnotesize{$(1,3)$}};
\node at (1.4,0) {\footnotesize{$(3,1)$}};

\draw[fill] (N) circle (0.05);
\draw[fill] (S) circle (0.05);
\draw[fill] (E) circle (0.05);
\draw[fill] (W) circle (0.05);

\node at (0,-2.3) {1};
\node at (-2.3,0) {2};
\node at (0,2.3) {3};
\node at (2.3,0) {4};

\draw[<->, thick] (3,0) -- (4,0);


\begin{scope}[xshift=8cm, thick, scale = 0.8]

\draw[thick](0,0) circle (0.5);

\coordinate (2c) at (-3,0);
\coordinate (2t) at (-4,1);
\coordinate (2b) at (-4,-1);
\node (r2) at ($(2c)+(-1,0)$) {2}; 

\coordinate (4c) at (3,0);
\coordinate (4t) at (4,1);
\coordinate (4b) at (4,-1);
\node (r4) at ($(4c)+(1,0)$) {4};

\coordinate (3l1) at (-3.5,1.5);
\coordinate (3l2) at (-2.5,0.5);
\coordinate (3l3) at (-1.1,0.5);

\coordinate (3r1) at (3.5,1.5);
\coordinate (3r2) at (2.5,0.5);
\coordinate (3r3) at (1.1,0.5);
\coordinate (3aux) at (0,1.25);

\node (r3) at ($(3aux)+(0,0.5)$) {3};

\coordinate (1l1) at (-3.5,-1.5);
\coordinate (1l2) at (-2.5,-0.5);
\coordinate (1l3) at (-1.1,-0.5);
\coordinate (1r1) at (3.5,-1.5);
\coordinate (1r2) at (2.5,-0.5);
\coordinate (1r3) at (1.1,-0.5);
\coordinate (1aux) at (0,-1.25);

\node (r1) at ($(1aux)+(0,-0.5)$) {1};

\coordinate (2ht) at ($(2c)!0.2!(2t)$);
\coordinate (2hb) at ($(2c)!0.2!(2b)$);
\coordinate (4hb) at ($(4c)!0.2!(4b)$);
\coordinate (4ht) at ($(4c)!0.2!(4t)$);

\coordinate (phrt) at ($0.5*(0.92388, 0.382683)$);
\coordinate (phrb) at ($0.5*(0.92388, -0.382683)$);

\coordinate (phlt) at ($0.5*(-0.92388, 0.382683)$);
\coordinate (phlb) at ($0.5*(-0.92388, -0.382683)$);

\coordinate (c1l) at ($(2ht)!0.5!(2hb)$);
\coordinate (c1r) at ($(phlt)!0.5!(phlb)$);

\coordinate (c2l) at ($(4ht)!0.5!(4hb)$);
\coordinate (c2r) at ($(phrt)!0.5!(phrb)$);




\draw (2c) to (2t);
\draw (2c) to (2b);

\draw (4c) to (4t);
\draw (4c) to (4b);
\draw (3l1) to (3l2);
\draw (3l2) to (3l3);
\draw (3r1) to (3r2);
\draw (3r2) to (3r3);
\draw (3r3) to[out=115, in=0] (3aux);
\draw (3aux) to[out=180, in=65] (3l3);

\draw (1l1) to (1l2);
\draw (1l2) to (1l3);
\draw (1r1) to (1r2);
\draw (1r2) to (1r3);
\draw (1r3) to[out=-115, in=0] (1aux);
\draw (1aux) to[out=180, in=-65] (1l3);
\node at (0,0) {\small $p$};
\node at ($(3l2)!0.5!(3l3) - (0,0.4)$) {\scriptsize $(1,3)$};
\node at ($(3r2)!0.5!(3r3) - (0,0.4)$) {\scriptsize $(3,1)$};

\node at ($(0,0.5)!0.55!(3aux)$) {\scriptsize $(3,p)$};
\node at ($(0,-0.5)!0.65!(1aux)$) {\scriptsize $(1,p)$};

\coordinate (p3l2) at (-2.5,0.5);
\coordinate (p3l3) at (-1.1,0.5);

\draw[->,thick] ($(3l2)+(0.2,0.2)$) -- ($(3l3)+(-0.2,0.2)$) node[midway, above, color=black] {$q_1^\mu$};
\draw[<-,thick] ($1.5*({cos(-150)},{sin(-150)})$) arc (-150:-115:1.5);
\node at ($1.80*({cos(-130)},{sin(-130)})$) {$\ell^\mu$};
\draw[Blue] ($(3l2)!0.7!(1l2)$) to ($(3l3)!0.7!(1l3) + (-0.5,0)$) to[out=0,in=-180] ($(3aux)!0.7!(0,0.5)$) to[out=0,in=170] ($0.3*(3r3)+0.7*(1r3)+(0.3,0)$) to[out=-10,in=180] ($(3r2)!0.7!(1r2)$);
\draw[Maroon] ($0.8*({cos(-160)},{sin(-160)})$) to[out=-50,in=180] ($(0,-0.8)$)to[out=0,in=-130] ($0.8*({cos(-20)},{sin(-20)})$);
\node[Blue] at ($(3r3)!0.5!(3r2) + (0,0.4)$) {\small $X_{2,4}$};
\node[Maroon] at ($(1r3)!0.5!(1r2) + (0,-0.4)$) {\small $X_{3,3}$};

\end{scope}
\end{tikzpicture}
\end{center}
    \caption{(Left) Triangulation of the punctured disk containing curves $\mathcal{T}=\{(1,p),(3,p),(1,3),(3,1)\}$, where $p$ is labelling the puncture. (Right) Dual fatgraph with the internal edges labelled accordingly. In blue we represent the core of curve $X_{2,4}$ and in red the one for the tadpole $X_{3,3}$.}
    \label{fig:BubbleFatGraph}
\end{figure}

\paragraph{One-loop bubble} The one-loop bubble graph is dual to the triangulation of the punctured disk with 4 marked points on the boundary represented in figure \ref{fig:BubbleFatGraph}. The respective kinematic variables associated to the curves in the triangulation are then: 
\begin{equation}
    X_{1,3}=(p_1+p_2)^2, \quad X_{3,1}=(p_3+p_4)^2, \quad X_{1,p}=l^2, \quad X_{3,p}=(l+p_1+p_2)^2, 
\end{equation}
which precisely match the momenta of the dual propagators on the bubble diagram, once we assign the loop-momentum $l^\mu$ to edge $(1,p)$. In addition, we see that (because of momentum conservation) $X_{1,3}=X_{3,1}$, even though $(1,3)$ and $(3,1)$ are different curves on the surface.
After setting these $X_{\mathcal{T}}=0$, on top of the closed curve, we are left with the following $X_\mathcal{C}$'s:
\begin{equation}
    X_{i,i} \equiv 0, \quad X_{2,4}=X_{4,2}=(p_2+p_3)^2,\quad  X_{2,p}=(l+p_1)^2, \quad X_{4,p}=(l+p_1+p_2+p_3)^2,
\end{equation}
where the $X_{i,i}$'s are called \textit{tadpole} curves starting at $i$, going around the puncture and ending back at $i$. These carry zero momentum and therefore don't contribute to the LS.

Let's now see how thinking about each monomial in the LS in terms of a collection of curves filling the fat graph let us automatically conclude that $X_{2,4}/X_{4,2}$ \textit{cannot} enter the bubble LS. Drawing $(2,4)$ on the fatgraph, we see that its core fills edges $\{(1,3),(3,p),(3,1)\}$. Therefore, we need one more core that covers \textit{only} edge $(1,p)$ -- this is the core of curve $(3,3)$, however $X_{3,3} \equiv0$ and therefore this term vanishes in the LS. Similarly the absence of $X_{4,2}$ is due to the vanishing of $X_{1,1}$. 

As for the remaining curves, using \eqref{eq:linearu}, we find the following linearized expansions
\begin{equation}
\begin{aligned}
    u_{2,p}^{X_{2,p}} &= 1- X_{2,p}(y_{1,3})\left[1-y_{1,p}(1+y_{3,p})\right] = 1- X_{2,p}(y_{1,3})\left[1+\eta_{2}y_{1,p}(1+y_{3,p})\right]\, ,\\
    u_{4,p}^{X_{4,p}} &= 1- X_{4,p}(y_{3,1})\left[1-y_{3,p}(1+y_{1,p})\right]= 1- X_{4,p}(y_{3,1})\left[1+\eta_{4}y_{3,p}(1+y_{1,p})\right] \, ,\\
     u_\Delta^\Delta &= 1-\Delta y_{1,p}y_{3,p} \, ,
\end{aligned}
\end{equation}
where we replaced the $-1$ on the RHS by $\eta_{2i}=-1$, so that in the final expression we can distinguish the terms coming from cores/extensions of each $X_{2i,p}$ curve. Finally, putting everything together, we obtain that, at linear order in all $y$-variables, 
\begin{equation}    \prod_\mathcal{C}u_{\mathcal{C}}^{X_{\mathcal{C}}} \rightarrow (-1)X_{2,p} X_{4,p} (\Delta - \underbracket[0.4pt]{(\eta_2 \eta_4 + \eta_2 + \eta_4)}_{\text{extensions}})\prod_{P\in \mathcal{T}} y_P  + \mathcal{O}(y^2)\,,\label{eq:bubble_ext}
\end{equation}
which after setting  $\Delta=1-D$ and $\eta_{2i} = -1$, yields 
\begin{equation}
\text{LS}_{\bub} = (D-2)X_{2,p} X_{4,p}.
    \label{eq:BubbleLS_SurfRes}  
\end{equation}
Note that this can be graphically encoded in the following form: 

\begin{equation}
\vcenter{\hbox{\begin{tikzpicture}[line width=0.6,line cap=round,every node/.style={font=\footnotesize}]
		\begin{scope}[scale = 0.4]
            
			\draw[thick](0,0) circle (0.5);
			\draw[thick,Green] (0,0) circle (0.85);
			\coordinate (2c) at (-3,0);
			\coordinate (2t) at (-4,1);
			\coordinate (2b) at (-4,-1);
			\node (r2) at ($(2c)+(-1,0)$) {2}; 

			\coordinate (4c) at (3,0);
			\coordinate (4t) at (4,1);
			\coordinate (4b) at (4,-1);
			\node (r4) at ($(4c)+(1,0)$) {4};

			\coordinate (3l1) at (-3.5,1.5);
			\coordinate (3l2) at (-2.5,0.5);
			\coordinate (3l3) at (-1.1,0.5);

			\coordinate (3r1) at (3.5,1.5);
			\coordinate (3r2) at (2.5,0.5);
			\coordinate (3r3) at (1.1,0.5);
			\coordinate (3aux) at (0,1.25);

			\node (r3) at ($(3aux)+(0,0.5)$) {3};

			\coordinate (1l1) at (-3.5,-1.5);
			\coordinate (1l2) at (-2.5,-0.5);
			\coordinate (1l3) at (-1.1,-0.5);
			\coordinate (1r1) at (3.5,-1.5);
			\coordinate (1r2) at (2.5,-0.5);
			\coordinate (1r3) at (1.1,-0.5);
			\coordinate (1aux) at (0,-1.25);

			\node (r1) at ($(1aux)+(0,-0.5)$) {1};

			\coordinate (2ht) at ($(2c)!0.2!(2t)$);
			\coordinate (2hb) at ($(2c)!0.2!(2b)$);
			\coordinate (4hb) at ($(4c)!0.2!(4b)$);
			\coordinate (4ht) at ($(4c)!0.2!(4t)$);

			\coordinate (phrt) at ($0.5*(0.92388, 0.382683)$);
			\coordinate (phrb) at ($0.5*(0.92388, -0.382683)$);

			\coordinate (phlt) at ($0.5*(-0.92388, 0.382683)$);
			\coordinate (phlb) at ($0.5*(-0.92388, -0.382683)$);

			\coordinate (c1l) at ($(2ht)!0.5!(2hb)$);
			\coordinate (c1r) at ($(phlt)!0.5!(phlb)$);

			\coordinate (c2l) at ($(4ht)!0.5!(4hb)$);
			\coordinate (c2r) at ($(phrt)!0.5!(phrb)$);




			\draw (2c) to (2t);
			\draw (2c) to (2b);

			\draw (4c) to (4t);
			\draw (4c) to (4b);
			\draw (3l1) to (3l2);
			\draw (3l2) to (3l3);
			\draw (3r1) to (3r2);
			\draw (3r2) to (3r3);
			\draw (3r3) to[out=115, in=0] (3aux);
			\draw (3aux) to[out=180, in=65] (3l3);

			\draw (1l1) to (1l2);
			\draw (1l2) to (1l3);
			\draw (1r1) to (1r2);
			\draw (1r2) to (1r3);
			\draw (1r3) to[out=-115, in=0] (1aux);
			\draw (1aux) to[out=180, in=-65] (1l3);
			\node at (0,0) {$p$};


            \node[Blue] at ($(3l2)!0.5!(3l3) + (0,0.7)$){\tiny $X_{2,p}$};
            \node[Maroon] at ($(3r2)!0.5!(3r3) + (0,0.7)$){\tiny $X_{4,p}$};
			
			\draw[Blue, thick] ($(3l2)!0.5!(1l2)$) to ($(3l3)!0.5!(1l3)$);
			\draw[Maroon, thick] ($(3r2)!0.5!(1r2)$) to ($(3r3)!0.5!(1r3)$);
			\node[Green] at ($1.65*({cos(-135)},{sin(-135)})$) {\scriptsize $\Delta$};
		\end{scope}
	\end{tikzpicture}
}}
-
\vcenter{\hbox{\begin{tikzpicture}[line width=0.6,line cap=round,every node/.style={font=\footnotesize}]
		\begin{scope}[scale = 0.4]

			\draw[thick](0,0) circle (0.5);

			\coordinate (2c) at (-3,0);
			\coordinate (2t) at (-4,1);
			\coordinate (2b) at (-4,-1);
			\node (r2) at ($(2c)+(-1,0)$) {2}; 

			\coordinate (4c) at (3,0);
			\coordinate (4t) at (4,1);
			\coordinate (4b) at (4,-1);
			\node (r4) at ($(4c)+(1,0)$) {4};

			\coordinate (3l1) at (-3.5,1.5);
			\coordinate (3l2) at (-2.5,0.5);
			\coordinate (3l3) at (-1.1,0.5);

			\coordinate (3r1) at (3.5,1.5);
			\coordinate (3r2) at (2.5,0.5);
			\coordinate (3r3) at (1.1,0.5);
			\coordinate (3aux) at (0,1.25);

			\node (r3) at ($(3aux)+(0,0.5)$) {3};

			\coordinate (1l1) at (-3.5,-1.5);
			\coordinate (1l2) at (-2.5,-0.5);
			\coordinate (1l3) at (-1.1,-0.5);
			\coordinate (1r1) at (3.5,-1.5);
			\coordinate (1r2) at (2.5,-0.5);
			\coordinate (1r3) at (1.1,-0.5);
			\coordinate (1aux) at (0,-1.25);

			\node (r1) at ($(1aux)+(0,-0.5)$) {1};

			\coordinate (2ht) at ($(2c)!0.2!(2t)$);
			\coordinate (2hb) at ($(2c)!0.2!(2b)$);
			\coordinate (4hb) at ($(4c)!0.2!(4b)$);
			\coordinate (4ht) at ($(4c)!0.2!(4t)$);

			\coordinate (phrt) at ($0.5*(0.92388, 0.382683)$);
			\coordinate (phrb) at ($0.5*(0.92388, -0.382683)$);

			\coordinate (phlt) at ($0.5*(-0.92388, 0.382683)$);
			\coordinate (phlb) at ($0.5*(-0.92388, -0.382683)$);

			\coordinate (c1l) at ($(2ht)!0.5!(2hb)$);
			\coordinate (c1r) at ($(phlt)!0.5!(phlb)$);

			\coordinate (c2l) at ($(4ht)!0.5!(4hb)$);
			\coordinate (c2r) at ($(phrt)!0.5!(phrb)$);




			\draw (2c) to (2t);
			\draw (2c) to (2b);

			\draw (4c) to (4t);
			\draw (4c) to (4b);
			\draw (3l1) to (3l2);
			\draw (3l2) to (3l3);
			\draw (3r1) to (3r2);
			\draw (3r2) to (3r3);
			\draw (3r3) to[out=115, in=0] (3aux);
			\draw (3aux) to[out=180, in=65] (3l3);

			\draw (1l1) to (1l2);
			\draw (1l2) to (1l3);
			\draw (1r1) to (1r2);
			\draw (1r2) to (1r3);
			\draw (1r3) to[out=-115, in=0] (1aux);
			\draw (1aux) to[out=180, in=-65] (1l3);
			\node at (0,0) {$p$};



			\draw[Blue, thick] ($(3l2)!0.5!(1l2)$) to ($(3l3)!0.5!(1l3)$); \draw[dotted, Blue, thick] ($(3l3)!0.5!(1l3)$) to[out=0,in=180] ($(3aux)!0.5!(0,0.5)$) to[out=0,in=135] ($0.8*({cos(20)},{sin(20)})$);
			\draw[Maroon, thick] ($(3r2)!0.5!(1r2)$) to ($(3r3)!0.5!(1r3)$);
			
			\draw[Maroon, thick, dotted] ($(3r3)!0.5!(1r3)$) to[out=180,in=0] ($(1aux)!0.5!(0,-0.5)$) to[out=180,in=-45] ($0.8*({cos(-160)},{sin(-160)})$);
            \node[Blue] at ($(3l2)!0.5!(3l3) + (0,0.7)$){\tiny $\eta_2X_{2,p}$};
            \node[Maroon] at ($(3r2)!0.5!(3r3) + (0,0.7)$){\tiny $\eta_4X_{4,p}$};
		\end{scope}
	\end{tikzpicture}
}}
-
\vcenter{\hbox{\begin{tikzpicture}[line width=0.6,line cap=round,every node/.style={font=\footnotesize}]
		\begin{scope}[scale = 0.4]

			\draw[thick](0,0) circle (0.5);

			\coordinate (2c) at (-3,0);
			\coordinate (2t) at (-4,1);
			\coordinate (2b) at (-4,-1);
			\node (r2) at ($(2c)+(-1,0)$) {2}; 

			\coordinate (4c) at (3,0);
			\coordinate (4t) at (4,1);
			\coordinate (4b) at (4,-1);
			\node (r4) at ($(4c)+(1,0)$) {4};

			\coordinate (3l1) at (-3.5,1.5);
			\coordinate (3l2) at (-2.5,0.5);
			\coordinate (3l3) at (-1.1,0.5);

			\coordinate (3r1) at (3.5,1.5);
			\coordinate (3r2) at (2.5,0.5);
			\coordinate (3r3) at (1.1,0.5);
			\coordinate (3aux) at (0,1.25);

			\node (r3) at ($(3aux)+(0,0.5)$) {3};

			\coordinate (1l1) at (-3.5,-1.5);
			\coordinate (1l2) at (-2.5,-0.5);
			\coordinate (1l3) at (-1.1,-0.5);
			\coordinate (1r1) at (3.5,-1.5);
			\coordinate (1r2) at (2.5,-0.5);
			\coordinate (1r3) at (1.1,-0.5);
			\coordinate (1aux) at (0,-1.25);

			\node (r1) at ($(1aux)+(0,-0.5)$) {1};

			\coordinate (2ht) at ($(2c)!0.2!(2t)$);
			\coordinate (2hb) at ($(2c)!0.2!(2b)$);
			\coordinate (4hb) at ($(4c)!0.2!(4b)$);
			\coordinate (4ht) at ($(4c)!0.2!(4t)$);

			\coordinate (phrt) at ($0.5*(0.92388, 0.382683)$);
			\coordinate (phrb) at ($0.5*(0.92388, -0.382683)$);

			\coordinate (phlt) at ($0.5*(-0.92388, 0.382683)$);
			\coordinate (phlb) at ($0.5*(-0.92388, -0.382683)$);

			\coordinate (c1l) at ($(2ht)!0.5!(2hb)$);
			\coordinate (c1r) at ($(phlt)!0.5!(phlb)$);

			\coordinate (c2l) at ($(4ht)!0.5!(4hb)$);
			\coordinate (c2r) at ($(phrt)!0.5!(phrb)$);




			\draw (2c) to (2t);
			\draw (2c) to (2b);

			\draw (4c) to (4t);
			\draw (4c) to (4b);
			\draw (3l1) to (3l2);
			\draw (3l2) to (3l3);
			\draw (3r1) to (3r2);
			\draw (3r2) to (3r3);
			\draw (3r3) to[out=115, in=0] (3aux);
			\draw (3aux) to[out=180, in=65] (3l3);

			\draw (1l1) to (1l2);
			\draw (1l2) to (1l3);
			\draw (1r1) to (1r2);
			\draw (1r2) to (1r3);
			\draw (1r3) to[out=-115, in=0] (1aux);
			\draw (1aux) to[out=180, in=-65] (1l3);
			\node at (0,0) {$p$};



			\draw[Blue, thick] ($(3l2)!0.5!(1l2)$) to ($(3l3)!0.5!(1l3)$);
            \draw[Blue, thick, dotted] ($(3l3)!0.5!(1l3)$) to[out=0,in=180] ($(3aux)!0.5!(0,0.5)$) to[out=0,in=90] ($0.8*({cos(0)},{sin(0)})$) to[out=-90,in=0]($0.8*({cos(-90)},{sin(-90)})$) to[out=180,in=-45] ($0.8*({cos(-160)},{sin(-160)})$);
			\draw[Maroon, thick] ($(3r2)!0.5!(1r2)$) to ($(3r3)!0.5!(1r3)$);
            \node[Blue] at ($(3l2)!0.5!(3l3) + (0,0.7)$){\tiny $\eta_2X_{2,p}$};
            \node[Maroon] at ($(3r2)!0.5!(3r3) + (0,0.7)$){\tiny $X_{4,p}$};
		\end{scope}
	\end{tikzpicture}
}}
-
\vcenter{\hbox{\begin{tikzpicture}[line width=0.6,line cap=round,every node/.style={font=\footnotesize}]
		\begin{scope}[scale = 0.4]

			\draw[thick](0,0) circle (0.5);

			\coordinate (2c) at (-3,0);
			\coordinate (2t) at (-4,1);
			\coordinate (2b) at (-4,-1);
			\node (r2) at ($(2c)+(-1,0)$) {2}; 

			\coordinate (4c) at (3,0);
			\coordinate (4t) at (4,1);
			\coordinate (4b) at (4,-1);
			\node (r4) at ($(4c)+(1,0)$) {4};

			\coordinate (3l1) at (-3.5,1.5);
			\coordinate (3l2) at (-2.5,0.5);
			\coordinate (3l3) at (-1.1,0.5);

			\coordinate (3r1) at (3.5,1.5);
			\coordinate (3r2) at (2.5,0.5);
			\coordinate (3r3) at (1.1,0.5);
			\coordinate (3aux) at (0,1.25);

			\node (r3) at ($(3aux)+(0,0.5)$) {3};

			\coordinate (1l1) at (-3.5,-1.5);
			\coordinate (1l2) at (-2.5,-0.5);
			\coordinate (1l3) at (-1.1,-0.5);
			\coordinate (1r1) at (3.5,-1.5);
			\coordinate (1r2) at (2.5,-0.5);
			\coordinate (1r3) at (1.1,-0.5);
			\coordinate (1aux) at (0,-1.25);

			\node (r1) at ($(1aux)+(0,-0.5)$) {1};

			\coordinate (2ht) at ($(2c)!0.2!(2t)$);
			\coordinate (2hb) at ($(2c)!0.2!(2b)$);
			\coordinate (4hb) at ($(4c)!0.2!(4b)$);
			\coordinate (4ht) at ($(4c)!0.2!(4t)$);

			\coordinate (phrt) at ($0.5*(0.92388, 0.382683)$);
			\coordinate (phrb) at ($0.5*(0.92388, -0.382683)$);

			\coordinate (phlt) at ($0.5*(-0.92388, 0.382683)$);
			\coordinate (phlb) at ($0.5*(-0.92388, -0.382683)$);

			\coordinate (c1l) at ($(2ht)!0.5!(2hb)$);
			\coordinate (c1r) at ($(phlt)!0.5!(phlb)$);

			\coordinate (c2l) at ($(4ht)!0.5!(4hb)$);
			\coordinate (c2r) at ($(phrt)!0.5!(phrb)$);




			\draw (2c) to (2t);
			\draw (2c) to (2b);

			\draw (4c) to (4t);
			\draw (4c) to (4b);
			\draw (3l1) to (3l2);
			\draw (3l2) to (3l3);
			\draw (3r1) to (3r2);
			\draw (3r2) to (3r3);
			\draw (3r3) to[out=115, in=0] (3aux);
			\draw (3aux) to[out=180, in=65] (3l3);

			\draw (1l1) to (1l2);
			\draw (1l2) to (1l3);
			\draw (1r1) to (1r2);
			\draw (1r2) to (1r3);
			\draw (1r3) to[out=-115, in=0] (1aux);
			\draw (1aux) to[out=180, in=-65] (1l3);
			\node at (0,0) {$p$};



			\draw[Blue, thick] ($(3l2)!0.5!(1l2)$) to ($(3l3)!0.5!(1l3)$);
			\draw[Maroon, thick] ($(3r2)!0.5!(1r2)$) to ($(3r3)!0.5!(1r3)$) ;
            \draw[Maroon, dotted, thick] ($(3r3)!0.5!(1r3)$) to[out=180,in=0] ($(1aux)!0.5!(0,-0.5)$) to[out=180,in=-90] ($0.8*({cos(180)},{sin(180)})$) to[out=90,in=180]($0.8*({cos(90)},{sin(90)})$) to[out=0,in=135] ($0.8*({cos(20)},{sin(20)})$);
            \node[Blue] at ($(3l2)!0.5!(3l3) + (0,0.7)$){\tiny $X_{2,p}$};
            \node[Maroon] at ($(3r2)!0.5!(3r3) + (0,0.7)$){\tiny $\eta_4X_{4,p}$};
		\end{scope}
	\end{tikzpicture}
}}
\end{equation}
where we see explicitly that each term gives a possible way of filling the fatgraph using curves/extensions in a way that each edge is covered once and only once. 

\section{Tree-level leading singularities from explicit gluing}
\label{sec:LSfromQFT}

A leading singularity for a process involving $n$ gluons can be computed by gluing 3 point amplitudes using the gluon polarization-sum tensor in the following way
 \begin{equation}
     \text{LS}_n = \mathcal{A}_3^\mu(p^\mu) \left[-\eta_{\mu\nu} +\frac{p_\mu q_\nu + p_\nu q_\mu}{p\cdot q}\right] \mathcal{A}_3^{\nu \alpha}(-p^\nu, k^\alpha)  \cdots \left[-\eta^{\beta \rho} +\frac{p^\prime_\beta q^\prime_\rho + p^\prime_\beta q^\prime_\rho}{p^\prime\cdot q^\prime}\right] \mathcal{A}_3^\rho(p^{\prime \, \rho}),
 \end{equation}
where $q_\mu/q^\prime_\mu$ are reference vectors that should drop out in the final answer, and $\mathcal{A}^3_\mu (p^\mu)$ is the 3-point vertex with the polarization vector stripped off from the leg with momentum $p^\mu$. At tree-level, since at each polarization sum is gluing two trees, both satisfying the Ward-identity $p^\mu \mathcal{A}_\mu = 0$, we can compute the spin sums using $\eta_{\mu\nu}$, and so the LS is given by
  \begin{equation}
     \text{LS}_n = \mathcal{A}_3^\mu \left(-\eta_{\mu\nu}\right) \mathcal{A}_3^{\nu \alpha}  \cdots \left(-\eta_{\beta \rho}\right) \mathcal{A}_3^{\rho}, 
     \label{eq:LSTreeGluing}
 \end{equation}
which is then simply a contraction of 3-point amplitudes. Let us now study what this contraction looks like when the gluons are scalar-scaffolded.

\subsection{Building blocks: 3-point amplitudes \label{sec:ContPic}}

In section \ref{sec:Dictionary}, we reviewed the momentum-space Feynman rules for the 3-point gluon interaction, and the interaction between a gluon and two scaffolding scalars, given in eqns. \eqref{eq:3ptGluonPol} and  \eqref{eq:GluonScalarMap}, respectively.
Having these expressions in momentum space, we now wish to represent each term in a graphical way that makes the Lorentz-contractions manifest. 

Let's start by drawing the three point vertex as a fat-graph -- which is equivalently the double-line representation of this color ordered vertex. We start by drawing a blue line on the edges of fat graph associated with gluons to represent the Lorentz index naturally associated to its polarization vector. This way we can represent contractions of two polarization vectors by connecting the blue lines of the respective edges. As for the contractions involving polarization and difference of momenta, such as $\epsilon_3.(q_2-q_1)$, we represent them by ending the polarization blue line on a red handle anchored in the two respective edges with momentum $q_2$ and $q_1$. So the full $3$-point amplitude can be represented as follows 
\begin{equation}
\begin{aligned}
\mathcal{A}_3[g_1,g_2,g_3] =& \,\epsilon_1 \cdot(q_2 - q_3) \epsilon_2 \cdot \epsilon_3 + \epsilon_2 \cdot(q_3 - q_1) \epsilon_1 \cdot \epsilon_3  + \epsilon_3 \cdot(q_1 - q_2) \epsilon_1 \cdot \epsilon_2 \\ 
=& \,\quad \quad \begin{gathered}
    \begin{tikzpicture}[line width=0.6,scale=0.27,line cap=round,every node/.style={font=\scriptsize}]
		\node (0) at (-2.75, 1.5) {};
		\node (1) at (-2, 2.25) {};
            \coordinate (01m) at ($(0)!0.5!(1)$);
		\node (2) at (-1, -0.25) {};
		\node (3) at (-0.25, 0.5) {};
		\node (4) at (-1, -3.25) {};
		\node(5) at (0, -3.25) {};
		\node (6) at (0, -0.5) {};
		\node   (7) at (2.75, -0.5) {};
		\node   (8) at (2.75, 0.5) {};		
		\node   (25) at (-2.25, -0.75) {1};
		\node   (27) at (0.5, 1.75) {2};
		\node   (29) at (1.25, -1.75) {3};

           \coordinate (45m) at ($(4)!0.5!(5)$);
         \coordinate (26m) at ($(2)!0.5!(6)$);
        \coordinate (36m) at ($(3)!0.5!(6)$);
        \coordinate (78m) at ($(7)!0.5!(8)$);

        \coordinate (Anch1) at ($(6)!0.25!(5)$);
        \coordinate (Anch2) at ($(6)!0.25!(7)$);
         \coordinate (mAnch) at ($(Anch1)!0.5!(Anch2)$);
		
		\draw[thick] (0.center) to (2.center);
		\draw[thick] (2.center) to (4.center);
		\draw[thick] (6.center) to (5.center);
		\draw[thick] [in=135, out=-45] (1.center) to (3.center);
		\draw[thick] (3.center) to (8.center);
		\draw[thick]  (6.center) to (7.center);
		\draw[thick][style=Blue] (01m.center) to (mAnch);
        \draw[thick][style=Blue] (45m.center) to (26m.center);
        \draw[thick][style=Blue] (36m.center) to (78m.center);
         \draw[Blue,thick] (26m) to[out=90,in=180] (36m);
         \draw[Maroon,thick] (Anch1) to[out=0,in=-90] (Anch2);
\end{tikzpicture}
\end{gathered} \quad \, \, + \quad \, \, \, \, \,  \begin{gathered}
    \begin{tikzpicture}[line width=0.6,scale=0.27,line cap=round,every node/.style={font=\scriptsize}]
		\node (0) at (-2.75, 1.5) {};
		\node (1) at (-2, 2.25) {};
            \coordinate (01m) at ($(0)!0.5!(1)$);
		\node (2) at (-1, -0.25) {};
		\node (3) at (-0.25, 0.5) {};
		\node (4) at (-1, -3.25) {};
		\node(5) at (0, -3.25) {};
		\node (6) at (0, -0.5) {};
		\node   (7) at (2.75, -0.5) {};
		\node   (8) at (2.75, 0.5) {};		
		\node   (25) at (-2.25, -0.75) {1};
		\node   (27) at (0.5, 1.75) {2};
		\node   (29) at (1.25, -1.75) {3};

           \coordinate (45m) at ($(4)!0.5!(5)$);
         \coordinate (26m) at ($(2)!0.5!(6)$);
        \coordinate (36m) at ($(3)!0.5!(6)$);
        \coordinate (78m) at ($(7)!0.5!(8)$);
        \coordinate (23m) at ($(2)!0.5!(3)$);

        \coordinate (Anch1) at ($(2)!0.2!(0)$);
        \coordinate (Anch2) at ($(2)!0.2!(4)$);
         \coordinate (mAnch) at ($(Anch1)!0.5!(Anch2)$);
		
		\draw[thick] (0.center) to (2.center);
		\draw[thick] (2.center) to (4.center);
		\draw[thick] (6.center) to (5.center);
		\draw[thick] [in=135, out=-45] (1.center) to (3.center);
		\draw[thick] (3.center) to (8.center);
		\draw[thick]  (6.center) to (7.center);
		\draw[thick][style=Blue] (01m.center) to (23m.center);
        \draw[thick][style=Blue] (45m.center) to (26m.center);
        \draw[thick][style=Blue] (36m.center) to (78m.center);
         \draw[Blue,thick] (26m) to[out=90,in=-45] (23m);
         \draw[Blue,thick] (36m) to[out=180,in=45] (-1.25,-0.4);
         \draw[Maroon,thick] (Anch1) to[out=-135,in=180] (Anch2);
\end{tikzpicture}
\end{gathered} \quad \, \, + \quad \, \, \, \, \, \begin{gathered}
    \begin{tikzpicture}[line width=0.6,scale=0.27,line cap=round,every node/.style={font=\scriptsize}]
		\node (0) at (-2.75, 1.5) {};
		\node (1) at (-2, 2.25) {};
            \coordinate (01m) at ($(0)!0.5!(1)$);
		\node (2) at (-1, -0.25) {};
		\node (3) at (-0.25, 0.5) {};
		\node (4) at (-1, -3.25) {};
		\node(5) at (0, -3.25) {};
		\node (6) at (0, -0.5) {};
		\node   (7) at (2.75, -0.5) {};
		\node   (8) at (2.75, 0.5) {};		
		\node   (25) at (-2.25, -0.75) {1};
		\node   (27) at (0.5, 1.75) {2};
		\node   (29) at (1.25, -1.75) {3};

           \coordinate (45m) at ($(4)!0.5!(5)$);
         \coordinate (26m) at ($(2)!0.5!(6)$);
        \coordinate (36m) at ($(3)!0.5!(6)$);
        \coordinate (78m) at ($(7)!0.5!(8)$);
        \coordinate (23m) at ($(2)!0.5!(3)$);

        \coordinate (Anch1) at ($(3)!0.2!(1)$);
        \coordinate (Anch2) at ($(3)!0.2!(8)$);
         \coordinate (mAnch) at ($(Anch1)!0.5!(Anch2)$);
		
		\draw[thick] (0.center) to (2.center);
		\draw[thick] (2.center) to (4.center);
		\draw[thick] (6.center) to (5.center);
		\draw[thick] [in=135, out=-45] (1.center) to (3.center);
		\draw[thick] (3.center) to (8.center);
		\draw[thick]  (6.center) to (7.center);
		\draw[thick][style=Blue] (01m.center) to (23m.center);
        \draw[thick][style=Blue] (45m.center) to (26m.center);
        \draw[thick][style=Blue] (36m.center) to (78m.center);
         \draw[Blue,thick] (36m) to[out=180,in=-45] (23m);
         \draw[Blue,thick] (26m) to[out=90,in=-125] (-0.1, 0.75);
         \draw[Maroon,thick] (Anch1) to[out=45,in=90] (Anch2);
\end{tikzpicture}
\end{gathered}
\end{aligned}
\label{eq:3ptGluonRep}
\end{equation}
where the index of a given edge can be read via the region on its left. The scalar-scalar-gluon interaction, entering the scaffolding vertices, is given by a single contraction which we can similarly represent as
\begin{equation}
\mathcal{A}_3[g_1,\phi_2,\phi_3] =\,\epsilon_1 \cdot(p_2 - p_3) 
= \quad\begin{gathered}
    \begin{tikzpicture}[line width=0.6,scale=0.31,line cap=round,every node/.style={font=\scriptsize}]
		\node (0) at (-2.75, 1.5) {};
		\node (1) at (-2, 2.25) {};
            \coordinate (01m) at ($(0)!0.5!(1)$);
		\node (2) at (-1, -0.25) {};
		\node (3) at (-0.25, 0.5) {};
		\node (4) at (-1, -3.25) {};
		\node(5) at (0, -3.25) {};
		\node (6) at (0, -0.5) {};
		\node   (7) at (2.75, -0.5) {};
		\node   (8) at (2.75, 0.5) {};		
		\node   (25) at (-2.25, -0.75) {1};
		\node   (27) at (0.5, 1.75) {2};
		\node   (29) at (1.25, -1.75) {3};
           \coordinate (45m) at ($(4)!0.5!(5)$);
         \coordinate (26m) at ($(2)!0.5!(6)$);
        \coordinate (36m) at ($(3)!0.5!(6)$);
        \coordinate (78m) at ($(7)!0.5!(8)$);

        \coordinate (Anch1) at ($(6)!0.25!(5)$);
        \coordinate (Anch2) at ($(6)!0.25!(7)$);
         \coordinate (mAnch) at ($(Anch1)!0.5!(Anch2)$);
		
		\draw[thick] (0.center) to (2.center);
		\draw[thick] (2.center) to (4.center);
		\draw[thick] (6.center) to (5.center);
		\draw[thick] [in=135, out=-45] (1.center) to (3.center);
		\draw[thick] (3.center) to (8.center);
		\draw[thick]  (6.center) to (7.center);
		\draw[thick][style=Blue] (01m.center) to (mAnch);
         \draw[Maroon,thick] (Anch1) to[out=0,in=-90] (Anch2);

\end{tikzpicture}
\end{gathered} \quad \equiv 2 \epsilon_1 \cdot p_2
\label{eq:3ptGluonScalar}
\end{equation}
where $p_2^\mu$ and $p_3^\mu$ are the momenta of the scalar edges $2$ and $3$, respectively. Since these edges correspond to scalar particles we don't have any polarization contraction associated to them, and thus no blue curve. Note that in both \eqref{eq:3ptGluonRep} and \eqref{eq:3ptGluonScalar}, once we fix the place of the red handles the blue lines/contraction curves are completely fixed. Henceforth, we call this graphical rule for polarization contractions the \textit{Lorentz-contraction picture}.

Due to on-shell gauge-invariance, we have some freedom in defining the rule for what the blue curve ending on the red handle means, since  
\begin{equation}
    \epsilon_1 \cdot (p_2 - p_3 ) =2 \epsilon_1 \cdot p_2 =-2 \epsilon_1 \cdot p_3 \, .
    \label{eq:FeynRuleOnShell}
\end{equation}

For the rest of the paper, we take the second expression in \eqref{eq:FeynRuleOnShell} to represent the blue-red contraction (as emphasized in \eqref{eq:3ptGluonScalar}), which effectively means that the Lorentz index carried by a blue line ending on a red handle is contracted with the momentum of the edge to the \textit{left} of the handle\footnote{For simplicity we will drop the factors of $2$, as these can be easily reinstate afterwards by a factor of $2^{\#V}$, where $\#V$ number of vertices $\#V$ on the diagram.}. As it turns out this choice will be important, as it will make the translation of the contractions into scalar-scaffolidng variables, $X_{i,j}$, follow a simple graphical interpretation. 

Let's then see what happens when we scalar-scaffold already in the simplest example at $3$-points. In this case, we can scaffold each gluon by attaching a scalar-scalar-gluon vertex to each edge in \eqref{eq:3ptGluonScalar}. Note that this produces a blue line entering each edge of the $3$-point interaction, and by contracting these in all possible three ways, we get the full $3$-point amplitude, but now scalar-scaffolded. 

One particular such contraction, resulting from placing the internal red handle in region $3$ is the following
\begin{equation}
    \begin{gathered}
    \begin{tikzpicture}[line width=0.6,scale=0.3,line cap=round,every node/.style={font=\scriptsize}]
		\node (0) at (-2.75, 1.5) {};
		\node (1) at (-2, 2.25) {};
		\node (2) at (-1, -0.25) {};
		\node (3) at (-0.25, 0.5) {};
		\node (4) at (-1, -3.25) {};
		\node(5) at (0, -3.25) {};
		\node (6) at (0, -0.5) {};
		\node   (7) at (2.75, -0.5) {};
		\node   (8) at (2.75, 0.5) {};
		\node   (9) at (-4.25, 1.5) {};
		\node   (10) at (-2.5, 4.25) {};
		\node   (11) at (-4.25, 2.5) {};
		\node   (12) at (-3.25, 3.75) {};
		\node   (13) at (-3, 2.5) {};
		\node   (14) at (-2.25, -5) {};
		\node   (15) at (-1.25, -5) {};
		\node   (16) at (-0.5, -4) {};
		\node   (17) at (0.25, -5) {};
		\node   (18) at (1.25, -5) {};
		\node   (19) at (4.75, -2) {};
		\node   (20) at (4.75, -0.75) {};
		\node   (21) at (3.75, 0) {};
		\node   (22) at (4.75, 0.75) {};
		\node   (23) at (4.75, 2) {};
		\node   (24) at (-2.75, -0.75) {};
		\node   (25) at (-2.75, -0.75) {1};
		\node   (27) at (1, 2.25) {3};
		\node   (28) at (4.75, 0) {4};
		\node   (29) at (1.25, -2) {5};
		\node   (30) at (-0.5, -4.95) {6};
		\node   (31) at (-0.25, 0) {};
		\node   (32) at (3.25, 0) {};
		\node   (33) at (-2.5, 2) {};
		\node   (34) at (-0.5, 0) {};
		\node   (35) at (-0.5, -3.25) {};
		\node   (36) at (-3.75, 3) {};
		\node   (37) at (-3.75, 3.2) {2};
		\draw[thick] (0.center) to (2.center);
		\draw[thick] (2.center) to (4.center);
		\draw[thick] (6.center) to (5.center);
		\draw[thick] [in=135, out=-45] (1.center) to (3.center);
		\draw[thick] (3.center) to (8.center);
		\draw[thick]  (6.center) to (7.center);
		\draw[thick]  (9.center) to (0.center);
		\draw[thick]  (11.center) to (13.center);
		\draw[thick] (13.center) to (12.center);
		\draw[thick] (10.center) to (1.center);
		\draw[thick](14.center) to (4.center);
		\draw[thick](15.center) to (16.center);
		\draw[thick](16.center) to (17.center);
		\draw[thick](5.center) to (18.center);
	\draw[thick] (7.center) to (19.center);
		\draw[thick](8.center) to (23.center);
	\draw[thick](21.center) to (22.center);
		\draw[thick](21.center) to (20.center);

           \coordinate (45m) at ($(4)!0.5!(5)$);
         \coordinate (26m) at ($(2)!0.5!(6)$);
        \coordinate (36m) at ($(3)!0.5!(6)$);
        \coordinate (78m) at ($(7)!0.5!(8)$);
        \coordinate (23m) at ($(2)!0.5!(3)$);

         \coordinate (1617m) at ($(16)!0.4!(17)$);
          \coordinate (1615m) at ($(16)!0.4!(15)$);
          \coordinate (Anch6m) at ($(1617m)!0.5!(1615m)$);

           \coordinate (1312m) at ($(13)!0.4!(12)$);
          \coordinate (1311m) at ($(13)!0.4!(11)$);
          \coordinate (Anch2m) at ($(1312m)!0.5!(1311m)$);
          
          \coordinate (2122m) at ($(21)!0.4!(22)$);
          \coordinate (2120m) at ($(21)!0.4!(20)$);
          \coordinate (Anch4m) at ($(2122m)!0.5!(2120m)$);

          \coordinate (Anch1) at ($(3)!0.2!(1)$);
          \coordinate (Anch2) at ($(3)!0.2!(8)$);
          \coordinate (mAnch) at ($(Anch1)!0.5!(Anch2)$);
		
		\draw[thick] (0.center) to (2.center);
		\draw[thick] (2.center) to (4.center);
		\draw[thick] (6.center) to (5.center);
		\draw[thick] [in=135, out=-45] (1.center) to (3.center);
		\draw[thick] (3.center) to (8.center);
		\draw[thick]  (6.center) to (7.center);
		\draw[thick][style=Blue] (Anch2m) to (23m.center);
        \draw[thick][style=Blue] (Anch6m.center) to (26m.center);
        \draw[thick][style=Blue] (36m.center) to (Anch4m.center);
         \draw[Blue,thick] (36m) to[out=180,in=-45] (23m);
         \draw[Blue,thick] (26m) to[out=90,in=-125] (-0.1, 0.75);
         \draw[Maroon,thick] (Anch1) to[out=45,in=90] (Anch2);
         \draw[Maroon,thick] (1617m) to[out=-135,in=-45] (1615m);
         \draw[Maroon,thick] (1312m) to[out=185,in=90] (1311m);
          \draw[Maroon,thick] (2122m) to[out=-35,in=45] (2120m);
\end{tikzpicture}
\end{gathered}\quad  \begin{aligned}
    = &\left[-(x_3-x_1)\cdot(x_6-x_5) \right]\, \left[-(x_2-x_1)\cdot(x_4-x_3) \right] \\
    = & \left[X_{3,6}\right]\left[X_{2,4} - X_{1,4} \right]
\end{aligned}
\label{eq:3ptContraction}
\end{equation}
where we have used the dual coordinates, $x_i^\mu$, to write the  momentum of the edge between regions $i$ and $j$ as $(x_{j}^\mu - x_i^\mu)$, and the overall minus sign is because we are gluing with $-\eta_{\mu,\nu}$. This allows us efficiently read off the translation of the contractions into the planar variables, $X_{i,j}$, using 
\begin{equation}
    (x_b - x_a) \cdot (x_d - x_c) = X_{b,c} + X_{a,d}-X_{b,d} - X_{a,c} \, ,
    \label{eq:xdotxToXij}
\end{equation}

So evaluating each term on the scaffolding locus, $X_{1,3}=X_{3,5}=X_{1,5}=0$, we have that contraction \eqref{eq:3ptContraction} gives $X_{3,6}(X_{2,4}-X_{1,4})$. Performing cyclic transformations (by two) to this term, we produce the two other contractions, and by adding all three together we recover the scalar-scaffolded 3-point amplitude given in \eqref{eq:3ptScaff}. Note that just by keeping track of the Lorentz contractions in this pictorial way we generated a picture (in \eqref{eq:3ptContraction}) very similar to the one found in \eqref{eq:3ptLS_Surf}! More precisely, here we see that a single pattern for index contractions produces multiple monomials which correspond to two different ways of covering the fat-graph with cores/extensions. 

Another aspect to point out is that while in this simple example the Lorentz-contraction picture yields three different terms (six monomials) which add up to the final result for $\text{LS}_3 $, this is, in general, not the case. Instead, when we add different contractions some monomials might cancel in the final result. It is then natural to ask whether from this perspective there is a simple way to predict the sign of a given monomial and whether it survives after we add everything together. As we explain in section \ref{sec:MonPicture}, we can start answering this question by thinking about each monomial separately, and reverse engineering which contractions give rise to it. In section \ref{sec:CancellationsVRule}, we give a general rule for the cancellations both at tree- and loop-level. For now, let's quickly look at the simplest case in which this cancellations happens: the $s$-channel LS$_4$.

\paragraph{4-point s-channel LS} This LS is given by the gluing of two $\mathcal{A}_3$'s and, therefore, there are a total of 9 contractions each corresponding to the different configurations of red-handle markings (on the two vertices). Two of these, as well as their respective $X_{i,j}$ contractions, are 
\begin{equation}
\begin{aligned}
&\quad \, \, \,\begin{gathered}
	\begin{tikzpicture}[line width=0.6,scale=0.3,line cap=round,every node/.style={font=\normalsize
}]
		\node (1l) at (-2, -1) {};
		\node (1r) at (2, -1) {};
		\node (5l) at (-2, 1) {};
		\node (5r) at (2, 1) {};
		
		\node (2) at (-5, -3) {};
		\node (2d) at (-5, -5) {};
		\node (2l) at (-7, -3) {};
		\node (2r) at (-4, -3) {};
		\node (2dr) at (-4, -5) {};
		\node (2u) at (-5, -2) {};
		\node (2lu) at (-7, -2) {};

		\node (4) at (-5, 3) {};
		\node (4u) at (-5, 5) {};
		\node (4l) at (-7, 3) {};
		\node (4r) at (-4, 3) {};
		\node (4ur) at (-4, 5) {};
		\node (4d) at (-5, 2) {};
		\node (4ld) at (-7, 2) {};

		\node (6) at (5, 3) {};
		\node (6u) at (5, 5) {};
		\node (6r) at (7, 3) {};
		\node (6l) at (4, 3) {};
		\node (6ul) at (4, 5) {};
		\node (6d) at (5, 2) {};
		\node (6dr) at (7, 2) {};
		
		\node (8) at (5, -3) {};
		\node (8r) at (7, -3) {};
		\node (8d) at (5, -5) {};
		\node (8dl) at (4, -5) {};
		\node (8l) at (4, -3) {};
		\node (8ur) at (7, -2) {};
		\node (8u) at (5, -2) {};;
		
		\node (3) at (-3, 0) {};
		\node (7) at (3, 0) {};

		\draw[thick] (2l.center) to (2.center);
		\draw[thick] (2.center) to (2d.center);
		
		\draw[thick] (4u.center) to (4.center);
		\draw[thick] (4.center) to (4l.center);

		\draw[thick] (6u.center) to (6.center);
		\draw[thick] (6.center) to (6r.center);
		
		\draw[thick] (8r.center) to (8.center);
		\draw[thick] (8.center) to (8d.center);
		
		\draw[thick] (2lu.center) to (2u.center);
		\draw[thick] (2u.center) to (3.center);
		\draw[thick] (3.center) to (4d.center);
		\draw[thick] (4d.center) to (4ld.center);

		\draw[thick] (6dr.center) to (6d.center);
		\draw[thick] (6d.center) to (7.center);
		\draw[thick] (7.center) to (8u.center);
		\draw[thick] (8u.center) to (8ur.center);

		\draw[thick] (4ur.center) to (4r.center);
		\draw[thick] (4r.center) to (5l.center);
		\draw[thick] (5l.center) to (5r.center);
		\draw[thick] (5r.center) to (6l.center);
		\draw[thick] (6l.center) to (6ul.center);
		
		\draw[thick] (2dr.center) to (2r.center);
		\draw[thick] (2r.center) to (1l.center);
		\draw[thick] (1l.center) to (1r.center);
		\draw[thick] (1r.center) to (8l.center);
		\draw[thick] (8l.center) to (8dl.center);
		\coordinate (2md) at ($(2)!0.3!(2d)$);
        \coordinate (2ml) at ($(2)!0.3!(2l)$);
		\coordinate (2hm) at ($(2md)!0.5!(2ml)$);
        
 		\coordinate (4mu) at ($(4)!0.3!(4u)$);
        \coordinate (4ml) at ($(4)!0.3!(4l)$);
		\coordinate (4hm) at ($(4mu)!0.5!(4ml)$);
        
 		\coordinate (6mu) at ($(6)!0.3!(6u)$);
        \coordinate (6mr) at ($(6)!0.3!(6r)$);
		\coordinate (6hm) at ($(6mu)!0.5!(6mr)$);
        
  		\coordinate (8mr) at ($(8)!0.3!(8r)$);
        \coordinate (8md) at ($(8)!0.3!(8d)$);
		\coordinate (8hm) at ($(8mr)!0.5!(8md)$);
        
        \coordinate (5hr) at ($(5r)!0.2!(6l)$);
       	\coordinate (5hl) at ($(5r)!0.15!(5l)$);
		\coordinate (5hm) at ($(5hl)!0.5!(5hr)$);

        \coordinate (3hu) at ($(3)!0.25!(4d)$);
       	\coordinate (3hd) at ($(3)!0.25!(2u)$);
		\coordinate (3hm) at ($(3hu)!0.5!(3hd)$);

        \draw[Maroon,very thick] (2md) to[out=180,in=-90] (2ml);
        \draw[Maroon,very thick] (4ml) to[out=90,in=180] (4mu);
        \draw[Maroon,very thick] (6mu) to[out=0,in=90] (6mr);
        \draw[Maroon,very thick] (8mr) to[out=-90,in=0] (8md);
        \draw[Maroon,very thick] (5hl) to[out=90,in=135] (5hr);
        \draw[Maroon,very thick] (3hu) to[out=-135,in=135] (3hd);

        \draw[Blue,very thick] (4hm) to (-3,1);
        \draw[Blue,very thick] (-3,1) to[out=-45,in=45] (-3,-1);
        \draw[Blue,very thick] (-3,-1) to (2hm);

        \draw[Blue,very thick] (3hm) to (0.5,0);
        \draw[Blue,very thick] (0.5,0) to[out=0,in=-135] (6hm);
       	
       	\draw[Blue,very thick] (8hm) to (3,-1);
        \draw[Blue,very thick] (3,-1) to[out=135,in=-70] (5hm);

        \node[] at (0,-4) {$1$};
        \node[] at (0,4) {$5$};
        \node[] at (-6,0) {$3$};
        \node[] at (6,0) {$7$};
        \node[] at (-6.5,-4.5) {$2$};
        \node[] at (-6.5,4.5) {$4$};
        \node[] at (6.5,4.5) {$6$};
        \node[] at (6.5,-4.5) {$8$};
	\end{tikzpicture}
\end{gathered} \\
& =(X_{2,4}-X_{1,4})(X_{3,6}-X_{1,6})X_{5,8}
\end{aligned}
\quad \quad
\vline
\quad \quad
\begin{aligned}
    &\quad \, \, 
\begin{gathered}
	\begin{tikzpicture}[line width=0.6,scale=0.3,line cap=round,every node/.style={font=\normalsize}]
		\node (1l) at (-2, -1) {};
		\node (1r) at (2, -1) {};
		\node (5l) at (-2, 1) {};
		\node (5r) at (2, 1) {};
		
		\node (2) at (-5, -3) {};
		\node (2d) at (-5, -5) {};
		\node (2l) at (-7, -3) {};
		\node (2r) at (-4, -3) {};
		\node (2dr) at (-4, -5) {};
		\node (2u) at (-5, -2) {};
		\node (2lu) at (-7, -2) {};

		\node (4) at (-5, 3) {};
		\node (4u) at (-5, 5) {};
		\node (4l) at (-7, 3) {};
		\node (4r) at (-4, 3) {};
		\node (4ur) at (-4, 5) {};
		\node (4d) at (-5, 2) {};
		\node (4ld) at (-7, 2) {};

		\node (6) at (5, 3) {};
		\node (6u) at (5, 5) {};
		\node (6r) at (7, 3) {};
		\node (6l) at (4, 3) {};
		\node (6ul) at (4, 5) {};
		\node (6d) at (5, 2) {};
		\node (6dr) at (7, 2) {};
		
		\node (8) at (5, -3) {};
		\node (8r) at (7, -3) {};
		\node (8d) at (5, -5) {};
		\node (8dl) at (4, -5) {};
		\node (8l) at (4, -3) {};
		\node (8ur) at (7, -2) {};
		\node (8u) at (5, -2) {};;
		
		\node (3) at (-3, 0) {};
		\node (7) at (3, 0) {};

		\draw[thick] (2l.center) to (2.center);
		\draw[thick] (2.center) to (2d.center);
		
		\draw[thick] (4u.center) to (4.center);
		\draw[thick] (4.center) to (4l.center);

		\draw[thick] (6u.center) to (6.center);
		\draw[thick] (6.center) to (6r.center);
		
		\draw[thick] (8r.center) to (8.center);
		\draw[thick] (8.center) to (8d.center);
		
		\draw[thick] (2lu.center) to (2u.center);
		\draw[thick] (2u.center) to (3.center);
		\draw[thick] (3.center) to (4d.center);
		\draw[thick] (4d.center) to (4ld.center);

		\draw[thick] (6dr.center) to (6d.center);
		\draw[thick] (6d.center) to (7.center);
		\draw[thick] (7.center) to (8u.center);
		\draw[thick] (8u.center) to (8ur.center);

		\draw[thick] (4ur.center) to (4r.center);
		\draw[thick] (4r.center) to (5l.center);
		\draw[thick] (5l.center) to (5r.center);
		\draw[thick] (5r.center) to (6l.center);
		\draw[thick] (6l.center) to (6ul.center);
		
		\draw[thick] (2dr.center) to (2r.center);
		\draw[thick] (2r.center) to (1l.center);
		\draw[thick] (1l.center) to (1r.center);
		\draw[thick] (1r.center) to (8l.center);
		\draw[thick] (8l.center) to (8dl.center);
		\coordinate (2md) at ($(2)!0.3!(2d)$);
        \coordinate (2ml) at ($(2)!0.3!(2l)$);
		\coordinate (2hm) at ($(2md)!0.5!(2ml)$);
        
 		\coordinate (4mu) at ($(4)!0.3!(4u)$);
        \coordinate (4ml) at ($(4)!0.3!(4l)$);
		\coordinate (4hm) at ($(4mu)!0.5!(4ml)$);
        
 		\coordinate (6mu) at ($(6)!0.3!(6u)$);
        \coordinate (6mr) at ($(6)!0.3!(6r)$);
		\coordinate (6hm) at ($(6mu)!0.5!(6mr)$);
        
  		\coordinate (8mr) at ($(8)!0.3!(8r)$);
        \coordinate (8md) at ($(8)!0.3!(8d)$);
		\coordinate (8hm) at ($(8mr)!0.5!(8md)$);
        
        \coordinate (5hr) at ($(5r)!0.2!(6l)$);
       	\coordinate (5hl) at ($(5r)!0.15!(5l)$);
		\coordinate (5hm) at ($(5hl)!0.5!(5hr)$);

        \coordinate (1hl) at ($(1l)!0.2!(2r)$);
       	\coordinate (1hr) at ($(1l)!0.15!(1r)$);
		\coordinate (1hm) at ($(1hl)!0.5!(1hr)$);

        \draw[Maroon,very thick] (2md) to[out=180,in=-90] (2ml);
        \draw[Maroon,very thick] (4ml) to[out=90,in=180] (4mu);
        \draw[Maroon,very thick] (6mu) to[out=0,in=90] (6mr);
        \draw[Maroon,very thick] (8mr) to[out=-90,in=0] (8md);
        \draw[Maroon,very thick] (5hl) to[out=90,in=135] (5hr);
        \draw[Maroon,very thick] (1hl) to[out=-45,in=-90] (1hr);

        \draw[Blue,very thick] (4hm) to (-3,1);
        \draw[Blue,very thick] (-3,1) to[out=-45,in=100] (1hm);
   		\draw[Blue,very thick] (8hm) to (3,-1);
        \draw[Blue,very thick] (3,-1) to[out=135,in=-70] (5hm);

        \draw[Blue,very thick] (2hm) to[out=45,in=180] (-1,0);
        \draw[Blue,very thick] (-1,0) to (1,0);
        \draw[Blue,very thick] (1,0) to[out=0,in=-135] (6hm);

        \node[] at (0,-4) {$1$};
        \node[] at (0,4) {$5$};
        \node[] at (-6,0) {$3$};
        \node[] at (6,0) {$7$};
        \node[] at (-6.5,-4.5) {$2$};
        \node[] at (-6.5,4.5) {$4$};
        \node[] at (6.5,4.5) {$6$};
        \node[] at (6.5,-4.5) {$8$};
	\end{tikzpicture}
\end{gathered} \\
&=X_{1,4} X_{5,8} ( X_{2,6}-X_{2,5}-X_{1,6})
\end{aligned}
\label{eq:SchannContract}
\end{equation}
It follows from the structure of the contraction picture that, whatever set of handles we choose, the fatgraph will be filled with \textit{non-overlapping blue curves}, as illustrated in the two examples in \eqref{eq:SchannContract} -- again this is closely reproducing the picture for leading singularities from the surface integral, by recasting the polarization contractions as curves on the surface we are led to collections of curves that fully cover the fatgraph!

As we already saw at $3$-points, each blue line contributes, in general, with more than one $X_{i,j}$. In particular, from \eqref{eq:SchannContract}, we see that both contraction pictures generate the monomial $X_{1,4}X_{1,6}X_{5,8}$, however, with opposite signs and so it cancels in the final answer. To understand these cancellations on general grounds let's start by finding a simple way of encoding all monomials generated by a given contraction curve.

\subsection{From contractions to curves and cores}
\label{sec:MonPicture}

So far, we have that given a contraction -- represented by a blue line -- we get a factor of the form \eqref{eq:xdotxToXij}. So now, we would like to understand how we can go from the blue contraction curve inside the fat-graph to the $4$-curves, entering in the $r.h.s.$ of \eqref{eq:xdotxToXij}. In general, we have that a contraction curve takes the form:
\begin{equation}
    \begin{gathered}
    \begin{tikzpicture}[line width=0.6,scale=0.5,line cap=round]
		\node (3) at (0, 0.5) {};
		\node (6) at (0, -0.5) {};
		\node   (7) at (2.75, -0.5) {};
            \node (6m) at ($(6)!0.5!(7)$) {\ldots};
            \node (61) at (0.75, -0.5) {};
            \node (62) at (2, -0.5) {};
		\node   (8) at (2.75, 0.5) {};
		\node   (19) at (4.75, -2) {};
		\node   (20) at (4.75, -0.75) {};
		\node   (21) at (3.75, 0) {};
		\node   (22) at (4.75, 0.75) {};
		\node   (23) at (4.75, 2) {};
		\node   (28) at (4.75, 0) {};
		\node   (31) at (-0.25, 0) {};
		\node   (32) at (3.25, 0) {};
            \node (3u) at (0, 2) {};
            \node (2u) at (-1, 2) {};
            \node (2) at (-1, 0.5) {};
            \node (1) at (-1.75, 0.5) {};
            \node (1p) at (-1.75, -0.5) {};
            \node (mt) at (-2.25, 0.5) {\ldots};
            \node (mb) at (-2.25, -0.5) {\ldots};
            \node (4) at (-2.9, 0.5) {};
            \node (4p) at (-2.9, -0.5) {};
             \node (5p) at (-3.75, -0.5) {};
             \node (5pb) at (-3.75, -2) {};
             \node (9p) at (-4.75, -0.5) {};
             \node (9pb) at (-4.75, -2) {};
             \node (10p) at (-7.5, -0.5) {};
             \node (10) at (-7.5, 0.5) {};
             \node (101) at (-5.4, 0.5) {};
             \node (102) at (-6, 0.5) {\ldots};
              \node (103) at (-6.7, 0.5) {};
		\draw[thick] (3.center) to (8.center);
            \draw[thick] (3.center) to (3u.center);
             \draw[thick] (2.center) to (2u.center);
            \draw[thick]  (6.center) to (61.center);
            \draw[thick]  (62.center) to (7.center);
	    \draw[thick] (7.center) to (19.center);
		\draw[thick](8.center) to (23.center);
	    \draw[thick](21.center) to (22.center);
		\draw[thick](21.center) to (20.center);
            \draw[thick](2.center) to (1.center);
		\draw[thick](1p.center) to (6.center);
            \draw[thick](4p.center) to (5p.center);
             \draw[thick](5p.center) to (5pb.center);
             \draw[thick](9p.center) to (9pb.center);
             \draw[thick](9p.center) to (10p.center);
             \draw[thick](4.center) to (101.center);
             \draw[thick](103.center) to (10.center);
             \node  (12) at (-8.5, 0.) {};
             \node  (12u) at (-9.5, 0.75) {};
             \node  (12b) at (-9.5, -0.75) {};
             \draw[thick](12.center) to (12u.center);
             \draw[thick](12.center) to (12b.center);
             \node  (10u) at (-9.5, 2.) {};
             \node  (10pb) at (-9.5, -2.) {};
             \draw[thick](10.center) to (10u.center);
             \draw[thick](10p.center) to (10pb.center);
             \coordinate (12mb) at ($(12)!0.3!(12b)$);
             \coordinate (12mu) at ($(12)!0.3!(12u)$);
             \coordinate (21mb) at ($(21)!0.3!(20)$);
             \coordinate (21mu) at ($(21)!0.3!(22)$);
             \coordinate (12a) at ($(12mu)!0.5!(12mb)$);
             \coordinate (21a) at ($(21mu)!0.5!(21mb)$);
            
             \coordinate (bd1) at ($(10u)!0.5!(12u)$);
             \coordinate (bu1) at (-10.5, 2.);
             \path (bu1) -- (bd1) node [font=\small, midway, sloped] {$\dots$};

             \coordinate (bpu1) at ($(12b)!0.5!(10pb)$);
             \coordinate (bpd1) at (-10.5, -2.);
             \path (bpu1) -- (bpd1) node [font=\small, midway, sloped] {$\dots$};
            
             \coordinate (du1) at ($(23)!0.5!(22)$);
             \coordinate (dd1) at (5.75, 2);
             \path (du1) -- (dd1) node [font=\small, midway, sloped] {$\dots$};

             \coordinate (dpu1) at ($(20)!0.5!(19)$);
             \coordinate (dpd1) at (5.75, -2);
             \path (dpu1) -- (dpd1) node [font=\small, midway, sloped] {$\dots$};

             \coordinate (up1) at ($(2u)!0.5!(3u)$);
             \coordinate (up2) at (-0.5, 3);
             \path (up1) -- (up2) node [font=\small, midway,sloped] {$\dots$};

             \coordinate (down1) at ($(9pb)!0.5!(5pb)$);
             \coordinate (down2) at (-4.25, -3);
             \path (down1) -- (down2) node [font=\small, midway,sloped] {$\dots$};

            \draw[Maroon,very thick] (12mb) to[out=135,in=-135] (12mu);
            \draw[Maroon,very thick] (21mb) to[out=45,in=-45] (21mu);
            \draw[Blue,very thick] (12a) -- (21a);
            \node[] at (-6.5,-1.25) {$a$};
            \node[] at (-9.5,0) {$b$};
            \node[] at (1.5,1.3) {$c$};
            \node[] at (4.7,0) {$d$};
\end{tikzpicture}
\end{gathered} \quad = X_{b,d}+X_{a,c}-X_{b,c}-X_{a,d}.
\label{eq:contractCurve}
\end{equation}

Quite nicely, we can generate these four different curves by simple operations on the contraction curve: we start at either end of the contraction curve and either turn right and then left until exiting the fatgraph, or just turn left continuously. For example, if at both ends we turn right and then left until exiting, we get curve $X_{b,d}$ -- which is precisely the curve whose \textit{core} agrees with the contraction curve. If instead we turn left at both ends, then we get curve $X_{a,c}$, and the two other mixed cases lead to $X_{b,c}/X_{a,d}$. 

Now to make connection with the picture we got from the surface integral, let's represent each one of these four curves via the respective \textit{cores}. As noted above, the core for curve $X_{b,d}$ precisely agrees with the starting contraction curve, but for the remaining we find:
\begin{align}
    &\begin{gathered}
    \begin{tikzpicture}[line width=0.6,scale=0.5,line cap=round]
		\node (3) at (0, 0.5) {};
		\node (6) at (0, -0.5) {};
		\node   (7) at (2.75, -0.5) {};
            \node (6m) at ($(6)!0.5!(7)$) {\ldots};
            \node (61) at (0.75, -0.5) {};
            \node (62) at (2, -0.5) {};
		\node   (8) at (2.75, 0.5) {};
		\node   (19) at (4.75, -2) {};
		\node   (20) at (4.75, -0.75) {};
		\node   (21) at (3.75, 0) {};
		\node   (22) at (4.75, 0.75) {};
		\node   (23) at (4.75, 2) {};
		\node   (28) at (4.75, 0) {};
		\node   (31) at (-0.25, 0) {};
		\node   (32) at (3.25, 0) {};
            \node (3u) at (0, 2) {};
            \node (2u) at (-1, 2) {};
            \node (2) at (-1, 0.5) {};
            \node (1) at (-1.75, 0.5) {};
            \node (1p) at (-1.75, -0.5) {};
            \node (mt) at (-2.25, 0.5) {\ldots};
            \node (mb) at (-2.25, -0.5) {\ldots};
            \node (4) at (-2.9, 0.5) {};
            \node (4p) at (-2.9, -0.5) {};
             \node (5p) at (-3.75, -0.5) {};
             \node (5pb) at (-3.75, -2) {};
             \node (9p) at (-4.75, -0.5) {};
             \node (9pb) at (-4.75, -2) {};
             \node (10p) at (-7.5, -0.5) {};
             \node (10) at (-7.5, 0.5) {};
             \node (101) at (-5.4, 0.5) {};
             \node (102) at (-6, 0.5) {\ldots};
              \node (103) at (-6.7, 0.5) {};
		\draw[thick] (3.center) to (8.center);
            \draw[thick] (3.center) to (3u.center);
             \draw[thick] (2.center) to (2u.center);
            \draw[thick]  (6.center) to (61.center);
            \draw[thick]  (62.center) to (7.center);
	    \draw[thick] (7.center) to (19.center);
		\draw[thick](8.center) to (23.center);
	    \draw[thick](21.center) to (22.center);
		\draw[thick](21.center) to (20.center);
            \draw[thick](2.center) to (1.center);
		\draw[thick](1p.center) to (6.center);
            \draw[thick](4p.center) to (5p.center);
             \draw[thick](5p.center) to (5pb.center);
             \draw[thick](9p.center) to (9pb.center);
             \draw[thick](9p.center) to (10p.center);
             \draw[thick](4.center) to (101.center);
             \draw[thick](103.center) to (10.center);
             \node  (12) at (-8.5, 0.) {};
             \node  (12u) at (-9.5, 0.75) {};
             \node  (12b) at (-9.5, -0.75) {};
             \draw[thick](12.center) to (12u.center);
             \draw[thick](12.center) to (12b.center);
             \node  (10u) at (-9.5, 2.) {};
             \node  (10pb) at (-9.5, -2.) {};
             \draw[thick](10.center) to (10u.center);
             \draw[thick](10p.center) to (10pb.center);
             \coordinate (12mb) at ($(12)!0.3!(12b)$);
             \coordinate (12mu) at ($(12)!0.3!(12u)$);
             \coordinate (21mb) at ($(21)!0.3!(20)$);
             \coordinate (21mu) at ($(21)!0.3!(22)$);
             \coordinate (12a) at ($(12mu)!0.5!(12mb)$);
             \coordinate (21a) at ($(21mu)!0.5!(21mb)$);
            \draw[Maroon,very thick] (12mb) to[out=135,in=-135] (12mu);
            \draw[Maroon,very thick] (21mb) to[out=45,in=-45] (21mu);
            \draw[Green,very thick] (-4.25, 0) -- (-0.5, 0);
            \node[] at (-6.5,-1.25) {$a$};
            \node[] at (-9.5,0) {$b$};
            \node[] at (1.5,1.3) {$c$};
            \node[] at (4.7,0) {$d$};
\end{tikzpicture}
\end{gathered} \quad =X_{a,c}, \label{eq:curveInt}\\
&\begin{gathered}
    \begin{tikzpicture}[line width=0.6,scale=0.5,line cap=round]
		\node (3) at (0, 0.5) {};
		\node (6) at (0, -0.5) {};
		\node   (7) at (2.75, -0.5) {};
            \node (6m) at ($(6)!0.5!(7)$) {\ldots};
            \node (61) at (0.75, -0.5) {};
            \node (62) at (2, -0.5) {};
		\node   (8) at (2.75, 0.5) {};
		\node   (19) at (4.75, -2) {};
		\node   (20) at (4.75, -0.75) {};
		\node   (21) at (3.75, 0) {};
		\node   (22) at (4.75, 0.75) {};
		\node   (23) at (4.75, 2) {};
		\node   (28) at (4.75, 0) {};
		\node   (31) at (-0.25, 0) {};
		\node   (32) at (3.25, 0) {};
            \node (3u) at (0, 2) {};
            \node (2u) at (-1, 2) {};
            \node (2) at (-1, 0.5) {};
            \node (1) at (-1.75, 0.5) {};
            \node (1p) at (-1.75, -0.5) {};
            \node (mt) at (-2.25, 0.5) {\ldots};
            \node (mb) at (-2.25, -0.5) {\ldots};
            \node (4) at (-2.9, 0.5) {};
            \node (4p) at (-2.9, -0.5) {};
             \node (5p) at (-3.75, -0.5) {};
             \node (5pb) at (-3.75, -2) {};
             \node (9p) at (-4.75, -0.5) {};
             \node (9pb) at (-4.75, -2) {};
             \node (10p) at (-7.5, -0.5) {};
             \node (10) at (-7.5, 0.5) {};
             \node (101) at (-5.4, 0.5) {};
             \node (102) at (-6, 0.5) {\ldots};
              \node (103) at (-6.7, 0.5) {};
		\draw[thick] (3.center) to (8.center);
            \draw[thick] (3.center) to (3u.center);
             \draw[thick] (2.center) to (2u.center);
            \draw[thick]  (6.center) to (61.center);
            \draw[thick]  (62.center) to (7.center);
	    \draw[thick] (7.center) to (19.center);
		\draw[thick](8.center) to (23.center);
	    \draw[thick](21.center) to (22.center);
		\draw[thick](21.center) to (20.center);
            \draw[thick](2.center) to (1.center);
		\draw[thick](1p.center) to (6.center);
            \draw[thick](4p.center) to (5p.center);
             \draw[thick](5p.center) to (5pb.center);
             \draw[thick](9p.center) to (9pb.center);
             \draw[thick](9p.center) to (10p.center);
             \draw[thick](4.center) to (101.center);
             \draw[thick](103.center) to (10.center);
             \node  (12) at (-8.5, 0.) {};
             \node  (12u) at (-9.5, 0.75) {};
             \node  (12b) at (-9.5, -0.75) {};
             \draw[thick](12.center) to (12u.center);
             \draw[thick](12.center) to (12b.center);
             \node  (10u) at (-9.5, 2.) {};
             \node  (10pb) at (-9.5, -2.) {};
             \draw[thick](10.center) to (10u.center);
             \draw[thick](10p.center) to (10pb.center);
             \coordinate (12mb) at ($(12)!0.3!(12b)$);
             \coordinate (12mu) at ($(12)!0.3!(12u)$);
             \coordinate (21mb) at ($(21)!0.3!(20)$);
             \coordinate (21mu) at ($(21)!0.3!(22)$);
             \coordinate (12a) at ($(12mu)!0.5!(12mb)$);
             \coordinate (21a) at ($(21mu)!0.5!(21mb)$);
            \draw[Maroon,very thick] (12mb) to[out=135,in=-135] (12mu);
            \draw[Maroon,very thick] (21mb) to[out=45,in=-45] (21mu);
            \draw[Green,very thick] (-8, 0) -- (-0.5, 0);
            \node[] at (-6.5,-1.25) {$a$};
            \node[] at (-9.5,0) {$b$};
            \node[] at (1.5,1.3) {$c$};
            \node[] at (4.7,0) {$d$};
\end{tikzpicture}
\end{gathered} \quad =X_{b,c}, \label{eq:curveL}\\
&\begin{gathered}
    \begin{tikzpicture}[line width=0.6,scale=0.5,line cap=round]
		\node (3) at (0, 0.5) {};
		\node (6) at (0, -0.5) {};
		\node   (7) at (2.75, -0.5) {};
            \node (6m) at ($(6)!0.5!(7)$) {\ldots};
            \node (61) at (0.75, -0.5) {};
            \node (62) at (2, -0.5) {};
		\node   (8) at (2.75, 0.5) {};
		\node   (19) at (4.75, -2) {};
		\node   (20) at (4.75, -0.75) {};
		\node   (21) at (3.75, 0) {};
		\node   (22) at (4.75, 0.75) {};
		\node   (23) at (4.75, 2) {};
		\node   (28) at (4.75, 0) {};
		\node   (31) at (-0.25, 0) {};
		\node   (32) at (3.25, 0) {};
            \node (3u) at (0, 2) {};
            \node (2u) at (-1, 2) {};
            \node (2) at (-1, 0.5) {};
            \node (1) at (-1.75, 0.5) {};
            \node (1p) at (-1.75, -0.5) {};
            \node (mt) at (-2.25, 0.5) {\ldots};
            \node (mb) at (-2.25, -0.5) {\ldots};
            \node (4) at (-2.9, 0.5) {};
            \node (4p) at (-2.9, -0.5) {};
             \node (5p) at (-3.75, -0.5) {};
             \node (5pb) at (-3.75, -2) {};
             \node (9p) at (-4.75, -0.5) {};
             \node (9pb) at (-4.75, -2) {};
             \node (10p) at (-7.5, -0.5) {};
             \node (10) at (-7.5, 0.5) {};
             \node (101) at (-5.4, 0.5) {};
             \node (102) at (-6, 0.5) {\ldots};
              \node (103) at (-6.7, 0.5) {};
		\draw[thick] (3.center) to (8.center);
            \draw[thick] (3.center) to (3u.center);
             \draw[thick] (2.center) to (2u.center);
            \draw[thick]  (6.center) to (61.center);
            \draw[thick]  (62.center) to (7.center);
	    \draw[thick] (7.center) to (19.center);
		\draw[thick](8.center) to (23.center);
	    \draw[thick](21.center) to (22.center);
		\draw[thick](21.center) to (20.center);
            \draw[thick](2.center) to (1.center);
		\draw[thick](1p.center) to (6.center);
            \draw[thick](4p.center) to (5p.center);
             \draw[thick](5p.center) to (5pb.center);
             \draw[thick](9p.center) to (9pb.center);
             \draw[thick](9p.center) to (10p.center);
             \draw[thick](4.center) to (101.center);
             \draw[thick](103.center) to (10.center);
             \node  (12) at (-8.5, 0.) {};
             \node  (12u) at (-9.5, 0.75) {};
             \node  (12b) at (-9.5, -0.75) {};
             \draw[thick](12.center) to (12u.center);
             \draw[thick](12.center) to (12b.center);
             \node  (10u) at (-9.5, 2.) {};
             \node  (10pb) at (-9.5, -2.) {};
             \draw[thick](10.center) to (10u.center);
             \draw[thick](10p.center) to (10pb.center);
             \coordinate (12mb) at ($(12)!0.3!(12b)$);
             \coordinate (12mu) at ($(12)!0.3!(12u)$);
             \coordinate (21mb) at ($(21)!0.3!(20)$);
             \coordinate (21mu) at ($(21)!0.3!(22)$);
             \coordinate (12a) at ($(12mu)!0.5!(12mb)$);
             \coordinate (21a) at ($(21mu)!0.5!(21mb)$);
            \draw[Maroon,very thick] (12mb) to[out=135,in=-135] (12mu);
            \draw[Maroon,very thick] (21mb) to[out=45,in=-45] (21mu);
            \draw[Green,very thick] (-4.25, 0) -- (3.25, 0);
            \node[] at (-6.5,-1.25) {$a$};
            \node[] at (-9.5,0) {$b$};
            \node[] at (1.5,1.3) {$c$};
            \node[] at (4.7,0) {$d$};
\end{tikzpicture}
\end{gathered} \quad =X_{a,d},\label{eq:curveR}
\end{align}
so we have that each core is a \textit{subcurve} of the original contraction curve! Of course, not any subcurve of the contraction appears in the answer, as we get at most the four terms in \eqref{eq:contractCurve}. Starting from the contraction curve, $\mathcal{C}$, we can identify the subcurves associated to cores of $X_{i,j}$ produce by this contraction as follows:
\begin{enumerate}
    \item Start from one end of the contraction curve and identify the intersection at which it makes one right turn followed by only left turns until the end of the curve, and consider the subcurve that goes from the starting point until this intersection. 
    \item By doing this operation we generate two subcurves: one coming from starting at the left end of $\mathcal{C}$, we call $C_L$; and another from starting at the right end of $\mathcal{C}$, $C_R$. For example in \eqref{eq:contractCurve},  $C_L$ is the curve in \eqref{eq:curveL} and $C_R$ that in \eqref{eq:curveR}.
    \item In addition to these two subcurves, from the contraction we also get the cores corresponding to the full contraction curve, which we call $C^\cup$, together with the subcurve correspoding to the intersection, $C^\cap = C_L \cap C_R $, which in the example above is given by the core in \eqref{eq:curveInt}.
\end{enumerate}

Note that in certain cases the intersection between $C_L$ and $C_R$ is empty (see example below), and therefore we have that $\mathcal{C}$ only gives rise to $C^\cup$, $C_L$ and $C_R$, separately.

Finally, from \eqref{eq:contractCurve}, we have that the cores $C^\cup$ and $C^\cap$ contribute to the LS with a plus sign ($+$), while $C_L$ and $C_R$ come with a minus sign ($-$). As we will see momentarily, the fact that this simple picture lets us read off the cores directly by identifying subcurves of the contraction curve, will be useful to go in the reverse direction: given $X_{i,j}$ we can determine which contraction curves generate it, as well as the respective sign.

\paragraph{Example} Let us look at the contraction curve from regions $2$ to $6$, entering the contraction on the right in \eqref{eq:SchannContract}. In this case, following the procedure outlined above we get that $C_L$ is the core associated to curve $X_{2,5}$, while $C_R$ is the core of curve $X_{1,6}$:
\begin{equation}
	\begin{gathered}
		\begin{tikzpicture}[line width=0.6,scale=0.25,line cap=round,every node/.style={font=\normalsize}]
		\node (1l) at (-2, -1) {};
		\node (1r) at (2, -1) {};
		\node (5l) at (-2, 1) {};
		\node (5r) at (2, 1) {};
		
		\node (2) at (-5, -3) {};
		\node (2d) at (-5, -5) {};
		\node (2l) at (-7, -3) {};
		\node (2r) at (-4, -3) {};
		\node (2dr) at (-4, -5) {};
		\node (2u) at (-5, -2) {};
		\node (2lu) at (-7, -2) {};

		\node (4) at (-5, 3) {};
		\node (4u) at (-5, 5) {};
		\node (4l) at (-7, 3) {};
		\node (4r) at (-4, 3) {};
		\node (4ur) at (-4, 5) {};
		\node (4d) at (-5, 2) {};
		\node (4ld) at (-7, 2) {};

		\node (6) at (5, 3) {};
		\node (6u) at (5, 5) {};
		\node (6r) at (7, 3) {};
		\node (6l) at (4, 3) {};
		\node (6ul) at (4, 5) {};
		\node (6d) at (5, 2) {};
		\node (6dr) at (7, 2) {};
		
		\node (8) at (5, -3) {};
		\node (8r) at (7, -3) {};
		\node (8d) at (5, -5) {};
		\node (8dl) at (4, -5) {};
		\node (8l) at (4, -3) {};
		\node (8ur) at (7, -2) {};
		\node (8u) at (5, -2) {};;
		
		\node (3) at (-3, 0) {};
		\node (7) at (3, 0) {};

		\draw[thick] (2l.center) to (2.center);
		\draw[thick] (2.center) to (2d.center);
		
		\draw[thick] (4u.center) to (4.center);
		\draw[thick] (4.center) to (4l.center);

		\draw[thick] (6u.center) to (6.center);
		\draw[thick] (6.center) to (6r.center);
		
		\draw[thick] (8r.center) to (8.center);
		\draw[thick] (8.center) to (8d.center);
		
		\draw[thick] (2lu.center) to (2u.center);
		\draw[thick] (2u.center) to (3.center);
		\draw[thick] (3.center) to (4d.center);
		\draw[thick] (4d.center) to (4ld.center);

		\draw[thick] (6dr.center) to (6d.center);
		\draw[thick] (6d.center) to (7.center);
		\draw[thick] (7.center) to (8u.center);
		\draw[thick] (8u.center) to (8ur.center);

		\draw[thick] (4ur.center) to (4r.center);
		\draw[thick] (4r.center) to (5l.center);
		\draw[thick] (5l.center) to (5r.center);
		\draw[thick] (5r.center) to (6l.center);
		\draw[thick] (6l.center) to (6ul.center);
		
		\draw[thick] (2dr.center) to (2r.center);
		\draw[thick] (2r.center) to (1l.center);
		\draw[thick] (1l.center) to (1r.center);
		\draw[thick] (1r.center) to (8l.center);
		\draw[thick] (8l.center) to (8dl.center);
		\coordinate (2md) at ($(2)!0.3!(2d)$);
        \coordinate (2ml) at ($(2)!0.3!(2l)$);
		\coordinate (2hm) at ($(2md)!0.5!(2ml)$);
        
 		\coordinate (4mu) at ($(4)!0.3!(4u)$);
        \coordinate (4ml) at ($(4)!0.3!(4l)$);
		\coordinate (4hm) at ($(4mu)!0.5!(4ml)$);
        
 		\coordinate (6mu) at ($(6)!0.3!(6u)$);
        \coordinate (6mr) at ($(6)!0.3!(6r)$);
		\coordinate (6hm) at ($(6mu)!0.5!(6mr)$);
        
  		\coordinate (8mr) at ($(8)!0.3!(8r)$);
        \coordinate (8md) at ($(8)!0.3!(8d)$);
		\coordinate (8hm) at ($(8mr)!0.5!(8md)$);
        
        \coordinate (5hr) at ($(5r)!0.2!(6l)$);
       	\coordinate (5hl) at ($(5r)!0.15!(5l)$);
		\coordinate (5hm) at ($(5hl)!0.5!(5hr)$);

        \coordinate (1hl) at ($(1l)!0.2!(2r)$);
       	\coordinate (1hr) at ($(1l)!0.15!(1r)$);
		\coordinate (1hm) at ($(1hl)!0.5!(1hr)$);

        \draw[Maroon,very thick] (2md) to[out=180,in=-90] (2ml);
        \draw[Maroon,very thick] (4ml) to[out=90,in=180] (4mu);
        \draw[Maroon,very thick] (6mu) to[out=0,in=90] (6mr);
        \draw[Maroon,very thick] (8mr) to[out=-90,in=0] (8md);
        \draw[Maroon,very thick] (5hl) to[out=90,in=135] (5hr);
        \draw[Maroon,very thick] (1hl) to[out=-45,in=-90] (1hr);

        \draw[Blue,very thick] (2hm) to[out=45,in=180] (-1,0);
        \draw[Blue,very thick] (-1,0) to (1,0);
        \draw[Blue,very thick] (1,0) to[out=0,in=-135] (6hm);

        \node[] at (0,-4) {$1$};
        \node[] at (0,4) {$5$};
        \node[] at (-6,0) {$3$};
        \node[] at (6,0) {$7$};
        \node[] at (-6.5,-4.5) {$2$};
        \node[] at (-6.5,4.5) {$4$};
        \node[] at (6.5,4.5) {$6$};
        \node[] at (6.5,-4.5) {$8$};
	\end{tikzpicture}
	\end{gathered}
\quad  \longrightarrow \quad \begin{aligned}
	&\begin{gathered}
		\begin{tikzpicture}[line width=0.6,scale=0.2,line cap=round,every node/.style={font=\footnotesize}]
		\node (1l) at (-2, -1) {};
		\node (1r) at (2, -1) {};
		\node (5l) at (-2, 1) {};
		\node (5r) at (2, 1) {};
		
		\node (2) at (-5, -3) {};
		\node (2d) at (-5, -5) {};
		\node (2l) at (-7, -3) {};
		\node (2r) at (-4, -3) {};
		\node (2dr) at (-4, -5) {};
		\node (2u) at (-5, -2) {};
		\node (2lu) at (-7, -2) {};

		\node (4) at (-5, 3) {};
		\node (4u) at (-5, 5) {};
		\node (4l) at (-7, 3) {};
		\node (4r) at (-4, 3) {};
		\node (4ur) at (-4, 5) {};
		\node (4d) at (-5, 2) {};
		\node (4ld) at (-7, 2) {};

		\node (6) at (5, 3) {};
		\node (6u) at (5, 5) {};
		\node (6r) at (7, 3) {};
		\node (6l) at (4, 3) {};
		\node (6ul) at (4, 5) {};
		\node (6d) at (5, 2) {};
		\node (6dr) at (7, 2) {};
		
		\node (8) at (5, -3) {};
		\node (8r) at (7, -3) {};
		\node (8d) at (5, -5) {};
		\node (8dl) at (4, -5) {};
		\node (8l) at (4, -3) {};
		\node (8ur) at (7, -2) {};
		\node (8u) at (5, -2) {};;
		
		\node (3) at (-3, 0) {};
		\node (7) at (3, 0) {};

		\draw[thick] (2l.center) to (2.center);
		\draw[thick] (2.center) to (2d.center);
		
		\draw[thick] (4u.center) to (4.center);
		\draw[thick] (4.center) to (4l.center);

		\draw[thick] (6u.center) to (6.center);
		\draw[thick] (6.center) to (6r.center);
		
		\draw[thick] (8r.center) to (8.center);
		\draw[thick] (8.center) to (8d.center);
		
		\draw[thick] (2lu.center) to (2u.center);
		\draw[thick] (2u.center) to (3.center);
		\draw[thick] (3.center) to (4d.center);
		\draw[thick] (4d.center) to (4ld.center);

		\draw[thick] (6dr.center) to (6d.center);
		\draw[thick] (6d.center) to (7.center);
		\draw[thick] (7.center) to (8u.center);
		\draw[thick] (8u.center) to (8ur.center);

		\draw[thick] (4ur.center) to (4r.center);
		\draw[thick] (4r.center) to (5l.center);
		\draw[thick] (5l.center) to (5r.center);
		\draw[thick] (5r.center) to (6l.center);
		\draw[thick] (6l.center) to (6ul.center);
		
		\draw[thick] (2dr.center) to (2r.center);
		\draw[thick] (2r.center) to (1l.center);
		\draw[thick] (1l.center) to (1r.center);
		\draw[thick] (1r.center) to (8l.center);
		\draw[thick] (8l.center) to (8dl.center);
		\coordinate (2md) at ($(2)!0.3!(2d)$);
        \coordinate (2ml) at ($(2)!0.3!(2l)$);
		\coordinate (2hm) at ($(2md)!0.5!(2ml)$);
        
 		\coordinate (4mu) at ($(4)!0.3!(4u)$);
        \coordinate (4ml) at ($(4)!0.3!(4l)$);
		\coordinate (4hm) at ($(4mu)!0.5!(4ml)$);
        
 		\coordinate (6mu) at ($(6)!0.3!(6u)$);
        \coordinate (6mr) at ($(6)!0.3!(6r)$);
		\coordinate (6hm) at ($(6mu)!0.5!(6mr)$);
        
  		\coordinate (8mr) at ($(8)!0.3!(8r)$);
        \coordinate (8md) at ($(8)!0.3!(8d)$);
		\coordinate (8hm) at ($(8mr)!0.5!(8md)$);
        
        \coordinate (5hl) at ($(5l)!0.2!(4r)$);
       	\coordinate (5hr) at ($(5l)!0.15!(5r)$);
		\coordinate (5hm) at ($(5hl)!0.5!(5hr)$);

        \coordinate (1hr) at ($(1r)!0.2!(8l)$);
       	\coordinate (1hl) at ($(1r)!0.15!(1l)$);
		\coordinate (1hm) at ($(1hl)!0.5!(1hr)$);


        \draw[ForestGreen,very thick] (-4.5,-2.5) to (-2,0);

        \node[] at (0,-4) {$1$};
        \node[] at (0,4) {$5$};
        \node[] at (-6,0) {$3$};
        \node[] at (6,0) {$7$};
        \node[] at (-6.5,-4.5) {$2$};
        \node[] at (-6.5,4.5) {$4$};
        \node[] at (6.5,4.5) {$6$};
        \node[] at (6.5,-4.5) {$8$};
	\end{tikzpicture}
	\end{gathered} \\[-0.5em]
    & \quad \quad \quad \, \,\, C_L
\end{aligned}
\quad , \quad 
 \begin{aligned}
&\begin{gathered}
		\begin{tikzpicture}[line width=0.6,scale=0.2,line cap=round,every node/.style={font=\footnotesize}]
		\node (1l) at (-2, -1) {};
		\node (1r) at (2, -1) {};
		\node (5l) at (-2, 1) {};
		\node (5r) at (2, 1) {};
		
		\node (2) at (-5, -3) {};
		\node (2d) at (-5, -5) {};
		\node (2l) at (-7, -3) {};
		\node (2r) at (-4, -3) {};
		\node (2dr) at (-4, -5) {};
		\node (2u) at (-5, -2) {};
		\node (2lu) at (-7, -2) {};

		\node (4) at (-5, 3) {};
		\node (4u) at (-5, 5) {};
		\node (4l) at (-7, 3) {};
		\node (4r) at (-4, 3) {};
		\node (4ur) at (-4, 5) {};
		\node (4d) at (-5, 2) {};
		\node (4ld) at (-7, 2) {};

		\node (6) at (5, 3) {};
		\node (6u) at (5, 5) {};
		\node (6r) at (7, 3) {};
		\node (6l) at (4, 3) {};
		\node (6ul) at (4, 5) {};
		\node (6d) at (5, 2) {};
		\node (6dr) at (7, 2) {};
		
		\node (8) at (5, -3) {};
		\node (8r) at (7, -3) {};
		\node (8d) at (5, -5) {};
		\node (8dl) at (4, -5) {};
		\node (8l) at (4, -3) {};
		\node (8ur) at (7, -2) {};
		\node (8u) at (5, -2) {};;
		
		\node (3) at (-3, 0) {};
		\node (7) at (3, 0) {};

		\draw[thick] (2l.center) to (2.center);
		\draw[thick] (2.center) to (2d.center);
		
		\draw[thick] (4u.center) to (4.center);
		\draw[thick] (4.center) to (4l.center);

		\draw[thick] (6u.center) to (6.center);
		\draw[thick] (6.center) to (6r.center);
		
		\draw[thick] (8r.center) to (8.center);
		\draw[thick] (8.center) to (8d.center);
		
		\draw[thick] (2lu.center) to (2u.center);
		\draw[thick] (2u.center) to (3.center);
		\draw[thick] (3.center) to (4d.center);
		\draw[thick] (4d.center) to (4ld.center);

		\draw[thick] (6dr.center) to (6d.center);
		\draw[thick] (6d.center) to (7.center);
		\draw[thick] (7.center) to (8u.center);
		\draw[thick] (8u.center) to (8ur.center);

		\draw[thick] (4ur.center) to (4r.center);
		\draw[thick] (4r.center) to (5l.center);
		\draw[thick] (5l.center) to (5r.center);
		\draw[thick] (5r.center) to (6l.center);
		\draw[thick] (6l.center) to (6ul.center);
		
		\draw[thick] (2dr.center) to (2r.center);
		\draw[thick] (2r.center) to (1l.center);
		\draw[thick] (1l.center) to (1r.center);
		\draw[thick] (1r.center) to (8l.center);
		\draw[thick] (8l.center) to (8dl.center);
		\coordinate (2md) at ($(2)!0.3!(2d)$);
        \coordinate (2ml) at ($(2)!0.3!(2l)$);
		\coordinate (2hm) at ($(2md)!0.5!(2ml)$);
        
 		\coordinate (4mu) at ($(4)!0.3!(4u)$);
        \coordinate (4ml) at ($(4)!0.3!(4l)$);
		\coordinate (4hm) at ($(4mu)!0.5!(4ml)$);
        
 		\coordinate (6mu) at ($(6)!0.3!(6u)$);
        \coordinate (6mr) at ($(6)!0.3!(6r)$);
		\coordinate (6hm) at ($(6mu)!0.5!(6mr)$);
        
  		\coordinate (8mr) at ($(8)!0.3!(8r)$);
        \coordinate (8md) at ($(8)!0.3!(8d)$);
		\coordinate (8hm) at ($(8mr)!0.5!(8md)$);
        
        \coordinate (5hl) at ($(5l)!0.2!(4r)$);
       	\coordinate (5hr) at ($(5l)!0.15!(5r)$);
		\coordinate (5hm) at ($(5hl)!0.5!(5hr)$);

        \coordinate (1hr) at ($(1r)!0.2!(8l)$);
       	\coordinate (1hl) at ($(1r)!0.15!(1l)$);
		\coordinate (1hm) at ($(1hl)!0.5!(1hr)$);


        \draw[ForestGreen,very thick] (4.5,2.5) to (2,0);

        \node[] at (0,-4) {$1$};
        \node[] at (0,4) {$5$};
        \node[] at (-6,0) {$3$};
        \node[] at (6,0) {$7$};
        \node[] at (-6.5,-4.5) {$2$};
        \node[] at (-6.5,4.5) {$4$};
        \node[] at (6.5,4.5) {$6$};
        \node[] at (6.5,-4.5) {$8$};
	\end{tikzpicture}
	\end{gathered}\\[-0.5em]
    & \quad \quad \quad \, \,\, C_R
\end{aligned}
\end{equation}
and therefore, in this case, $C_L \cap C_R$ is empty. Hence, we have that this contraction only produces the terms $X_{2,6}$, corresponding to the full contraction curve, $X_{2,5}$ and $X_{1,6}$.

\subsection{From cores to contractions}

From the rule described above we have that given $X_{i,j}$ is generated by some contraction curve $\mathcal{C}$ \textit{iff} the core of $X_{i,j}$ is one of the four special subcurves of $\mathcal{C}$. Proceeding backwards we also have that given an $X_{i,j}$, we can tell all possible contractions that give rise to it by \textit{extending} its core according with the moves described in the subcurve-operation. 

Practically, starting with a core of a given curve, $X_{i,j}$, we can extend it into a full contraction curve by starting at each end of the core and turning right once and then left as many times as we want -- we call these $L/R$ \textit{extensions} of the core, and we will graphically represent these extensions by dashed lines. The number of left turns we make on a given extension determines which contraction curve we end up with, but for any extension we obtain a contraction curve that generates $X_{i,j}$.

From the sign rule derived in the previous section, we have that the sign of $X_{i,j}$ in a given contraction curve is determined by how many extensions ($L$, $R$ or both), we have to perform to go from the core of $X_{i,j}$ to the contraction curve under consideration. So in summary, we have that $X_{i,j}$ always comes with a factor of $(-1)^{N^\mathcal{C}_e}$ with $N^\mathcal{C}_e$ being the number extensions needed to get $\mathcal{C}$: $N^\mathcal{C}_e=0$ if no extensions, $N^\mathcal{C}_e=1$ if we extend on L or R and $N^\mathcal{C}_e=2$ if we extend on both L and R. 

For example, consider the  core $X_{3,7}$ in the $s$-channel LS$_4$. In this case, there are four different ways of extending the core into contractions that generate $X_{3,7}$, which are 
\begin{equation}
\begin{aligned}
	 &\begin{gathered}
	 	\begin{tikzpicture}[line width=0.6,scale=0.2,line cap=round,every node/.style={font=\footnotesize}]
		\node (1l) at (-2, -1) {};
		\node (1r) at (2, -1) {};
		\node (5l) at (-2, 1) {};
		\node (5r) at (2, 1) {};
		
		\node (2) at (-5, -3) {};
		\node (2d) at (-5, -5) {};
		\node (2l) at (-7, -3) {};
		\node (2r) at (-4, -3) {};
		\node (2dr) at (-4, -5) {};
		\node (2u) at (-5, -2) {};
		\node (2lu) at (-7, -2) {};

		\node (4) at (-5, 3) {};
		\node (4u) at (-5, 5) {};
		\node (4l) at (-7, 3) {};
		\node (4r) at (-4, 3) {};
		\node (4ur) at (-4, 5) {};
		\node (4d) at (-5, 2) {};
		\node (4ld) at (-7, 2) {};

		\node (6) at (5, 3) {};
		\node (6u) at (5, 5) {};
		\node (6r) at (7, 3) {};
		\node (6l) at (4, 3) {};
		\node (6ul) at (4, 5) {};
		\node (6d) at (5, 2) {};
		\node (6dr) at (7, 2) {};
		
		\node (8) at (5, -3) {};
		\node (8r) at (7, -3) {};
		\node (8d) at (5, -5) {};
		\node (8dl) at (4, -5) {};
		\node (8l) at (4, -3) {};
		\node (8ur) at (7, -2) {};
		\node (8u) at (5, -2) {};;
		
		\node (3) at (-3, 0) {};
		\node (7) at (3, 0) {};

		\draw[thick] (2l.center) to (2.center);
		\draw[thick] (2.center) to (2d.center);
		
		\draw[thick] (4u.center) to (4.center);
		\draw[thick] (4.center) to (4l.center);

		\draw[thick] (6u.center) to (6.center);
		\draw[thick] (6.center) to (6r.center);
		
		\draw[thick] (8r.center) to (8.center);
		\draw[thick] (8.center) to (8d.center);
		
		\draw[thick] (2lu.center) to (2u.center);
		\draw[thick] (2u.center) to (3.center);
		\draw[thick] (3.center) to (4d.center);
		\draw[thick] (4d.center) to (4ld.center);

		\draw[thick] (6dr.center) to (6d.center);
		\draw[thick] (6d.center) to (7.center);
		\draw[thick] (7.center) to (8u.center);
		\draw[thick] (8u.center) to (8ur.center);

		\draw[thick] (4ur.center) to (4r.center);
		\draw[thick] (4r.center) to (5l.center);
		\draw[thick] (5l.center) to (5r.center);
		\draw[thick] (5r.center) to (6l.center);
		\draw[thick] (6l.center) to (6ul.center);
		
		\draw[thick] (2dr.center) to (2r.center);
		\draw[thick] (2r.center) to (1l.center);
		\draw[thick] (1l.center) to (1r.center);
		\draw[thick] (1r.center) to (8l.center);
		\draw[thick] (8l.center) to (8dl.center);
		\coordinate (2md) at ($(2)!0.3!(2d)$);
        \coordinate (2ml) at ($(2)!0.3!(2l)$);
		\coordinate (2hm) at ($(2md)!0.5!(2ml)$);
        
 		\coordinate (4mu) at ($(4)!0.3!(4u)$);
        \coordinate (4ml) at ($(4)!0.3!(4l)$);
		\coordinate (4hm) at ($(4mu)!0.5!(4ml)$);
        
 		\coordinate (6mu) at ($(6)!0.3!(6u)$);
        \coordinate (6mr) at ($(6)!0.3!(6r)$);
		\coordinate (6hm) at ($(6mu)!0.5!(6mr)$);
        
  		\coordinate (8mr) at ($(8)!0.3!(8r)$);
        \coordinate (8md) at ($(8)!0.3!(8d)$);
		\coordinate (8hm) at ($(8mr)!0.5!(8md)$);
        
        \coordinate (5hr) at ($(5r)!0.2!(6l)$);
       	\coordinate (5hl) at ($(5r)!0.15!(5l)$);
		\coordinate (5hm) at ($(5hl)!0.5!(5hr)$);

        \coordinate (3hu) at ($(3)!0.25!(4d)$);
       	\coordinate (3hd) at ($(3)!0.25!(2u)$);
		\coordinate (3hm) at ($(3hu)!0.5!(3hd)$);

		\draw[ForestGreen, very thick] (-2,0) to (2,0);
        \node[] at (0,-4) {$1$};
        \node[] at (0,4) {$5$};
        \node[] at (-6,0) {$3$};
        \node[] at (6,0) {$7$};
        \node[] at (-6.5,-4.5) {$2$};
        \node[] at (-6.5,4.5) {$4$};
        \node[] at (6.5,4.5) {$6$};
        \node[] at (6.5,-4.5) {$8$};
	\end{tikzpicture}
	 \end{gathered}\\[-0.5em]
     & \quad \quad  \quad \, \,  X_{3,7}       
\end{aligned}
\quad \vline \quad 
\begin{aligned}
	&\begin{gathered}
	\begin{tikzpicture}[line width=0.6,scale=0.2,line cap=round,every node/.style={font=\footnotesize}]
		\node (1l) at (-2, -1) {};
		\node (1r) at (2, -1) {};
		\node (5l) at (-2, 1) {};
		\node (5r) at (2, 1) {};
		
		\node (2) at (-5, -3) {};
		\node (2d) at (-5, -5) {};
		\node (2l) at (-7, -3) {};
		\node (2r) at (-4, -3) {};
		\node (2dr) at (-4, -5) {};
		\node (2u) at (-5, -2) {};
		\node (2lu) at (-7, -2) {};

		\node (4) at (-5, 3) {};
		\node (4u) at (-5, 5) {};
		\node (4l) at (-7, 3) {};
		\node (4r) at (-4, 3) {};
		\node (4ur) at (-4, 5) {};
		\node (4d) at (-5, 2) {};
		\node (4ld) at (-7, 2) {};

		\node (6) at (5, 3) {};
		\node (6u) at (5, 5) {};
		\node (6r) at (7, 3) {};
		\node (6l) at (4, 3) {};
		\node (6ul) at (4, 5) {};
		\node (6d) at (5, 2) {};
		\node (6dr) at (7, 2) {};
		
		\node (8) at (5, -3) {};
		\node (8r) at (7, -3) {};
		\node (8d) at (5, -5) {};
		\node (8dl) at (4, -5) {};
		\node (8l) at (4, -3) {};
		\node (8ur) at (7, -2) {};
		\node (8u) at (5, -2) {};;
		
		\node (3) at (-3, 0) {};
		\node (7) at (3, 0) {};

		\draw[thick] (2l.center) to (2.center);
		\draw[thick] (2.center) to (2d.center);
		
		\draw[thick] (4u.center) to (4.center);
		\draw[thick] (4.center) to (4l.center);

		\draw[thick] (6u.center) to (6.center);
		\draw[thick] (6.center) to (6r.center);
		
		\draw[thick] (8r.center) to (8.center);
		\draw[thick] (8.center) to (8d.center);
		
		\draw[thick] (2lu.center) to (2u.center);
		\draw[thick] (2u.center) to (3.center);
		\draw[thick] (3.center) to (4d.center);
		\draw[thick] (4d.center) to (4ld.center);

		\draw[thick] (6dr.center) to (6d.center);
		\draw[thick] (6d.center) to (7.center);
		\draw[thick] (7.center) to (8u.center);
		\draw[thick] (8u.center) to (8ur.center);

		\draw[thick] (4ur.center) to (4r.center);
		\draw[thick] (4r.center) to (5l.center);
		\draw[thick] (5l.center) to (5r.center);
		\draw[thick] (5r.center) to (6l.center);
		\draw[thick] (6l.center) to (6ul.center);
		
		\draw[thick] (2dr.center) to (2r.center);
		\draw[thick] (2r.center) to (1l.center);
		\draw[thick] (1l.center) to (1r.center);
		\draw[thick] (1r.center) to (8l.center);
		\draw[thick] (8l.center) to (8dl.center);
		\coordinate (2md) at ($(2)!0.3!(2d)$);
        \coordinate (2ml) at ($(2)!0.3!(2l)$);
		\coordinate (2hm) at ($(2md)!0.5!(2ml)$);
        
 		\coordinate (4mu) at ($(4)!0.3!(4u)$);
        \coordinate (4ml) at ($(4)!0.3!(4l)$);
		\coordinate (4hm) at ($(4mu)!0.5!(4ml)$);
        
 		\coordinate (6mu) at ($(6)!0.3!(6u)$);
        \coordinate (6mr) at ($(6)!0.3!(6r)$);
		\coordinate (6hm) at ($(6mu)!0.5!(6mr)$);
        
  		\coordinate (8mr) at ($(8)!0.3!(8r)$);
        \coordinate (8md) at ($(8)!0.3!(8d)$);
		\coordinate (8hm) at ($(8mr)!0.5!(8md)$);
        
        \coordinate (5hr) at ($(5r)!0.2!(6l)$);
       	\coordinate (5hl) at ($(5r)!0.15!(5l)$);
		\coordinate (5hm) at ($(5hl)!0.5!(5hr)$);

        \coordinate (3hu) at ($(3)!0.25!(4d)$);
       	\coordinate (3hd) at ($(3)!0.25!(2u)$);
		\coordinate (3hm) at ($(3hu)!0.5!(3hd)$);
		\draw[ForestGreen, very thick] (-2,0) to (2,0);
		\draw[ForestGreen, very thick,dashed] (-2,0) to (-4.5,2.5);
        \node[] at (0,-4) {$1$};
        \node[] at (0,4) {$5$};
        \node[] at (-6,0) {$3$};
        \node[] at (6,0) {$7$};
        \node[] at (-6.5,-4.5) {$2$};
        \node[] at (-6.5,4.5) {$4$};
        \node[] at (6.5,4.5) {$6$};
        \node[] at (6.5,-4.5) {$8$};
	\end{tikzpicture}
	\end{gathered}\\[-0.5em]
& \quad \quad  \, \,  -X_{3,7}       
\end{aligned}
\quad \vline \quad 
\begin{aligned}
	&\begin{gathered}
	\begin{tikzpicture}[line width=0.6,scale=0.2,line cap=round,every node/.style={font=\footnotesize}]
		\node (1l) at (-2, -1) {};
		\node (1r) at (2, -1) {};
		\node (5l) at (-2, 1) {};
		\node (5r) at (2, 1) {};
		
		\node (2) at (-5, -3) {};
		\node (2d) at (-5, -5) {};
		\node (2l) at (-7, -3) {};
		\node (2r) at (-4, -3) {};
		\node (2dr) at (-4, -5) {};
		\node (2u) at (-5, -2) {};
		\node (2lu) at (-7, -2) {};

		\node (4) at (-5, 3) {};
		\node (4u) at (-5, 5) {};
		\node (4l) at (-7, 3) {};
		\node (4r) at (-4, 3) {};
		\node (4ur) at (-4, 5) {};
		\node (4d) at (-5, 2) {};
		\node (4ld) at (-7, 2) {};

		\node (6) at (5, 3) {};
		\node (6u) at (5, 5) {};
		\node (6r) at (7, 3) {};
		\node (6l) at (4, 3) {};
		\node (6ul) at (4, 5) {};
		\node (6d) at (5, 2) {};
		\node (6dr) at (7, 2) {};
		
		\node (8) at (5, -3) {};
		\node (8r) at (7, -3) {};
		\node (8d) at (5, -5) {};
		\node (8dl) at (4, -5) {};
		\node (8l) at (4, -3) {};
		\node (8ur) at (7, -2) {};
		\node (8u) at (5, -2) {};;
		
		\node (3) at (-3, 0) {};
		\node (7) at (3, 0) {};

		\draw[thick] (2l.center) to (2.center);
		\draw[thick] (2.center) to (2d.center);
		
		\draw[thick] (4u.center) to (4.center);
		\draw[thick] (4.center) to (4l.center);

		\draw[thick] (6u.center) to (6.center);
		\draw[thick] (6.center) to (6r.center);
		
		\draw[thick] (8r.center) to (8.center);
		\draw[thick] (8.center) to (8d.center);
		
		\draw[thick] (2lu.center) to (2u.center);
		\draw[thick] (2u.center) to (3.center);
		\draw[thick] (3.center) to (4d.center);
		\draw[thick] (4d.center) to (4ld.center);

		\draw[thick] (6dr.center) to (6d.center);
		\draw[thick] (6d.center) to (7.center);
		\draw[thick] (7.center) to (8u.center);
		\draw[thick] (8u.center) to (8ur.center);

		\draw[thick] (4ur.center) to (4r.center);
		\draw[thick] (4r.center) to (5l.center);
		\draw[thick] (5l.center) to (5r.center);
		\draw[thick] (5r.center) to (6l.center);
		\draw[thick] (6l.center) to (6ul.center);
		
		\draw[thick] (2dr.center) to (2r.center);
		\draw[thick] (2r.center) to (1l.center);
		\draw[thick] (1l.center) to (1r.center);
		\draw[thick] (1r.center) to (8l.center);
		\draw[thick] (8l.center) to (8dl.center);
		\coordinate (2md) at ($(2)!0.3!(2d)$);
        \coordinate (2ml) at ($(2)!0.3!(2l)$);
		\coordinate (2hm) at ($(2md)!0.5!(2ml)$);
        
 		\coordinate (4mu) at ($(4)!0.3!(4u)$);
        \coordinate (4ml) at ($(4)!0.3!(4l)$);
		\coordinate (4hm) at ($(4mu)!0.5!(4ml)$);
        
 		\coordinate (6mu) at ($(6)!0.3!(6u)$);
        \coordinate (6mr) at ($(6)!0.3!(6r)$);
		\coordinate (6hm) at ($(6mu)!0.5!(6mr)$);
        
  		\coordinate (8mr) at ($(8)!0.3!(8r)$);
        \coordinate (8md) at ($(8)!0.3!(8d)$);
		\coordinate (8hm) at ($(8mr)!0.5!(8md)$);
        
        \coordinate (5hr) at ($(5r)!0.2!(6l)$);
       	\coordinate (5hl) at ($(5r)!0.15!(5l)$);
		\coordinate (5hm) at ($(5hl)!0.5!(5hr)$);

        \coordinate (3hu) at ($(3)!0.25!(4d)$);
       	\coordinate (3hd) at ($(3)!0.25!(2u)$);
		\coordinate (3hm) at ($(3hu)!0.5!(3hd)$);

		\draw[ForestGreen, very thick] (-2,0) to (2,0);
		\draw[ForestGreen, very thick,dashed] (2,0) to (4.5,-2.5);
        \node[] at (0,-4) {$1$};
        \node[] at (0,4) {$5$};
        \node[] at (-6,0) {$3$};
        \node[] at (6,0) {$7$};
        \node[] at (-6.5,-4.5) {$2$};
        \node[] at (-6.5,4.5) {$4$};
        \node[] at (6.5,4.5) {$6$};
        \node[] at (6.5,-4.5) {$8$};
	\end{tikzpicture}
	\end{gathered}
    \\[-0.5em]
& \quad \quad  \, \,  -X_{3,7} 
\end{aligned}
\quad \vline \quad 
\begin{aligned}
	&\begin{gathered}
		\begin{tikzpicture}[line width=0.6,scale=0.2,line cap=round,every node/.style={font=\footnotesize}]
		\node (1l) at (-2, -1) {};
		\node (1r) at (2, -1) {};
		\node (5l) at (-2, 1) {};
		\node (5r) at (2, 1) {};
		
		\node (2) at (-5, -3) {};
		\node (2d) at (-5, -5) {};
		\node (2l) at (-7, -3) {};
		\node (2r) at (-4, -3) {};
		\node (2dr) at (-4, -5) {};
		\node (2u) at (-5, -2) {};
		\node (2lu) at (-7, -2) {};

		\node (4) at (-5, 3) {};
		\node (4u) at (-5, 5) {};
		\node (4l) at (-7, 3) {};
		\node (4r) at (-4, 3) {};
		\node (4ur) at (-4, 5) {};
		\node (4d) at (-5, 2) {};
		\node (4ld) at (-7, 2) {};

		\node (6) at (5, 3) {};
		\node (6u) at (5, 5) {};
		\node (6r) at (7, 3) {};
		\node (6l) at (4, 3) {};
		\node (6ul) at (4, 5) {};
		\node (6d) at (5, 2) {};
		\node (6dr) at (7, 2) {};
		
		\node (8) at (5, -3) {};
		\node (8r) at (7, -3) {};
		\node (8d) at (5, -5) {};
		\node (8dl) at (4, -5) {};
		\node (8l) at (4, -3) {};
		\node (8ur) at (7, -2) {};
		\node (8u) at (5, -2) {};;
		
		\node (3) at (-3, 0) {};
		\node (7) at (3, 0) {};

		\draw[thick] (2l.center) to (2.center);
		\draw[thick] (2.center) to (2d.center);
		
		\draw[thick] (4u.center) to (4.center);
		\draw[thick] (4.center) to (4l.center);

		\draw[thick] (6u.center) to (6.center);
		\draw[thick] (6.center) to (6r.center);
		
		\draw[thick] (8r.center) to (8.center);
		\draw[thick] (8.center) to (8d.center);
		
		\draw[thick] (2lu.center) to (2u.center);
		\draw[thick] (2u.center) to (3.center);
		\draw[thick] (3.center) to (4d.center);
		\draw[thick] (4d.center) to (4ld.center);

		\draw[thick] (6dr.center) to (6d.center);
		\draw[thick] (6d.center) to (7.center);
		\draw[thick] (7.center) to (8u.center);
		\draw[thick] (8u.center) to (8ur.center);

		\draw[thick] (4ur.center) to (4r.center);
		\draw[thick] (4r.center) to (5l.center);
		\draw[thick] (5l.center) to (5r.center);
		\draw[thick] (5r.center) to (6l.center);
		\draw[thick] (6l.center) to (6ul.center);
		
		\draw[thick] (2dr.center) to (2r.center);
		\draw[thick] (2r.center) to (1l.center);
		\draw[thick] (1l.center) to (1r.center);
		\draw[thick] (1r.center) to (8l.center);
		\draw[thick] (8l.center) to (8dl.center);
		\coordinate (2md) at ($(2)!0.3!(2d)$);
        \coordinate (2ml) at ($(2)!0.3!(2l)$);
		\coordinate (2hm) at ($(2md)!0.5!(2ml)$);
        
 		\coordinate (4mu) at ($(4)!0.3!(4u)$);
        \coordinate (4ml) at ($(4)!0.3!(4l)$);
		\coordinate (4hm) at ($(4mu)!0.5!(4ml)$);
        
 		\coordinate (6mu) at ($(6)!0.3!(6u)$);
        \coordinate (6mr) at ($(6)!0.3!(6r)$);
		\coordinate (6hm) at ($(6mu)!0.5!(6mr)$);
        
  		\coordinate (8mr) at ($(8)!0.3!(8r)$);
        \coordinate (8md) at ($(8)!0.3!(8d)$);
		\coordinate (8hm) at ($(8mr)!0.5!(8md)$);
        
        \coordinate (5hr) at ($(5r)!0.2!(6l)$);
       	\coordinate (5hl) at ($(5r)!0.15!(5l)$);
		\coordinate (5hm) at ($(5hl)!0.5!(5hr)$);

        \coordinate (3hu) at ($(3)!0.25!(4d)$);
       	\coordinate (3hd) at ($(3)!0.25!(2u)$);
		\coordinate (3hm) at ($(3hu)!0.5!(3hd)$);

		\draw[ForestGreen, very thick] (-2,0) to (2,0);
		\draw[ForestGreen, very thick,dashed] (2,0) to (4.5,-2.5);
		\draw[ForestGreen, very thick,dashed] (-2,0) to (-4.5,2.5);

        \node[] at (0,-4) {$1$};
        \node[] at (0,4) {$5$};
        \node[] at (-6,0) {$3$};
        \node[] at (6,0) {$7$};
        \node[] at (-6.5,-4.5) {$2$};
        \node[] at (-6.5,4.5) {$4$};
        \node[] at (6.5,4.5) {$6$};
        \node[] at (6.5,-4.5) {$8$};
	\end{tikzpicture}
	\end{gathered}\\[-0.5em]
     & \quad \quad  \quad \, \,  X_{3,7} 
\end{aligned}
\end{equation}
where, since this example is small the extensions consist in simply turning right once at either end, with the respective signs given by the rule, $(-1)^{N_e}$. 

Finally, we can use the same ideas to determine which contractions produce a full monomial, $M=\prod_C X_C$. Just like for a single $X_C$, with start with the collection of cores in $M$ and consider all possible extensions of each core into contraction curves, but keeping in mind that  the only allowed sets of contractions corresponded to non-crossing blue curves that filled all of the edges in the fat graph -- such that each edge is filled by \textit{one and only one} contraction curve. In fact, by itself, this is already a very powerful constraint as it forbids some combinations of monomials from the get go. For example, for the $s$-channel LS$_4$, this constraint tells us that the monomial $X_{2,5}X_{1,4}X_{1,6}$ can never appear. As for the sign of a given monomial, $M$, in set of contractions it is simply captured by a factor of $(-1)^{N^{\text{tot}}_e}$, with $N^{\text{tot}}_e$ being the total number of extensions needed to go from the cores to the set of contractions covering all edges of the fatgraph. 

So we have fully discovered the formulation of LS from the surface integral from standard gluing of polarizations! Once we scalar-scaffold the gluons, the different terms on the LS correspond to collections curves (taken as \textit{cores} or \textit{extensions}) which completely cover the fatgraph (once and only once), and the sign of a given monomial depends solely on the number of extensions! So simply following the simplest path to encode the Lorentz contractions graphically we are naturally led to formulation of LS from the linearized $u$'s and the residues of the surface integral; producing exactly the same monomials with the same signs and cancellations!  

\subsection{Tr $F^3$ leading singularities}
To finish the tree-level discussion, we would like discuss the case of leading singularities with purely Tr $F^3$ vertices, as the picture above highly simplifies in this case. As understood from the computation of leading singularities via the surface integral, already for the $3$-point amplitude described in \eqref{eq:3ptLS_Surf}, the Tr $F^3$ vertex has the highest unit term in the LS, corresponds to the case where each internal edge of the fatgraph is filled with a core (and no extensions are used). 

This picture trivially generalizes for any graph at any number of points! This is, given a graph, the respective LS with all internal Tr $F^3$ vertices can be determine by drawing a core inside each internal edge of the fat graph and directly reading the respective monomial from it -- so the full answer is given by this single term. 

A simple way to argue that contributions coming from extensions don't enter is by noting that if we cover the fatgraph using cores and one extension, then we can always have one higher order LS in which we replace the extensions by a core. Since the pure Tr $F^3$ LS is the highest units LS, this must be it. From the contraction picture, this translates into the fact that the Tr $F^3$ vertex can be represented by a $3$-point fatgraph with $3$ red handles: 
\begin{equation}
\mathcal{A}^{F^3}_3[g_1,g_2,g_3] =\,\epsilon_1 \cdot(q_2 - q_3) \, \epsilon_2 \cdot(q_3 - q_1) \, \epsilon_3 \cdot(q_1 - q_2)
= \, \, \begin{gathered}
    \begin{tikzpicture}[line width=0.6,scale=0.27,line cap=round,every node/.style={font=\scriptsize}]
		\node (0) at (-2.75, 1.5) {};
		\node (1) at (-2, 2.25) {};
            \coordinate (01m) at ($(0)!0.5!(1)$);
		\node (2) at (-1, -0.25) {};
		\node (3) at (-0.25, 0.5) {};
		\node (4) at (-1, -3.25) {};
		\node(5) at (0, -3.25) {};
		\node (6) at (0, -0.5) {};
		\node   (7) at (2.75, -0.5) {};
		\node   (8) at (2.75, 0.5) {};		
		\node   (25) at (-2.25, -0.75) {1};
		\node   (27) at (0.5, 1.75) {2};
		\node   (29) at (1.25, -1.75) {3};
           \coordinate (45m) at ($(4)!0.5!(5)$);
         \coordinate (26m) at ($(2)!0.5!(6)$);
        \coordinate (36m) at ($(3)!0.5!(6)$);
        \coordinate (78m) at ($(7)!0.5!(8)$);

        \coordinate (Anch1) at ($(6)!0.25!(5)$);
        \coordinate (Anch2) at ($(6)!0.25!(7)$);
         \coordinate (mAnch) at ($(Anch1)!0.5!(Anch2)$);
		
		\draw[thick] (0.center) to (2.center);
		\draw[thick] (2.center) to (4.center);
		\draw[thick] (6.center) to (5.center);
		\draw[thick] [in=135, out=-45] (1.center) to (3.center);
		\draw[thick] (3.center) to (8.center);
		\draw[thick]  (6.center) to (7.center);
		\draw[thick][style=Blue] (01m.center) to (mAnch);
         \draw[Maroon,thick] (Anch1) to[out=0,in=-90] (Anch2);

          \coordinate (Anch1p) at ($(2)!0.2!(0)$);
        \coordinate (Anch2p) at ($(2)!0.2!(4)$);
         \coordinate (mAnchp) at ($(Anch1)!0.5!(Anch2)$);
		
        \draw[thick][style=Blue] (36m.center) to (78m.center);
         \draw[Blue,thick] (36m) to[out=180,in=45] (-1.25,-0.4);
         
         \draw[Maroon,thick] (Anch1p) to[out=-135,in=180] (Anch2p);
         
         \coordinate (Anch1pp) at ($(3)!0.2!(1)$);
        \coordinate (Anch2pp) at ($(3)!0.2!(8)$);
         \coordinate (mAnchpp) at ($(Anch1)!0.5!(Anch2)$);
        \draw[thick][style=Blue] (45m.center) to (26m.center);

         \draw[Blue,thick] (26m) to[out=90,in=-125] (-0.1, 0.75);
         \draw[Maroon,thick] (Anch1pp) to[out=45,in=90] (Anch2pp);

\end{tikzpicture}
\end{gathered},
\label{eq:3ptF3}
\end{equation}
and so to compute a pure Tr $F^3$ LS, we should place $3$ red handles in all internal vertices of the fatgraph, which automatically determines all the contractions. For example in the $s$-channel LS we get:
\begin{equation}
	\begin{gathered}
		\begin{tikzpicture}[line width=0.6,scale=0.3,line cap=round,every node/.style={font=\normalsize}]
		\node (1l) at (-2, -1) {};
		\node (1r) at (2, -1) {};
		\node (5l) at (-2, 1) {};
		\node (5r) at (2, 1) {};
		
		\node (2) at (-5, -3) {};
		\node (2d) at (-5, -5) {};
		\node (2l) at (-7, -3) {};
		\node (2r) at (-4, -3) {};
		\node (2dr) at (-4, -5) {};
		\node (2u) at (-5, -2) {};
		\node (2lu) at (-7, -2) {};

		\node (4) at (-5, 3) {};
		\node (4u) at (-5, 5) {};
		\node (4l) at (-7, 3) {};
		\node (4r) at (-4, 3) {};
		\node (4ur) at (-4, 5) {};
		\node (4d) at (-5, 2) {};
		\node (4ld) at (-7, 2) {};

		\node (6) at (5, 3) {};
		\node (6u) at (5, 5) {};
		\node (6r) at (7, 3) {};
		\node (6l) at (4, 3) {};
		\node (6ul) at (4, 5) {};
		\node (6d) at (5, 2) {};
		\node (6dr) at (7, 2) {};
		
		\node (8) at (5, -3) {};
		\node (8r) at (7, -3) {};
		\node (8d) at (5, -5) {};
		\node (8dl) at (4, -5) {};
		\node (8l) at (4, -3) {};
		\node (8ur) at (7, -2) {};
		\node (8u) at (5, -2) {};;
		
		\node (3) at (-3, 0) {};
		\node (7) at (3, 0) {};

		\draw[thick] (2l.center) to (2.center);
		\draw[thick] (2.center) to (2d.center);
		
		\draw[thick] (4u.center) to (4.center);
		\draw[thick] (4.center) to (4l.center);

		\draw[thick] (6u.center) to (6.center);
		\draw[thick] (6.center) to (6r.center);
		
		\draw[thick] (8r.center) to (8.center);
		\draw[thick] (8.center) to (8d.center);
		
		\draw[thick] (2lu.center) to (2u.center);
		\draw[thick] (2u.center) to (3.center);
		\draw[thick] (3.center) to (4d.center);
		\draw[thick] (4d.center) to (4ld.center);

		\draw[thick] (6dr.center) to (6d.center);
		\draw[thick] (6d.center) to (7.center);
		\draw[thick] (7.center) to (8u.center);
		\draw[thick] (8u.center) to (8ur.center);

		\draw[thick] (4ur.center) to (4r.center);
		\draw[thick] (4r.center) to (5l.center);
		\draw[thick] (5l.center) to (5r.center);
		\draw[thick] (5r.center) to (6l.center);
		\draw[thick] (6l.center) to (6ul.center);
		
		\draw[thick] (2dr.center) to (2r.center);
		\draw[thick] (2r.center) to (1l.center);
		\draw[thick] (1l.center) to (1r.center);
		\draw[thick] (1r.center) to (8l.center);
		\draw[thick] (8l.center) to (8dl.center);
		\coordinate (2md) at ($(2)!0.3!(2d)$);
        \coordinate (2ml) at ($(2)!0.3!(2l)$);
		\coordinate (2hm) at ($(2md)!0.5!(2ml)$);
        
 		\coordinate (4mu) at ($(4)!0.3!(4u)$);
        \coordinate (4ml) at ($(4)!0.3!(4l)$);
		\coordinate (4hm) at ($(4mu)!0.5!(4ml)$);
        
 		\coordinate (6mu) at ($(6)!0.3!(6u)$);
        \coordinate (6mr) at ($(6)!0.3!(6r)$);
		\coordinate (6hm) at ($(6mu)!0.5!(6mr)$);
        
  		\coordinate (8mr) at ($(8)!0.3!(8r)$);
        \coordinate (8md) at ($(8)!0.3!(8d)$);
		\coordinate (8hm) at ($(8mr)!0.5!(8md)$);
        
        \coordinate (5hl) at ($(5l)!0.2!(4r)$);
       	\coordinate (5hr) at ($(5l)!0.15!(5r)$);
		\coordinate (5hm) at ($(5hl)!0.5!(5hr)$);

        \coordinate (1hr) at ($(1r)!0.2!(8l)$);
       	\coordinate (1hl) at ($(1r)!0.15!(1l)$);
		\coordinate (1hm) at ($(1hl)!0.5!(1hr)$);

		\coordinate (5hrp) at ($(5r)!0.2!(6l)$);
       	\coordinate (5hlp) at ($(5r)!0.15!(5l)$);
		\coordinate (5hmp) at ($(5hlp)!0.5!(5hrp)$);

        \coordinate (1hlp) at ($(1l)!0.2!(2r)$);
       	\coordinate (1hrp) at ($(1l)!0.15!(1r)$);
		\coordinate (1hmp) at ($(1hlp)!0.5!(1hrp)$);
		
		\coordinate (3hu) at ($(3)!0.25!(4d)$);
       	\coordinate (3hd) at ($(3)!0.25!(2u)$);
		\coordinate (3hm) at ($(3hu)!0.5!(3hd)$);
		
		\coordinate (7hu) at ($(7)!0.25!(6d)$);
       	\coordinate (7hd) at ($(7)!0.25!(8u)$);
		\coordinate (7hm) at ($(7hu)!0.5!(7hd)$);
        
        \draw[Maroon,very thick] (2md) to[out=180,in=-90] (2ml);
        \draw[Maroon,very thick] (4ml) to[out=90,in=180] (4mu);
        \draw[Maroon,very thick] (6mu) to[out=0,in=90] (6mr);
        \draw[Maroon,very thick] (8mr) to[out=-90,in=0] (8md);
        \draw[Maroon,very thick] (5hl) to[out=45,in=90] (5hr);
        \draw[Maroon,very thick] (1hl) to[out=-90,in=-135] (1hr);
        
		\draw[Maroon,very thick] (5hlp) to[out=90,in=135] (5hrp);
        \draw[Maroon,very thick] (1hlp) to[out=-45,in=-90] (1hrp);

		\draw[Maroon,very thick] (7hu) to[out=-45,in=45] (7hd);
		\draw[Maroon,very thick] (3hu) to[out=-135,in=135] (3hd);

		
		\draw[Blue,very thick] (4hm) to (-3,1);
        \draw[Blue,very thick] (-3,1) to[out=-45,in=100] (1hmp);
   		\draw[Blue,very thick] (8hm) to (3,-1);
        \draw[Blue,very thick] (3,-1) to[out=135,in=-70] (5hmp);
		
		\draw[Blue,very thick] (3hm) to (7hm);
		
		\draw[Blue, very thick] (2hm) to (-3,-1);
		\draw[Blue, very thick] (-3,-1) to[out=45,in=-100] (5hm);
		
		\draw[Blue, very thick] (6hm) to (3,1);
		\draw[Blue, very thick] (3,1) to[out=-135,in=80] (1hm);

        \node[] at (0,-4) {$1$};
        \node[] at (0,4) {$5$};
        \node[] at (-6,0) {$3$};
        \node[] at (6,0) {$7$};
        \node[] at (-6.5,-4.5) {$2$};
        \node[] at (-6.5,4.5) {$4$};
        \node[] at (6.5,4.5) {$6$};
        \node[] at (6.5,-4.5) {$8$};
	\end{tikzpicture} 
	\end{gathered}
	\quad \Leftrightarrow \quad 
	\begin{gathered}
		\begin{tikzpicture}[line width=0.6,scale=0.3,line cap=round,every node/.style={font=\normalsize}]
		\node (1l) at (-2, -1) {};
		\node (1r) at (2, -1) {};
		\node (5l) at (-2, 1) {};
		\node (5r) at (2, 1) {};
		
		\node (2) at (-5, -3) {};
		\node (2d) at (-5, -5) {};
		\node (2l) at (-7, -3) {};
		\node (2r) at (-4, -3) {};
		\node (2dr) at (-4, -5) {};
		\node (2u) at (-5, -2) {};
		\node (2lu) at (-7, -2) {};

		\node (4) at (-5, 3) {};
		\node (4u) at (-5, 5) {};
		\node (4l) at (-7, 3) {};
		\node (4r) at (-4, 3) {};
		\node (4ur) at (-4, 5) {};
		\node (4d) at (-5, 2) {};
		\node (4ld) at (-7, 2) {};

		\node (6) at (5, 3) {};
		\node (6u) at (5, 5) {};
		\node (6r) at (7, 3) {};
		\node (6l) at (4, 3) {};
		\node (6ul) at (4, 5) {};
		\node (6d) at (5, 2) {};
		\node (6dr) at (7, 2) {};
		
		\node (8) at (5, -3) {};
		\node (8r) at (7, -3) {};
		\node (8d) at (5, -5) {};
		\node (8dl) at (4, -5) {};
		\node (8l) at (4, -3) {};
		\node (8ur) at (7, -2) {};
		\node (8u) at (5, -2) {};;
		
		\node (3) at (-3, 0) {};
		\node (7) at (3, 0) {};

		\draw[thick] (2l.center) to (2.center);
		\draw[thick] (2.center) to (2d.center);
		
		\draw[thick] (4u.center) to (4.center);
		\draw[thick] (4.center) to (4l.center);

		\draw[thick] (6u.center) to (6.center);
		\draw[thick] (6.center) to (6r.center);
		
		\draw[thick] (8r.center) to (8.center);
		\draw[thick] (8.center) to (8d.center);
		
		\draw[thick] (2lu.center) to (2u.center);
		\draw[thick] (2u.center) to (3.center);
		\draw[thick] (3.center) to (4d.center);
		\draw[thick] (4d.center) to (4ld.center);

		\draw[thick] (6dr.center) to (6d.center);
		\draw[thick] (6d.center) to (7.center);
		\draw[thick] (7.center) to (8u.center);
		\draw[thick] (8u.center) to (8ur.center);

		\draw[thick] (4ur.center) to (4r.center);
		\draw[thick] (4r.center) to (5l.center);
		\draw[thick] (5l.center) to (5r.center);
		\draw[thick] (5r.center) to (6l.center);
		\draw[thick] (6l.center) to (6ul.center);
		
		\draw[thick] (2dr.center) to (2r.center);
		\draw[thick] (2r.center) to (1l.center);
		\draw[thick] (1l.center) to (1r.center);
		\draw[thick] (1r.center) to (8l.center);
		\draw[thick] (8l.center) to (8dl.center);
		\coordinate (2md) at ($(2)!0.3!(2d)$);
        \coordinate (2ml) at ($(2)!0.3!(2l)$);
		\coordinate (2hm) at ($(2md)!0.5!(2ml)$);
        
 		\coordinate (4mu) at ($(4)!0.3!(4u)$);
        \coordinate (4ml) at ($(4)!0.3!(4l)$);
		\coordinate (4hm) at ($(4mu)!0.5!(4ml)$);
        
 		\coordinate (6mu) at ($(6)!0.3!(6u)$);
        \coordinate (6mr) at ($(6)!0.3!(6r)$);
		\coordinate (6hm) at ($(6mu)!0.5!(6mr)$);
        
  		\coordinate (8mr) at ($(8)!0.3!(8r)$);
        \coordinate (8md) at ($(8)!0.3!(8d)$);
		\coordinate (8hm) at ($(8mr)!0.5!(8md)$);
        
        \coordinate (5hl) at ($(5l)!0.2!(4r)$);
       	\coordinate (5hr) at ($(5l)!0.15!(5r)$);
		\coordinate (5hm) at ($(5hl)!0.5!(5hr)$);

        \coordinate (1hr) at ($(1r)!0.2!(8l)$);
       	\coordinate (1hl) at ($(1r)!0.15!(1l)$);
		\coordinate (1hm) at ($(1hl)!0.5!(1hr)$);

		\coordinate (5hrp) at ($(5r)!0.2!(6l)$);
       	\coordinate (5hlp) at ($(5r)!0.15!(5l)$);
		\coordinate (5hmp) at ($(5hlp)!0.5!(5hrp)$);

        \coordinate (1hlp) at ($(1l)!0.2!(2r)$);
       	\coordinate (1hrp) at ($(1l)!0.15!(1r)$);
		\coordinate (1hmp) at ($(1hlp)!0.5!(1hrp)$);
		
		\coordinate (3hu) at ($(3)!0.25!(4d)$);
       	\coordinate (3hd) at ($(3)!0.25!(2u)$);
		\coordinate (3hm) at ($(3hu)!0.5!(3hd)$);
		
		\coordinate (7hu) at ($(7)!0.25!(6d)$);
       	\coordinate (7hd) at ($(7)!0.25!(8u)$);
		\coordinate (7hm) at ($(7hu)!0.5!(7hd)$);
        
        \draw[ForestGreen,very thick] (-4.5,-2.5) to (-2.5,-0.5);
        
        \draw[ForestGreen,very thick] (-4.5,2.5) to (-2.5,0.5);
        
        \draw[ForestGreen,very thick] (-2,0) to (2,0);

        \draw[ForestGreen,very thick] (4.5,2.5) to (2.5,0.5);
        
        \draw[ForestGreen,very thick] (4.5,-2.5) to (2.5,-0.5);

        \node[] at (0,-4) {$1$};
        \node[] at (0,4) {$5$};
        \node[] at (-6,0) {$3$};
        \node[] at (6,0) {$7$};
        \node[] at (-6.5,-4.5) {$2$};
        \node[] at (-6.5,4.5) {$4$};
        \node[] at (6.5,4.5) {$6$};
        \node[] at (6.5,-4.5) {$8$};
	\end{tikzpicture} 
	\end{gathered}
\end{equation}
which when mapped into cores, precisely agrees with the rule described above: we get a core per internal edge of the graph. 

In summary, Tr $F^3$ LS are extremely simple from this perspective. Note in particular that starting from the pure Tr $F^3$ LS we can go down in units to produce mixed LS with both pure YM and Tr $F^3$ vertices by turning some cores into extensions -- this amounts to changing the number of red handles in the internal vertices from $3$ down to one. 

\section{Leading singularities at one-loop}
\label{sec:loop-level}
At loop-level and, in particular, at one-loop, we can compute a $D$-dimensional LS almost in the same way as for tree-level. The only difference comes from the last gluing step where we glue two legs of the same object. 

While at tree level, the spin-sum in \eqref{eq:LSTreeGluing} can be replaced by $-\eta_{\mu\nu}$ by virtue of the ward identities, the same is no longer true for loop-level LS. This is if we glue using only $\eta_{\mu\nu}$, the final result would match what we get by taking the respective maximal residue of the loop-integrand, where the gluon propagators are proportional to $\eta_{\mu \nu}$. However, as it is well known, to correctly define a loop-integrand, which yields a gauge-invariant answer post loop integration, one needs to introduce \textit{ghosts}. The contributions from the ghost field also change the LS, but this change is automatically taken into account when we consider the correct spin-sum gluing rule.  

Concretely, when computing  an $n$-point 1-loop LS, if we first glue all tree-like propagators, in the last step we need to glue two external legs of an $(n+2)$-tree LS, this is
\begin{equation}
    \text{LS}_\text{1-loop}^{(n)} = \left(-\eta^{\mu\nu} + \frac{p^\mu q^\nu + p^\nu q^\mu}{p \cdot q}\right) (\text{LS}^{(n+2)}_\text{tree})_{\mu \nu} \,,
    \label{eq:LoopGlue1}
\end{equation}
where the two legs that we are gluing have momentum $p^\mu$ and $-p^\mu$. As opposed to what happens at tree-level, we now have that $p^\mu(\text{LS}^{(n+2)}_\text{tree})_{\mu \nu} \neq 0$, and therefore the second term in \eqref{eq:LoopGlue1} is no longer zero. Instead, a leading singularity with two free indices,  $\text{LS}^{\mu\nu}$, obeys $p_{\mu} p_\nu \text{LS}^{\mu\nu} = 0$, which implies that
\begin{equation}
    p_{\mu} \text{LS}^{\mu\nu} = \mathcal{N} p^\nu, \quad 
     \text{LS}^{\mu\nu}  p_{\nu}= \mathcal{N}^\prime p^\mu
\, ,  
\label{eq:loopGlueNLNR}    
\end{equation}
where we have intentionally distinguished between $\mathcal{N}$ and $\mathcal{N}^\prime$ since under the convention chosen in \eqref{eq:3ptGluonScalar} these are different functions, as we will show momentarily. In practice, this means that when we are computing any one-loop LS, we get a sum of three contributions, 
\begin{equation}
    \text{LS}_\text{1-loop} = -\mathcal{M}_{\mu\nu}\eta^{\mu\nu}+ \mathcal{N} + \mathcal{N}^\prime \, ,
    \label{eq:NCorect1loop}
\end{equation}
where the correction to the naive gluing with $\eta_{\mu \nu}$ is precisely $\mathcal{N} + \mathcal{N}^\prime$. For the naive gluing part, the story at loop-level is exactly the same as at tree-level: we draw the respective fat graph (where each gluon is scalar-scaffolded) and given a choice of red handles, we can extract all the monomials (and signs) by drawing the contractions curves in a way such that each edge of the fat graph is covered by a \textit{single} curve. From such a collection of contraction curves we can read the respective monomials/signs by considering the different subcurves of the contraction curve -- everything just like at tree-level.

Quite nicely, already simply from this picture, we see that there are a couple of novelties at one-loop. First of all we find that we can have contraction curves that \textit{self-intersect} as they go around the loop: 
\begin{equation}
\vcenter{\hbox{\begin{tikzpicture}[line width=0.6 ,scale=0.6,line cap=round,every node/.style={font=\normalsize},rotate=90]
        	\coordinate (str) at ($0.5*({cos(30)},{sin(30)})$);
            \coordinate (sb) at ($0.5*({cos(-90)},{sin(-90)})$);
        	\coordinate (stl) at ($0.5*({cos(150)},{sin(150)})$);
        	
        	\coordinate (tr) at ($1.5*({cos(30)},{sin(30)})$);
            \coordinate (b) at ($1.5*({cos(-90)},{sin(-90)})$);
        	\coordinate (tl) at ($1.5*({cos(150)},{sin(150)})$);
        	
        	\coordinate (trv) at ($3*({cos(30)},{sin(30)})$);
        	\coordinate (tlv) at ($3*({cos(150)},{sin(150)})$);
        	\coordinate (bv) at ($3*({cos(-90)},{sin(-90)})$);
        	
        	\coordinate (tr1) at ($(tr)!0.15!(tl)$);
        	\coordinate (tr1v) at ($(tr1)+1.5*({cos(30)},{sin(30)})$);
		      \coordinate (tr2) at ($(tr)!0.15!(b)$);
        	\coordinate (tr2v) at ($(tr2)+1.5*({cos(30)},{sin(30)})$);
        	
        	\coordinate (tl1) at ($(tl)!0.15!(tr)$);
        	\coordinate (tl1v) at ($(tl1)+1.5*({cos(150)},{sin(150)})$);
		      \coordinate (tl2) at ($(tl)!0.15!(b)$);
        	\coordinate (tl2v) at ($(tl2)+1.5*({cos(150)},{sin(150)})$);
        	
        	\coordinate (b1) at ($(b)!0.15!(tl)$);
        	\coordinate (b1v) at ($(b1)+1.5*({cos(-90)},{sin(-90)})$);
		\coordinate (b2) at ($(b)!0.15!(tr)$);
        	\coordinate (b2v) at ($(b2)+1.5*({cos(-90)},{sin(-90)})$);
        	\coordinate (br1) at ($(b2v)+({cos(-45)},{sin(-45)})$);
        	\coordinate (br2) at ($(bv)+({cos(-45)},{sin(-45)})$);
      	\coordinate (bl1) at ($(b1v)+({cos(-135)},{sin(-135)})$);
      	\coordinate (bl2) at ($(bv)+({cos(-135)},{sin(-135)})$);

        \draw[thick] (str) to (sb) to (stl) to (str);

        \draw[thick] (tr1) to (tr1v);
        \draw[thick] (tr2) to (tr2v);
        \draw[thick] (tl1) to (tl1v);
        \draw[thick] (tl2) to (tl2v);

		\draw[thick] (b2) to (b2v);
        \draw[thick] (b1) to (b1v);
        	\draw[thick] (b2v) to (br1);        	
        	\draw[thick] (bv) to (br2); 
        	\draw[thick] (b1v) to (bl1);        	
        	\draw[thick] (bv) to (bl2);

        	\draw[thick] (b1) to (tl2);
        	\draw[thick] (tl1) to (tr1);
        \draw[thick] (tr2) to (b2);
        
        	\coordinate (ht1) at ($(b2)!0.1!(tr2)$);
        	\coordinate (ht2) at ($(b2)!0.1!(b2v)$);
        	\coordinate (htm) at ($(ht1)!0.5!(ht2)$);
        	\draw[Maroon, thick] (ht1) to[out=-30,in=10] (ht2);
		\coordinate (p1) at ($(b1)!0.5!(sb)$);
		\coordinate (p21) at ($(stl)!0.4!(tl2)$);
		\coordinate (p22) at ($(stl)!0.4!(tl1)$);
		\coordinate (p31) at ($(str)!0.4!(tr1)$);
		\coordinate (p32) at ($(str)!0.4!(tr2)$);
		\coordinate (p3) at ($(str)!0.4!(tr)$);
        
		\coordinate (p4) at ($(b1)!0.5!(b2)$);
        
		\coordinate (p5) at ($(b1v)!0.5!(b2v)+(0,0.3)$);
		\coordinate (p6) at ($(bl1)!0.5!(bl2)$);
		\draw[Blue, thick] (htm) to[out=130, in=-70] (p21);
		\draw[Blue, thick] (p21) to[out=100,in=-160] (p22) to[out=20, in=160] (p31) to[out=-20,in=80] (p32) to [out=-110,in=90] (p4) to[out=-90, in=90] (p5) to[out=-90, in=45] (p6); 

		\coordinate (hi1t) at ($(stl)!0.2!(str)$);
		\coordinate (hi1b) at ($(stl)!0.2!(sb)$);
		\coordinate (hi2t) at ($(str)!0.2!(stl)$);
		\coordinate (hi2b) at ($(str)!0.2!(sb)$);
		\coordinate (hot) at ($(b1v)!0.1!(b1)$);
		\coordinate (hob) at ($(b1v)!0.1!(bl1)$);	
		
		\draw[Maroon, thick] (hi1t) to[out=-90,in=20] (hi1b);
		\draw[Maroon, thick] (hi2t) to[out=-90,in=160] (hi2b);
		\draw[Maroon, thick] (hot) to[out=180,in=135] (hob);
        \node (reg1) at ($(bl2)!0.5!(br2)$) {$1$};
        \node (reg3) at ($(b1)-(1,0)$) {$3$};
        \node (reg7) at ($(b2)+(1,0)$) {$7$};
        \node (reg7) at ($(tl1)!0.5!(tr1)+(0,0.5)$) {$5$};
    \end{tikzpicture}}}
    \quad 
    \longrightarrow
     \quad\vcenter{\hbox{\begin{tikzpicture}[line width=0.6 ,scale=0.6,line cap=round,every node/.style={font=\normalsize},rotate=90]
        	\coordinate (str) at ($0.5*({cos(30)},{sin(30)})$);
            \coordinate (sb) at ($0.5*({cos(-90)},{sin(-90)})$);
        	\coordinate (stl) at ($0.5*({cos(150)},{sin(150)})$);
        	
        	\coordinate (tr) at ($1.5*({cos(30)},{sin(30)})$);
            \coordinate (b) at ($1.5*({cos(-90)},{sin(-90)})$);
        	\coordinate (tl) at ($1.5*({cos(150)},{sin(150)})$);
        	
        	\coordinate (trv) at ($3*({cos(30)},{sin(30)})$);
        	\coordinate (tlv) at ($3*({cos(150)},{sin(150)})$);
        	\coordinate (bv) at ($3*({cos(-90)},{sin(-90)})$);
        	
        	\coordinate (tr1) at ($(tr)!0.15!(tl)$);
        	\coordinate (tr1v) at ($(tr1)+1.5*({cos(30)},{sin(30)})$);
		      \coordinate (tr2) at ($(tr)!0.15!(b)$);
        	\coordinate (tr2v) at ($(tr2)+1.5*({cos(30)},{sin(30)})$);
        	
        	\coordinate (tl1) at ($(tl)!0.15!(tr)$);
        	\coordinate (tl1v) at ($(tl1)+1.5*({cos(150)},{sin(150)})$);
		      \coordinate (tl2) at ($(tl)!0.15!(b)$);
        	\coordinate (tl2v) at ($(tl2)+1.5*({cos(150)},{sin(150)})$);
        	
        	\coordinate (b1) at ($(b)!0.15!(tl)$);
        	\coordinate (b1v) at ($(b1)+1.5*({cos(-90)},{sin(-90)})$);
		\coordinate (b2) at ($(b)!0.15!(tr)$);
        	\coordinate (b2v) at ($(b2)+1.5*({cos(-90)},{sin(-90)})$);
        	\coordinate (br1) at ($(b2v)+({cos(-45)},{sin(-45)})$);
        	\coordinate (br2) at ($(bv)+({cos(-45)},{sin(-45)})$);
      	\coordinate (bl1) at ($(b1v)+({cos(-135)},{sin(-135)})$);
      	\coordinate (bl2) at ($(bv)+({cos(-135)},{sin(-135)})$);

        \draw[thick] (str) to (sb) to (stl) to (str);

        \draw[thick] (tr1) to (tr1v);
        \draw[thick] (tr2) to (tr2v);
        \draw[thick] (tl1) to (tl1v);
        \draw[thick] (tl2) to (tl2v);

		\draw[thick] (b2) to (b2v);
        \draw[thick] (b1) to (b1v);
        	\draw[thick] (b2v) to (br1);        	
        	\draw[thick] (bv) to (br2); 
        	\draw[thick] (b1v) to (bl1);        	
        	\draw[thick] (bv) to (bl2);

        	\draw[thick] (b1) to (tl2);
        	\draw[thick] (tl1) to (tr1);
        \draw[thick] (tr2) to (b2);
        
        \coordinate (ht1) at ($(b2)!0.1!(tr2)$);
        \coordinate (ht2) at ($(b2)!0.1!(b2v)$);
        \coordinate (htm) at ($(ht1)!0.5!(ht2)$);
        \draw[Maroon, thick] (ht1) to[out=-30,in=10] (ht2);
		\coordinate (p1) at ($(b1)!0.5!(sb)$);
		\coordinate (p21) at ($(stl)!0.4!(tl2)$);
		\coordinate (p22) at ($(stl)!0.4!(tl1)$);
		\coordinate (p31) at ($(str)!0.4!(tr1)$);
		\coordinate (p32) at ($(str)!0.4!(tr2)$);
		\coordinate (p3) at ($(str)!0.4!(tr)$);
		\coordinate (p41) at ($(b1)!0.3!(b2)$);
		\coordinate (p42) at ($(b1)!0.6!(b2)$);
        
        \coordinate (brm) at ($(br1)!0.5!(br2)$);
		\coordinate (p51) at ($(b1v)!0.3!(b2v)+(0,0.3)$);
        \coordinate (p52) at ($(b1v)!0.6!(b2v)+(0,0.3)$);

		\coordinate (p6) at ($(bl1)!0.5!(bl2)$);
		\draw[Blue, thick] (p42) to[out=90, in=-70] (p21);
        \draw[Blue, dashed] (p42) to (p52) to[out=-90, in=135] (brm);
		\draw[Blue, thick] (p21) to[out=100,in=-160] (p22) to[out=20, in=160] (p31) to[out=-20,in=80] (p32) to [out=-110,in=90] (p41) to[out=-90, in=90] (p51) to[out=-90, in=45] (p6); 

		\coordinate (hi1t) at ($(stl)!0.2!(str)$);
		\coordinate (hi1b) at ($(stl)!0.2!(sb)$);
		\coordinate (hi2t) at ($(str)!0.2!(stl)$);
		\coordinate (hi2b) at ($(str)!0.2!(sb)$);
		\coordinate (hot) at ($(b1v)!0.1!(b1)$);
		\coordinate (hob) at ($(b1v)!0.1!(bl1)$);	
		
		\draw[Maroon, thick] (hi1t) to[out=-90,in=20] (hi1b);
		\draw[Maroon, thick] (hi2t) to[out=-90,in=160] (hi2b);
		\draw[Maroon, thick] (hot) to[out=180,in=135] (hob);
        \node (reg1) at ($(bl2)!0.5!(br2)$) {$1$};
        \node (reg3) at ($(b1)-(1,0)$) {$3$};
        \node (reg7) at ($(b2)+(1,0)$) {$7$};
        \node (reg7) at ($(tl1)!0.5!(tr1)+(0,0.5)$) {$5$};
    \end{tikzpicture}}}
\end{equation}
and more than that we see that we can \textit{at most} have curves that self intersect one time while looping around the puncture, as any higher self-intersection would lead to a contraction curve that goes from a given edge twice. 

Another new feature is the fact that from the $\eta_{\mu,\nu}$ gluing of the $3$-points around the loop, there will be a piece where all $\eta_{\mu,\nu}$ are glued to each other which gives a contribution proportional to $D$! In the contraction picture this means that there is a \textit{closed} contraction curve which is precisely proportional to $D$! For example, in the case of gluing a 4-point tree-level into a bubble LS, by gluing the contraction pattern that connects legs A and B, we obtain the $D$ contribution directly, graphically this is represented by
\begin{equation}
    \begin{tikzpicture}[line width=0.6 ,scale=0.4,line cap=round,every node/.style={font=\normalsize},baseline=(O.base)]
	\begin{scope}[scale = 0.8]
        \node (O) at (0,0) {};
		\node (1l) at (-2, -1) {};
		\node (1r) at (2, -1) {};
		\node (5l) at (-2, 1) {};
		\node (5r) at (2, 1) {};
		
		\node (2) at (-5, -3) {};
		\node (2d) at (-5, -5) {};
		\node (2l) at (-7, -3) {};
		\node (2r) at (-4, -3) {};
		\node (2dr) at (-4, -5) {};
		\node (2u) at (-5, -2) {};
		\node (2lu) at (-7, -2) {};

		\node (4) at (-5, 3) {};
		\node (4u) at (-5, 5) {};
		\node (4l) at (-7, 3) {};
		\node (4r) at (-4, 3) {};
		\node (4ur) at (-4, 5) {};
		\node (4d) at (-5, 2) {};
		\node (4ld) at (-7, 2) {};

		\node (6) at (5, 3) {};
		\node (6u) at (5, 5) {};
		\node (6r) at (7, 3) {};
		\node (6l) at (4, 3) {};
		\node (6ul) at (4, 5) {};
		\node (6d) at (5, 2) {};
		\node (6dr) at (7, 2) {};
		
		\node (8) at (5, -3) {};
		\node (8r) at (7, -3) {};
		\node (8d) at (5, -5) {};
		\node (8dl) at (4, -5) {};
		\node (8l) at (4, -3) {};
		\node (8ur) at (7, -2) {};
		\node (8u) at (5, -2) {};;
		
		\node (3) at (-3, 0) {};
		\node (7) at (3, 0) {};

		\draw[thick] (2u.center) to (3.center);
		\draw[thick] (3.center) to (4d.center);

		\draw[thick] (6d.center) to (7.center);
		\draw[thick] (7.center) to (8u.center);

		\draw[thick] (4r.center) to (5l.center);
		\draw[thick] (5l.center) to (5r.center);
		\draw[thick] (5r.center) to (6l.center);
		
		\draw[thick] (2r.center) to (1l.center);
		\draw[thick] (1l.center) to (1r.center);
		\draw[thick] (1r.center) to (8l.center);

		\coordinate (2md) at ($(2)!0.3!(2d)$);
        \coordinate (2ml) at ($(2)!0.3!(2l)$);
		\coordinate (2hm) at ($(2md)!0.5!(2ml)$);
        
 		\coordinate (4mu) at ($(4)!0.3!(4u)$);
        \coordinate (4ml) at ($(4)!0.3!(4l)$);
		\coordinate (4hm) at ($(4mu)!0.5!(4ml)$);
        
 		\coordinate (6mu) at ($(6)!0.3!(6u)$);
        \coordinate (6mr) at ($(6)!0.3!(6r)$);
		\coordinate (6hm) at ($(6mu)!0.5!(6mr)$);
        
  		\coordinate (8mr) at ($(8)!0.3!(8r)$);
        \coordinate (8md) at ($(8)!0.3!(8d)$);
		\coordinate (8hm) at ($(8mr)!0.5!(8md)$);
        
       \coordinate (1hl) at ($(1l)!0.2!(2r)$);
       	\coordinate (1hr) at ($(1l)!0.15!(1r)$);
		\coordinate (1hm) at ($(1hl)!0.5!(1hr)$);
		
		\coordinate (1hpl) at ($(1r)!0.15!(1l)$);
       	\coordinate (1hpr) at ($(1r)!0.2!(8l)$);
		\coordinate (1hpm) at ($(1hpl)!0.5!(1hpr)$);
		
		\coordinate (r35) at ($(4d)!0.5!(4r)$);
		\coordinate (r57) at ($(6l)!0.5!(6d)$);

        	\coordinate (r13) at ($(2u)!0.5!(2r)$);
        	\coordinate (r13aux) at ($(1l)!0.6!(3)$);
        	\coordinate (r17) at ($(8u)!0.5!(8l)$);
		\coordinate (r17aux) at ($(7)!0.4!(1r)$);
       	
       	\draw[Maroon,very thick] (1hl) to[out=-45,in=-90] (1hr);
       	\draw[Maroon,very thick] (1hpl) to[out=-90,in=-135] (1hpr);
       	
		\draw[Blue, very thick] (r35) to (-3,1) to[out=-45,in=100] (1hm);
		\draw [Blue, very thick] (r57) to (3,1) to[out= -135, in = 80] (1hpm);

		\draw[Blue, very thick] (r13.center) to[out=45, in=-135] (r13aux.center) to[out=45, in=180] (0,0.2) to[out=0,in=135] (r17aux) to[out=-45, in=135] (r17) ;
        \node at ($(2u)+0.5*(-1.3,-0)$) {$A$};
        \node at ($(8u)+0.7*(+1.3,-0.2)$) {$B$};

        \coordinate (2mur) at ($(2u)!0.5!(2r)$);
\end{scope}
\node (marrow) at (8,1) {$\eta_{\mu\nu}\eta^{\mu\nu} = D$};
\draw[->, thick] (6,0) -- (10,0);

\begin{scope}[xshift=18cm, thick, scale = 1.7]

\draw[thick](0,0) circle (0.5);

\coordinate (2c) at (-3,0);
\coordinate (2t) at (-4,1);
\coordinate (2b) at (-4,-1);
\node (r2) at ($(2c)+(-1,0)$) {2}; 

\coordinate (4c) at (3,0);
\coordinate (4t) at (4,1);
\coordinate (4b) at (4,-1);
\node (r4) at ($(4c)+(1,0)$) {4};

\coordinate (3l1) at (-3.5,1.5);
\coordinate (3l2) at (-2.5,0.5);
\coordinate (3l3) at (-1.1,0.5);

\coordinate (3r1) at (3.5,1.5);
\coordinate (3r2) at (2.5,0.5);
\coordinate (3r3) at (1.1,0.5);
\coordinate (3aux) at (0,1.25);

\node (r3) at ($(3aux)+(0,0.5)$) {3};

\coordinate (1l1) at (-3.5,-1.5);
\coordinate (1l2) at (-2.5,-0.5);
\coordinate (1l3) at (-1.1,-0.5);
\coordinate (1r1) at (3.5,-1.5);
\coordinate (1r2) at (2.5,-0.5);
\coordinate (1r3) at (1.1,-0.5);
\coordinate (1aux) at (0,-1.25);

\node (r1) at ($(1aux)+(0,-0.5)$) {1};

\coordinate (2ht) at ($(2c)!0.2!(2t)$);
\coordinate (2hb) at ($(2c)!0.2!(2b)$);
\coordinate (4hb) at ($(4c)!0.2!(4b)$);
\coordinate (4ht) at ($(4c)!0.2!(4t)$);

\coordinate (phrt) at ($0.5*(0.92388, 0.382683)$);
\coordinate (phrb) at ($0.5*(0.92388, -0.382683)$);

\coordinate (phlt) at ($0.5*(-0.92388, 0.382683)$);
\coordinate (phlb) at ($0.5*(-0.92388, -0.382683)$);

\coordinate (c1l) at ($(2ht)!0.5!(2hb)$);
\coordinate (c1r) at ($(phlt)!0.5!(phlb)$);

\coordinate (c2l) at ($(4ht)!0.5!(4hb)$);
\coordinate (c2r) at ($(phrt)!0.5!(phrb)$);

\draw[Maroon, very thick] (2hb) to[out=135, in=-135] (2ht);
\draw[Maroon, very thick] (4hb) to[out=45, in=-45] (4ht);

\draw[Maroon, very thick] (phrt) to[out=-160, in=90] (0.32,0) to[out=-90, in=160] (phrb);
\draw[Maroon, very thick] (phlt) to[out=-20, in=90] (-0.32,0) to[out=-90, in=20] (phlb);

\draw[Blue, very thick] (c1l) to (c1r);
\draw[Blue, very thick] (c2l) to (c2r);

\draw (2c) to (2t);
\draw (2c) to (2b);

\draw (4c) to (4t);
\draw (4c) to (4b);
\draw (3l1) to (3l2);
\draw (3l2) to (3l3);
\draw (3r1) to (3r2);
\draw (3r2) to (3r3);
\draw (3r3) to[out=115, in=0] (3aux);
\draw (3aux) to[out=180, in=65] (3l3);

\draw (1l1) to (1l2);
\draw (1l2) to (1l3);
\draw (1r1) to (1r2);
\draw (1r2) to (1r3);
\draw (1r3) to[out=-115, in=0] (1aux);
\draw (1aux) to[out=180, in=-65] (1l3);
\node at (0,0) {$p$};


\draw[very thick, Blue](0,0) circle (0.85);
\end{scope}
\end{tikzpicture} 
\end{equation}
So once again, just by recasting Lorentz contractions in terms of curves on surfaces we are inevitably led to self-intersecting curves (with precisely one self-intersection) as well as to closed curves associated to the spacetime dimension $D$, exactly the new one-loop ingredients in the surface integral! Recall, however, that in the surface integral the precise exponent of the closed curve is not simply $D$, and, as we will show next, this is precisely because it will receive non-trivial contributions from $\mathcal{N}$ and $\mathcal{N}^\prime$.

Having understood how the naive gluing mimics the tree-level results, the entire focus of this part will be on understanding the correction terms, $\mathcal{N}$ and $\mathcal{N}^\prime$, and see how these can also be interpreted graphically within the established formalism, and in particular how these let us determine the exponents of the closed curves in the surface integral.

\subsection{Corrections to loop-level gluing}
\label{sec:Corrections_LoopGluing}

As it turns out, the corrections to the $\eta_{\mu,\nu}$ gluing, $\mathcal{N}/  \mathcal{N}^\prime$, follow a simple recursive structure. Let us consider a general object obtained from gluing 3-point on-shell amplitudes from which we stripped off the polarization vectors of gluon $A$ (index $\alpha$) and $B$ (index $\beta$), $\mathcal{M}^{\alpha\beta}$, then we want to show that
\begin{equation}
\begin{cases}
   ( p_A)_\alpha\mathcal{M}^{\alpha\beta} = \mathcal{N}_A (p_B)^\beta \, , \\
    \mathcal{M}^{\alpha\beta}(p_B)_\beta = \mathcal{N}_B (p_A)^\alpha, \, 
\end{cases} \label{eq:N-identities}
\end{equation}
so that when we further glue $\alpha$ and $\beta$, we take $p_A^\mu = - p_B^\mu$, and the expression turns into \eqref{eq:loopGlueNLNR} with $\mathcal{N}_A= - \mathcal{N}$ and $\mathcal{N}_B=-\mathcal{N}^\prime$. 
For the 3-point case under the convention in \eqref{eq:3ptGluonScalar},  stripping off the polarization vectors yields
\begin{equation}
    \mathcal{A}_3(1^\alpha, 2^\beta, 3^\gamma) = 2p_{2}^{\alpha} \eta^{\beta,\gamma} + 2p_{3}^{\beta} \eta^{\alpha,\gamma} +2 p_{1}^{\gamma} \eta^{\alpha,\beta} \, ,
\end{equation}
and thus we can verify the identity in \eqref{eq:N-identities} for $A=1$ and $B=2$, 
\begin{equation}
\begin{cases}
    (\epsilon_3)_\gamma(p_1)_\alpha \mathcal{A}_3^{\alpha\beta\gamma} = (\mathcal{A}_3)^{\alpha \beta} (p_1)_\alpha = (2\epsilon_3 \cdot p_1) (- p_2)^\beta\, \\ 
    (\epsilon_3)_\gamma(p_2)_\beta \mathcal{A}_3^{\alpha\beta\gamma} = (\mathcal{A}_3)^{\alpha \beta} (p_2)_\beta  = 0\,  
\end{cases} \, , \label{eq:Nidentities3pt}
\end{equation}
which implies that $\mathcal{N} \equiv -\mathcal{N}_1 = 2\epsilon_3 \cdot p_1$ while $\mathcal{N}^\prime \equiv\mathcal{N}_2 = 0$, and so it is clear that the asymmetry between $\mathcal{N}$ and $\mathcal{N}^\prime$ ties back to choice of the $3$-point in \eqref{eq:3ptGluonScalar}. 

Now by induction, assuming $\mathcal{M}_n$ is an object obtained from on-shell gluing which satisfies \eqref{eq:N-identities}, we can build an object with 1 more external leg, $\mathcal{M}_{n+1}$, by gluing a 3-point to one of the external legs, $i.e.$
\begin{equation}
   \mathcal{M}_{n+1}(A^\alpha,B^\nu) = \mathcal{M}_n(A^\alpha,C^\beta) [-\eta_{\beta \mu}] \mathcal{A}_3(c^\mu,B^\nu) \, , 
\end{equation}
which in turn also satisfies
\begin{equation}
\begin{aligned}
    (p_A)_\alpha \mathcal{M}_{n+1}(A^\alpha,B^\nu) &= (p_A)_\alpha \mathcal{M}_n(A^\alpha,C^\beta) [-\eta_{\beta \mu}] \mathcal{A}_3(c^\mu,B^\nu)|_{p_c=-p_C}  \\
    &= \mathcal{N}_A (-p_C)_\mu \,\mathcal{A}_3(c^\mu,B^\nu)|_{p_c=-p_C} = (\mathcal{N}_A \mathcal{N}_c ) p_B^\nu, \\[6pt]
    (p_B)_\nu \mathcal{M}_{n+1}(A^\alpha,B^\nu) &=  \mathcal{M}_n(A^\alpha,C^\beta)[-\eta_{\beta \mu}]  \mathcal{A}_3(c^\mu, B^\nu) (p_B)_\nu |_{p_c=-p_C}  = 0, 
\end{aligned}
\label{eq:InductiveN}
\end{equation}
proving the desired identity. In addition, from \eqref{eq:InductiveN}, we learn that any object build from on-shell gluing of $3$-point amplitudes satisfies $\mathcal{N}_B=0$, and therefore in \eqref{eq:NCorect1loop}, we have $\mathcal{N}^\prime=0$. As for $\mathcal{N}_A = \mathcal{N} \neq 0$, it can be built by taking the product of all the $\mathcal{N}$ for the lower point on-shell functions, $i.e.$ all the $3$-point $\mathcal{N}$'s.

Quite remarkably, the Lorentz contractions producing $\mathcal{N}$ also follow a simple graphical picture. Starting with the $3$-point case described above, we have  $\mathcal{N} = - \mathcal{N}_1 = 2\epsilon_3 \cdot p_1$ which is simply the Feynman rule for the scalar-scalar-gluon vertex. So if we are gluing leg $1$ of this $3$-point vertex using the spin sum \eqref{eq:LoopGlue1}, the correction term can be interpreted as the following effective contraction

\begin{equation}
    (p_1)_\mu \mathcal{A}_3(1^\mu,2^\nu,3^\rho) =2 p_1^\rho (-p_2)^\nu \quad \Leftrightarrow \quad \vcenter{\hbox{\begin{tikzpicture}[line width=0.6 ,scale=0.35,line cap=round,every node/.style={font=\normalsize}]
		\node (1l) at (-2, -0.6) {};
		\node (1r) at (1, -0.6) {};
		\node (5l) at (-2, 0.6) {};
		\node (5r) at (1, 0.6) {};
		\node (3) at (-3, 0) {};
		\node (7) at (3, 0) {};
		  \coordinate (4r) at ($(5l)+2*(-1,1)$);
        \coordinate (2r) at ($(1l)+2*(-1,-1)$);
        \coordinate (4d) at ($(3)+1.8*(-1,1)$);
        \coordinate (2u) at ($(3)+1.8*(-1,-1)$);

		\draw[thick] (2u.center) to (3.center);
		\draw[thick] (3.center) to (4d.center);
		\draw[thick] (4r.center) to (5l.center);
		\draw[thick] (5l.center) to (5r.center);
		\draw[thick] (2r.center) to (1l.center);
		\draw[thick] (1l.center) to (1r.center);
        \draw[thick,Green] (4d) to (4r);
        \node (g1) at ($(4r)+0.5*(-1,1.5)$) {$1^\mu$};
        \draw[->,Green] ($(4r)!0.5!(4d)+0.5*(-1,1)$) to ($(4r)!0.5!(4d)$);
        \node (g2) at ($(1r)+0.5*(1.2,-0.5)$) {$2^\nu$};
        
        \node (g3) at ($(2r)+0.5*(-1,-1)$) {$3^\rho$};

	\end{tikzpicture} }}=
    \vcenter{\hbox{\begin{tikzpicture}[line width=0.6 ,scale=0.35,line cap=round,every node/.style={font=\normalsize}]
		\node (1l) at (-2, -0.6) {};
		\node (1r) at (1, -0.6) {};
		\node (5l) at (-2, 0.6) {};
		\node (5r) at (1, 0.6) {};
		\node (3) at (-3, 0) {};
		\node (7) at (3, 0) {};
		  \coordinate (4r) at ($(5l)+2*(-1,1)$);
        \coordinate (2r) at ($(1l)+2*(-1,-1)$);
        \coordinate (4d) at ($(3)+1.8*(-1,1)$);
        \coordinate (2u) at ($(3)+1.8*(-1,-1)$);

		\draw[thick] (2u.center) to (3.center);
		\draw[thick] (3.center) to (4d.center);
		\draw[thick] (4r.center) to (5l.center);
		\draw[thick] (5l.center) to (5r.center);
		\draw[thick] (2r.center) to (1l.center);
		\draw[thick] (1l.center) to (1r.center);
        \draw[thick,Green] (4d) to (4r);
        \draw[thick,Green] (5r.center) to (1r.center);

        \node (g1) at ($(4r)+0.5*(-1,1.5)$) {$1^\mu$};
        \node (p2) at ($(5r)+0.5*(1.5,+1)$) {\scriptsize{$(-p_2^\nu)$}};
        \node (g2) at ($(1r)+0.5*(1.2,-0.5)$) {$2^\nu$};
        \draw[->,Green] ($(5r)!0.5!(1r)$) to ($(5r)!0.5!(1r)+0.8*(1,0)$);
        \node (g3) at ($(2r)+0.5*(-1,-1)$) {$3^\rho$};

        \draw[dashed, Green] ($(4d)!0.5!(4r)$) to ($(5l)!0.5!(3)$) to[out=-45, in=180] ($(5l)!0.5!(1l)$) to ($(5r)!0.5!(1r)$);
        \coordinate(hl) at ($(5l)!0.2!(4r)$);
        \coordinate (hr) at ($(5l)!0.2!(5r)$);
        \draw[thick, Maroon] (hl) to[out=45,in=90] (hr) ;
        \draw[Blue, thick] ($(2u)!0.5!(2r)$) to[out=45,in=-120] ($(hl)!0.5!(hr)$);
	\end{tikzpicture} }} 
    \label{eq:NGraphRule}
\end{equation}
so that the polarization of the leg on the \textit{left} of $1$, leg $3$, is contracted with $p_1^\mu$ (as represented by the blue line + red handle); the new contraction curve represented in dashed green goes along the leg to the \textit{right} of $1$ and is proportional to its own momentum, $(-p_2)^\mu$ -- so if it contracts with an on-shell amplitude we get zero from on-shell gauge-invariance.

We now proceed to compute the bubble leading singularity from explicit gluing, so that we can see what the $\mathcal{N}$ correction looks like in this case. This will also let us illustrate how using the rule above we can trivially read off $\mathcal{N}$ in terms of the loop scalar variables $X_{i,j}$ just like we do for the rest of the gluing. From this one-loop analysis, it will also be clear under what conditions $\mathcal{N}$ vanishes. These conditions are general and hold at arbitrary loop order, and are therefore crucial to determine the correct exponents of the closed curves.

\paragraph{The bubble LS: from gluing to the surface integral}

We can obtain the bubble LS by gluing legs $A$ and $B$ of the 4-point $t$-channel LS. In this way, taking \eqref{eq:loopGlueNLNR}  with $ q_A^\mu = - q_B^\mu = l^\mu$, in addition to the naive gluing we get  $\mathcal{N}_A$ and $\mathcal{N}_B$ (where $\mathcal{N}_B$ should vanish), which we can compute using the graphical rule as follows
\begin{equation}
\begin{aligned}
    \mathcal{N}_A &: 
    \begin{tikzpicture}[line width=0.6 ,scale=0.3,line cap=round,every node/.style={font=\normalsize},baseline=(O.base)]
	\begin{scope}[scale = 0.8]
        \node (O) at (0,0) {};
		\node (1l) at (-2, -1) {};
		\node (1r) at (2, -1) {};
		\node (5l) at (-2, 1) {};
		\node (5r) at (2, 1) {};
		
		\node (2) at (-5, -3) {};
		\node (2d) at (-5, -5) {};
		\node (2l) at (-7, -3) {};
		\node (2r) at (-4, -3) {};
		\node (2dr) at (-4, -5) {};
		\node (2u) at (-5, -2) {};
		\node (2lu) at (-7, -2) {};

		\node (4) at (-5, 3) {};
		\node (4u) at (-5, 5) {};
		\node (4l) at (-7, 3) {};
		\node (4r) at (-4, 3) {};
		\node (4ur) at (-4, 5) {};
		\node (4d) at (-5, 2) {};
		\node (4ld) at (-7, 2) {};

		\node (6) at (5, 3) {};
		\node (6u) at (5, 5) {};
		\node (6r) at (7, 3) {};
		\node (6l) at (4, 3) {};
		\node (6ul) at (4, 5) {};
		\node (6d) at (5, 2) {};
		\node (6dr) at (7, 2) {};
		
		\node (8) at (5, -3) {A};
		\node (8r) at (7, -3) {};
		\node (8d) at (5, -5) {};
		\node (8dl) at (4, -5) {};
		\node (8l) at (4, -3) {};
		\node (8ur) at (7, -2) {};
		\node (8u) at (5, -2) {};;
		
		\node (3) at (-3, 0) {};
		\node (7) at (3, 0) {};

		\draw[thick] (2u.center) to (3.center);
		\draw[thick] (3.center) to (4d.center);

		\draw[thick] (6d.center) to (7.center);
		\draw[thick] (7.center) to (8u.center);

		\draw[thick] (4r.center) to (5l.center);
		\draw[thick] (5l.center) to (5r.center);
		\draw[thick] (5r.center) to (6l.center);
		
		\draw[thick] (2r.center) to (1l.center);
		\draw[thick] (1l.center) to (1r.center);
		\draw[thick] (1r.center) to (8l.center);

		\coordinate (2md) at ($(2)!0.3!(2d)$);
        \coordinate (2ml) at ($(2)!0.3!(2l)$);
		\coordinate (2hm) at ($(2md)!0.5!(2ml)$);
        
 		\coordinate (4mu) at ($(4)!0.3!(4u)$);
        \coordinate (4ml) at ($(4)!0.3!(4l)$);
		\coordinate (4hm) at ($(4mu)!0.5!(4ml)$);
        
 		\coordinate (6mu) at ($(6)!0.3!(6u)$);
        \coordinate (6mr) at ($(6)!0.3!(6r)$);
		\coordinate (6hm) at ($(6mu)!0.5!(6mr)$);
        
  		\coordinate (8mr) at ($(8)!0.3!(8r)$);
        \coordinate (8md) at ($(8)!0.3!(8d)$);
		\coordinate (8hm) at ($(8mr)!0.5!(8md)$);
        
       \coordinate (1hl) at ($(1l)!0.2!(2r)$);
       	\coordinate (1hr) at ($(1l)!0.15!(1r)$);
		\coordinate (1hm) at ($(1hl)!0.5!(1hr)$);
		
		\coordinate (1hpl) at ($(1r)!0.15!(1l)$);
       	\coordinate (1hpr) at ($(1r)!0.2!(8l)$);
		\coordinate (1hpm) at ($(1hpl)!0.5!(1hpr)$);
		
		\coordinate (r35) at ($(4d)!0.5!(4r)$);
		\coordinate (r57) at ($(6l)!0.5!(6d)$);

        	\coordinate (r13) at ($(2u)!0.5!(2r)$);
        	\coordinate (r13aux) at ($(1l)!0.6!(3)$);
        	\coordinate (r17) at ($(8u)!0.5!(8l)$);
		\coordinate (r17aux) at ($(7)!0.4!(1r)$);
       	
       	\draw[Maroon,very thick] (1hl) to[out=-45,in=-90] (1hr);
       	\draw[Maroon,very thick] (1hpl) to[out=-90,in=-135] (1hpr);
       	
		\draw[Blue, very thick] (r35) to (-3,1) to[out=-45,in=100] (1hm);
		\draw [Blue, very thick] (r57) to (3,1) to[out= -135, in = 80] (1hpm);

		\draw[Green, very thick, dashed] (r13.center) to[out=45, in=-135] (r13aux.center) to[out=45, in=180] (0,0.2) to[out=0,in=135] (r17aux) to[out=-45, in=135] (r17) ;
        \node at ($(2u)+0.5*(-1.3,-0)$) {B};
        \draw[Green, thick] (2u.center) to (2r.center);
        \draw[Green, thick] (8u.center) to (8l.center);
        \coordinate (2mur) at ($(2u)!0.5!(2r)$);
        \draw[Green, ->, thick] (2mur) to ($(2mur)+(-1,-1)$);
        \node (p4) at ($(2r)+(0.3,-1)$) {\scriptsize{$(-p_B^\mu)$}};
\end{scope}
\draw[->, thick] (4.5,0) -- (7.5,0);

\begin{scope}[xshift=16cm, thick, scale = 1.7]

\draw[thick](0,0) circle (0.5);

\coordinate (2c) at (-3,0);
\coordinate (2t) at (-4,1);
\coordinate (2b) at (-4,-1);
\node (r2) at ($(2c)+(-1,0)$) {2}; 

\coordinate (4c) at (3,0);
\coordinate (4t) at (4,1);
\coordinate (4b) at (4,-1);
\node (r4) at ($(4c)+(1,0)$) {4};

\coordinate (3l1) at (-3.5,1.5);
\coordinate (3l2) at (-2.5,0.5);
\coordinate (3l3) at (-1.1,0.5);

\coordinate (3r1) at (3.5,1.5);
\coordinate (3r2) at (2.5,0.5);
\coordinate (3r3) at (1.1,0.5);
\coordinate (3aux) at (0,1.25);

\node (r3) at ($(3aux)+(0,0.5)$) {3};

\coordinate (1l1) at (-3.5,-1.5);
\coordinate (1l2) at (-2.5,-0.5);
\coordinate (1l3) at (-1.1,-0.5);
\coordinate (1r1) at (3.5,-1.5);
\coordinate (1r2) at (2.5,-0.5);
\coordinate (1r3) at (1.1,-0.5);
\coordinate (1aux) at (0,-1.25);

\node (r1) at ($(1aux)+(0,-0.5)$) {1};

\coordinate (2ht) at ($(2c)!0.2!(2t)$);
\coordinate (2hb) at ($(2c)!0.2!(2b)$);
\coordinate (4hb) at ($(4c)!0.2!(4b)$);
\coordinate (4ht) at ($(4c)!0.2!(4t)$);

\coordinate (phrt) at ($0.5*(0.92388, 0.382683)$);
\coordinate (phrb) at ($0.5*(0.92388, -0.382683)$);

\coordinate (phlt) at ($0.5*(-0.92388, 0.382683)$);
\coordinate (phlb) at ($0.5*(-0.92388, -0.382683)$);

\coordinate (c1l) at ($(2ht)!0.5!(2hb)$);
\coordinate (c1r) at ($(phlt)!0.5!(phlb)$);

\coordinate (c2l) at ($(4ht)!0.5!(4hb)$);
\coordinate (c2r) at ($(phrt)!0.5!(phrb)$);

\draw[Maroon, very thick] (2hb) to[out=135, in=-135] (2ht);
\draw[Maroon, very thick] (4hb) to[out=45, in=-45] (4ht);

\draw[Maroon, very thick] (phrt) to[out=-160, in=90] (0.32,0) to[out=-90, in=160] (phrb);
\draw[Maroon, very thick] (phlt) to[out=-20, in=90] (-0.32,0) to[out=-90, in=20] (phlb);

\draw[Blue, very thick] (c1l) to (c1r);
\draw[Blue, very thick] (c2l) to (c2r);

\draw (2c) to (2t);
\draw (2c) to (2b);

\draw (4c) to (4t);
\draw (4c) to (4b);
\draw (3l1) to (3l2);
\draw (3l2) to (3l3);
\draw (3r1) to (3r2);
\draw (3r2) to (3r3);
\draw (3r3) to[out=115, in=0] (3aux);
\draw (3aux) to[out=180, in=65] (3l3);

\draw (1l1) to (1l2);
\draw (1l2) to (1l3);
\draw (1r1) to (1r2);
\draw (1r2) to (1r3);
\draw (1r3) to[out=-115, in=0] (1aux);
\draw (1aux) to[out=180, in=-65] (1l3);
\node at (0,0) {$p$};

\draw[->,thick] ($(3l2)+(0.2,0.2)$) -- ($(3l3)+(-0.2,0.2)$) node[midway, above, color=black] {$q_1^\mu$};
\draw[<-,thick] ($1.5*({cos(-150)},{sin(-150)})$) arc (-150:-115:1.5);
\node at ($1.9*({cos(-130)},{sin(-130)})$) {$\ell^\mu$};


\draw[thick, dashed, Green](0,0) circle (0.85);
\end{scope}
	\end{tikzpicture} \, \propto (\epsilon_1\cdot \ell)(\epsilon_2\cdot (q_1+\ell))
 \\
    \mathcal{N}_B &:  \begin{tikzpicture}[line width=0.6 ,scale=0.25,line cap=round,every node/.style={font=\normalsize}, baseline=(O.base)]
            \node (O) at (0,0) {};
		\node (1l) at (-2, -1) {};
		\node (1r) at (2, -1) {};
		\node (5l) at (-2, 1) {};
		\node (5r) at (2, 1) {};
		
		\node (2) at (-5, -3) {};
		\node (2d) at (-5, -5) {};
		\node (2l) at (-7, -3) {};
		\node (2r) at (-4, -3) {};
		\node (2dr) at (-4, -5) {};
		\node (2u) at (-5, -2) {};
		\node (2lu) at (-7, -2) {};

		\node (4) at (-5, 3) {};
		\node (4u) at (-5, 5) {};
		\node (4l) at (-7, 3) {};
		\node (4r) at (-4, 3) {};
		\node (4ur) at (-4, 5) {};
		\node (4d) at (-5, 2) {};
		\node (4ld) at (-7, 2) {};

		\node (6) at (5, 3) {};
		\node (6u) at (5, 5) {};
		\node (6r) at (7, 3) {};
		\node (6l) at (4, 3) {};
		\node (6ul) at (4, 5) {};
		\node (6d) at (5, 2) {};
		\node (6dr) at (7, 2) {};
		
		\node (8) at (5, -3) {A};
		\node (8r) at (7, -3) {};
		\node (8d) at (5, -5) {};
		\node (8dl) at (4, -5) {};
		\node (8l) at (4, -3) {};
		\node (8ur) at (7, -2) {};
		\node (8u) at (5, -2) {};;
		
		\node (3) at (-3, 0) {};
		\node (7) at (3, 0) {};

		\draw[thick] (2u.center) to (3.center);
		\draw[thick] (3.center) to (4d.center);

		\draw[thick] (6d.center) to (7.center);
		\draw[thick] (7.center) to (8u.center);

		\draw[thick] (4r.center) to (5l.center);
		\draw[thick] (5l.center) to (5r.center);
		\draw[thick] (5r.center) to (6l.center);
		
		\draw[thick] (2r.center) to (1l.center);
		\draw[thick] (1l.center) to (1r.center);
		\draw[thick] (1r.center) to (8l.center);

		\coordinate (2md) at ($(2)!0.3!(2d)$);
        \coordinate (2ml) at ($(2)!0.3!(2l)$);
		\coordinate (2hm) at ($(2md)!0.5!(2ml)$);
        
 		\coordinate (4mu) at ($(4)!0.3!(4u)$);
        \coordinate (4ml) at ($(4)!0.3!(4l)$);
		\coordinate (4hm) at ($(4mu)!0.5!(4ml)$);
        
 		\coordinate (6mu) at ($(6)!0.3!(6u)$);
        \coordinate (6mr) at ($(6)!0.3!(6r)$);
		\coordinate (6hm) at ($(6mu)!0.5!(6mr)$);
        
  		\coordinate (8mr) at ($(8)!0.3!(8r)$);
        \coordinate (8md) at ($(8)!0.3!(8d)$);
		\coordinate (8hm) at ($(8mr)!0.5!(8md)$);
        
        \coordinate (5hl) at ($(5l)!0.2!(4r)$);
       	\coordinate (5hr) at ($(5l)!0.15!(5r)$);
		\coordinate (5hm) at ($(5hl)!0.5!(5hr)$);

        \coordinate (1hr) at ($(1r)!0.2!(8l)$);
       	\coordinate (1hl) at ($(1r)!0.15!(1l)$);
		\coordinate (1hm) at ($(1hl)!0.5!(1hr)$);

		\coordinate (5hrp) at ($(5r)!0.2!(6l)$);
       	\coordinate (5hlp) at ($(5r)!0.15!(5l)$);
		\coordinate (5hmp) at ($(5hlp)!0.5!(5hrp)$);

        \coordinate (1hlp) at ($(1l)!0.2!(2r)$);
       	\coordinate (1hrp) at ($(1l)!0.15!(1r)$);
		\coordinate (1hmp) at ($(1hlp)!0.5!(1hrp)$);
		
		\coordinate (3hu) at ($(3)!0.25!(4d)$);
       	\coordinate (3hd) at ($(3)!0.25!(2u)$);
		\coordinate (3hm) at ($(3hu)!0.5!(3hd)$);
		
		\coordinate (7hu) at ($(7)!0.25!(6d)$);
       	\coordinate (7hd) at ($(7)!0.25!(8u)$);
		\coordinate (7hm) at ($(7hu)!0.5!(7hd)$);
        	
        	\coordinate (r13) at ($(2u)!0.5!(2r)$);
        	\coordinate (r35) at ($(4d)!0.5!(4r)$);
        	\coordinate (r13aux) at ($(1l)!0.5!(3)$);
        	\coordinate (r35aux) at ($(3)!0.5!(5l)$);

		\coordinate (3ht) at ($(3)!0.2!(4d)$);
		\coordinate (3hb) at ($(3)!0.2!(2u)$);
		\coordinate (3hm) at ($(3ht)!0.5!(3hb)$);
		\coordinate (caux) at ($(5r)!0.5!(1r)$);
		\draw[Green, very thick, dashed] (r13) to[out=45, in=-135] (r13aux) to[out=45, in=-45] (r35aux) to[out=135, in=-45] (r35) ;
		\draw[Maroon, very thick] (3ht) to[out=-135,in=135] (3hb);
		\draw[Blue, very thick] (3hm) to (caux);
        \draw[Green, thick] (4d.center) to (4r.center);
        \coordinate (4dmr) at ($(4d)!0.5!(4r)$);
        \node at ($(2u)!0.5!(2r)+0.7*(-1,-1)$) {$B$};

        \draw[Green, ->, thick] (4dmr.center) to ($(4dmr)+(-1,1)$);
        \node[Green] at ($(4dmr)+(1,1.3)$) {\scriptsize $(-q_1^\mu)$};
        \draw[Green, thick] (2u.center) to (2r.center);
	\end{tikzpicture} \propto q_1.\epsilon_1  = 0 
\end{aligned}
\label{eq:N3}
\end{equation}
where for $\mathcal{N}_A$ we show the final contraction curve when we scalar-scaffold gluons $1$ and $2$.
So we obtain, as expected, $\mathcal{N}_A \neq 0$,  $\mathcal{N}_B=0$. So we see that the non-zero contribution comes from the green contraction curve that connects $A$ to $B$ by turning \textit{always} left. If such a path does not exist, then by turning consistently left with the green dashed curve (as dictated by the rule \eqref{eq:NGraphRule}), we eventually reach an external edge and the contribution vanishes due to on-shell gauge-invariance. So in general, we have:
\begin{center}
    \textit{When gluing two legs of the same on-shell object, the $\mathcal{N}$ correction is non-vanishing if and only if there is an exclusively left-turning path connecting the two legs.}
\end{center} 

By construction, such a path will always be one that forms a closed curve around the loop, and thus the $\mathcal{N}$ corrections to the loop gluing will simply redefine the factor that we associate with the closed curves! This is the monomials entering in $\mathcal{N}$ are the same as those that in the naive gluing come with a factor of $D$,  signalling the presence of the closed contraction curve. To see this explicitly, let's compute the full gluing in scalar-scaffolding, by adding $\mathcal{N}_A$ to the $\eta_{\mu,\nu}$ piece,
\begin{equation}
    \text{LS}_{\bub} = -\text{LS}_4(A^\mu,B^\nu)(\eta_{\mu\nu}) + \mathcal{N}_A \, ,
\end{equation}
where from \eqref{eq:N3}, we have that $\mathcal{N}_A =X_{2,p}X_{4,p}$; and the $\eta_{\mu,\nu}$ piece we compute by looking at all possible ways of filling each road of the graph with cores and extensions. Recall from Sec. \ref{sec:examples_LS_SurfInt} that for this 2-point process, the only non-vanishing invariants are $X_{2,4}$, $X_{4,2}$, $X_{2,p}$ and $X_{4,p}$, since the tadpoles $X_{i,i}=0$ in physical kinematics. 

Using exactly the same analysis as in Sec. \ref{sec:examples_LS_SurfInt}, we see that starting  with the core of $X_{2,4}$/$X_{4,2}$, to fill the graph we can only  add the core of $X_{3,3}$/$X_{1,1}$, respectively. And therefore, $X_{2,4}$/$X_{4,2}$ do not enter LS$_{\bub}$. Therefore, we are left with only one combination: $X_{2,p}X_{4,p}$. These can fully cover the graph if we consider their extensions, or we can keep only their cores but add the  closed curve, which captures the contraction giving $-\eta_{\mu\nu} \eta^{\mu\nu} = -D$. Let us focus on the latter, in this case the full contraction picture looks exactly like that of $\mathcal{N}_A$ in \eqref{eq:N3}. Adding these two pieces together we find
\begin{equation}
    \text{LS}^{\text{closed}}_{\bub} = X_{2,p}X_{4,p}(1-D) = 	\begin{tikzpicture}[scale = 0.4, baseline=(O.base)]
            \node (O) at (6,0) {$+$};
		
    \begin{scope}[thick, scale = 1.2]

\draw[thick](0,0) circle (0.5);

\coordinate (2c) at (-3,0);
\coordinate (2t) at (-4,1);
\coordinate (2b) at (-4,-1);
\coordinate (4c) at (3,0);
\coordinate (4t) at (4,1);
\coordinate (4b) at (4,-1);

\coordinate (3l1) at (-3.5,1.5);
\coordinate (3l2) at (-2.5,0.5);
\coordinate (3l3) at (-1.1,0.5);

\coordinate (3r1) at (3.5,1.5);
\coordinate (3r2) at (2.5,0.5);
\coordinate (3r3) at (1.1,0.5);
\coordinate (3aux) at (0,1.25);

\coordinate (1l1) at (-3.5,-1.5);
\coordinate (1l2) at (-2.5,-0.5);
\coordinate (1l3) at (-1.1,-0.5);
\coordinate (1r1) at (3.5,-1.5);
\coordinate (1r2) at (2.5,-0.5);
\coordinate (1r3) at (1.1,-0.5);
\coordinate (1aux) at (0,-1.25);

\coordinate (2ht) at ($(2c)!0.2!(2t)$);
\coordinate (2hb) at ($(2c)!0.2!(2b)$);
\coordinate (4hb) at ($(4c)!0.2!(4b)$);
\coordinate (4ht) at ($(4c)!0.2!(4t)$);

\coordinate (phrt) at ($0.5*(0.92388, 0.382683)$);
\coordinate (phrb) at ($0.5*(0.92388, -0.382683)$);

\coordinate (phlt) at ($0.5*(-0.92388, 0.382683)$);
\coordinate (phlb) at ($0.5*(-0.92388, -0.382683)$);

\coordinate (c1l) at ($(2ht)!0.5!(2hb)$);
\coordinate (c1r) at ($(phlt)!0.5!(phlb)$);

\coordinate (c2l) at ($(4ht)!0.5!(4hb)$);
\coordinate (c2r) at ($(phrt)!0.5!(phrb)$);

\draw[Maroon, very thick] (2hb) to[out=135, in=-135] (2ht);
\draw[Maroon, very thick] (4hb) to[out=45, in=-45] (4ht);

\draw[Maroon, very thick] (phrt) to[out=-160, in=90] (0.32,0) to[out=-90, in=160] (phrb);
\draw[Maroon, very thick] (phlt) to[out=-20, in=90] (-0.32,0) to[out=-90, in=20] (phlb);

\draw[Blue, very thick] (c1l) to (c1r);
\draw[Blue, very thick] (c2l) to (c2r);

\draw (2c) to (2t);
\draw (2c) to (2b);

\draw (4c) to (4t);
\draw (4c) to (4b);
\draw (3l1) to (3l2);
\draw (3l2) to (3l3);
\draw (3r1) to (3r2);
\draw (3r2) to (3r3);
\draw (3r3) to[out=115, in=0] (3aux);
\draw (3aux) to[out=180, in=65] (3l3);

\draw (1l1) to (1l2);
\draw (1l2) to (1l3);
\draw (1r1) to (1r2);
\draw (1r2) to (1r3);
\draw (1r3) to[out=-115, in=0] (1aux);
\draw (1aux) to[out=180, in=-65] (1l3);


\draw[thick, dashed, Green](0,0) circle (0.85);
\end{scope}
\begin{scope}[xshift=12cm, thick, scale = 1.2]

\draw[thick](0,0) circle (0.5);

\coordinate (2c) at (-3,0);
\coordinate (2t) at (-4,1);
\coordinate (2b) at (-4,-1);
\coordinate (4c) at (3,0);
\coordinate (4t) at (4,1);
\coordinate (4b) at (4,-1);

\coordinate (3l1) at (-3.5,1.5);
\coordinate (3l2) at (-2.5,0.5);
\coordinate (3l3) at (-1.1,0.5);

\coordinate (3r1) at (3.5,1.5);
\coordinate (3r2) at (2.5,0.5);
\coordinate (3r3) at (1.1,0.5);
\coordinate (3aux) at (0,1.25);

\coordinate (1l1) at (-3.5,-1.5);
\coordinate (1l2) at (-2.5,-0.5);
\coordinate (1l3) at (-1.1,-0.5);
\coordinate (1r1) at (3.5,-1.5);
\coordinate (1r2) at (2.5,-0.5);
\coordinate (1r3) at (1.1,-0.5);
\coordinate (1aux) at (0,-1.25);

\coordinate (2ht) at ($(2c)!0.2!(2t)$);
\coordinate (2hb) at ($(2c)!0.2!(2b)$);
\coordinate (4hb) at ($(4c)!0.2!(4b)$);
\coordinate (4ht) at ($(4c)!0.2!(4t)$);

\coordinate (phrt) at ($0.5*(0.92388, 0.382683)$);
\coordinate (phrb) at ($0.5*(0.92388, -0.382683)$);

\coordinate (phlt) at ($0.5*(-0.92388, 0.382683)$);
\coordinate (phlb) at ($0.5*(-0.92388, -0.382683)$);

\coordinate (c1l) at ($(2ht)!0.5!(2hb)$);
\coordinate (c1r) at ($(phlt)!0.5!(phlb)$);

\coordinate (c2l) at ($(4ht)!0.5!(4hb)$);
\coordinate (c2r) at ($(phrt)!0.5!(phrb)$);

\draw[Maroon, very thick] (2hb) to[out=135, in=-135] (2ht);
\draw[Maroon, very thick] (4hb) to[out=45, in=-45] (4ht);

\draw[Maroon, very thick] (phrt) to[out=-160, in=90] (0.32,0) to[out=-90, in=160] (phrb);
\draw[Maroon, very thick] (phlt) to[out=-20, in=90] (-0.32,0) to[out=-90, in=20] (phlb);

\draw[Blue, very thick] (c1l) to (c1r);
\draw[Blue, very thick] (c2l) to (c2r);

\draw (2c) to (2t);
\draw (2c) to (2b);

\draw (4c) to (4t);
\draw (4c) to (4b);
\draw (3l1) to (3l2);
\draw (3l2) to (3l3);
\draw (3r1) to (3r2);
\draw (3r2) to (3r3);
\draw (3r3) to[out=115, in=0] (3aux);
\draw (3aux) to[out=180, in=65] (3l3);

\draw (1l1) to (1l2);
\draw (1l2) to (1l3);
\draw (1r1) to (1r2);
\draw (1r2) to (1r3);
\draw (1r3) to[out=-115, in=0] (1aux);
\draw (1aux) to[out=180, in=-65] (1l3);


\draw[very thick, Blue](0,0) circle (0.85);

\end{scope}
\end{tikzpicture}\, .
\end{equation}
Matching this to the piece of the LS from the surface integral proportional to the closed curve exponent $\Delta$, let us learn that $\Delta=(1-D)$, exactly as expected!

Finally for the full $\text{LS}_{\bub}$ we also get contributions \textit{without} the closed contraction curve, where instead we fill the graph from 
extensions of $X_{2,p}$ and $X_{4,p}$. There are three possible ways of doing the extensions -- we can extend either core to fill the whole loop (two possibilities with $N_e=1$) or we can extend them both once (one possibility with $N_e=2$), so that the final result reads
\begin{equation}
    \text{LS}_{\bub} = X_{2,p} X_{4,p} (\underbracket[0.4pt]{(1-D)}_{\text{closed}} - \underbracket[0.4pt]{((-1)^2 + (-1)^1 + (-1)^1)}_{\text{extensions}} ) =(2-D)X_{2,p}X_{4,p} \, ,
    \label{eq:LSBubbleQFT}
\end{equation}
note that once again the picture of the extension precisely agrees with the different $\eta_{2i}$ terms in \eqref{eq:bubble_ext}, and we get exactly the same result!

Even without asking for it, the core picture for contractions at loop-level automatically creates a distinction between extending cores around the loop and the contribution from closed-curves -- allowing for the perfect matching of the two pictures!

\subsection{$\Delta$'s at 1-loop}

From the previous discussion it is clear that since the monomials generated by the correction $\mathcal{N}$ are exactly those multiplying $D$, all this correction does is to effectively shift $D$ as follows: 
\begin{equation}
\begin{aligned}
    \text{LS}_{\text{1-loop}} &= \mathcal{N} +  (-D)\left[\sum_{M_C} c_{M_C} \prod_{i\in M_C} X_i\right] +\sum_M c_M\prod_{i\in M} X_i, \\
    &=  (\widetilde{\mathcal{N}}-D)\left[\sum_{M_C} c_{M_C} \prod_{i\in M_C} X_i\right] +\sum_M c_M\prod_{i\in M} X_i ,
\end{aligned}
 \label{eq:LS1loopQFT}
\end{equation}
where $M_C$ stands for the set of monomials containing the closed contraction curve, while $M$ are those that don't. $c_{M}/c_{M_C}$ are the $c$-numbers corresponding to the coefficient of each monomial, and $\widetilde{\mathcal{N}}$ is simply the overall numerical coefficient multiplying the monomials in the correction terms $\mathcal{N}$. 

Note that, from the point of view of the final gluing result, this separation of monomials might seem somewhat artificial -- for example, already in the bubble LS, we can see 
that some of the monomials appearing in $M_C$ also enter in $M$. However, when we think of the contractions in this graphical way, via cores and extensions this separation is completely meaningful and so are the coefficients/signs -- this is what lets us join the $\mathcal{N}$ and $(-D)$ pieces. Most importantly this picture is also what let's us perfectly match this gluing result to the maximal residue of the surface integral, which we can also write as follows:
\begin{equation}
    \text{LS}_{\text{1-loop}} = \Delta_1 \left[\sum_{M_C} c_{M_C} \prod_{i\in M_C} X_i\right] +\sum_M c_M\prod_{i\in M} X_i \, ,
    \label{eq:LS1loopGen}
\end{equation}
where $\Delta_1$ is the exponent of the closed curve. And since we know that not only the monomials but also the coefficients $c_{M_C}$ and $c_M$ are the same in \eqref{eq:LS1loopGen} and in \eqref{eq:LS1loopQFT}, we can use this to fix $\Delta_1$ to be 
\begin{empheq}[box=\boxed]{equation}
 \Delta_1 = \widetilde{\mathcal{N}} - D 
    \label{eq:delta1loop}
\end{empheq}

For the bubble we saw that $\widetilde{\mathcal{N}} =1$ and that gave us the expected $\Delta_1=1-D$. So from here we learn that to determine the exponent of the closed curve at one-loop, one just needs to evaluate the $\mathcal{N}$. 

\paragraph{$\Delta$ for planar 1-loop:  the n-Gon LS}
In a 1-loop planar LS, the most general structure we can have is the n-Gon LS, which we can obtain from gluing a $(n+2)$-gluon tree LS. As it was previously explained, we can compute $\mathcal{N}$ by applying the graphical rule to the legs being glued together:
\begin{equation}
    \begin{tikzpicture}[line width=0.6 ,scale=0.4,line cap=round,every node/.style={font=\normalsize},baseline=(O.base)]
	\begin{scope}[xshift=-10cm,scale = 0.6]
        \node (O) at (0,0) {};
		\coordinate (r1c) at (-9,0);
        \coordinate (3cl) at ($(r1c)+1.5*({cos(30)},{sin(30)})$);
        \coordinate (2np5l) at ($(r1c)+1.5*({cos(30)},-{sin(30)})$);
        \coordinate (1u) at ($(r1c)+2*(-1,1)$);        
        \coordinate (1d) at ($(r1c)+2*(-1,-1)$);
        \coordinate (3l) at ($(3cl)+2.2*(-1,1)$);
        \coordinate (3u) at ($(3l)+2*(0,1)$);
        \coordinate (1ur) at ($(1u)+2*(-1,0)$);
        \coordinate (2c) at ($(1u) + (0,1)$);
        \coordinate (2r) at ($(2c) + 2*(-1,0)$);        
        \coordinate (2u) at ($(2c) + 2*(0,1)$);
        \coordinate (2np3c) at (7,0);
        \coordinate (2np5ld) at ($(2np5l)+2.2*(-1,-1)$);
        \coordinate (3cr) at ($(3cl) + 1.5*(1,0)$);
        \coordinate (3cru) at ($(3cr) + 2.2*(0,1)$);
        \coordinate (2np3c) at (+9,0);
        \coordinate (2np1cr) at ($(2np3c)+1.5*(-{cos(30)},{sin(30)})$);
        \coordinate (2np5r) at ($(2np3c)+1.5*(-{cos(30)},-{sin(30)})$);
        \coordinate (2np3u) at ($(2np3c)+2*(1,1)$);        
        \coordinate (2np3d) at ($(2np3c)+2*(1,-1)$);
        \coordinate (2np1r) at ($(2np1cr)+2.2*(1,1)$);
        \coordinate (2np1u) at ($(2np1r)+2*(0,1)$);
        \coordinate (2np3ur) at ($(2np3u)+2*(1,0)$);
        \coordinate (2np2c) at ($(2np3u) + (0,1)$);
        \coordinate (2np2r) at ($(2np2c) + 2*(1,0)$);        
        \coordinate (2np2u) at ($(2np2c) + 2*(0,1)$);
        \coordinate (2np5rd) at ($(2np5r)+2.2*(1,-1)$);
        \coordinate (2np1l) at ($(2np1cr) + 1.5*(-1,0)$);
        \coordinate (2np1lu) at ($(2np1l) + 2.2*(0,1)$);
        \draw[very thick] (1d) to (r1c) to (1u) to (1ur);
        \draw[very thick] (3cru) to (3cr) to (3cl) to (3l) to (3u);
        \draw [very thick] (2r) to (2c) to (2u);
        \draw[very thick] (2np5ld) to (2np5l);
        \draw[very thick] (2np3d) to (2np3c) to (2np3u) to (2np3ur);
        \draw[very thick] (2np1lu) to (2np1l) to (2np1cr) to (2np1r) to (2np1u);
        \draw [very thick] (2np2r) to (2np2c) to (2np2u);
        \draw[very thick] (2np5rd) to (2np5r);
        \draw[very thick] (2np5l) to (2np5r);
        \node (dotsl) at ($(3cr)+(1.5,0)$) {\dots};
        \node (dotsr) at ($(2np1l)+(-1.5,0)$) {\dots};
        \coordinate (2km1r) at ($(-0.7,{1.5*sin(30)})$);
        \coordinate (2km1l) at ($(2km1r)+2.2*(-1,0)$);
        \coordinate (2km1lu) at ($(2km1l) + 2.2*(0,1)$);
        \coordinate (2km1ru) at ($(2km1r)+2.2*(0,1)$);
        \coordinate (2km1rdi) at ($(2km1ru)+1.8*(-1,1)$);
        \draw [very thick] (2km1lu) to (2km1l) to (2km1r) to (2km1ru) to (2km1rdi);
        \coordinate (2kp1l) at ($(0.7,{1.5*sin(30)})$);
        \coordinate (2kp1r) at ($(2kp1l)+2.2*(1,0)$);
        \coordinate (2kp1ru) at ($(2kp1r) + 2.2*(0,1)$);
        \coordinate (2kp1lu) at ($(2kp1l)+2.2*(0,1)$);
        \coordinate (2kp1ldi) at ($(2kp1lu)+1.8*(1,1)$);
        \draw [very thick] (2kp1ru) to (2kp1r) to (2kp1l) to (2kp1lu) to (2kp1ldi);
        \coordinate (2kc) at ($(2km1ru)+1*({cos(60)},{sin(60)})$);
        \coordinate (2kl) at ($(2kc)+1.5*(-1,1)$);
        \coordinate (2kr) at ($(2kc)+1.5*(1,1)$);
        \draw[very thick] (2kl) to (2kc) to (2kr);

        \draw[very thick, Green, dashed] ($(2np5ld)!0.5!(1d)$) to[out=45,in=180] ($(2np5l)!0.5!(3cl)$) to[out=0,in=180] ($(2np5r)!0.5!(2np1cr)$) to[out=-10,in=135] ($(2np5rd)!0.5!(2np3d)$);
        \draw[<->,dashed] ($(2np5ld) + 0.5*(-1,-1)$) to[out=-45, in=-135] ($(2np5rd) + 0.5*(1,-1)$); 
        \node at (0,-7) {\small glue};
    \end{scope}
    \draw[thick, ->] (-1.7,0) -- (1,0);
    \begin{scope}[xshift= 7cm, scale=0.6, rotate = 90]
        \draw[very thick](0,0) circle (3);
        \draw[very thick, Green, dashed](0,0) circle (3.75);

        \coordinate (r1lauxvec) at ($({cos(240)},{sin(240)})$);
        \draw[very thick] ($4.5*({cos(306)},{sin(306)})$) arc (-54:-20:4.5) ;
        \draw[very thick] ($4.5*({cos(20)},{sin(20)})$) arc (20:84:4.5) ;

        \draw[very thick] ($4.5*({cos(96)},{sin(96)})$) arc (96:160:4.5);
        \draw[very thick] ($4.5*({cos(200)},{sin(200)})$) arc (200:234:4.5);
        \draw[very thick] ($4.5*({cos(246)},{sin(246)})$) arc (246:294:4.5);
        \draw[very thick, decoration={text along path,
       text={...},text align={center}},decorate] ($4.5*({cos(-20)},{sin(-20)})$) arc (-20:20:4.5);

       \draw[very thick, decoration={text along path,
       text={...},text align={center}},decorate] ($4.5*({cos(160)},{sin(160)})$) arc (160:200:4.5);

        \begin{scope}[rotate=90]
            \coordinate (h1) at ($3*({cos(10)},{sin(10)})$);
        		\coordinate (h2) at ($3*({cos(-10)},{sin(-10)})$);
            \draw[very thick, Maroon] (h1) to[out=-170,in=90](2.5,0) to[out=-90, in=170] (h2);
            \coordinate (rtmp) at ($4.5*({cos(6)},{sin(6)})$) {};
            \coordinate (rtmp2) at ($(rtmp) + 2*(1,0)$){};
            \coordinate (rtmp3) at ($4.5*({cos(-6)},{sin(-6)})$);
            \coordinate (rtmp4) at ($(rtmp3) + 2*(1,0)$);
            \coordinate (rtmp5) at ($(rtmp2) + 1.5*(1,1)$){};
            \coordinate (rtmp6) at ($(rtmp4) + 1.5*(1,-1)$);
    
            \coordinate (r2tmp) at ($7.5*(1,0)$);
            \coordinate (r2tmp1) at ($(r2tmp) + (1,1)$);
            \coordinate (r2tmp2) at ($(r2tmp) + (1,-1)$);
           	
           	\coordinate (b1) at ($(rtmp2)!0.5!(rtmp4)$);
           	\coordinate (b2) at ($(h1)!0.5!(h2) + (-0.15,0)$);
           	\draw[very thick, Blue] (b1) to (b2);
            \draw[very thick] (rtmp) -- (rtmp2) -- (rtmp5);
            \draw[very thick] (rtmp3) -- (rtmp4) -- (rtmp6);
            \draw[very thick] (r2tmp2) -- (r2tmp) -- (r2tmp1);
        \end{scope}
        
        \begin{scope}[rotate=240]
            \coordinate (h1) at ($3*({cos(10)},{sin(10)})$);
        		\coordinate (h2) at ($3*({cos(-10)},{sin(-10)})$);
            \draw[very thick, Maroon] (h1) to[out=-170,in=90](2.5,0) to[out=-90, in=170] (h2);
            \coordinate (rtmp) at ($4.5*({cos(6)},{sin(6)})$) {};
            \coordinate (rtmp2) at ($(rtmp) + 2*(1,0)$){};
            \coordinate (rtmp3) at ($4.5*({cos(-6)},{sin(-6)})$);
            \coordinate (rtmp4) at ($(rtmp3) + 2*(1,0)$);
            \coordinate (rtmp5) at ($(rtmp2) + 1.5*(1,1)$){};
            \coordinate (rtmp6) at ($(rtmp4) + 1.5*(1,-1)$);
    
            \coordinate (r2tmp) at ($7.5*(1,0)$);
            \coordinate (r2tmp1) at ($(r2tmp) + (1,1)$);
            \coordinate (r2tmp2) at ($(r2tmp) + (1,-1)$);
           	
           	\coordinate (b1) at ($(rtmp2)!0.5!(rtmp4)$);
           	\coordinate (b2) at ($(h1)!0.5!(h2) + (-0.15,0)$);
           	\draw[very thick, Blue] (b1) to (b2);
            \draw[very thick] (rtmp) -- (rtmp2) -- (rtmp5);
            \draw[very thick] (rtmp3) -- (rtmp4) -- (rtmp6);
            \draw[very thick] (r2tmp2) -- (r2tmp) -- (r2tmp1);
        \end{scope}
        
        \begin{scope}[rotate=300]
            \coordinate (h1) at ($3*({cos(10)},{sin(10)})$);
        		\coordinate (h2) at ($3*({cos(-10)},{sin(-10)})$);
            \draw[very thick, Maroon] (h1) to[out=-170,in=90](2.5,0) to[out=-90, in=170] (h2);
            \coordinate (rtmp) at ($4.5*({cos(6)},{sin(6)})$) {};
            \coordinate (rtmp2) at ($(rtmp) + 2*(1,0)$){};
            \coordinate (rtmp3) at ($4.5*({cos(-6)},{sin(-6)})$);
            \coordinate (rtmp4) at ($(rtmp3) + 2*(1,0)$);
            \coordinate (rtmp5) at ($(rtmp2) + 1.5*(1,1)$){};
            \coordinate (rtmp6) at ($(rtmp4) + 1.5*(1,-1)$);
    
            \coordinate (r2tmp) at ($7.5*(1,0)$);
            \coordinate (r2tmp1) at ($(r2tmp) + (1,1)$);
            \coordinate (r2tmp2) at ($(r2tmp) + (1,-1)$);
           	
           	\coordinate (b1) at ($(rtmp2)!0.5!(rtmp4)$);
           	\coordinate (b2) at ($(h1)!0.5!(h2) + (-0.15,0)$);
           	\draw[very thick, Blue] (b1) to (b2);
            \draw[very thick] (rtmp) -- (rtmp2) -- (rtmp5);
            \draw[very thick] (rtmp3) -- (rtmp4) -- (rtmp6);
            \draw[very thick] (r2tmp2) -- (r2tmp) -- (r2tmp1);
        \end{scope}
    \end{scope}
	\end{tikzpicture} 
    \label{eq:NGon}
\end{equation}
as in the bubble case, the only non-vanishing contribution to $\mathcal{N}$ comes from the curve which goes from one leg to another by turning exclusively left, which in the glued surface precisely gives the closed curve. In addition, as emphasized on the right, the remaining contractions take exactly the form of what we would get by actually having the closed contraction curve. Since there is only one such path we get  $\widetilde{\mathcal{N}} =1$, and therefore: 
\begin{empheq}[box=\boxed]{equation}
 \Delta_1^{\text{planar}} = 1 - D 
    \label{eq:delta1loop_Pl}
\end{empheq}

If we are literally computing the $n$-gon LS then from \eqref{eq:NGon}, we can automatically read the single monomial entering $\mathcal{N}$: $\prod_i X_{2i,p}$.
But of course, by virtue of the structure of $\mathcal{N}$, this result is independent of the objects which are glued to the external legs of the $n$-gon, as long as these are on-shell objects. Therefore, the exponent in \eqref{eq:delta1loop_Pl} is valid for all 1-loop planar graphs.

\paragraph{$\Delta$ for 1-loop non-planar}
In the case of 1-loop non-planar leading singularities, the result is even simpler. For non-planar diagrams it is easy to see that there is \textit{no} curve going between the two last glued legs that turns exclusively left, as even the closed curve around the loop has to turn right (at least once)! For example, take the following representative 1-loop non-planar diagram: 
\begin{equation}
    \begin{tikzpicture}[line width=0.6 ,scale=0.4,line cap=round,every node/.style={font=\normalsize},baseline=(O.base)]
        \begin{scope}[scale=0.6, rotate = 90]
        \draw[very thick, Green, dashed](0,0) circle (4.25);

        
        \draw[very thick] ($5*({cos(306)},{sin(306)})$) arc (-54:-20:5) ;
        \draw[very thick, decoration={text along path,
       text={...},text align={center}},decorate] ($5*({cos(-20)},{sin(-20)})$) arc (-20:20:5);
        \draw[very thick] ($5*({cos(20)},{sin(20)})$) arc (20:84:5) ;
        \draw[very thick] ($5*({cos(96)},{sin(96)})$) arc (96:130:5) ;
        \draw[very thick, decoration={text along path,
       text={...},text align={center}},decorate] ($5*({cos(130)},{sin(130)})$) arc (130:160:5);
        \draw[very thick] ($5*({cos(160)},{sin(160)})$) arc (160:200:5);
        \draw[very thick] ($3.5*({cos(188)},{sin(188)})$) arc (-172:172:3.5);

        \draw[very thick, decoration={text along path,
       text={...},text align={center}},decorate] ($5*({cos(200)},{sin(200)})$) arc (200:230:5);
        \draw[very thick] ($5*({cos(230)},{sin(230)})$) arc (230:244:5);
        \draw[very thick] ($5*({cos(256)},{sin(256)})$) arc (256:294:5);

        
        
        \begin{scope}[rotate=250]
            \coordinate (rtmp) at ($5*({cos(6)},{sin(6)})$) {};
            \coordinate (rtmp2) at ($(rtmp) + 2*(1,0)$){};
            \coordinate (rtmp3) at ($5*({cos(-6)},{sin(-6)})$);
            \coordinate (rtmp4) at ($(rtmp3) + 2*(1,0)$);
            \coordinate (rtmp5) at ($(rtmp2) + 1.5*(1,1)$){};
            \coordinate (rtmp6) at ($(rtmp4) + 1.5*(1,-1)$);
    
            \coordinate (r2tmp) at ($(5,0)+3*(1,0)$);
            \coordinate (r2tmp1) at ($(r2tmp) + (1,1)$);
            \coordinate (r2tmp2) at ($(r2tmp) + (1,-1)$);
            
            \draw[very thick] (rtmp) -- (rtmp2) -- (rtmp5);
            \draw[very thick] (rtmp3) -- (rtmp4) -- (rtmp6);
            \draw[very thick] (r2tmp2) -- (r2tmp) -- (r2tmp1);
            
            \coordinate (h1) at ($3.5*({cos(10)},{sin(10)})$);
        		\coordinate (h2) at ($3.5*({cos(-10)},{sin(-10)})$);
            \draw[very thick, Maroon] (h1) to[out=-170,in=90](2.8,0) to[out=-90, in=170] (h2);
           	\coordinate (b1) at ($(rtmp2)!0.5!(rtmp4)$);
           	\coordinate (b2) at ($(h1)!0.5!(h2) + (-0.15,0)$);
           	\draw[very thick, Blue] (b1) to (b2);

        \end{scope}
        \begin{scope}[rotate=300]
            \coordinate (rtmp) at ($5*({cos(6)},{sin(6)})$) {};
            \coordinate (rtmp2) at ($(rtmp) + 2*(1,0)$){};
            \coordinate (rtmp3) at ($5*({cos(-6)},{sin(-6)})$);
            \coordinate (rtmp4) at ($(rtmp3) + 2*(1,0)$);
            \coordinate (rtmp5) at ($(rtmp2) + 1.5*(1,1)$){};
            \coordinate (rtmp6) at ($(rtmp4) + 1.5*(1,-1)$);
    
            \coordinate (r2tmp) at ($(5,0)+3*(1,0)$);
            \coordinate (r2tmp1) at ($(r2tmp) + (1,1)$);
            \coordinate (r2tmp2) at ($(r2tmp) + (1,-1)$);
            \draw[very thick] (rtmp) -- (rtmp2) -- (rtmp5);
            \draw[very thick] (rtmp3) -- (rtmp4) -- (rtmp6);
            \draw[very thick] (r2tmp2) -- (r2tmp) -- (r2tmp1);
            
            \coordinate (h1) at ($3.5*({cos(10)},{sin(10)})$);
        		\coordinate (h2) at ($3.5*({cos(-10)},{sin(-10)})$);
            \draw[very thick, Maroon] (h1) to[out=-170,in=90](2.8,0) to[out=-90, in=170] (h2);
           	\coordinate (b1) at ($(rtmp2)!0.5!(rtmp4)$);
           	\coordinate (b2) at ($(h1)!0.5!(h2) + (-0.15,0)$);
           	\draw[very thick, Blue] (b1) to (b2);

        \end{scope}
        \begin{scope}[rotate=90]
            \coordinate (rtmp) at ($5*({cos(6)},{sin(6)})$) {};
            \coordinate (rtmp2) at ($(rtmp) + 2*(1,0)$){};
            \coordinate (rtmp3) at ($5*({cos(-6)},{sin(-6)})$);
            \coordinate (rtmp4) at ($(rtmp3) + 2*(1,0)$);
            \coordinate (rtmp5) at ($(rtmp2) + 1.5*(1,1)$){};
            \coordinate (rtmp6) at ($(rtmp4) + 1.5*(1,-1)$);
    
            \coordinate (r2tmp) at ($(5,0)+3*(1,0)$);
            \coordinate (r2tmp1) at ($(r2tmp) + (1,1)$);
            \coordinate (r2tmp2) at ($(r2tmp) + (1,-1)$);
            \draw[very thick] (rtmp) -- (rtmp2) -- (rtmp5);
            \draw[very thick] (rtmp3) -- (rtmp4) -- (rtmp6);
            \draw[very thick] (r2tmp2) -- (r2tmp) -- (r2tmp1);
            
            \coordinate (h1) at ($3.5*({cos(10)},{sin(10)})$);
        		\coordinate (h2) at ($3.5*({cos(-10)},{sin(-10)})$);
            \draw[very thick, Maroon] (h1) to[out=-170,in=90](2.8,0) to[out=-90, in=170] (h2);
           	\coordinate (b1) at ($(rtmp2)!0.5!(rtmp4)$);
           	\coordinate (b2) at ($(h1)!0.5!(h2) + (-0.15,0)$);
           	\draw[very thick, Blue] (b1) to (b2);
        \end{scope}
        \begin{scope}[rotate=180]
            \coordinate (rtmp) at ($3.5*({cos(8)},{sin(8)})$) {};
            \coordinate (rtmp2) at ($(rtmp) + 2*(-1,0)$){};
            \coordinate (rtmp3) at ($3.5*({cos(-8)},{sin(-8)})$);
            \coordinate (rtmp4) at ($(rtmp3) + 2*(-1,0)$);
            \coordinate (rtmp5) at ($(rtmp2) + 1.5*(-1,1)$){};
            \coordinate (rtmp6) at ($(rtmp4) + 1.5*(-1,-1)$);
    
            \coordinate (r2tmp) at ($(3.5,0)+ 3*(-1,0)$);
            \coordinate (r2tmp1) at ($(r2tmp) + (-1,1)$);
            \coordinate (r2tmp2) at ($(r2tmp) + (-1,-1)$);
            \draw[very thick] (rtmp) -- (rtmp2) -- (rtmp5);
            \draw[very thick] (rtmp3) -- (rtmp4) -- (rtmp6);
            \draw[very thick] (r2tmp2) -- (r2tmp) -- (r2tmp1);
            
            	\coordinate (h1) at ($5*({cos(7)},{sin(7)})$);
        		\coordinate (h2) at ($5*({cos(-7)},{sin(-7)})$);
            \draw[very thick, Maroon] (h1) to[out=-15,in=90](5.6,0) to[out=-90,in=15] (h2);
           	\coordinate (b1) at ($(rtmp2)!0.5!(rtmp4)$);
           	\coordinate (b2) at ($(h1)!0.5!(h2) + (0.15,0)$);
           	\draw[very thick, Blue] (b1) to (b2);
        \end{scope}

    \end{scope}
    \end{tikzpicture}
\end{equation}

we can trivially see that a closed curve around the loop is not an \textit{exclusively} left-turning curve. As such, all the $\mathcal{N}$ corrections have to vanish! Therefore for non-planar surfaces we find
\begin{empheq}[box=\boxed]{equation}
 \Delta_1^{\text{non-planar}} =- D 
    \label{eq:delta1loop_NPl}
\end{empheq}

So, already at 1-loop, we learn that the ``turn-left'' rule for the $\mathcal{N}$ corrections essentially differentiates between two types of closed curves: those that are \textit{homotopic} to an internal boundary of the graph, whose $\Delta = 1-D$; from those which are not, and thus $\Delta$ is simply given by the naive gluing rules to be $-D$. As we will see next, this distinction carries over to LS beyond one-loop order and it will be crucial in determining the exponents of closed curves at all loop orders.

\section{Higher loops}
\label{sec:HigherLoops}

Before getting to the new features of gluing higher-loop LS, already at the level of the surface integral there are a few new features worth highlighting:

\paragraph{Mapping class group} For higher genus surfaces entering at higher-loops, the infinitely many curves on the surface are organized into orbits of the mapping class group (MCG) -- which identifies curves that are the same up to a motion of the punctures around the surface. While there is no MCG for planar one-loop surfaces, already for one-loop non-planar there is a non-trivial MCG action which rotates the internal boundary. To extract the a single copy of the amplitude from the surface integral one needs to understand how to properly mod out the action of the MCG on the surface \cite{Mirzakhani:2006fta,Arkani-Hamed:2023lbd}. However, extracting the maximal residue for the LS automatically selects the curves from a given MCG class -- those that are allowed to enter in collections of curves that cover the fatgraph \textit{once and only once}! This is similar to what we saw for the case of the infinite self-intersecting curves, where only those that self-intersect at most once around a puncture have a chance to contribute to the LS. This is something we also consistently ``discover'' from recasting the gluing via curves. 

\paragraph{Closed curves} As mentioned in Sec. \ref{sec:Dictionary}, for higher genus surfaces we have more types of closed curves that in general can be labelled by the set of punctures they enclose, $\Delta_{J}$. Once $|J|>1$ then we start having closed curves that self intersect themselves, which we label $\Delta^{(q)}_{J}$, with $q$ the number of self-intersections. However, it is easy to see that closed curves with $q>1$ do not contribute for the maximal residues giving the LS. Even so, matching with the explicit gluing we will be able to determine the exponents of non-self-intersecting curves which contribute to the LS. 
\vspace{2mm}

Let us now study the new features of gluing at higher-loops. As it is clear from the one-loop discussion, higher-loop LS are computed in exactly the same way with the extra feature that as the number of loops increases there are increasingly more corrections to the naive gluing. Concretely, to obtain an $(n+1)$-loop LS we glue two legs of an $n$-loop object, which already contains $\mathcal{N}$ corrections. Hence, the higher the loop order we go, the more nested $\mathcal{N}$-type corrections to naive gluing we get. For example, starting at 2-loops, any LS can be computed from gluing a 1-loop LS as
\begin{equation}
\begin{aligned}
    \text{LS}_{\text{2-loop}} &= \left(-\eta^{\alpha\beta} + \frac{p^{\alpha}q^{\beta} + p^{\beta}q^{\alpha}}{p \cdot q} \right) (\text{LS}_{\text{1-loop}})_{\alpha,\beta} \\
    &= \left(-\eta^{\alpha\beta} + \frac{p^{\alpha}q^{\beta} + p^{\beta}q^{\alpha}}{p \cdot q} \right) (\mathcal{M}_{\alpha,\beta} + \mathcal{N}_{\alpha,\beta} ) \, \\
     &= -\underbracket[0.4pt]{\eta^{\alpha\beta} \mathcal{M}_{\alpha,\beta}}_{(1)} -\underbracket[0.4pt]{\eta^{\alpha\beta}\mathcal{N}_{\alpha,\beta}}_{(2)}  + \underbracket[0.4pt]{\frac{p^{\alpha}q^{\beta} + p^{\beta}q^{\alpha}}{p \cdot q}  \mathcal{M}_{\alpha,\beta}}_{(3)}+ \underbracket[0.4pt]{\frac{p^{\alpha}q^{\beta} + p^{\beta}q^{\alpha}}{p \cdot q} \mathcal{N}_{\alpha,\beta}}_{(4)}   \, 
\end{aligned}
    \label{eq:2loopN}
\end{equation}
where in the second line, we split $\text{LS}_{\text{1-loop}}$ into the naive-gluing, $\mathcal{M}$, and the $\mathcal{N}$ correction, so that the total two-loop correction is given terms $(2)$, $(3)$ and $(4)$. Among these, the $(2)$ and $(3)$ terms are just the corrections to each one-loop, for which we already have a simple picture from the previous section, so now we want to understand the new nested $\mathcal{N}$ correction, given by $(4)$. Let's start by looking at a simple $2$-loop example.

\subsection*{2-loop bubble example}

Let's consider the 2-loop bubble LS, which can be obtained by gluing legs A and B in the 4-point 1-loop LS as follows:
\begin{equation}
\begin{aligned}
    &\vcenter{\hbox{\begin{tikzpicture}[line width=0.6 ,scale=1,line cap=round,every node/.style={font=\normalsize}]
        \begin{scope}[rotate=90,scale =0.55]
            \coordinate (str) at ($0.5*({cos(30)},{sin(30)})$);
            \coordinate (sb) at ($0.5*({cos(-90)},{sin(-90)})$);
        	\coordinate (stl) at ($0.5*({cos(150)},{sin(150)})$);
        	
        	\coordinate (tr) at ($1.5*({cos(30)},{sin(30)})$);
            \coordinate (b) at ($1.5*({cos(-90)},{sin(-90)})$);
        	\coordinate (tl) at ($1.5*({cos(150)},{sin(150)})$);
        	
        	\coordinate (trv) at ($3*({cos(30)},{sin(30)})$);
        	\coordinate (tlv) at ($3*({cos(150)},{sin(150)})$);
        	\coordinate (bv) at ($2.5*({cos(-90)},{sin(-90)})$);
        	
        	\coordinate (tr1) at ($(tr)!0.15!(tl)$);
        	\coordinate (tr1v) at ($(tr1)+1*({cos(30)},{sin(30)})$);
		      \coordinate (tr2) at ($(tr)!0.15!(b)$);
        	\coordinate (tr2v) at ($(tr2)+1*({cos(30)},{sin(30)})$);
        	
        	\coordinate (tl1) at ($(tl)!0.15!(tr)$);
        	\coordinate (tl1v) at ($(tl1)+1*({cos(150)},{sin(150)})$);
		      \coordinate (tl2) at ($(tl)!0.15!(b)$);
        	\coordinate (tl2v) at ($(tl2)+1*({cos(150)},{sin(150)})$);
            
            \coordinate (b1) at ($(b)!0.15!(tl)$);
        	\coordinate (b1v) at ($(b1)+1*({cos(-90)},{sin(-90)})$);
		      \coordinate (b2) at ($(b)!0.15!(tr)$);
        	\coordinate (b2v) at ($(b2)+1*({cos(-90)},{sin(-90)})$);
        	\coordinate (br1) at ($(b2v)+({cos(-45)},{sin(-45)})$);
        	\coordinate (br2) at ($(bv)+({cos(-45)},{sin(-45)})$);
      	\coordinate (bl1) at ($(b1v)+({cos(-135)},{sin(-135)})$);
      	\coordinate (bl2) at ($(bv)+({cos(-135)},{sin(-135)})$);
        	\node (g1) at ($(bl1)!0.5!(bl2)+0.3*({cos(-135)},{sin(-135)})$) {};
        	\node (g2) at ($(tr1v)!0.5!(tr2v)+0.4*({cos(35)},{sin(35)})$) {A$^\mu$};
        	\node (g4) at ($(br1)!0.5!(br2)+0.4*({cos(-45)},{sin(-45)})$) {B$^\nu$};
            \draw[<->,dashed] ($(tr1v)!0.5!(tr2v)+0.4*({cos(35)},{sin(35)}) + (0.4,0)$) to[out=-70, in=30] ($(br1)!0.5!(br2)+0.4*({cos(-45)},{sin(-45)})+ (0.35,0)$);  
            \node at ($0.5*(g2) + 0.5*(g4) + 0.5*({cos(-20)},{sin(-20)})$) {glue};
            
        \draw[very thick] (str) to (sb) to (stl) to (str);

        \draw[very thick] (tr1) to (tr1v);
        \draw[very thick] (tr2) to (tr2v);
        \draw[very thick] (tl1) to (tl1v);
        \draw[very thick] (tl2) to (tl2v);

		\draw[very thick] (b2) to (b2v);
        \draw[very thick] (b1) to (b1v);
        	\draw[very thick] (b2v) to (br1);        	
        	\draw[very thick] (bv) to (br2); 
        	\draw[very thick] (b1v) to (bl1);        	
        	\draw[very thick] (bv) to (bl2);

        	\draw[very thick] (b1) to (tl2);
        	\draw[very thick] (tl1) to (tr1);
        \draw[very thick] (tr2) to (b2);
        
        	\coordinate (ht1) at ($(b2)!0.1!(tr2)$);
        	\coordinate (ht2) at ($(b2)!0.1!(b2v)$);
        	\coordinate (htm) at ($(ht1)!0.5!(ht2)$);
		\coordinate (p1) at ($(b1)!0.5!(sb)$);
		\coordinate (p21) at ($(stl)!0.4!(tl2)$);
		\coordinate (p22) at ($(stl)!0.4!(tl1)$);
		\coordinate (p31) at ($(str)!0.4!(tr1)$);
		\coordinate (p32) at ($(str)!0.4!(tr2)$);
		\coordinate (p3) at ($(str)!0.4!(tr)$);
        	
        	\coordinate (p12) at ($(sb)!0.4!(b1)$);
		\coordinate (p11) at ($(sb)!0.4!(b2)$);
        	
		\coordinate (p4) at ($(b1)!0.5!(b2)$);
        
		\coordinate (p5) at ($(b1v)!0.5!(b2v)+(0,0.3)$);
		\coordinate (p6) at ($(bl1)!0.5!(bl2)$);

		\coordinate (hi1t) at ($(stl)!0.25!(str)$);
		\coordinate (hi1b) at ($(stl)!0.25!(sb)$);
		\coordinate (hi2t) at ($(str)!0.25!(stl)$);
		\coordinate (hi2b) at ($(str)!0.25!(sb)$);
		\coordinate (hbr) at ($(sb)!0.25!(str)$);
		\coordinate (hbl) at ($(sb)!0.25!(stl)$);	
		

		
        \end{scope}
    \end{tikzpicture}}}
    \, + \, 
    \vcenter{\hbox{\begin{tikzpicture}[line width=0.6 ,scale=1,line cap=round,every node/.style={font=\normalsize}]
        \begin{scope}[rotate=90,scale=0.55]
            \coordinate (str) at ($0.5*({cos(30)},{sin(30)})$);
            \coordinate (sb) at ($0.5*({cos(-90)},{sin(-90)})$);
        	\coordinate (stl) at ($0.5*({cos(150)},{sin(150)})$);
        	
        	\coordinate (tr) at ($1.5*({cos(30)},{sin(30)})$);
            \coordinate (b) at ($1.5*({cos(-90)},{sin(-90)})$);
        	\coordinate (tl) at ($1.5*({cos(150)},{sin(150)})$);
        	
        	\coordinate (trv) at ($3*({cos(30)},{sin(30)})$);
        	\coordinate (tlv) at ($3*({cos(150)},{sin(150)})$);
        	\coordinate (bv) at ($2.5*({cos(-90)},{sin(-90)})$);
        	
        	\coordinate (tr1) at ($(tr)!0.15!(tl)$);
        	\coordinate (tr1v) at ($(tr1)+1*({cos(30)},{sin(30)})$);
		      \coordinate (tr2) at ($(tr)!0.15!(b)$);
        	\coordinate (tr2v) at ($(tr2)+1*({cos(30)},{sin(30)})$);
        	
        	\coordinate (tl1) at ($(tl)!0.15!(tr)$);
        	\coordinate (tl1v) at ($(tl1)+1*({cos(150)},{sin(150)})$);
		      \coordinate (tl2) at ($(tl)!0.15!(b)$);
        	\coordinate (tl2v) at ($(tl2)+1*({cos(150)},{sin(150)})$);
            
            \coordinate (b1) at ($(b)!0.15!(tl)$);
        	\coordinate (b1v) at ($(b1)+1*({cos(-90)},{sin(-90)})$);
		      \coordinate (b2) at ($(b)!0.15!(tr)$);
        	\coordinate (b2v) at ($(b2)+1*({cos(-90)},{sin(-90)})$);
        	\coordinate (br1) at ($(b2v)+({cos(-45)},{sin(-45)})$);
        	\coordinate (br2) at ($(bv)+({cos(-45)},{sin(-45)})$);
      	\coordinate (bl1) at ($(b1v)+({cos(-135)},{sin(-135)})$);
      	\coordinate (bl2) at ($(bv)+({cos(-135)},{sin(-135)})$);
        	\node (g1) at ($(bl1)!0.5!(bl2)+0.3*({cos(-135)},{sin(-135)})$) {};
        	\node (g2) at ($(tr1v)!0.5!(tr2v)+0.4*({cos(35)},{sin(35)})$) {A$^\mu$};
        	\node (g4) at ($(br1)!0.5!(br2)+0.4*({cos(-45)},{sin(-45)})$) {B$^\nu$};
            \draw[<->,dashed] ($(tr1v)!0.5!(tr2v)+0.4*({cos(35)},{sin(35)}) + (0.4,0)$) to[out=-70, in=30] ($(br1)!0.5!(br2)+0.4*({cos(-45)},{sin(-45)})+ (0.35,0)$);  
            \node at ($0.5*(g2) + 0.5*(g4) + 0.5*({cos(-20)},{sin(-20)})$) {glue};
        	
        \draw[very thick] (str) to (sb) to (stl) to (str);

        \draw[very thick] (tr1) to (tr1v);
        \draw[very thick] (tr2) to (tr2v);
        \draw[very thick] (tl1) to (tl1v);
        \draw[very thick] (tl2) to (tl2v);

		\draw[very thick] (b2) to (b2v);
        \draw[very thick] (b1) to (b1v);
        	\draw[very thick] (b2v) to (br1);        	
        	\draw[very thick] (bv) to (br2); 
        	\draw[very thick] (b1v) to (bl1);        	
        	\draw[very thick] (bv) to (bl2);

        	\draw[very thick] (b1) to (tl2);
        	\draw[very thick] (tl1) to (tr1);
        \draw[very thick] (tr2) to (b2);
        
        	\coordinate (ht1) at ($(b2)!0.1!(tr2)$);
        	\coordinate (ht2) at ($(b2)!0.1!(b2v)$);
        	\coordinate (htm) at ($(ht1)!0.5!(ht2)$);
		\coordinate (p1) at ($(b1)!0.5!(sb)$);
		\coordinate (p21) at ($(stl)!0.4!(tl2)$);
		\coordinate (p22) at ($(stl)!0.4!(tl1)$);
		\coordinate (p31) at ($(str)!0.4!(tr1)$);
		\coordinate (p32) at ($(str)!0.4!(tr2)$);
		\coordinate (p3) at ($(str)!0.4!(tr)$);
        	
        	\coordinate (p12) at ($(sb)!0.4!(b1)$);
		\coordinate (p11) at ($(sb)!0.4!(b2)$);
        	
		\coordinate (p4) at ($(b1)!0.5!(b2)$);
        
		\coordinate (p5) at ($(b1v)!0.5!(b2v)+(0,0.3)$);
		\coordinate (p6) at ($(bl1)!0.5!(bl2)$);
		\draw[Green, thick, dashed] (p21) to[out=100,in=-160] (p22) to[out=20, in=160] (p31) to[out=-20,in=80] (p32) to[out=-100,in=50] (p11) to[out=-130,in=-40] (p12) to[out=130,in=-80] (p21);

		\coordinate (hi1t) at ($(stl)!0.25!(str)$);
		\coordinate (hi1b) at ($(stl)!0.25!(sb)$);
		\coordinate (hi2t) at ($(str)!0.25!(stl)$);
		\coordinate (hi2b) at ($(str)!0.25!(sb)$);
		\coordinate (hbr) at ($(sb)!0.25!(str)$);
		\coordinate (hbl) at ($(sb)!0.25!(stl)$);	
		
		\draw[Maroon, thick] (hi1t) to[out=-90,in=20] (hi1b);
		\draw[Maroon, thick] (hi2t) to[out=-90,in=160] (hi2b);
		\draw[Maroon, thick] (hbl) to[out=50,in=130] (hbr);
		\draw[Blue, thick] ($(hbl)!0.5!(hbr)$) to (p5);
		\draw[Blue, thick] ($(hi2t)!0.5!(hi2b)$) to ($(tr1v)!0.5!(tr2v)$);
		\draw[Blue, thick] ($(hi1t)!0.5!(hi1b)$) to ($(tl1v)!0.5!(tl2v)$);

		
        \end{scope}
    \end{tikzpicture}}} = \\
    & = \vcenter{\hbox{\begin{tikzpicture}[line width=0.6,line cap=round,every node/.style={font=\normalsize},baseline=(O.base)]
    \begin{scope}[scale=0.22]
        \node at (-5,3) {\small $(1)$};
        \draw[very thick] ($2.5*({cos(-75)},{sin(-75)})$) arc (-75:75:2.5);
        \draw[very thick] ($2.5*({cos(105)},{sin(105)})$) arc (105:255:2.5);
        \draw[very thick] ($2.5*({cos(105)},{sin(105)})$) to ($2.5*({cos(255)},{sin(255)})$);
        \draw[very thick] ($2.5*({cos(75)},{sin(75)})$) to ($2.5*({cos(-75)},{sin(-75)})$);
        
        \draw[very thick] ($4*({cos(8)},{sin(8)})$) arc (8:172:4);
        \draw[very thick] ($4*({cos(188)},{sin(188)})$) arc (188:352:4);

        	\coordinate (auxp2) at ($2.5*({cos(75)},{sin(75)})$);
        	\coordinate (aux2p2) at ($2.5*({cos(-75)},{sin(-75)})$);
        	\coordinate (auxp1) at ($2.5*({cos(180-75)},{sin(180-75)})$);
        	\coordinate (aux2p1) at ($2.5*({cos(-180+75)},{sin(-180+75)})$);

        \begin{scope}
            \coordinate (rtmp) at ($4*({cos(8)},{sin(8)})$) {};
            \coordinate (rtmp2) at ($(rtmp) + 2*(1,0)$){};
            \coordinate (rtmp3) at ($4*({cos(-8)},{sin(-8)})$);
            \coordinate (rtmp4) at ($(rtmp3) + 2*(1,0)$);
            \coordinate (rtmp5) at ($(rtmp2) + 1.5*(1,1)$){};
            \coordinate (rtmp6) at ($(rtmp4) + 1.5*(1,-1)$);
    
            \draw[very thick] (rtmp) -- (rtmp2) ;
            \draw[very thick] (rtmp3) -- (rtmp4) ;
        \end{scope}

        \begin{scope}[rotate=180]
            \coordinate (rtmp) at ($4*({cos(8)},{sin(8)})$) {};
            \coordinate (rtmp2) at ($(rtmp) + 2*(1,0)$){};
            \coordinate (rtmp3) at ($4*({cos(-8)},{sin(-8)})$);
            \coordinate (rtmp4) at ($(rtmp3) + 2*(1,0)$);
            \coordinate (rtmp5) at ($(rtmp2) + 1.5*(1,1)$){};
            \coordinate (rtmp6) at ($(rtmp4) + 1.5*(1,-1)$);
    
            \draw[very thick] (rtmp) -- (rtmp2) ;
            \draw[very thick] (rtmp3) -- (rtmp4) ;
        \end{scope}
    \end{scope} 
    \end{tikzpicture}}}
    +
    \vcenter{\hbox{\begin{tikzpicture}[line width=0.6,line cap=round,every node/.style={font=\normalsize},baseline=(O.base)]
    \begin{scope}[scale=0.22]
            \node at (-5,3) {\small $(2)$};
        \draw[very thick] ($2.5*({cos(-75)},{sin(-75)})$) arc (-75:75:2.5);
        \draw[very thick] ($2.5*({cos(105)},{sin(105)})$) arc (105:255:2.5);
        \draw[very thick] ($2.5*({cos(105)},{sin(105)})$) to ($2.5*({cos(255)},{sin(255)})$);
        \draw[very thick] ($2.5*({cos(75)},{sin(75)})$) to ($2.5*({cos(-75)},{sin(-75)})$);
        
        \draw[very thick] ($4*({cos(8)},{sin(8)})$) arc (8:172:4);
        \draw[very thick] ($4*({cos(188)},{sin(188)})$) arc (188:352:4);

        	\coordinate (auxp2) at ($2.5*({cos(75)},{sin(75)})$);
        	\coordinate (aux2p2) at ($2.5*({cos(-75)},{sin(-75)})$);
        	\coordinate (auxp1) at ($2.5*({cos(180-75)},{sin(180-75)})$);
        	\coordinate (aux2p1) at ($2.5*({cos(-180+75)},{sin(-180+75)})$);
        	

%
%

		\coordinate (hp1t1) at ($2.5*({cos(180-62)},{sin(180-62)})$);
        	\coordinate (hp1t2) at ($(auxp1)!0.15!(aux2p1)$);
        	\draw[Maroon, thick] (hp1t2) to[out=180,in=-80] (hp1t1);
        	
        	\coordinate (hp1b1) at ($2.5*({cos(-180+62)},{sin(-180+62)})$);
        	\coordinate (hp1b2) at ($(aux2p1)!0.15!(auxp1)$);
        	\draw[Maroon, thick] (hp1b2) to[out=180,in=80] (hp1b1);
		\coordinate (hp1m1) at ($2.5*({cos(180-10)},{sin(180-10)})$);
		\coordinate (hp1m2) at ($2.5*({cos(-180+10)},{sin(-180+10)})$);
		\draw[Maroon, thick] (hp1m1) to[out=-20,in=20] (hp1m2);
		
		\coordinate (bctl) at ($3.25*({cos(80)},{sin(80)})$);
		\draw[Blue, very thick] ($(hp1t1)!0.5!(hp1t2)$) to[out=75, in=180] (bctl);
		\draw[Blue, very thick]  (bctl) arc (80:50:3.25);
		
		\coordinate (bcbl) at ($3.25*({cos(-80)},{sin(-80)})$);
		\draw[Blue, very thick] ($(hp1b1)!0.5!(hp1b2)$) to[out=-75, in=180] (bcbl);
		\draw[Blue, very thick]  (bcbl) arc (-80:-50:3.25);
		
		\draw[Blue, very thick] ($(hp1m1)!0.5!(hp1m2)$) to (-6,0);
		
		
		

%
%
		
		\draw[Green, dashed] (0,-3.25) to (0,3.25) arc (90:270:3.25);

        \begin{scope}
            \coordinate (rtmp) at ($4*({cos(8)},{sin(8)})$) {};
            \coordinate (rtmp2) at ($(rtmp) + 2*(1,0)$){};
            \coordinate (rtmp3) at ($4*({cos(-8)},{sin(-8)})$);
            \coordinate (rtmp4) at ($(rtmp3) + 2*(1,0)$);
            \coordinate (rtmp5) at ($(rtmp2) + 1.5*(1,1)$){};
            \coordinate (rtmp6) at ($(rtmp4) + 1.5*(1,-1)$);
    
            \draw[very thick] (rtmp) -- (rtmp2) ;
            \draw[very thick] (rtmp3) -- (rtmp4) ;
        \end{scope}

        \begin{scope}[rotate=180]
           \coordinate (rtmp) at ($4*({cos(8)},{sin(8)})$) {};
            \coordinate (rtmp2) at ($(rtmp) + 2*(1,0)$){};
            \coordinate (rtmp3) at ($4*({cos(-8)},{sin(-8)})$);
            \coordinate (rtmp4) at ($(rtmp3) + 2*(1,0)$);
            \coordinate (rtmp5) at ($(rtmp2) + 1.5*(1,1)$){};
            \coordinate (rtmp6) at ($(rtmp4) + 1.5*(1,-1)$);
    
            \draw[very thick] (rtmp) -- (rtmp2) ;
            \draw[very thick] (rtmp3) -- (rtmp4) ;
        \end{scope}
    \end{scope} 
    \end{tikzpicture}}}
    \, + \, 
    \vcenter{\hbox{\begin{tikzpicture}[line width=0.6,line cap=round,every node/.style={font=\normalsize}]
    \begin{scope}[scale=0.22,yshift=-5cm]
            \node at (-5,3) {\small $(3)$};
        \draw[very thick] ($2.5*({cos(-75)},{sin(-75)})$) arc (-75:75:2.5);
        \draw[very thick] ($2.5*({cos(105)},{sin(105)})$) arc (105:255:2.5);
        \draw[very thick] ($2.5*({cos(105)},{sin(105)})$) to ($2.5*({cos(255)},{sin(255)})$);
        \draw[very thick] ($2.5*({cos(75)},{sin(75)})$) to ($2.5*({cos(-75)},{sin(-75)})$);
        
        \draw[very thick] ($4*({cos(8)},{sin(8)})$) arc (8:172:4);
        \draw[very thick] ($4*({cos(188)},{sin(188)})$) arc (188:352:4);

        	\coordinate (auxp2) at ($2.5*({cos(75)},{sin(75)})$);
        	\coordinate (aux2p2) at ($2.5*({cos(-75)},{sin(-75)})$);
        	\coordinate (auxp1) at ($2.5*({cos(180-75)},{sin(180-75)})$);
        	\coordinate (aux2p1) at ($2.5*({cos(-180+75)},{sin(-180+75)})$);
        	

        	\coordinate (hp2t1) at ($2.5*({cos(62)},{sin(62)})$);
        	\coordinate (hp2t2) at ($(auxp2)!0.15!(aux2p2)$);
        	\draw[Maroon, thick] (hp2t2) to[out=0,in=-100] (hp2t1);
        	
        	\coordinate (hp2b1) at ($2.5*({cos(-62)},{sin(-62)})$);
        	\coordinate (hp2b2) at ($(aux2p2)!0.15!(auxp2)$);
        	\draw[Maroon, thick] (hp2b2) to[out=0,in=100] (hp2b1);

		\coordinate (hp2m1) at ($2.5*({cos(10)},{sin(10)})$);
		\coordinate (hp2m2) at ($2.5*({cos(-10)},{sin(-10)})$);
		\draw[Maroon, thick] (hp2m1) to[out=-160,in=160] (hp2m2);

		\coordinate (bctr) at ($3.25*({cos(100)},{sin(100)})$);
		\draw[Blue, very thick] ($(hp2t1)!0.5!(hp2t2)$) to[out=105, in=0] (bctr);
		\draw[Blue, very thick]  (bctr) arc (100:130:3.25);
		
		\coordinate (bcbl) at ($3.25*({cos(-105)},{sin(-105)})$);
		\draw[Blue, very thick] ($(hp2b1)!0.5!(hp2b2)$) to[out=-105, in=0] (bcbl);
		\draw[Blue, very thick]  (bcbl) arc (-105:-130:3.25);
		
		\draw[Blue, very thick] ($(hp2m1)!0.5!(hp2m2)$) to (6,0);
		
		\draw[Green, dashed] (0,-3.25) to (0,3.25) arc (90:-90:3.25);

        \begin{scope}
            \coordinate (rtmp) at ($4*({cos(8)},{sin(8)})$) {};
            \coordinate (rtmp2) at ($(rtmp) + 2*(1,0)$){};
            \coordinate (rtmp3) at ($4*({cos(-8)},{sin(-8)})$);
            \coordinate (rtmp4) at ($(rtmp3) + 2*(1,0)$);
            \coordinate (rtmp5) at ($(rtmp2) + 1.5*(1,1)$){};
            \coordinate (rtmp6) at ($(rtmp4) + 1.5*(1,-1)$);
    
            \draw[very thick] (rtmp) -- (rtmp2) ;
            \draw[very thick] (rtmp3) -- (rtmp4) ;
        \end{scope}

        \begin{scope}[rotate=180]
            \coordinate (rtmp) at ($4*({cos(8)},{sin(8)})$) {};
            \coordinate (rtmp2) at ($(rtmp) + 2*(1,0)$){};
            \coordinate (rtmp3) at ($4*({cos(-8)},{sin(-8)})$);
            \coordinate (rtmp4) at ($(rtmp3) + 2*(1,0)$);
            \coordinate (rtmp5) at ($(rtmp2) + 1.5*(1,1)$){};
            \coordinate (rtmp6) at ($(rtmp4) + 1.5*(1,-1)$);
    
            \draw[very thick] (rtmp) -- (rtmp2) ;
            \draw[very thick] (rtmp3) -- (rtmp4) ;
        \end{scope}
    \end{scope} 
    \end{tikzpicture}}}
    + \hspace{-1.1cm}
    \vcenter{\hbox{\begin{tikzpicture}[line width=0.6,line cap=round,every node/.style={font=\normalsize}]
    \begin{scope}[scale=0.22,xshift=-20cm]
        \node at (-5,3) {\small $(4)$};
        \draw[very thick] ($2.5*({cos(-75)},{sin(-75)})$) arc (-75:75:2.5);
        \draw[very thick] ($2.5*({cos(105)},{sin(105)})$) arc (105:255:2.5);
        \draw[very thick] ($2.5*({cos(105)},{sin(105)})$) to ($2.5*({cos(255)},{sin(255)})$);
        \draw[very thick] ($2.5*({cos(75)},{sin(75)})$) to ($2.5*({cos(-75)},{sin(-75)})$);
        
        \draw[very thick] ($4*({cos(8)},{sin(8)})$) arc (8:172:4);
        \draw[very thick] ($4*({cos(188)},{sin(188)})$) arc (188:352:4);

        	\coordinate (auxp2) at ($2.5*({cos(75)},{sin(75)})$);
        	\coordinate (aux2p2) at ($2.5*({cos(-75)},{sin(-75)})$);
        	\coordinate (auxp1) at ($2.5*({cos(180-75)},{sin(180-75)})$);
        	\coordinate (aux2p1) at ($2.5*({cos(-180+75)},{sin(-180+75)})$);
        	

%
%

		\coordinate (hp1t1) at ($2.5*({cos(180-62)},{sin(180-62)})$);
        	\coordinate (hp1t2) at ($(auxp1)!0.15!(aux2p1)$);
        	\draw[Maroon, thick] (hp1t2) to[out=180,in=-80] (hp1t1);
        	
        	\coordinate (hp1b1) at ($2.5*({cos(-180+62)},{sin(-180+62)})$);
        	\coordinate (hp1b2) at ($(aux2p1)!0.15!(auxp1)$);
        	\draw[Maroon, thick] (hp1b2) to[out=180,in=80] (hp1b1);
		\coordinate (hp1m1) at ($2.5*({cos(180-10)},{sin(180-10)})$);
		\coordinate (hp1m2) at ($2.5*({cos(-180+10)},{sin(-180+10)})$);
		\draw[Maroon, thick] (hp1m1) to[out=-20,in=20] (hp1m2);
		
		\coordinate (bctl) at ($3.25*({cos(80)},{sin(80)})$);
		\draw[Blue, very thick] ($(hp1t1)!0.5!(hp1t2)$) to[out=75, in=180] (bctl);
		\draw[Blue, very thick]  (bctl) arc (80:50:3.25);
		
		\coordinate (bcbl) at ($3.25*({cos(-80)},{sin(-80)})$);
		\draw[Blue, very thick] ($(hp1b1)!0.5!(hp1b2)$) to[out=-75, in=180] (bcbl);
		\draw[Blue, very thick]  (bcbl) arc (-80:-50:3.25);
		
		\draw[Blue, very thick] ($(hp1m1)!0.5!(hp1m2)$) to (-6,0);
		\draw[Green, very thick] ($2.5*({cos(50)},{sin(50)})$) to ($4*({cos(50)},{sin(50)})$);
		\draw[->, Green, thick] ($3.25*({cos(25)},{sin(25)})$) arc (25:50:3.25);
		
		
		\draw[->, thick] ($4.5*({cos(20)},{sin(20)})$) arc (15:40:4.5);
		\node at ($6*({cos(30)},{sin(30)})$) {\scriptsize $q_i^\mu$};
		\draw[->, thick] (1,-0.5) to (1,1.5);
		\node at (1.8,0.1) { \tiny $q_j^\mu$};
		
		\draw[->,Green, very thick] ($5*(1,1)$) to ($(-7.25,-4)$);
		\node[Green] at ($(-6,-4.5)$) {\small $\propto q_i \cdot q_j =0$};

%
%
		
		\draw[Green, dashed] (0,-3.25) to (0,3.25) arc (90:270:3.25);

        \begin{scope}
            \coordinate (rtmp) at ($4*({cos(8)},{sin(8)})$) {};
            \coordinate (rtmp2) at ($(rtmp) + 2*(1,0)$){};
            \coordinate (rtmp3) at ($4*({cos(-8)},{sin(-8)})$);
            \coordinate (rtmp4) at ($(rtmp3) + 2*(1,0)$);
            \coordinate (rtmp5) at ($(rtmp2) + 1.5*(1,1)$){};
            \coordinate (rtmp6) at ($(rtmp4) + 1.5*(1,-1)$);
    
            \draw[very thick] (rtmp) -- (rtmp2) ;
            \draw[very thick] (rtmp3) -- (rtmp4) ;
        \end{scope}

        \begin{scope}[rotate=180]
            \coordinate (rtmp) at ($4*({cos(8)},{sin(8)})$) {};
            \coordinate (rtmp2) at ($(rtmp) + 2*(1,0)$){};
            \coordinate (rtmp3) at ($4*({cos(-8)},{sin(-8)})$);
            \coordinate (rtmp4) at ($(rtmp3) + 2*(1,0)$);
            \coordinate (rtmp5) at ($(rtmp2) + 1.5*(1,1)$){};
            \coordinate (rtmp6) at ($(rtmp4) + 1.5*(1,-1)$);
    
            \draw[very thick] (rtmp) -- (rtmp2) ;
            \draw[very thick] (rtmp3) -- (rtmp4) ;
        \end{scope}
    \end{scope} 
    \end{tikzpicture}}}
\end{aligned}
    \label{eq:2loopBub_Gluing}
\end{equation}
where in the first line we explicitly separated the $\mathcal{M}_{\alpha,\beta}$ and $\mathcal{N}_{\alpha,\beta}$ (gluing of the correction term). Now, in both cases we glue legs $3$ and $4$ with the $\eta_{\mu,\nu}$ tensor, and also find their respective $\mathcal{N}$ corrections using the rules of the previous sections, leading to the four contributions in the second line. Quite nicely, out of all these contributions, the one which corresponds to a nested $\mathcal{N}$ trivially vanishes, since it is proportional to $q_i \cdot q_j$, where $q_i^\mu$ and $q_j^\mu$ are momenta entering in the same on-shell vertex. Thus, we are left with three non-vanishing contributions!

Note that it is crucial for this nested correction to vanish; otherwise, we would not be able to interpret this term within our current picture, since it would permit a configuration with a closed curve around each loop -- a situation that would not only create an impossible contraction pattern but also violate the LS rule requiring the graph to be filled with non-overlapping curves. Therefore, given the vanishing of this nested correction, the graphical essence of the rules found at 1-loop order remain essentially unchanged at higher loops, as all remaining ingredients are fully understood within the one-loop world!

In summary, for a given monomial  $\prod_C X_C$ (which might include closed curves $\Delta$), we find that it contributes to the LS if the corresponding cores plus extensions fill the graph with non-overlapping curves. In addition, the roads around loop regions can then be filled by two types of curves: either contraction curves, whose contribution is $-D$, or $\mathcal{N}$ corrections if and only if the corresponding loop can be enclosed by a purely left-turning curve.

Sticking to this 2-loop bubble example, in addition to the closed curves enclosing a single puncture, $\Delta_1$ and $\Delta_2$ (whose exponents are automatically fixed by 1-loop consistency), we find a new closed curve, $\Delta_{1,2}$, enclosing both punctures: 
\begin{equation}
    \begin{tikzpicture}[line width=0.6 ,scale=0.35,line cap=round,every node/.style={font=\normalsize},baseline=(O.base)]
    \begin{scope}[scale=0.7]
        \draw[very thick] ($3*({cos(-75)},{sin(-75)})$) arc (-75:75:3);
        \draw[very thick] ($3*({cos(105)},{sin(105)})$) arc (105:255:3);
        \draw[very thick] ($3*({cos(105)},{sin(105)})$) to ($3*({cos(255)},{sin(255)})$);
        \draw[very thick] ($3*({cos(75)},{sin(75)})$) to ($3*({cos(-75)},{sin(-75)})$);
        
        \draw[very thick] ($4.5*({cos(8)},{sin(8)})$) arc (8:172:4.5);
        \draw[very thick] ($4.5*({cos(188)},{sin(188)})$) arc (188:352:4.5);

        \node (r1) at (0,-5.5) {\large{$1$}};
        \node (r2) at (-10,0) {\large{$2$}};
        \node (r3) at (0,5.5) {\large{$3$}};
        \node (r4) at (10,0) {\large{$4$}};
        \node (p1) at (1.75,0) {\large{$p_2$}};
        \node (p2) at (-1.75,0) {\large{$p_1$}};
        	
        	\draw[Maroon, very thick] (0,0) circle (3.75);
        	\draw[Blue, very thick] (0,3) to (0,-3);
        	\draw[Blue, very thick] (4.5,0) to (7.5,0);
        \draw[Blue, very thick] (-4.5,0) to (-7.5,0);
        \node[Maroon] at ($6*({cos(35)},{sin(35)})$) {$\Delta_{1,2}$};

        \begin{scope}
            \coordinate (rtmp) at ($4.5*({cos(8)},{sin(8)})$) {};
            \coordinate (rtmp2) at ($(rtmp) + 3*(1,0)$){};
            \coordinate (rtmp3) at ($4.5*({cos(-8)},{sin(-8)})$);
            \coordinate (rtmp4) at ($(rtmp3) + 3*(1,0)$);
            \coordinate (rtmp5) at ($(rtmp2) + 2*(1,1)$){};
            \coordinate (rtmp6) at ($(rtmp4) + 2*(1,-1)$);
    
            \coordinate (r2tmp) at ($(4.5,0)+4*(1,0)$);
            \coordinate (r2tmp1) at ($(r2tmp) + 1.5*(1,1)$);
            \coordinate (r2tmp2) at ($(r2tmp) + 1.5*(1,-1)$);
            \draw[very thick] (rtmp) -- (rtmp2) -- (rtmp5);
            \draw[very thick] (rtmp3) -- (rtmp4) -- (rtmp6);
            \draw[very thick] (r2tmp2) -- (r2tmp) -- (r2tmp1);
        \end{scope}

        \begin{scope}[rotate=180]
            \coordinate (rtmp) at ($4.5*({cos(8)},{sin(8)})$) {};
            \coordinate (rtmp2) at ($(rtmp) + 3*(1,0)$){};
            \coordinate (rtmp3) at ($4.5*({cos(-8)},{sin(-8)})$);
            \coordinate (rtmp4) at ($(rtmp3) + 3*(1,0)$);
            \coordinate (rtmp5) at ($(rtmp2) + 2*(1,1)$){};
            \coordinate (rtmp6) at ($(rtmp4) + 2*(1,-1)$);
    
            \coordinate (r2tmp) at ($(4.5,0)+4*(1,0)$);
            \coordinate (r2tmp1) at ($(r2tmp) + 1.5*(1,1)$);
            \coordinate (r2tmp2) at ($(r2tmp) + 1.5*(1,-1)$);
            \draw[very thick] (rtmp) -- (rtmp2) -- (rtmp5);
            \draw[very thick] (rtmp3) -- (rtmp4) -- (rtmp6);
            \draw[very thick] (r2tmp2) -- (r2tmp) -- (r2tmp1);
        \end{scope}
    \end{scope}
    \end{tikzpicture} \quad \propto \quad \Delta_{1,2} X_{1,3} X_{2,p} X_{4,p}
\end{equation}
in particular, from the contraction picture, this contraction curve comes with a $(-D)$ factor and it automatically fixes all remaining contractions around the loop, yielding the monomial $X_{1,3}X_{2,p_1}X_{4,p_2}$, as represented above. Moreover, this curve does not receive any $\mathcal{N}$ corrections since it is not a purely left-turning curve. Therefore its contribution to the LS is exactly $(-D)X_{1,3}X_{2,p_1}X_{4,p_2}$. 

Since the higher-loop gluing picture and surface integral residue are completely in the same footing, we can immediately use this result to fix the exponent of the closed curve $\Delta_{1,2} = -D$. Let us proceed to study the exponents of closed curves in generality.

\subsection{$\Delta$'s at all-loops}

Having understood the vanishing of nested corrections to the gluing at higher loops, we find, remarkably, perfect agreement between the LS monomials (and their corresponding coefficients) generated by the contraction picture and those obtained from the residue of the surface integral. Therefore, by requiring consistency with the explicit gluing, we can fully determine the exponents of all closed curves entering the LS.

From the contraction picture, any given closed curve will enter the LS as long as it does \textit{not} cross any edge of the fatgraph more than once. In turn, the exponent, $\Delta_J$, depends on whether this curve receives contributions from $\mathcal{N}$, but all such cases are summarized by the following rule

\begin{center}
Given a closed curve, $\Delta_J$, if it is \textit{homotopic} to an internal boundary of the graph ($i.e.$ it is \textit{purely left-turning}), then its exponent is $\Delta_J=1-D$; otherwise $\Delta_J=-D$.
\end{center}

Note that in particular, if we have a planar graph at any number of loops, the rule above implies that the curves that enclose a single puncture have exponent $\Delta_{|J|=1}=(1-D)$, while those enclosing more than one puncture $\Delta_{|J|>1}=-D$. 

While in the planar case we rephrase the rule above in terms of the number of punctures a given closed curve encloses, this is not the case when we have non-planar diagrams. For example consider the following non-planar graph: 
\begin{equation}
    \begin{tikzpicture}[line width=0.6 ,scale=0.4,line cap=round,every node/.style={font=\normalsize}]
    \begin{scope}[xshift=5cm,scale=0.6]
        \draw[very thick] (0,0) circle (6);

        \coordinate (l1r) at ($4.5*({cos(8)},{sin(8)})$);
        \coordinate (l2r) at ($4.5*({cos(-8)},{sin(-8)})$);
        \coordinate (l1l) at ($4.5*({cos(172)},{sin(172)})$);
        \coordinate (l2l) at ($4.5*({cos(188)},{sin(188)})$);
        \draw[dashed,Green, very thick] (0,0) circle (5.25);
        	\node at (0,-7.5) {$\Delta_{1,2} = 1-D$};

        
        \coordinate (p1v1) at ($(l1r)+(-2,0)$);
        \coordinate (p2v1) at ($(p1v1)+(-1,0)$);
        \coordinate (p3v1) at ($(p1v1)+(0,0.8)$);
        \coordinate (p4v1) at ($(p2v1)+(0,0.8)$);

        \draw[very thick] (p1v1) to (p3v1);
        \draw[very thick] (p2v1) to (p4v1);

        \coordinate (p1v2) at ($(l1r)+(-6,0)$);
        \coordinate (p2v2) at ($(p1v2)+(-1,0)$);
        \coordinate (p3v2) at ($(p1v2)+(0,0.8)$);
        \coordinate (p4v2) at ($(p2v2)+(0,0.8)$);

        \draw[very thick] (p1v2) to (p3v2);
        \draw[very thick] (p2v2) to (p4v2);

        \coordinate (p1v3) at ($(l2r)+(-3.75,0)$);
        \coordinate (p2v3) at ($(p1v3)+(-1,0)$);
        \coordinate (p3v3) at ($(p1v3)+(0,-0.8)$);
        \coordinate (p4v3) at ($(p2v3)+(0,-0.8)$);

        \draw[very thick] (p1v3) to (p3v3);
        \draw[very thick] (p2v3) to (p4v3);

        \draw[very thick] (l1r) to (p1v1);
        \draw[very thick] (p1v2) to (p2v1);
        \draw[very thick] (p2v2) to (l1l);
        \draw[very thick] (l2r) to (p1v3);
        \draw[very thick] (p2v3) to (l2l);

        \draw [very thick] ($4.5*({cos(8)},{sin(8)})$) arc (8:44:4.5);
        \draw [very thick] ($4.5*({cos(56)},{sin(56)})$) arc (56:124:4.5);
        \draw [very thick] ($4.5*({cos(136)},{sin(136)})$) arc (136:172:4.5);

        \draw [very thick] ($4.5*({cos(188)},{sin(188)})$) arc (188:214:4.5);
        \draw [very thick] ($4.5*({cos(226)},{sin(226)})$) arc (226:264:4.5);
        \draw [very thick] ($4.5*({cos(276)},{sin(276)})$) arc (276:314:4.5);
        \draw [very thick] ($4.5*({cos(326)},{sin(326)})$) arc (326:352:4.5);

        \begin{scope}[rotate=50]
            \coordinate (ac1) at ($4.5*({cos(6)},{sin(6)})$);
            \coordinate (ac1v) at ($(ac1)+(-0.8,0)$);
            \coordinate (ac2) at ($4.5*({cos(-6)},{sin(-6)})$);
            \coordinate (ac2v) at ($(ac2)+(-0.8,0)$);
            \draw[very thick] (ac1) to (ac1v);
            \draw[very thick] (ac2) to (ac2v);
        \end{scope}

        \begin{scope}[rotate=130]
            \coordinate (ac1) at ($4.5*({cos(6)},{sin(6)})$);
            \coordinate (ac1v) at ($(ac1)+(-0.8,0)$);
            \coordinate (ac2) at ($4.5*({cos(-6)},{sin(-6)})$);
            \coordinate (ac2v) at ($(ac2)+(-0.8,0)$);
            \draw[very thick] (ac1) to (ac1v);
            \draw[very thick] (ac2) to (ac2v);
        \end{scope}

        \begin{scope}[rotate=220]
            \coordinate (ac1) at ($4.5*({cos(6)},{sin(6)})$);
            \coordinate (ac1v) at ($(ac1)+(-0.8,0)$);
            \coordinate (ac2) at ($4.5*({cos(-6)},{sin(-6)})$);
            \coordinate (ac2v) at ($(ac2)+(-0.8,0)$);
            \draw[very thick] (ac1) to (ac1v);
            \draw[very thick] (ac2) to (ac2v);
        \end{scope}

        \begin{scope}[rotate=270]
            \coordinate (ac1) at ($4.5*({cos(6)},{sin(6)})$);
            \coordinate (ac1v) at ($(ac1)+(-0.8,0)$);
            \coordinate (ac2) at ($4.5*({cos(-6)},{sin(-6)})$);
            \coordinate (ac2v) at ($(ac2)+(-0.8,0)$);
            \draw[very thick] (ac1) to (ac1v);
            \draw[very thick] (ac2) to (ac2v);
        \end{scope}

        \begin{scope}[rotate=320]
            \coordinate (ac1) at ($4.5*({cos(6)},{sin(6)})$);
            \coordinate (ac1v) at ($(ac1)+(-0.8,0)$);
            \coordinate (ac2) at ($4.5*({cos(-6)},{sin(-6)})$);
            \coordinate (ac2v) at ($(ac2)+(-0.8,0)$);
            \draw[very thick] (ac1) to (ac1v);
            \draw[very thick] (ac2) to (ac2v);
        \end{scope}
           
    \end{scope}
    \end{tikzpicture}
        \quad 
    \begin{tikzpicture}[line width=0.6 ,scale=0.4,line cap=round,every node/.style={font=\normalsize}]
    \begin{scope}[xshift=5cm,scale=0.6]
        \draw[very thick] (0,0) circle (6);

        \coordinate (l1r) at ($4.5*({cos(8)},{sin(8)})$);
        \coordinate (l2r) at ($4.5*({cos(-8)},{sin(-8)})$);
        \coordinate (l1l) at ($4.5*({cos(172)},{sin(172)})$);
        \coordinate (l2l) at ($4.5*({cos(188)},{sin(188)})$);
        \draw[dashed,Green, very thick] (5.25,0) arc (0:180:5.25) to (5.25,0);
        
		\node at (0,-7.5) {$\Delta_1 = -D$};

        
        \coordinate (p1v1) at ($(l1r)+(-2,0)$);
        \coordinate (p2v1) at ($(p1v1)+(-1,0)$);
        \coordinate (p3v1) at ($(p1v1)+(0,0.8)$);
        \coordinate (p4v1) at ($(p2v1)+(0,0.8)$);

        \draw[very thick] (p1v1) to (p3v1);
        \draw[very thick] (p2v1) to (p4v1);

        \coordinate (p1v2) at ($(l1r)+(-6,0)$);
        \coordinate (p2v2) at ($(p1v2)+(-1,0)$);
        \coordinate (p3v2) at ($(p1v2)+(0,0.8)$);
        \coordinate (p4v2) at ($(p2v2)+(0,0.8)$);

        \draw[very thick] (p1v2) to (p3v2);
        \draw[very thick] (p2v2) to (p4v2);

        \coordinate (p1v3) at ($(l2r)+(-3.75,0)$);
        \coordinate (p2v3) at ($(p1v3)+(-1,0)$);
        \coordinate (p3v3) at ($(p1v3)+(0,-0.8)$);
        \coordinate (p4v3) at ($(p2v3)+(0,-0.8)$);

        \draw[very thick] (p1v3) to (p3v3);
        \draw[very thick] (p2v3) to (p4v3);

        \draw[very thick] (l1r) to (p1v1);
        \draw[very thick] (p1v2) to (p2v1);
        \draw[very thick] (p2v2) to (l1l);
        \draw[very thick] (l2r) to (p1v3);
        \draw[very thick] (p2v3) to (l2l);

        \draw [very thick] ($4.5*({cos(8)},{sin(8)})$) arc (8:44:4.5);
        \draw [very thick] ($4.5*({cos(56)},{sin(56)})$) arc (56:124:4.5);
        \draw [very thick] ($4.5*({cos(136)},{sin(136)})$) arc (136:172:4.5);

        \draw [very thick] ($4.5*({cos(188)},{sin(188)})$) arc (188:214:4.5);
        \draw [very thick] ($4.5*({cos(226)},{sin(226)})$) arc (226:264:4.5);
        \draw [very thick] ($4.5*({cos(276)},{sin(276)})$) arc (276:314:4.5);
        \draw [very thick] ($4.5*({cos(326)},{sin(326)})$) arc (326:352:4.5);

        \begin{scope}[rotate=50]
            \coordinate (ac1) at ($4.5*({cos(6)},{sin(6)})$);
            \coordinate (ac1v) at ($(ac1)+(-0.8,0)$);
            \coordinate (ac2) at ($4.5*({cos(-6)},{sin(-6)})$);
            \coordinate (ac2v) at ($(ac2)+(-0.8,0)$);
            \draw[very thick] (ac1) to (ac1v);
            \draw[very thick] (ac2) to (ac2v);
        \end{scope}

        \begin{scope}[rotate=130]
            \coordinate (ac1) at ($4.5*({cos(6)},{sin(6)})$);
            \coordinate (ac1v) at ($(ac1)+(-0.8,0)$);
            \coordinate (ac2) at ($4.5*({cos(-6)},{sin(-6)})$);
            \coordinate (ac2v) at ($(ac2)+(-0.8,0)$);
            \draw[very thick] (ac1) to (ac1v);
            \draw[very thick] (ac2) to (ac2v);
        \end{scope}

        \begin{scope}[rotate=220]
            \coordinate (ac1) at ($4.5*({cos(6)},{sin(6)})$);
            \coordinate (ac1v) at ($(ac1)+(-0.8,0)$);
            \coordinate (ac2) at ($4.5*({cos(-6)},{sin(-6)})$);
            \coordinate (ac2v) at ($(ac2)+(-0.8,0)$);
            \draw[very thick] (ac1) to (ac1v);
            \draw[very thick] (ac2) to (ac2v);
        \end{scope}

        \begin{scope}[rotate=270]
            \coordinate (ac1) at ($4.5*({cos(6)},{sin(6)})$);
            \coordinate (ac1v) at ($(ac1)+(-0.8,0)$);
            \coordinate (ac2) at ($4.5*({cos(-6)},{sin(-6)})$);
            \coordinate (ac2v) at ($(ac2)+(-0.8,0)$);
            \draw[very thick] (ac1) to (ac1v);
            \draw[very thick] (ac2) to (ac2v);
        \end{scope}

        \begin{scope}[rotate=320]
            \coordinate (ac1) at ($4.5*({cos(6)},{sin(6)})$);
            \coordinate (ac1v) at ($(ac1)+(-0.8,0)$);
            \coordinate (ac2) at ($4.5*({cos(-6)},{sin(-6)})$);
            \coordinate (ac2v) at ($(ac2)+(-0.8,0)$);
            \draw[very thick] (ac1) to (ac1v);
            \draw[very thick] (ac2) to (ac2v);
        \end{scope}
           
    \end{scope}
    \end{tikzpicture}
    \quad 
    \begin{tikzpicture}[line width=0.6 ,scale=0.4,line cap=round,every node/.style={font=\normalsize}]
    \begin{scope}[xshift=5cm,scale=0.6]
        \draw[very thick] (0,0) circle (6);

        \coordinate (l1r) at ($4.5*({cos(8)},{sin(8)})$);
        \coordinate (l2r) at ($4.5*({cos(-8)},{sin(-8)})$);
        \coordinate (l1l) at ($4.5*({cos(172)},{sin(172)})$);
        \coordinate (l2l) at ($4.5*({cos(188)},{sin(188)})$);
        \draw[dashed,Green, very thick] (5.25,0) arc (0:-180:5.25) to (5.25,0);
        
		\node at (0,-7.5) {$\Delta_2 = -D$};
        
        \coordinate (p1v1) at ($(l1r)+(-2,0)$);
        \coordinate (p2v1) at ($(p1v1)+(-1,0)$);
        \coordinate (p3v1) at ($(p1v1)+(0,0.8)$);
        \coordinate (p4v1) at ($(p2v1)+(0,0.8)$);

        \draw[very thick] (p1v1) to (p3v1);
        \draw[very thick] (p2v1) to (p4v1);

        \coordinate (p1v2) at ($(l1r)+(-6,0)$);
        \coordinate (p2v2) at ($(p1v2)+(-1,0)$);
        \coordinate (p3v2) at ($(p1v2)+(0,0.8)$);
        \coordinate (p4v2) at ($(p2v2)+(0,0.8)$);

        \draw[very thick] (p1v2) to (p3v2);
        \draw[very thick] (p2v2) to (p4v2);

        \coordinate (p1v3) at ($(l2r)+(-3.75,0)$);
        \coordinate (p2v3) at ($(p1v3)+(-1,0)$);
        \coordinate (p3v3) at ($(p1v3)+(0,-0.8)$);
        \coordinate (p4v3) at ($(p2v3)+(0,-0.8)$);

        \draw[very thick] (p1v3) to (p3v3);
        \draw[very thick] (p2v3) to (p4v3);

        \draw[very thick] (l1r) to (p1v1);
        \draw[very thick] (p1v2) to (p2v1);
        \draw[very thick] (p2v2) to (l1l);
        \draw[very thick] (l2r) to (p1v3);
        \draw[very thick] (p2v3) to (l2l);

        \draw [very thick] ($4.5*({cos(8)},{sin(8)})$) arc (8:44:4.5);
        \draw [very thick] ($4.5*({cos(56)},{sin(56)})$) arc (56:124:4.5);
        \draw [very thick] ($4.5*({cos(136)},{sin(136)})$) arc (136:172:4.5);

        \draw [very thick] ($4.5*({cos(188)},{sin(188)})$) arc (188:214:4.5);
        \draw [very thick] ($4.5*({cos(226)},{sin(226)})$) arc (226:264:4.5);
        \draw [very thick] ($4.5*({cos(276)},{sin(276)})$) arc (276:314:4.5);
        \draw [very thick] ($4.5*({cos(326)},{sin(326)})$) arc (326:352:4.5);

        \begin{scope}[rotate=50]
            \coordinate (ac1) at ($4.5*({cos(6)},{sin(6)})$);
            \coordinate (ac1v) at ($(ac1)+(-0.8,0)$);
            \coordinate (ac2) at ($4.5*({cos(-6)},{sin(-6)})$);
            \coordinate (ac2v) at ($(ac2)+(-0.8,0)$);
            \draw[very thick] (ac1) to (ac1v);
            \draw[very thick] (ac2) to (ac2v);
        \end{scope}

        \begin{scope}[rotate=130]
            \coordinate (ac1) at ($4.5*({cos(6)},{sin(6)})$);
            \coordinate (ac1v) at ($(ac1)+(-0.8,0)$);
            \coordinate (ac2) at ($4.5*({cos(-6)},{sin(-6)})$);
            \coordinate (ac2v) at ($(ac2)+(-0.8,0)$);
            \draw[very thick] (ac1) to (ac1v);
            \draw[very thick] (ac2) to (ac2v);
        \end{scope}

        \begin{scope}[rotate=220]
            \coordinate (ac1) at ($4.5*({cos(6)},{sin(6)})$);
            \coordinate (ac1v) at ($(ac1)+(-0.8,0)$);
            \coordinate (ac2) at ($4.5*({cos(-6)},{sin(-6)})$);
            \coordinate (ac2v) at ($(ac2)+(-0.8,0)$);
            \draw[very thick] (ac1) to (ac1v);
            \draw[very thick] (ac2) to (ac2v);
        \end{scope}

        \begin{scope}[rotate=270]
            \coordinate (ac1) at ($4.5*({cos(6)},{sin(6)})$);
            \coordinate (ac1v) at ($(ac1)+(-0.8,0)$);
            \coordinate (ac2) at ($4.5*({cos(-6)},{sin(-6)})$);
            \coordinate (ac2v) at ($(ac2)+(-0.8,0)$);
            \draw[very thick] (ac1) to (ac1v);
            \draw[very thick] (ac2) to (ac2v);
        \end{scope}

        \begin{scope}[rotate=320]
            \coordinate (ac1) at ($4.5*({cos(6)},{sin(6)})$);
            \coordinate (ac1v) at ($(ac1)+(-0.8,0)$);
            \coordinate (ac2) at ($4.5*({cos(-6)},{sin(-6)})$);
            \coordinate (ac2v) at ($(ac2)+(-0.8,0)$);
            \draw[very thick] (ac1) to (ac1v);
            \draw[very thick] (ac2) to (ac2v);
        \end{scope}
           
    \end{scope}
    \end{tikzpicture}
\end{equation}
so we see that question of how many ``punctures/regions'' it encloses is ill-defined, but the correct form of the rule still holds: $\Delta^\prime$ is homotopic to a boundary of the surface, or equivalently, a purely left-turning curve, and therefore it's exponent is $(1-D)$, while the same is not true for $\Delta_{1}$ or $\Delta_{2}$, and thus their exponent is simply $(-D)$.

\section{Cancellations and the V-Rule}
\label{sec:CancellationsVRule}

Throughout the text we saw that we can easily produce all the terms entering a given LS by drawing all the possible contraction curves filling the fat graph. However, each contraction curve produces different $X_{i,j}$'s from its cores and extensions, with different signs, and therefore when we add all the contributions together there are some monomials that \textit{cancel}. We now present a systematic way of checking whether a given monomial survives or cancels out in the full answer that we call the \textit{V-rule}. We also discuss a different sort of cancellations pointed out in eqn. \eqref{eq:LSBubbleQFT}, coming from the extensions of curves around the puncture.  

Let us start by revisiting the first cancellation we saw in the 4-point $s$-channel LS, where the monomial $X_{1,4}X_{1,6}X_{5,8}$ cancelled when we added the two contractions of equation \eqref{eq:SchannContract}. To understand this cancellation, we start by drawing the cores of $X_{1,4},\,X_{1,6},$ and $X_{5,8}$ as well as the respective extensions to fill the whole fatgraph with non overlapping curves. This produces two different pictures, represented in  \eqref{eq:4ptVrule}, which exactly match the two contractions patterns of \eqref{eq:SchannContract}. 

\begin{equation}
\label{eq:4ptVrule}
    \begin{aligned}
	 &\begin{gathered}
	 	\begin{tikzpicture}[line width=0.6,scale=0.3,line cap=round,every node/.style={font=\footnotesize}]
		\node (1l) at (-2, -1) {};
		\node (1r) at (2, -1) {};
		\node (5l) at (-2, 1) {};
		\node (5r) at (2, 1) {};
		
		\node (2) at (-5, -3) {};
		\node (2d) at (-5, -5) {};
		\node (2l) at (-7, -3) {};
		\node (2r) at (-4, -3) {};
		\node (2dr) at (-4, -5) {};
		\node (2u) at (-5, -2) {};
		\node (2lu) at (-7, -2) {};

		\node (4) at (-5, 3) {};
		\node (4u) at (-5, 5) {};
		\node (4l) at (-7, 3) {};
		\node (4r) at (-4, 3) {};
		\node (4ur) at (-4, 5) {};
		\node (4d) at (-5, 2) {};
		\node (4ld) at (-7, 2) {};

		\node (6) at (5, 3) {};
		\node (6u) at (5, 5) {};
		\node (6r) at (7, 3) {};
		\node (6l) at (4, 3) {};
		\node (6ul) at (4, 5) {};
		\node (6d) at (5, 2) {};
		\node (6dr) at (7, 2) {};
		
		\node (8) at (5, -3) {};
		\node (8r) at (7, -3) {};
		\node (8d) at (5, -5) {};
		\node (8dl) at (4, -5) {};
		\node (8l) at (4, -3) {};
		\node (8ur) at (7, -2) {};
		\node (8u) at (5, -2) {};;
		
		\node (3) at (-3, 0) {};
		\node (7) at (3, 0) {};

		\draw[thick] (2l.center) to (2.center);
		\draw[thick] (2.center) to (2d.center);
		
		\draw[thick] (4u.center) to (4.center);
		\draw[thick] (4.center) to (4l.center);

		\draw[thick] (6u.center) to (6.center);
		\draw[thick] (6.center) to (6r.center);
		
		\draw[thick] (8r.center) to (8.center);
		\draw[thick] (8.center) to (8d.center);
		
		\draw[thick] (2lu.center) to (2u.center);
		\draw[thick] (2u.center) to (3.center);
		\draw[thick] (3.center) to (4d.center);
		\draw[thick] (4d.center) to (4ld.center);

		\draw[thick] (6dr.center) to (6d.center);
		\draw[thick] (6d.center) to (7.center);
		\draw[thick] (7.center) to (8u.center);
		\draw[thick] (8u.center) to (8ur.center);

		\draw[thick] (4ur.center) to (4r.center);
		\draw[thick] (4r.center) to (5l.center);
		\draw[thick] (5l.center) to (5r.center);
		\draw[thick] (5r.center) to (6l.center);
		\draw[thick] (6l.center) to (6ul.center);
		
		\draw[thick] (2dr.center) to (2r.center);
		\draw[thick] (2r.center) to (1l.center);
		\draw[thick] (1l.center) to (1r.center);
		\draw[thick] (1r.center) to (8l.center);
		\draw[thick] (8l.center) to (8dl.center);
		\coordinate (2md) at ($(2)!0.3!(2d)$);
        \coordinate (2ml) at ($(2)!0.3!(2l)$);
		\coordinate (2hm) at ($(2md)!0.5!(2ml)$);
        
 		\coordinate (4mu) at ($(4)!0.3!(4u)$);
        \coordinate (4ml) at ($(4)!0.3!(4l)$);
		\coordinate (4hm) at ($(4mu)!0.5!(4ml)$);
        
 		\coordinate (6mu) at ($(6)!0.3!(6u)$);
        \coordinate (6mr) at ($(6)!0.3!(6r)$);
		\coordinate (6hm) at ($(6mu)!0.5!(6mr)$);
        
  		\coordinate (8mr) at ($(8)!0.3!(8r)$);
        \coordinate (8md) at ($(8)!0.3!(8d)$);
		\coordinate (8hm) at ($(8mr)!0.5!(8md)$);
        
        \coordinate (5hr) at ($(5r)!0.2!(6l)$);
       	\coordinate (5hl) at ($(5r)!0.15!(5l)$);
		\coordinate (5hm) at ($(5hl)!0.5!(5hr)$);

        \coordinate (3hu) at ($(3)!0.25!(4d)$);
       	\coordinate (3hd) at ($(3)!0.25!(2u)$);
		\coordinate (3hm) at ($(3hu)!0.5!(3hd)$);

        \draw[ForestGreen, very thick] (-4.5,2.5) to (-2.5,0.5);
        \draw[ForestGreen, very thick] (4.5,2.5) to (2,0);
        \draw[ForestGreen, very thick] (4.5,-2.5) to (2.5,-0.5);
        
        \draw[ForestGreen, very thick,dashed] (2,0) to (-2,0);
        \draw[ForestGreen, very thick,dashed] (-2,0) to (-4.5,-2.5);

        \node[] at (0,-4) {$1$};
        \node[] at (0,4) {$5$};
        \node[] at (-6,0) {$3$};
        \node[] at (6,0) {$7$};
        \node[] at (-6.5,-4.5) {$2$};
        \node[] at (-6.5,4.5) {$4$};
        \node[] at (6.5,4.5) {$6$};
        \node[] at (6.5,-4.5) {$8$};
	\end{tikzpicture}
	 \end{gathered}\\[-0.5em]
     &  \quad \, \,  -X_{1,4}X_{1,6}X_{5,8}       
\end{aligned}
\quad \vline \quad 
 \begin{aligned}
	 &\begin{gathered}
	 	\begin{tikzpicture}[line width=0.6,scale=0.3,line cap=round,every node/.style={font=\footnotesize}]
		\node (1l) at (-2, -1) {};
		\node (1r) at (2, -1) {};
		\node (5l) at (-2, 1) {};
		\node (5r) at (2, 1) {};
		
		\node (2) at (-5, -3) {};
		\node (2d) at (-5, -5) {};
		\node (2l) at (-7, -3) {};
		\node (2r) at (-4, -3) {};
		\node (2dr) at (-4, -5) {};
		\node (2u) at (-5, -2) {};
		\node (2lu) at (-7, -2) {};

		\node (4) at (-5, 3) {};
		\node (4u) at (-5, 5) {};
		\node (4l) at (-7, 3) {};
		\node (4r) at (-4, 3) {};
		\node (4ur) at (-4, 5) {};
		\node (4d) at (-5, 2) {};
		\node (4ld) at (-7, 2) {};

		\node (6) at (5, 3) {};
		\node (6u) at (5, 5) {};
		\node (6r) at (7, 3) {};
		\node (6l) at (4, 3) {};
		\node (6ul) at (4, 5) {};
		\node (6d) at (5, 2) {};
		\node (6dr) at (7, 2) {};
		
		\node (8) at (5, -3) {};
		\node (8r) at (7, -3) {};
		\node (8d) at (5, -5) {};
		\node (8dl) at (4, -5) {};
		\node (8l) at (4, -3) {};
		\node (8ur) at (7, -2) {};
		\node (8u) at (5, -2) {};;
		
		\node (3) at (-3, 0) {};
		\node (7) at (3, 0) {};

		\draw[thick] (2l.center) to (2.center);
		\draw[thick] (2.center) to (2d.center);
		
		\draw[thick] (4u.center) to (4.center);
		\draw[thick] (4.center) to (4l.center);

		\draw[thick] (6u.center) to (6.center);
		\draw[thick] (6.center) to (6r.center);
		
		\draw[thick] (8r.center) to (8.center);
		\draw[thick] (8.center) to (8d.center);
		
		\draw[thick] (2lu.center) to (2u.center);
		\draw[thick] (2u.center) to (3.center);
		\draw[thick] (3.center) to (4d.center);
		\draw[thick] (4d.center) to (4ld.center);

		\draw[thick] (6dr.center) to (6d.center);
		\draw[thick] (6d.center) to (7.center);
		\draw[thick] (7.center) to (8u.center);
		\draw[thick] (8u.center) to (8ur.center);

		\draw[thick] (4ur.center) to (4r.center);
		\draw[thick] (4r.center) to (5l.center);
		\draw[thick] (5l.center) to (5r.center);
		\draw[thick] (5r.center) to (6l.center);
		\draw[thick] (6l.center) to (6ul.center);
		
		\draw[thick] (2dr.center) to (2r.center);
		\draw[thick] (2r.center) to (1l.center);
		\draw[thick] (1l.center) to (1r.center);
		\draw[thick] (1r.center) to (8l.center);
		\draw[thick] (8l.center) to (8dl.center);
		\coordinate (2md) at ($(2)!0.3!(2d)$);
        \coordinate (2ml) at ($(2)!0.3!(2l)$);
		\coordinate (2hm) at ($(2md)!0.5!(2ml)$);
        
 		\coordinate (4mu) at ($(4)!0.3!(4u)$);
        \coordinate (4ml) at ($(4)!0.3!(4l)$);
		\coordinate (4hm) at ($(4mu)!0.5!(4ml)$);
        
 		\coordinate (6mu) at ($(6)!0.3!(6u)$);
        \coordinate (6mr) at ($(6)!0.3!(6r)$);
		\coordinate (6hm) at ($(6mu)!0.5!(6mr)$);
        
  		\coordinate (8mr) at ($(8)!0.3!(8r)$);
        \coordinate (8md) at ($(8)!0.3!(8d)$);
		\coordinate (8hm) at ($(8mr)!0.5!(8md)$);
        
        \coordinate (5hr) at ($(5r)!0.2!(6l)$);
       	\coordinate (5hl) at ($(5r)!0.15!(5l)$);
		\coordinate (5hm) at ($(5hl)!0.5!(5hr)$);

        \coordinate (3hu) at ($(3)!0.25!(4d)$);
       	\coordinate (3hd) at ($(3)!0.25!(2u)$);
		\coordinate (3hm) at ($(3hu)!0.5!(3hd)$);

        \draw[ForestGreen, very thick] (-4.5,2.5) to (-2,0);
        \draw[ForestGreen, very thick] (4.5,2.5) to (2,0);
        \draw[ForestGreen, very thick] (4.5,-2.5) to (2.5,-0.5);
        
        \draw[ForestGreen, very thick,dashed] (2,0) to (-1,0);
        \draw[ForestGreen, very thick,dashed] (-4.5,-2.5) to (-2,0);

        \node[] at (0,-4) {$1$};
        \node[] at (0,4) {$5$};
        \node[] at (-6,0) {$3$};
        \node[] at (6,0) {$7$};
        \node[] at (-6.5,-4.5) {$2$};
        \node[] at (-6.5,4.5) {$4$};
        \node[] at (6.5,4.5) {$6$};
        \node[] at (6.5,-4.5) {$8$};
	\end{tikzpicture}
	 \end{gathered}\\[-0.5em]
     & \quad \, \,  +X_{1,4}X_{1,6}X_{5,8}       
\end{aligned}
\end{equation}

Looking at the two pictures, we recognize that the total number of extensions differs by one, and so the monomial is produced with opposite sign in each case, therefore cancelling in the final expression. Crucially, this difference in the total number of extensions comes from the conflicting extensions of $X_{1,4}$ and $X_{1,6}$, both of which had to fill road $(1,3)$. On top of that, for this conflict to result in a change in the number of extensions, it is also important that the conflicting extension of $X_{1,4}$ was a right turn, because it implies that its alternative -- only having the extension of $X_{1,6}$ -- had, overall, one less extension. 

As it turns out, this cancellation pattern can be made into a completely general rule that probes cancellations, which we call the \textit{V-Rule}. 

\paragraph{V-Rule} Given a monomial $\prod_C X_c$ and their possible extensions, if two cores have overlapping extensions, then the corresponding monomial does not appear in the leading singularity. We call it V-rule because graphically when two monomials have overlapping extensions, they converge in a V-Shape.

This rule comes from the simple fact that extensions are always done in accordance with the ``turn right once and left forever'' rule, and thus, whenever we have two overlapping extensions this means that we have two alternatives of filling the graph with non-overlapping curves which differ by one in the total number of extensions $N_e$. 

\paragraph{Cancellations around a planar loop} Let us now discuss a different type of cancellation arising at loop-level. Already for the planar $n$-gon at one-loop, if we look at the monomial that contains all the curves ending on the puncture and that does not contain any closed curve, $\prod_{i}X_{2i,p}$, from the contraction/surface picture this monomial can be generated in different ways and with different signs depending on exactly what extensions we pick for each curve (as pointed out in \eqref{eq:LSBubbleQFT})\footnote{this is of course true for any monomial containing more than one curve ending on the puncture}. In the cases discussed above, we saw that after summing over all such possibilities we got that the coefficient of $\prod_{i}X_{2i,p}$ was exactly $1$. Let us now show that this is always the case.

Say we are dealing with a monomial where there are $n$ curves ending on the puncture. Then we want to account for all the ways to  extend these $n$ curves around the loop, with any number of extensions $N_e \leq n$, where a configuration with $N_e$ extensions comes with a factor $(-1)^{N_e}$. Up to three extensions, such contributions would look like the following picture 
\begin{equation}
\begin{tikzpicture}[line width=0.6 ,scale=0.4,line cap=round,every node/.style={font=\large}]
	\begin{scope}[xshift= 7cm, scale=0.6]
        \draw[very thick](0,0) circle (3);

        \coordinate (r1lauxvec) at ($({cos(240)},{sin(240)})$);

        \draw[very thick] ($4.5*({cos(-96)},{sin(-96)})$) arc (-96:-129:4.5);
        \draw[very thick] ($4.5*({cos(-141)},{sin(-141)})$) arc (-141:-200:4.5);
        \draw[very thick] ($4.5*({cos(-84)},{sin(-84)})$) arc (-84:-51:4.5);
        \draw[very thick] ($4.5*({cos(-39)},{sin(-39)})$) arc (-39:20:4.5);
		\draw[dotted] ($4.5*({cos(150)},{sin(150)})$) arc (150:30:4.5);
		\coordinate (cont1) at ($3.75*({cos(-110)},{sin(-110)})$);
		\node at (0,6) {\normalsize $N_e=1$};
		\coordinate (cont2) at ($3.75*({cos(-150)},{sin(-150)})$);
		\draw[dotted, Blue, very thick] ($4.5*({cos(-135)},{sin(-135)})$) to[out=45, in=-55] (cont2);
		\draw[dotted, Blue, very thick] (cont2) arc (210:-130:3.75);

		\coordinate (cont3) at ($3.75*({cos(-55)},{sin(-55)})$);

        


        \begin{scope}[rotate=-90]
            \coordinate (rtmp) at ($4.5*({cos(6)},{sin(6)})$) {};
            \coordinate (rtmp2) at ($(rtmp) + 2*(1,0)$){};
            \coordinate (rtmp3) at ($4.5*({cos(-6)},{sin(-6)})$);
            \coordinate (rtmp4) at ($(rtmp3) + 2*(1,0)$);
            \coordinate (rtmp5) at ($(rtmp2) + 1.5*(1,1)$){};
            \coordinate (rtmp6) at ($(rtmp4) + 1.5*(1,-1)$);
    
            \coordinate (r2tmp) at ($7.5*(1,0)$);
            \coordinate (r2tmp1) at ($(r2tmp) + (1,1)$);
            \coordinate (r2tmp2) at ($(r2tmp) + (1,-1)$);
           	
           	\coordinate (b1) at ($(rtmp2)!0.5!(rtmp4)$);
           	\coordinate (b2) at ($(rtmp)!0.5!(rtmp3)$);
          	\draw[very thick, Green] (b1) to (b2);
            \draw[very thick] (rtmp) -- (rtmp2) -- (rtmp5);
            \draw[very thick] (rtmp3) -- (rtmp4) -- (rtmp6);
            \draw[very thick] (r2tmp2) -- (r2tmp) -- (r2tmp1);
        \end{scope}
        
        \begin{scope}[rotate=225]
            
            \coordinate (rtmp) at ($4.5*({cos(6)},{sin(6)})$) {};
            \coordinate (rtmp2) at ($(rtmp) + 2*(1,0)$){};
            \coordinate (rtmp3) at ($4.5*({cos(-6)},{sin(-6)})$);
            \coordinate (rtmp4) at ($(rtmp3) + 2*(1,0)$);
            \coordinate (rtmp5) at ($(rtmp2) + 1.5*(1,1)$){};
            \coordinate (rtmp6) at ($(rtmp4) + 1.5*(1,-1)$);
    		\coordinate (b1) at ($(rtmp2)!0.5!(rtmp4)$);
           	\coordinate (b2) at ($(rtmp)!0.5!(rtmp3)$);
          	\draw[very thick, Blue] (b1) to (b2);
            \coordinate (r2tmp) at ($7.5*(1,0)$);
            \coordinate (r2tmp1) at ($(r2tmp) + (1,1)$);
            \coordinate (r2tmp2) at ($(r2tmp) + (1,-1)$);
           	\coordinate (h1) at ($3*({cos(10)},{sin(10)})$);
        		\coordinate (h2) at ($3*({cos(-10)},{sin(-10)})$);
            \draw[very thick] (rtmp) -- (rtmp2) -- (rtmp5);
            \draw[very thick] (rtmp3) -- (rtmp4) -- (rtmp6);
            \draw[very thick] (r2tmp2) -- (r2tmp) -- (r2tmp1);
        \end{scope}
        
        \begin{scope}[rotate=-45]
            \coordinate (rtmp) at ($4.5*({cos(6)},{sin(6)})$) {};
            \coordinate (rtmp2) at ($(rtmp) + 2*(1,0)$){};
            \coordinate (rtmp3) at ($4.5*({cos(-6)},{sin(-6)})$);
            \coordinate (rtmp4) at ($(rtmp3) + 2*(1,0)$);
            \coordinate (rtmp5) at ($(rtmp2) + 1.5*(1,1)$){};
            \coordinate (rtmp6) at ($(rtmp4) + 1.5*(1,-1)$);
    		\coordinate (b1) at ($(rtmp2)!0.5!(rtmp4)$);
           	\coordinate (b2) at ($(rtmp)!0.5!(rtmp3)$);
          	\draw[very thick, Maroon] (b1) to (b2);
            \coordinate (r2tmp) at ($7.5*(1,0)$);
            \coordinate (r2tmp1) at ($(r2tmp) + (1,1)$);
            \coordinate (r2tmp2) at ($(r2tmp) + (1,-1)$);
           	
            \draw[very thick] (rtmp) -- (rtmp2) -- (rtmp5);
            \draw[very thick] (rtmp3) -- (rtmp4) -- (rtmp6);
            \draw[very thick] (r2tmp2) -- (r2tmp) -- (r2tmp1);
        \end{scope}
    \end{scope}
	\end{tikzpicture} 
	\quad \,
	\begin{tikzpicture}[line width=0.6 ,scale=0.4,line cap=round,every node/.style={font=\large}]
	\begin{scope}[xshift= 7cm, scale=0.6]
        \draw[very thick](0,0) circle (3);

        \coordinate (r1lauxvec) at ($({cos(240)},{sin(240)})$);

        \draw[very thick] ($4.5*({cos(-96)},{sin(-96)})$) arc (-96:-129:4.5);
        \draw[very thick] ($4.5*({cos(-141)},{sin(-141)})$) arc (-141:-200:4.5);
        \draw[very thick] ($4.5*({cos(-84)},{sin(-84)})$) arc (-84:-51:4.5);
        \draw[very thick] ($4.5*({cos(-39)},{sin(-39)})$) arc (-39:20:4.5);
		\draw[dotted] ($4.5*({cos(150)},{sin(150)})$) arc (150:30:4.5);
		\coordinate (cont1) at ($3.75*({cos(-110)},{sin(-110)})$);
		\draw[dotted, Green, very thick] (0,-4.5) to[out=90, in =-15] (cont1);
		\draw[dotted,Green, very thick] (cont1) arc (250:230:3.75);
		\node at (0,6) {\normalsize $N_e=2$};

		\coordinate (cont2) at ($3.75*({cos(-150)},{sin(-150)})$);
		\draw[dotted, Blue, very thick] ($4.5*({cos(-135)},{sin(-135)})$) to[out=45, in=-55] (cont2);
		\draw[dotted,Blue, very thick] (cont2) arc (210:-80:3.75);

		\coordinate (cont3) at ($3.75*({cos(-55)},{sin(-55)})$);

        


        \begin{scope}[rotate=-90]
            \coordinate (rtmp) at ($4.5*({cos(6)},{sin(6)})$) {};
            \coordinate (rtmp2) at ($(rtmp) + 2*(1,0)$){};
            \coordinate (rtmp3) at ($4.5*({cos(-6)},{sin(-6)})$);
            \coordinate (rtmp4) at ($(rtmp3) + 2*(1,0)$);
            \coordinate (rtmp5) at ($(rtmp2) + 1.5*(1,1)$){};
            \coordinate (rtmp6) at ($(rtmp4) + 1.5*(1,-1)$);
    
            \coordinate (r2tmp) at ($7.5*(1,0)$);
            \coordinate (r2tmp1) at ($(r2tmp) + (1,1)$);
            \coordinate (r2tmp2) at ($(r2tmp) + (1,-1)$);
           	
           	\coordinate (b1) at ($(rtmp2)!0.5!(rtmp4)$);
           	\coordinate (b2) at ($(rtmp)!0.5!(rtmp3)$);
          	\draw[very thick, Green] (b1) to (b2);
            \draw[very thick] (rtmp) -- (rtmp2) -- (rtmp5);
            \draw[very thick] (rtmp3) -- (rtmp4) -- (rtmp6);
            \draw[very thick] (r2tmp2) -- (r2tmp) -- (r2tmp1);
        \end{scope}
        
        \begin{scope}[rotate=225]
            
            \coordinate (rtmp) at ($4.5*({cos(6)},{sin(6)})$) {};
            \coordinate (rtmp2) at ($(rtmp) + 2*(1,0)$){};
            \coordinate (rtmp3) at ($4.5*({cos(-6)},{sin(-6)})$);
            \coordinate (rtmp4) at ($(rtmp3) + 2*(1,0)$);
            \coordinate (rtmp5) at ($(rtmp2) + 1.5*(1,1)$){};
            \coordinate (rtmp6) at ($(rtmp4) + 1.5*(1,-1)$);
    		\coordinate (b1) at ($(rtmp2)!0.5!(rtmp4)$);
           	\coordinate (b2) at ($(rtmp)!0.5!(rtmp3)$);
          	\draw[very thick, Blue] (b1) to (b2);
            \coordinate (r2tmp) at ($7.5*(1,0)$);
            \coordinate (r2tmp1) at ($(r2tmp) + (1,1)$);
            \coordinate (r2tmp2) at ($(r2tmp) + (1,-1)$);
           	\coordinate (h1) at ($3*({cos(10)},{sin(10)})$);
        		\coordinate (h2) at ($3*({cos(-10)},{sin(-10)})$);
            \draw[very thick] (rtmp) -- (rtmp2) -- (rtmp5);
            \draw[very thick] (rtmp3) -- (rtmp4) -- (rtmp6);
            \draw[very thick] (r2tmp2) -- (r2tmp) -- (r2tmp1);
        \end{scope}
        
        \begin{scope}[rotate=-45]
            \coordinate (rtmp) at ($4.5*({cos(6)},{sin(6)})$) {};
            \coordinate (rtmp2) at ($(rtmp) + 2*(1,0)$){};
            \coordinate (rtmp3) at ($4.5*({cos(-6)},{sin(-6)})$);
            \coordinate (rtmp4) at ($(rtmp3) + 2*(1,0)$);
            \coordinate (rtmp5) at ($(rtmp2) + 1.5*(1,1)$){};
            \coordinate (rtmp6) at ($(rtmp4) + 1.5*(1,-1)$);
    		\coordinate (b1) at ($(rtmp2)!0.5!(rtmp4)$);
           	\coordinate (b2) at ($(rtmp)!0.5!(rtmp3)$);
          	\draw[very thick, Maroon] (b1) to (b2);
            \coordinate (r2tmp) at ($7.5*(1,0)$);
            \coordinate (r2tmp1) at ($(r2tmp) + (1,1)$);
            \coordinate (r2tmp2) at ($(r2tmp) + (1,-1)$);
           	
            \draw[very thick] (rtmp) -- (rtmp2) -- (rtmp5);
            \draw[very thick] (rtmp3) -- (rtmp4) -- (rtmp6);
            \draw[very thick] (r2tmp2) -- (r2tmp) -- (r2tmp1);
        \end{scope}
    \end{scope}
	\end{tikzpicture}
	\quad \,
	\begin{tikzpicture}[line width=0.6 ,scale=0.4,line cap=round,every node/.style={font=\large}]
	\begin{scope}[xshift= 7cm, scale=0.6]
        \draw[very thick](0,0) circle (3);

        \coordinate (r1lauxvec) at ($({cos(240)},{sin(240)})$);

        \draw[very thick] ($4.5*({cos(-96)},{sin(-96)})$) arc (-96:-129:4.5);
        \draw[very thick] ($4.5*({cos(-141)},{sin(-141)})$) arc (-141:-200:4.5);
        \draw[very thick] ($4.5*({cos(-84)},{sin(-84)})$) arc (-84:-51:4.5);
        \draw[very thick] ($4.5*({cos(-39)},{sin(-39)})$) arc (-39:20:4.5);
		\draw[dotted] ($4.5*({cos(150)},{sin(150)})$) arc (150:30:4.5);
		\coordinate (cont1) at ($3.75*({cos(-110)},{sin(-110)})$);
		\draw[dotted,Green, very thick] (0,-4.5) to[out=90, in =-15] (cont1);
		\draw[dotted, Green, very thick] (cont1) arc (250:230:3.75);
		\node at (0,6) {\normalsize $N_e=3$};

		\coordinate (cont2) at ($3.75*({cos(-150)},{sin(-150)})$);
		\draw[dotted,Blue, very thick] ($4.5*({cos(-135)},{sin(-135)})$) to[out=45, in=-55] (cont2);
		\draw[dotted,Blue, very thick] (cont2) arc (210:-39:3.75);

		\coordinate (cont3) at ($3.75*({cos(-55)},{sin(-55)})$);
		\draw[dotted,Maroon, very thick] ($4.5*({cos(-45)},{sin(-45)})$) to[out=135, in=30] (cont3);
		\draw[dotted,Maroon, very thick] (cont3) arc (-55:-80:3.75);

        


        \begin{scope}[rotate=-90]
            \coordinate (rtmp) at ($4.5*({cos(6)},{sin(6)})$) {};
            \coordinate (rtmp2) at ($(rtmp) + 2*(1,0)$){};
            \coordinate (rtmp3) at ($4.5*({cos(-6)},{sin(-6)})$);
            \coordinate (rtmp4) at ($(rtmp3) + 2*(1,0)$);
            \coordinate (rtmp5) at ($(rtmp2) + 1.5*(1,1)$){};
            \coordinate (rtmp6) at ($(rtmp4) + 1.5*(1,-1)$);
    
            \coordinate (r2tmp) at ($7.5*(1,0)$);
            \coordinate (r2tmp1) at ($(r2tmp) + (1,1)$);
            \coordinate (r2tmp2) at ($(r2tmp) + (1,-1)$);
           	
           	\coordinate (b1) at ($(rtmp2)!0.5!(rtmp4)$);
           	\coordinate (b2) at ($(rtmp)!0.5!(rtmp3)$);
          	\draw[very thick, Green] (b1) to (b2);
            \draw[very thick] (rtmp) -- (rtmp2) -- (rtmp5);
            \draw[very thick] (rtmp3) -- (rtmp4) -- (rtmp6);
            \draw[very thick] (r2tmp2) -- (r2tmp) -- (r2tmp1);
        \end{scope}
        
        \begin{scope}[rotate=225]
            
            \coordinate (rtmp) at ($4.5*({cos(6)},{sin(6)})$) {};
            \coordinate (rtmp2) at ($(rtmp) + 2*(1,0)$){};
            \coordinate (rtmp3) at ($4.5*({cos(-6)},{sin(-6)})$);
            \coordinate (rtmp4) at ($(rtmp3) + 2*(1,0)$);
            \coordinate (rtmp5) at ($(rtmp2) + 1.5*(1,1)$){};
            \coordinate (rtmp6) at ($(rtmp4) + 1.5*(1,-1)$);
    		\coordinate (b1) at ($(rtmp2)!0.5!(rtmp4)$);
           	\coordinate (b2) at ($(rtmp)!0.5!(rtmp3)$);
          	\draw[very thick, Blue] (b1) to (b2);
            \coordinate (r2tmp) at ($7.5*(1,0)$);
            \coordinate (r2tmp1) at ($(r2tmp) + (1,1)$);
            \coordinate (r2tmp2) at ($(r2tmp) + (1,-1)$);
           	\coordinate (h1) at ($3*({cos(10)},{sin(10)})$);
        		\coordinate (h2) at ($3*({cos(-10)},{sin(-10)})$);
            \draw[very thick] (rtmp) -- (rtmp2) -- (rtmp5);
            \draw[very thick] (rtmp3) -- (rtmp4) -- (rtmp6);
            \draw[very thick] (r2tmp2) -- (r2tmp) -- (r2tmp1);
        \end{scope}
        
        \begin{scope}[rotate=-45]
            \coordinate (rtmp) at ($4.5*({cos(6)},{sin(6)})$) {};
            \coordinate (rtmp2) at ($(rtmp) + 2*(1,0)$){};
            \coordinate (rtmp3) at ($4.5*({cos(-6)},{sin(-6)})$);
            \coordinate (rtmp4) at ($(rtmp3) + 2*(1,0)$);
            \coordinate (rtmp5) at ($(rtmp2) + 1.5*(1,1)$){};
            \coordinate (rtmp6) at ($(rtmp4) + 1.5*(1,-1)$);
    		\coordinate (b1) at ($(rtmp2)!0.5!(rtmp4)$);
           	\coordinate (b2) at ($(rtmp)!0.5!(rtmp3)$);
          	\draw[very thick, Maroon] (b1) to (b2);
            \coordinate (r2tmp) at ($7.5*(1,0)$);
            \coordinate (r2tmp1) at ($(r2tmp) + (1,1)$);
            \coordinate (r2tmp2) at ($(r2tmp) + (1,-1)$);
           	
            \draw[very thick] (rtmp) -- (rtmp2) -- (rtmp5);
            \draw[very thick] (rtmp3) -- (rtmp4) -- (rtmp6);
            \draw[very thick] (r2tmp2) -- (r2tmp) -- (r2tmp1);
        \end{scope}
    \end{scope}
	\end{tikzpicture}
\end{equation}
Moreover, it follows from the cyclic symmetry of the problem that the number of configurations with $N_e$ total extensions is exactly $\binom{n}{N_e}$. Thus, we can write the total contribution as
\begin{equation}
   \text{LS} \supset (-1) \left( \sum_{N_e = 1}^{n} (-1)^{N_e}\binom{n}{N_e}\right) \underbracket[0.4pt]{\prod_i X_{i,p}}_{n \text{ curves}} \times  \prod_{(k,m)\neq p} X_{k,m} = \prod_i X_{i,p} \, ,
\end{equation}
where the product over $(k,m)$ corresponds to the remaining curves on the monomial that do not end on the puncture under consideration. The overall minus sign is inherited from the minus sign multiplying the naive gluing in equation \eqref{eq:NCorect1loop}. Putting this together with the contribution from the closed curves (as well as $\mathcal{N}$), we find  
\begin{equation}
    \text{LS} \supset ((1-D) +  1) \prod_i X_{i,p} \times  \prod_{(k,m)\neq p} X_{k,m}   = (2-D)  \prod_i X_{i,p} \times  \prod_{(k,m)\neq p} X_{k,m}  \, .
    \label{eq:FinalCancLoop}
\end{equation}

\section{A first look at fermion leading singularities}
\label{sec:FermLS}

After having seen how the structure of curves on surfaces naturally arises from standard momentum space gluing, it is an interesting question to ask whether a similar story holds for LS with fermions running in the loop -- as this is of course the case in QCD. We will now briefly discuss a simple way of capturing these fermionic LS in the same spirit of the rest of the paper, but leave further explorations for future work. 
For simplicity, let's stick to one-loop and consider the  $n$-gon LS diagram
\begin{equation}
\label{eq:FermionTrace}
    \vcenter{\hbox{\begin{tikzpicture}[line width=0.6 ,scale=0.9,line cap=round,every node/.style={font=\footnotesize}]
         \begin{feynman}
            \coordinate (p1) at ($3.5*({cos(250)},{sin(250)})$);
            \coordinate (p12) at ($1.5*({cos(250)},{sin(250)})$);

            \coordinate (p2) at ($3.5*({cos(200)},{sin(200)})$);
            \coordinate (p22) at ($1.5*({cos(200)},{sin(200)})$);

            \coordinate (p3) at ($3.5*({cos(150)},{sin(150)})$);
            \coordinate (p32) at ($1.5*({cos(150)},{sin(150)})$);
            \coordinate (p3aux) at ($1.5*({cos(100)},{sin(100)})$);
            \coordinate (pn) at ($3.5*({cos(300)},{sin(300)})$);
            \coordinate (pn2) at ($1.5*({cos(300)},{sin(300)})$);
            \coordinate (pnaux) at ($1.5*({cos(350)},{sin(350)})$);

            \draw[very thick, decoration={text along path, text={ ...},text align={center}},decorate] ($1.5*({cos(100)},{sin(100)})$) arc (100:-10:1.5);
            
            \diagram*{
                (p1) -- [gluon,edge label=\(\varepsilon_1^\mu \), momentum'={\(q_1^\mu\)}] (p12),
                (p12) --[fermion, edge label'={\(L_1^\mu\)}] (p22),
                (p22) --[fermion, edge label'={\(L_2^\mu\)}] (p32),
                (p32) --[fermion] (p3aux),
                (pn2) --[fermion, edge label'={\(L_n^\mu\)}] (p12),
                (pnaux) --[fermion, edge label'={\(L_{n-1}^\mu\)}] (pn2),

                (p3) -- [gluon,edge label=\(\varepsilon_3^\mu\), momentum'={\(q_3^\mu\)}] (p32),
                (p2) -- [gluon,edge label=\(\varepsilon_2^\mu\), momentum'={\(q_2^\mu\)}] (p22),
                (pn) -- [gluon,edge label=\(\varepsilon_n^\mu\), momentum'={\(q_n^\mu\)}] (pn2)
            };
         \end{feynman}
    \end{tikzpicture}}} = \Tr [\slashed{L}_1\slashed{\varepsilon}_1 \dots \slashed{L}_n\slashed{\varepsilon}_n ] = L_1^{\mu_1}\varepsilon_1^{\nu_1}\dots L_n^{\mu_n} \varepsilon_n^{\nu_n} \mathcal{L}_{\mu_1 \nu_1 \dots \mu_n \nu_n }\, ,
\end{equation}
where $\mathcal{L}_{\alpha_1\dots \alpha_{2n}} = \Tr [\gamma_{\alpha_1}\dots\gamma_{\alpha_{2n}}]$ and is related to the Pfaffian tensor\footnote{when contracted with an anti-symmetric matrix $M_{\mu \nu}$ we obtain the Pfaffian of that matrix Pf$(M) = $Pf$_{\mu_1 \nu_1 \dots \mu_n \nu_n} M_{\mu\nu} \dots M_{\mu_n\nu_n}$}, Pf, in the following way
\begin{equation}
    \mathcal{L}_{\alpha_1 \dots \alpha_{2n}} = \Tr [\gamma_{\alpha_1}\dots\gamma_{\alpha_{2n}}] = d_\gamma \left(\text{Pf}\right)_{\alpha_1 \dots \alpha_{2n}} \, , 
    \label{eq:LTensor}
\end{equation}
where $d_\gamma$ is the dimension of the gamma matrices, and 
\begin{equation}
	\text{Pf}_{\alpha_1\dots \alpha_{2n}} = \sum_{\sigma\in S_{2n}} (-1)^{sgn(\sigma)} \prod_{i=1}^{n} \eta_{\sigma(\alpha_i),\sigma(\alpha_{i+1})} \, ,
    \label{eq:Perm}
\end{equation}
with $S_{2n}$ being the group of permutations and $sgn(\sigma)$ is the sign of the permutation. As it turns out, each term in this sum follows a very simple geometric picture. In order to understand it, it is instructive to look at the two point example ($2n=4$). In this case, we use standard gamma matrix manipulations to find 
\begin{equation}
	\text{Pf}_{\alpha_1 \dots \alpha_4} = \eta_{\alpha_1 \alpha_2} \eta_{\alpha_3,\alpha_4} + \eta_{\alpha_4,\alpha_1}\eta_{\alpha_2,\alpha_3} - \eta_{\alpha_1,\alpha_3}\eta_{\alpha_2,\alpha_4} \, .
	\label{eq:Pf2pt}
\end{equation}
To keep track of the different Lorentzian contractions above, we draw lines between the $2n$ points on the boundary of a disk, and for collection of lines the overall sign of the respective contraction is given by the total number of intersections of the curves. So for the $2$ point case, the three terms in \eqref{eq:Pf2pt} are simply
\begin{equation}
\text{Pf}_{\alpha_1\dots\alpha_4} = 
	\vcenter{\hbox{\begin{tikzpicture}[line width=0.6 ,scale=1,line cap=round,every node/.style={font=\footnotesize},xshift=-6cm]
    		\begin{scope}[ scale=0.6]
    			\draw[very thick] (0,0) circle (2);
    			\coordinate (a1) at ($2*({cos(45)},{sin(45)})$);
    			\coordinate (a2) at ($2*({cos(135)},{sin(135)})$);
    			\coordinate (a3) at ($2*({cos(225)},{sin(225)})$);
    			\coordinate (a4) at ($2*({cos(-45)},{sin(-45)})$);
				
				\node (p1) at ($(a1)+(0.3,0.3)$) {$\alpha_2$};
				\node (p2) at ($(a2)+(-0.3,0.3)$) {$\alpha_1$};
				\node (p3) at ($(a3)+(-0.3,-0.3)$) {$\alpha_4$};
				\node (p4) at ($(a4)+(0.3,-0.3)$) {$\alpha_3$};

    			\draw[very thick,Blue] (a1)--(a2);
    			\draw[very thick,Blue] (a3)--(a4);

				\filldraw [very thick] (a1) circle (2pt);
    			\filldraw [very thick] (a2) circle (2pt);
    			\filldraw [very thick] (a3) circle (2pt);
    			\filldraw [very thick] (a4) circle (2pt);
    \end{scope}
    \end{tikzpicture}}} \quad + \quad  
		\vcenter{\hbox{\begin{tikzpicture}[line width=0.6 ,scale=1,line cap=round,every node/.style={font=\footnotesize}]
    		\begin{scope}[ scale=0.6]
    			\draw[very thick] (0,0) circle (2);
    			\coordinate (a1) at ($2*({cos(45)},{sin(45)})$);
    			\coordinate (a2) at ($2*({cos(135)},{sin(135)})$);
    			\coordinate (a3) at ($2*({cos(225)},{sin(225)})$);
    			\coordinate (a4) at ($2*({cos(-45)},{sin(-45)})$);
				
				\node (p1) at ($(a1)+(0.3,0.3)$) {$\alpha_2$};
				\node (p2) at ($(a2)+(-0.3,0.3)$) {$\alpha_1$};
				\node (p3) at ($(a3)+(-0.3,-0.3)$) {$\alpha_4$};
				\node (p4) at ($(a4)+(0.3,-0.3)$) {$\alpha_3$};

    			\draw[very thick,Blue] (a4)--(a1);
    			\draw[very thick,Blue] (a2)--(a3);

				\filldraw [very thick] (a1) circle (2pt);
    			\filldraw [very thick] (a2) circle (2pt);
    			\filldraw [very thick] (a3) circle (2pt);
    			\filldraw [very thick] (a4) circle (2pt);
    \end{scope}
    \end{tikzpicture}}} \quad + \quad
    \vcenter{\hbox{\begin{tikzpicture}[line width=0.6 ,scale=1,line cap=round,every node/.style={font=\footnotesize}]
    		\begin{scope}[ scale=0.6]
    			\draw[very thick] (0,0) circle (2);
    			\coordinate (a1) at ($2*({cos(45)},{sin(45)})$);
    			\coordinate (a2) at ($2*({cos(135)},{sin(135)})$);
    			\coordinate (a3) at ($2*({cos(225)},{sin(225)})$);
    			\coordinate (a4) at ($2*({cos(-45)},{sin(-45)})$);
				
				\node (p1) at ($(a1)+(0.3,0.3)$) {$\alpha_2$};
				\node (p2) at ($(a2)+(-0.3,0.3)$) {$\alpha_1$};
				\node (p3) at ($(a3)+(-0.3,-0.3)$) {$\alpha_4$};
				\node (p4) at ($(a4)+(0.3,-0.3)$) {$\alpha_3$};

    			\draw[very thick,Blue] (a1)--(a3);
    			\draw[very thick,Blue] (a2)--(a4);

				\filldraw [very thick] (a1) circle (2pt);
    			\filldraw [very thick] (a2) circle (2pt);
    			\filldraw [very thick] (a3) circle (2pt);
    			\filldraw [very thick] (a4) circle (2pt);
    \end{scope}
    \end{tikzpicture}}} \, .
    \label{eq:DiskContract}
  \end{equation}
Using this graphical representation for the permutations entering in \eqref{eq:Perm}, we can then replace the $(-1)^{\text{sgn}(\sigma)}$ by $(-1)^{\# \text{int}}$ where int is the total number of intersections between the contraction curves.  Note that given a contraction pattern entering $\mathcal{L}$, to extract the contribution to the LS we need to further contract it with the momenta of each loop propagator, $L_i^\mu $, and the polarizations, $\varepsilon_i^\mu $, as given in \eqref{eq:FermionTrace}.

One natural way of incorporating this fermion LS into our picture for pure gluon LS, is to consider the disk labelling the permutation contractions from \eqref{eq:DiskContract} to coincide with the internal boundary of the puncture in the fatgraph. In this way any curve contracted with  a red handle inside the puncture, is now extended along the relevant contraction curve inside the disk. When we do this extension we find that now curves are also allowed to end on loop propagators (and not only on the red handles at vertices) -- this is precisely capturing the contractions with $L_i^\mu$. For example, at two points, we get the following three different contraction pictures

\begin{equation}
	\vcenter{\hbox{\begin{tikzpicture}[line width=0.6,line cap=round,every node/.style={font=\footnotesize}]
		\begin{scope}[scale = 0.45]
            
\draw[thick](0,0) circle (0.7);

\coordinate (2c) at (-3,0);
\coordinate (2t) at (-4,1);
\coordinate (2b) at (-4,-1);
\node (r2) at ($(2c)+(-1,0)$) {2}; 

\coordinate (4c) at (3,0);
\coordinate (4t) at (4,1);
\coordinate (4b) at (4,-1);
\node (r4) at ($(4c)+(1,0)$) {4};

\coordinate (3l1) at (-3.5,1.5);
\coordinate (3l2) at (-2.5,0.5);
\coordinate (3l3) at (-1.1,0.5);

\coordinate (3r1) at (3.5,1.5);
\coordinate (3r2) at (2.5,0.5);
\coordinate (3r3) at (1.1,0.5);
\coordinate (3aux) at (0,1.25);

\node (r3) at ($(3aux)+(0,0.5)$) {3};

\coordinate (1l1) at (-3.5,-1.5);
\coordinate (1l2) at (-2.5,-0.5);
\coordinate (1l3) at (-1.1,-0.5);
\coordinate (1r1) at (3.5,-1.5);
\coordinate (1r2) at (2.5,-0.5);
\coordinate (1r3) at (1.1,-0.5);
\coordinate (1aux) at (0,-1.25);

\node (r1) at ($(1aux)+(0,-0.5)$) {1};

\coordinate (2ht) at ($(2c)!0.2!(2t)$);
\coordinate (2hb) at ($(2c)!0.2!(2b)$);
\coordinate (4hb) at ($(4c)!0.2!(4b)$);
\coordinate (4ht) at ($(4c)!0.2!(4t)$);

\coordinate (phrt) at ($0.5*(0.92388, 0.382683)$);
\coordinate (phrb) at ($0.5*(0.92388, -0.382683)$);

\coordinate (phlt) at ($0.5*(-0.92388, 0.382683)$);
\coordinate (phlb) at ($0.5*(-0.92388, -0.382683)$);

\coordinate (c1l) at ($(2ht)!0.5!(2hb)$);
\coordinate (c1r) at ($(phlt)!0.5!(phlb)$);

\coordinate (c2l) at ($(4ht)!0.5!(4hb)$);
\coordinate (c2r) at ($(phrt)!0.5!(phrb)$);

\draw[Maroon, very thick] (2hb) to[out=135, in=-135] (2ht);
\draw[Maroon, very thick] (4hb) to[out=45, in=-45] (4ht);



\draw (2c) to (2t);
\draw (2c) to (2b);

\draw (4c) to (4t);
\draw (4c) to (4b);
\draw (3l1) to (3l2);
\draw (3l2) to (3l3);
\draw (3r1) to (3r2);
\draw (3r2) to (3r3);
\draw (3r3) to[out=115, in=0] (3aux);
\draw (3aux) to[out=180, in=65] (3l3);

\draw (1l1) to (1l2);
\draw (1l2) to (1l3);
\draw (1r1) to (1r2);
\draw (1r2) to (1r3);
\draw (1r3) to[out=-115, in=0] (1aux);
\draw (1aux) to[out=180, in=-65] (1l3);
\node at (0,0) {$p$};

\draw[->,thick] ($(3l2)+(0.1,0.2)$) -- ($(3l3)+(-0.1,0.2)$) node[midway, above, color=black] {$q_1^\mu$};
\draw[<-,thick] ($1.5*({cos(-150)},{sin(-150)})$) arc (-150:-115:1.5);
\node at ($2*({cos(-130)},{sin(-130)})$) {$L^\mu$};

\node at (0,-4) {\footnotesize $-(\epsilon_1 \cdot \epsilon_2)(L\cdot(L+q_1))$};
\node at (0,-5) {\footnotesize $=0$};

			\coordinate (mh2) at ($(2hb)!0.5!(2ht)$);
			\coordinate (mh4) at ($(4hb)!0.5!(4ht)$);
			
			\coordinate (auxW) at (180:0.7);
			\coordinate (auxE) at (0:0.7);
			\coordinate (auxN) at (90:0.7);
			\coordinate (auxS) at (-90:0.7);
			
			\draw[Blue] (mh2) to (mh4);
            \draw[Blue] (auxN) to (auxS);

		\end{scope}
	\end{tikzpicture}
}}
\, 
\vcenter{\hbox{\begin{tikzpicture}[line width=0.6,line cap=round,every node/.style={font=\footnotesize}]
		\begin{scope}[scale = 0.45]
            
\draw[thick](0,0) circle (0.7);

\coordinate (2c) at (-3,0);
\coordinate (2t) at (-4,1);
\coordinate (2b) at (-4,-1);
\node (r2) at ($(2c)+(-1,0)$) {2}; 

\coordinate (4c) at (3,0);
\coordinate (4t) at (4,1);
\coordinate (4b) at (4,-1);
\node (r4) at ($(4c)+(1,0)$) {4};

\coordinate (3l1) at (-3.5,1.5);
\coordinate (3l2) at (-2.5,0.5);
\coordinate (3l3) at (-1.1,0.5);

\coordinate (3r1) at (3.5,1.5);
\coordinate (3r2) at (2.5,0.5);
\coordinate (3r3) at (1.1,0.5);
\coordinate (3aux) at (0,1.25);

\node (r3) at ($(3aux)+(0,0.5)$) {3};

\coordinate (1l1) at (-3.5,-1.5);
\coordinate (1l2) at (-2.5,-0.5);
\coordinate (1l3) at (-1.1,-0.5);
\coordinate (1r1) at (3.5,-1.5);
\coordinate (1r2) at (2.5,-0.5);
\coordinate (1r3) at (1.1,-0.5);
\coordinate (1aux) at (0,-1.25);

\node (r1) at ($(1aux)+(0,-0.5)$) {1};

\coordinate (2ht) at ($(2c)!0.2!(2t)$);
\coordinate (2hb) at ($(2c)!0.2!(2b)$);
\coordinate (4hb) at ($(4c)!0.2!(4b)$);
\coordinate (4ht) at ($(4c)!0.2!(4t)$);

\coordinate (phrt) at ($0.5*(0.92388, 0.382683)$);
\coordinate (phrb) at ($0.5*(0.92388, -0.382683)$);

\coordinate (phlt) at ($0.5*(-0.92388, 0.382683)$);
\coordinate (phlb) at ($0.5*(-0.92388, -0.382683)$);

\coordinate (c1l) at ($(2ht)!0.5!(2hb)$);
\coordinate (c1r) at ($(phlt)!0.5!(phlb)$);

\coordinate (c2l) at ($(4ht)!0.5!(4hb)$);
\coordinate (c2r) at ($(phrt)!0.5!(phrb)$);

\draw[Maroon, very thick] (2hb) to[out=135, in=-135] (2ht);
\draw[Maroon, very thick] (4hb) to[out=45, in=-45] (4ht);



\draw (2c) to (2t);
\draw (2c) to (2b);

\draw (4c) to (4t);
\draw (4c) to (4b);
\draw (3l1) to (3l2);
\draw (3l2) to (3l3);
\draw (3r1) to (3r2);
\draw (3r2) to (3r3);
\draw (3r3) to[out=115, in=0] (3aux);
\draw (3aux) to[out=180, in=65] (3l3);

\draw (1l1) to (1l2);
\draw (1l2) to (1l3);
\draw (1r1) to (1r2);
\draw (1r2) to (1r3);
\draw (1r3) to[out=-115, in=0] (1aux);
\draw (1aux) to[out=180, in=-65] (1l3);
\node at (0,0) {$p$};

\draw[->,thick] ($(3l2)+(0.1,0.2)$) -- ($(3l3)+(-0.1,0.2)$) node[midway, above, color=black] {$q_1^\mu$};
\draw[<-,thick] ($1.5*({cos(-150)},{sin(-150)})$) arc (-150:-115:1.5);
\node at ($2*({cos(-130)},{sin(-130)})$) {$L^\mu$};


			\coordinate (mh2) at ($(2hb)!0.5!(2ht)$);
			\coordinate (mh4) at ($(4hb)!0.5!(4ht)$);
			
			\coordinate (auxW) at (180:0.7);
			\coordinate (auxE) at (0:0.7);
			\coordinate (auxN) at (90:0.7);
			\coordinate (auxS) at (-90:0.7);
			
			\draw[Blue] (mh2) to (auxW) to[out=0, in=-90] (auxN);
			\draw[Blue] (mh4) to (auxE) to[out=180, in=90] (auxS);

            \node at (0,-4) {$(\epsilon_1\cdot(L+q_1))(\epsilon_2\cdot L)$};
            \node at (0,-5) {$ = (-X_{4,p})(-X_{2,p})$};
            
		\end{scope}
	\end{tikzpicture}
}}
\, 
\vcenter{\hbox{\begin{tikzpicture}[line width=0.6,line cap=round,every node/.style={font=\footnotesize}]
		\begin{scope}[scale = 0.45]
            
\draw[thick](0,0) circle (0.7);

\coordinate (2c) at (-3,0);
\coordinate (2t) at (-4,1);
\coordinate (2b) at (-4,-1);
\node (r2) at ($(2c)+(-1,0)$) {2}; 

\coordinate (4c) at (3,0);
\coordinate (4t) at (4,1);
\coordinate (4b) at (4,-1);
\node (r4) at ($(4c)+(1,0)$) {4};

\coordinate (3l1) at (-3.5,1.5);
\coordinate (3l2) at (-2.5,0.5);
\coordinate (3l3) at (-1.1,0.5);

\coordinate (3r1) at (3.5,1.5);
\coordinate (3r2) at (2.5,0.5);
\coordinate (3r3) at (1.1,0.5);
\coordinate (3aux) at (0,1.25);

\node (r3) at ($(3aux)+(0,0.5)$) {3};

\coordinate (1l1) at (-3.5,-1.5);
\coordinate (1l2) at (-2.5,-0.5);
\coordinate (1l3) at (-1.1,-0.5);
\coordinate (1r1) at (3.5,-1.5);
\coordinate (1r2) at (2.5,-0.5);
\coordinate (1r3) at (1.1,-0.5);
\coordinate (1aux) at (0,-1.25);

\node (r1) at ($(1aux)+(0,-0.5)$) {1};

\coordinate (2ht) at ($(2c)!0.2!(2t)$);
\coordinate (2hb) at ($(2c)!0.2!(2b)$);
\coordinate (4hb) at ($(4c)!0.2!(4b)$);
\coordinate (4ht) at ($(4c)!0.2!(4t)$);

\coordinate (phrt) at ($0.5*(0.92388, 0.382683)$);
\coordinate (phrb) at ($0.5*(0.92388, -0.382683)$);

\coordinate (phlt) at ($0.5*(-0.92388, 0.382683)$);
\coordinate (phlb) at ($0.5*(-0.92388, -0.382683)$);

\coordinate (c1l) at ($(2ht)!0.5!(2hb)$);
\coordinate (c1r) at ($(phlt)!0.5!(phlb)$);

\coordinate (c2l) at ($(4ht)!0.5!(4hb)$);
\coordinate (c2r) at ($(phrt)!0.5!(phrb)$);

\draw[Maroon, very thick] (2hb) to[out=135, in=-135] (2ht);
\draw[Maroon, very thick] (4hb) to[out=45, in=-45] (4ht);



\draw (2c) to (2t);
\draw (2c) to (2b);

\draw (4c) to (4t);
\draw (4c) to (4b);
\draw (3l1) to (3l2);
\draw (3l2) to (3l3);
\draw (3r1) to (3r2);
\draw (3r2) to (3r3);
\draw (3r3) to[out=115, in=0] (3aux);
\draw (3aux) to[out=180, in=65] (3l3);

\draw (1l1) to (1l2);
\draw (1l2) to (1l3);
\draw (1r1) to (1r2);
\draw (1r2) to (1r3);
\draw (1r3) to[out=-115, in=0] (1aux);
\draw (1aux) to[out=180, in=-65] (1l3);
\node at (0,0) {$p$};

\draw[->,thick] ($(3l2)+(0.1,0.2)$) -- ($(3l3)+(-0.1,0.2)$) node[midway, above, color=black] {$q_1^\mu$};
\draw[<-,thick] ($1.5*({cos(-150)},{sin(-150)})$) arc (-150:-115:1.5);
\node at ($2*({cos(-130)},{sin(-130)})$) {$L^\mu$};


			\coordinate (mh2) at ($(2hb)!0.5!(2ht)$);
			\coordinate (mh4) at ($(4hb)!0.5!(4ht)$);
			
			\coordinate (auxW) at (180:0.7);
			\coordinate (auxE) at (0:0.7);
			\coordinate (auxN) at (90:0.7);
			\coordinate (auxS) at (-90:0.7);
			
			\draw[Blue] (mh2) to (auxW) to[out=0, in=90] (auxS);
			\draw[Blue] (mh4) to (auxE) to[out=180, in=-90] (auxN);
			
			\node at (0,-4) {$(\epsilon_2\cdot(L+q_1))(\epsilon_1\cdot L)$};
            \node at (0,-5) {$=( - X_{2,p})(-X_{4,p})$};

		\end{scope}
	\end{tikzpicture}
}}
\end{equation}
%

Already from this example it is clear that this way of encoding the fermionic LS is different from the pure gluon case as it does not seem to ``care'' about filling the loop edges of the fatgraph -- instead the action is in filling the \textit{inside} puncture region with all possible contraction patterns giving the permutations in \eqref{eq:Perm}!

Nonetheless, it is still the case that we can read off the monomials associated with a given contraction pattern by an analogue of the cutting operation we found for pure gluon LS. To illustrate how the cutting works, we will work it out for the following two contraction curves
\begin{equation}
\label{eq:c1andc2Ferm}
\vcenter{\hbox{\begin{tikzpicture}[line width=0.6,line cap=round,every node/.style={font=\footnotesize}]
		\begin{scope}[scale = 0.5]

			\draw[thick](0,0) circle (0.7);
			\coordinate (2c) at (-3,0);
			\coordinate (2t) at (-4,1);
			\coordinate (2b) at (-4,-1);
			\node (r2) at ($(2c)+(-1,0)$) {$b$}; 
			
			\node at (0,-0.1) {\scriptsize $p$};
			
			\coordinate (4c) at (3,0);
			\coordinate (4t) at (4,1);
			\coordinate (4b) at (4,-1);
			\node (r4) at ($(4c)+(1,0)$) {$d$};

			\coordinate (3l1) at (-3.5,1.5);
			\coordinate (3l2) at (-2.5,0.5);
			\coordinate (3l3) at (-1.1,0.5);

			\coordinate (3r1) at (3.5,1.5);
			\coordinate (3r2) at (2.5,0.5);
			\coordinate (3r3) at (1.1,0.5);
			\coordinate (3aux) at (0,1.25);

			\node (r3) at ($(3r3)!0.5!(3r2)+(0,0.5)$) {$c$};

			\coordinate (1l1) at (-3.5,-1.5);
			\coordinate (1l2) at (-2.5,-0.5);
			\coordinate (1l3) at (-1.1,-0.5);
			\coordinate (1r1) at (3.5,-1.5);
			\coordinate (1r2) at (2.5,-0.5);
			\coordinate (1r3) at (1.1,-0.5);
			\coordinate (1aux) at (0,-1.25);

			\node (r1) at ($(1l3)!0.5!(1l2)+(0,-0.5)$) {$a$};

			\coordinate (2ht) at ($(2c)!0.2!(2t)$);
			\coordinate (2hb) at ($(2c)!0.2!(2b)$);
			\coordinate (4hb) at ($(4c)!0.2!(4b)$);
			\coordinate (4ht) at ($(4c)!0.2!(4t)$);

			\coordinate (phrt) at ($0.5*(0.92388, 0.382683)$);
			\coordinate (phrb) at ($0.5*(0.92388, -0.382683)$);

			\coordinate (phlt) at ($0.5*(-0.92388, 0.382683)$);
			\coordinate (phlb) at ($0.5*(-0.92388, -0.382683)$);

			\coordinate (c1l) at ($(2ht)!0.5!(2hb)$);
			\coordinate (c1r) at ($(phlt)!0.5!(phlb)$);

			\coordinate (c2l) at ($(4ht)!0.5!(4hb)$);
			\coordinate (c2r) at ($(phrt)!0.5!(phrb)$);



			\draw (2c) to (2t);
			\draw (2c) to (2b);

			\draw (4c) to (4t);
			\draw (4c) to (4b);
			\draw (3l1) to (3l2);
			\draw (3l2) to (3l3);
			\draw (3r1) to (3r2);
			\draw (3r2) to (3r3);
			\draw (3r3) to[out=115, in=0] (3aux);
			\draw[dotted] (3aux) to[out=180, in=65] (3l3);

			\draw (1l1) to (1l2);
			\draw (1l2) to (1l3);
			\draw (1r1) to (1r2);
			\draw (1r2) to (1r3);
			\draw[dotted] (1r3) to[out=-115, in=0] (1aux);
			\draw (1aux) to[out=180, in=-65] (1l3);
			\draw[Maroon, thick] (2hb) to[out=135, in=-135] (2ht);
			\draw[Maroon, thick] (4hb) to[out=45, in=-45] (4ht);
			
			\coordinate (mh2) at ($(2hb)!0.5!(2ht)$);
			\coordinate (mh4) at ($(4hb)!0.5!(4ht)$);
			
			\coordinate (auxW) at (180:0.7);
			\coordinate (auxE) at (0:0.7);
			\coordinate (auxN) at (90:0.7);
			\coordinate (auxS) at (-90:0.7);
			
			\draw[Blue, very thick] (mh4) to (mh2);
			\node[Blue] at ($(3l3)!0.5!(3l2) + (0,0.5)$) {$\mathcal{C}_1$};


			
		\end{scope}
	\end{tikzpicture}
}}
    \vcenter{\hbox{\begin{tikzpicture}[line width=0.6,line cap=round,every node/.style={font=\footnotesize}]
		\begin{scope}[scale = 0.5]

			\draw[thick](0,0) circle (0.7);
			\coordinate (2c) at (-3,0);
			\coordinate (2t) at (-4,1);
			\coordinate (2b) at (-4,-1);
			\node (r2) at ($(2c)+(-1,0)$) {$b$}; 
			
			\node at (0,-0.1) {\scriptsize $p$};
			
			\coordinate (4c) at (3,0);
			\coordinate (4t) at (4,1);
			\coordinate (4b) at (4,-1);
			\node (r4) at ($(4c)+(1,0)$) {$d$};

			\coordinate (3l1) at (-3.5,1.5);
			\coordinate (3l2) at (-2.5,0.5);
			\coordinate (3l3) at (-1.1,0.5);

			\coordinate (3r1) at (3.5,1.5);
			\coordinate (3r2) at (2.5,0.5);
			\coordinate (3r3) at (1.1,0.5);
			\coordinate (3aux) at (0,1.25);

			\node (r3) at ($(3r3)!0.5!(3r2)+(0,0.5)$) {$c$};

			\coordinate (1l1) at (-3.5,-1.5);
			\coordinate (1l2) at (-2.5,-0.5);
			\coordinate (1l3) at (-1.1,-0.5);
			\coordinate (1r1) at (3.5,-1.5);
			\coordinate (1r2) at (2.5,-0.5);
			\coordinate (1r3) at (1.1,-0.5);
			\coordinate (1aux) at (0,-1.25);

			\node (r1) at ($(1l3)!0.5!(1l2)+(0,-0.5)$) {$a$};

			\coordinate (2ht) at ($(2c)!0.2!(2t)$);
			\coordinate (2hb) at ($(2c)!0.2!(2b)$);
			\coordinate (4hb) at ($(4c)!0.2!(4b)$);
			\coordinate (4ht) at ($(4c)!0.2!(4t)$);

			\coordinate (phrt) at ($0.5*(0.92388, 0.382683)$);
			\coordinate (phrb) at ($0.5*(0.92388, -0.382683)$);

			\coordinate (phlt) at ($0.5*(-0.92388, 0.382683)$);
			\coordinate (phlb) at ($0.5*(-0.92388, -0.382683)$);

			\coordinate (c1l) at ($(2ht)!0.5!(2hb)$);
			\coordinate (c1r) at ($(phlt)!0.5!(phlb)$);

			\coordinate (c2l) at ($(4ht)!0.5!(4hb)$);
			\coordinate (c2r) at ($(phrt)!0.5!(phrb)$);



			\draw (2c) to (2t);
			\draw (2c) to (2b);

			\draw (4c) to (4t);
			\draw (4c) to (4b);
			\draw (3l1) to (3l2);
			\draw (3l2) to (3l3);
			\draw (3r1) to (3r2);
			\draw (3r2) to (3r3);
			\draw (3r3) to[out=115, in=0] (3aux);
			\draw[dotted] (3aux) to[out=180, in=65] (3l3);

			\draw (1l1) to (1l2);
			\draw (1l2) to (1l3);
			\draw (1r1) to (1r2);
			\draw (1r2) to (1r3);
			\draw[dotted] (1r3) to[out=-115, in=0] (1aux);
			\draw (1aux) to[out=180, in=-65] (1l3);
			\draw[Maroon, thick] (2hb) to[out=135, in=-135] (2ht);
			\draw[Maroon, thick] (4hb) to[out=45, in=-45] (4ht);
			
			\coordinate (mh2) at ($(2hb)!0.5!(2ht)$);
			\coordinate (mh4) at ($(4hb)!0.5!(4ht)$);
			
			\coordinate (auxW) at (180:0.7);
			\coordinate (auxE) at (0:0.7);
			\coordinate (auxN) at (90:0.7);
			\coordinate (auxS) at (-90:0.7);
			
			\draw[Blue, very thick] (mh2) to ($(3l3)!0.5!(1l3)$) to (auxW) to[out=0, in=-90] (auxN);
            \node[Blue] at ($(3l3)!0.5!(3l2) + (0,0.5)$) {$\mathcal{C}_2$};

			


			
		\end{scope}
	\end{tikzpicture}}}
\end{equation}
The cutting operation then goes as follows: 
\begin{enumerate}
\itemsep0em
	\item Start from each end of the contraction curve and identify the points at which it \textit{exits} the loop region. Consider then the subcurve that goes from the starting point until this intersection.
	\item As happened before, with this operation we generate two subcurves: one coming from starting at the left end of $\mathcal{C}$, we call $C_L$; and another from starting at the right end of $\mathcal{C}$, $C_R$. For example for $\mathcal{C}_1$ in \eqref{eq:c1andc2Ferm},  $C_L$ and $C_R$ are the following curves 
    \begin{equation}
	\vcenter{\hbox{\begin{tikzpicture}[line width=0.6,line cap=round,every node/.style={font=\footnotesize}]
		\begin{scope}[scale = 0.5]

			\draw[thick](0,0) circle (0.7);
			\coordinate (2c) at (-3,0);
			\coordinate (2t) at (-4,1);
			\coordinate (2b) at (-4,-1);
			\node (r2) at ($(2c)+(-1,0)$) {$b$}; 
			
			
			\coordinate (4c) at (3,0);
			\coordinate (4t) at (4,1);
			\coordinate (4b) at (4,-1);
			\node (r4) at ($(4c)+(1,0)$) {$d$};

			\coordinate (3l1) at (-3.5,1.5);
			\coordinate (3l2) at (-2.5,0.5);
			\coordinate (3l3) at (-1.1,0.5);

			\coordinate (3r1) at (3.5,1.5);
			\coordinate (3r2) at (2.5,0.5);
			\coordinate (3r3) at (1.1,0.5);
			\coordinate (3aux) at (0,1.25);

			\node (r3) at ($(3r3)!0.5!(3r2)+(0,0.5)$) {$c$};

			\coordinate (1l1) at (-3.5,-1.5);
			\coordinate (1l2) at (-2.5,-0.5);
			\coordinate (1l3) at (-1.1,-0.5);
			\coordinate (1r1) at (3.5,-1.5);
			\coordinate (1r2) at (2.5,-0.5);
			\coordinate (1r3) at (1.1,-0.5);
			\coordinate (1aux) at (0,-1.25);

			\node (r1) at ($(1l3)!0.5!(1l2)+(0,-0.5)$) {$a$};

			\coordinate (2ht) at ($(2c)!0.2!(2t)$);
			\coordinate (2hb) at ($(2c)!0.2!(2b)$);
			\coordinate (4hb) at ($(4c)!0.2!(4b)$);
			\coordinate (4ht) at ($(4c)!0.2!(4t)$);

			\coordinate (phrt) at ($0.5*(0.92388, 0.382683)$);
			\coordinate (phrb) at ($0.5*(0.92388, -0.382683)$);

			\coordinate (phlt) at ($0.5*(-0.92388, 0.382683)$);
			\coordinate (phlb) at ($0.5*(-0.92388, -0.382683)$);

			\coordinate (c1l) at ($(2ht)!0.5!(2hb)$);
			\coordinate (c1r) at ($(phlt)!0.5!(phlb)$);

			\coordinate (c2l) at ($(4ht)!0.5!(4hb)$);
			\coordinate (c2r) at ($(phrt)!0.5!(phrb)$);



			\draw (2c) to (2t);
			\draw (2c) to (2b);

			\draw (4c) to (4t);
			\draw (4c) to (4b);
			\draw (3l1) to (3l2);
			\draw (3l2) to (3l3);
			\draw (3r1) to (3r2);
			\draw (3r2) to (3r3);
			\draw (3r3) to[out=115, in=0] (3aux);
			\draw[dotted] (3aux) to[out=180, in=65] (3l3);

			\draw (1l1) to (1l2);
			\draw (1l2) to (1l3);
			\draw (1r1) to (1r2);
			\draw (1r2) to (1r3);
			\draw[dotted] (1r3) to[out=-115, in=0] (1aux);
			\draw (1aux) to[out=180, in=-65] (1l3);
			\draw[Maroon, thick] (2hb) to[out=135, in=-135] (2ht);
			\draw[Maroon, thick] (4hb) to[out=45, in=-45] (4ht);
			
			\coordinate (mh2) at ($(2hb)!0.5!(2ht)$);
			\coordinate (mh4) at ($(4hb)!0.5!(4ht)$);
			
			\coordinate (auxW) at (180:0.7);
			\coordinate (auxE) at (0:0.7);
			\coordinate (auxN) at (90:0.7);
			\coordinate (auxS) at (-90:0.7);
			
			\draw[Blue, very thick] (mh2) to (auxE);
			\node at ($(3l3)!0.5!(3l2) + (0,0.5)$) {$C_L$};


			
		\end{scope}
	\end{tikzpicture}
}}
\vcenter{\hbox{\begin{tikzpicture}[line width=0.6,line cap=round,every node/.style={font=\footnotesize}]
		\begin{scope}[scale = 0.5]

			\draw[thick](0,0) circle (0.7);
			\coordinate (2c) at (-3,0);
			\coordinate (2t) at (-4,1);
			\coordinate (2b) at (-4,-1);
			\node (r2) at ($(2c)+(-1,0)$) {$b$}; 
			
			
			\coordinate (4c) at (3,0);
			\coordinate (4t) at (4,1);
			\coordinate (4b) at (4,-1);
			\node (r4) at ($(4c)+(1,0)$) {$d$};

			\coordinate (3l1) at (-3.5,1.5);
			\coordinate (3l2) at (-2.5,0.5);
			\coordinate (3l3) at (-1.1,0.5);

			\coordinate (3r1) at (3.5,1.5);
			\coordinate (3r2) at (2.5,0.5);
			\coordinate (3r3) at (1.1,0.5);
			\coordinate (3aux) at (0,1.25);

			\node (r3) at ($(3r3)!0.5!(3r2)+(0,0.5)$) {$c$};

			\coordinate (1l1) at (-3.5,-1.5);
			\coordinate (1l2) at (-2.5,-0.5);
			\coordinate (1l3) at (-1.1,-0.5);
			\coordinate (1r1) at (3.5,-1.5);
			\coordinate (1r2) at (2.5,-0.5);
			\coordinate (1r3) at (1.1,-0.5);
			\coordinate (1aux) at (0,-1.25);

			\node (r1) at ($(1l3)!0.5!(1l2)+(0,-0.5)$) {$a$};

			\coordinate (2ht) at ($(2c)!0.2!(2t)$);
			\coordinate (2hb) at ($(2c)!0.2!(2b)$);
			\coordinate (4hb) at ($(4c)!0.2!(4b)$);
			\coordinate (4ht) at ($(4c)!0.2!(4t)$);

			\coordinate (phrt) at ($0.5*(0.92388, 0.382683)$);
			\coordinate (phrb) at ($0.5*(0.92388, -0.382683)$);

			\coordinate (phlt) at ($0.5*(-0.92388, 0.382683)$);
			\coordinate (phlb) at ($0.5*(-0.92388, -0.382683)$);

			\coordinate (c1l) at ($(2ht)!0.5!(2hb)$);
			\coordinate (c1r) at ($(phlt)!0.5!(phlb)$);

			\coordinate (c2l) at ($(4ht)!0.5!(4hb)$);
			\coordinate (c2r) at ($(phrt)!0.5!(phrb)$);



			\draw (2c) to (2t);
			\draw (2c) to (2b);

			\draw (4c) to (4t);
			\draw (4c) to (4b);
			\draw (3l1) to (3l2);
			\draw (3l2) to (3l3);
			\draw (3r1) to (3r2);
			\draw (3r2) to (3r3);
			\draw (3r3) to[out=115, in=0] (3aux);
			\draw[dotted] (3aux) to[out=180, in=65] (3l3);

			\draw (1l1) to (1l2);
			\draw (1l2) to (1l3);
			\draw (1r1) to (1r2);
			\draw (1r2) to (1r3);
			\draw[dotted] (1r3) to[out=-115, in=0] (1aux);
			\draw (1aux) to[out=180, in=-65] (1l3);
			\draw[Maroon, thick] (2hb) to[out=135, in=-135] (2ht);
			\draw[Maroon, thick] (4hb) to[out=45, in=-45] (4ht);
			
			\coordinate (mh2) at ($(2hb)!0.5!(2ht)$);
			\coordinate (mh4) at ($(4hb)!0.5!(4ht)$);
			
			\coordinate (auxW) at (180:0.7);
			\coordinate (auxE) at (0:0.7);
			\coordinate (auxN) at (90:0.7);
			\coordinate (auxS) at (-90:0.7);
			
			\draw[Blue, very thick] (auxW) to (mh4);
			
			\node at ($(1r3)!0.5!(1r2) + (0,-0.5)$) {$C_R$};


			
		\end{scope}
	\end{tikzpicture}
}}
\end{equation}
	\item Additionally, we also need to consider the full curve $C^\cup$, together with the subcurve correspoding to the intersection, $C^\cap = C_L \cap C_R $, which in the example above is given by the curves below
    \begin{equation}
	\vcenter{\hbox{\begin{tikzpicture}[line width=0.6,line cap=round,every node/.style={font=\footnotesize}]
		\begin{scope}[scale = 0.5]

			\draw[thick](0,0) circle (0.7);
			\coordinate (2c) at (-3,0);
			\coordinate (2t) at (-4,1);
			\coordinate (2b) at (-4,-1);
			\node (r2) at ($(2c)+(-1,0)$) {$b$}; 
			
			
			\coordinate (4c) at (3,0);
			\coordinate (4t) at (4,1);
			\coordinate (4b) at (4,-1);
			\node (r4) at ($(4c)+(1,0)$) {$d$};

			\coordinate (3l1) at (-3.5,1.5);
			\coordinate (3l2) at (-2.5,0.5);
			\coordinate (3l3) at (-1.1,0.5);

			\coordinate (3r1) at (3.5,1.5);
			\coordinate (3r2) at (2.5,0.5);
			\coordinate (3r3) at (1.1,0.5);
			\coordinate (3aux) at (0,1.25);

			\node (r3) at ($(3r3)!0.5!(3r2)+(0,0.5)$) {$c$};

			\coordinate (1l1) at (-3.5,-1.5);
			\coordinate (1l2) at (-2.5,-0.5);
			\coordinate (1l3) at (-1.1,-0.5);
			\coordinate (1r1) at (3.5,-1.5);
			\coordinate (1r2) at (2.5,-0.5);
			\coordinate (1r3) at (1.1,-0.5);
			\coordinate (1aux) at (0,-1.25);

			\node (r1) at ($(1l3)!0.5!(1l2)+(0,-0.5)$) {$a$};

			\coordinate (2ht) at ($(2c)!0.2!(2t)$);
			\coordinate (2hb) at ($(2c)!0.2!(2b)$);
			\coordinate (4hb) at ($(4c)!0.2!(4b)$);
			\coordinate (4ht) at ($(4c)!0.2!(4t)$);

			\coordinate (phrt) at ($0.5*(0.92388, 0.382683)$);
			\coordinate (phrb) at ($0.5*(0.92388, -0.382683)$);

			\coordinate (phlt) at ($0.5*(-0.92388, 0.382683)$);
			\coordinate (phlb) at ($0.5*(-0.92388, -0.382683)$);

			\coordinate (c1l) at ($(2ht)!0.5!(2hb)$);
			\coordinate (c1r) at ($(phlt)!0.5!(phlb)$);

			\coordinate (c2l) at ($(4ht)!0.5!(4hb)$);
			\coordinate (c2r) at ($(phrt)!0.5!(phrb)$);



			\draw (2c) to (2t);
			\draw (2c) to (2b);

			\draw (4c) to (4t);
			\draw (4c) to (4b);
			\draw (3l1) to (3l2);
			\draw (3l2) to (3l3);
			\draw (3r1) to (3r2);
			\draw (3r2) to (3r3);
			\draw (3r3) to[out=115, in=0] (3aux);
			\draw[dotted] (3aux) to[out=180, in=65] (3l3);

			\draw (1l1) to (1l2);
			\draw (1l2) to (1l3);
			\draw (1r1) to (1r2);
			\draw (1r2) to (1r3);
			\draw[dotted] (1r3) to[out=-115, in=0] (1aux);
			\draw (1aux) to[out=180, in=-65] (1l3);
			\draw[Maroon, thick] (2hb) to[out=135, in=-135] (2ht);
			\draw[Maroon, thick] (4hb) to[out=45, in=-45] (4ht);
			
			\coordinate (mh2) at ($(2hb)!0.5!(2ht)$);
			\coordinate (mh4) at ($(4hb)!0.5!(4ht)$);
			
			\coordinate (auxW) at (180:0.7);
			\coordinate (auxE) at (0:0.7);
			\coordinate (auxN) at (90:0.7);
			\coordinate (auxS) at (-90:0.7);
			
			\draw[Blue, very thick] (auxW) to (auxE);
			
			\node at ($(1aux)+ (0,-0.7)$) {$C^\cap$};


			
		\end{scope}
	\end{tikzpicture}
}}
\vcenter{\hbox{\begin{tikzpicture}[line width=0.6,line cap=round,every node/.style={font=\footnotesize}]
		\begin{scope}[scale = 0.5]

			\draw[thick](0,0) circle (0.7);
			\coordinate (2c) at (-3,0);
			\coordinate (2t) at (-4,1);
			\coordinate (2b) at (-4,-1);
			\node (r2) at ($(2c)+(-1,0)$) {$b$}; 
			
			
			\coordinate (4c) at (3,0);
			\coordinate (4t) at (4,1);
			\coordinate (4b) at (4,-1);
			\node (r4) at ($(4c)+(1,0)$) {$d$};

			\coordinate (3l1) at (-3.5,1.5);
			\coordinate (3l2) at (-2.5,0.5);
			\coordinate (3l3) at (-1.1,0.5);

			\coordinate (3r1) at (3.5,1.5);
			\coordinate (3r2) at (2.5,0.5);
			\coordinate (3r3) at (1.1,0.5);
			\coordinate (3aux) at (0,1.25);

			\node (r3) at ($(3r3)!0.5!(3r2)+(0,0.5)$) {$c$};

			\coordinate (1l1) at (-3.5,-1.5);
			\coordinate (1l2) at (-2.5,-0.5);
			\coordinate (1l3) at (-1.1,-0.5);
			\coordinate (1r1) at (3.5,-1.5);
			\coordinate (1r2) at (2.5,-0.5);
			\coordinate (1r3) at (1.1,-0.5);
			\coordinate (1aux) at (0,-1.25);

			\node (r1) at ($(1l3)!0.5!(1l2)+(0,-0.5)$) {$a$};

			\coordinate (2ht) at ($(2c)!0.2!(2t)$);
			\coordinate (2hb) at ($(2c)!0.2!(2b)$);
			\coordinate (4hb) at ($(4c)!0.2!(4b)$);
			\coordinate (4ht) at ($(4c)!0.2!(4t)$);

			\coordinate (phrt) at ($0.5*(0.92388, 0.382683)$);
			\coordinate (phrb) at ($0.5*(0.92388, -0.382683)$);

			\coordinate (phlt) at ($0.5*(-0.92388, 0.382683)$);
			\coordinate (phlb) at ($0.5*(-0.92388, -0.382683)$);

			\coordinate (c1l) at ($(2ht)!0.5!(2hb)$);
			\coordinate (c1r) at ($(phlt)!0.5!(phlb)$);

			\coordinate (c2l) at ($(4ht)!0.5!(4hb)$);
			\coordinate (c2r) at ($(phrt)!0.5!(phrb)$);



			\draw (2c) to (2t);
			\draw (2c) to (2b);

			\draw (4c) to (4t);
			\draw (4c) to (4b);
			\draw (3l1) to (3l2);
			\draw (3l2) to (3l3);
			\draw (3r1) to (3r2);
			\draw (3r2) to (3r3);
			\draw (3r3) to[out=115, in=0] (3aux);
			\draw[dotted] (3aux) to[out=180, in=65] (3l3);

			\draw (1l1) to (1l2);
			\draw (1l2) to (1l3);
			\draw (1r1) to (1r2);
			\draw (1r2) to (1r3);
			\draw[dotted] (1r3) to[out=-115, in=0] (1aux);
			\draw (1aux) to[out=180, in=-65] (1l3);
			\draw[Maroon, thick] (2hb) to[out=135, in=-135] (2ht);
			\draw[Maroon, thick] (4hb) to[out=45, in=-45] (4ht);
			
			\coordinate (mh2) at ($(2hb)!0.5!(2ht)$);
			\coordinate (mh4) at ($(4hb)!0.5!(4ht)$);
			
			\coordinate (auxW) at (180:0.7);
			\coordinate (auxE) at (0:0.7);
			\coordinate (auxN) at (90:0.7);
			\coordinate (auxS) at (-90:0.7);
			
			\draw[Blue, very thick] (mh4) to (mh2);
			
            \node at ($(1aux)+(0,-0.7)$) {$C^\cup$};



			
		\end{scope}
	\end{tikzpicture}
}}
\end{equation}
    \item Now there are two possibilities: \textbf{(1)} $C^\cap \neq C_L,C_R$, in which case the contribution from the contraction curve in terms of $X_\mathcal{C}$ is simply: 
    \begin{equation}
        X_{C^\cup} + X_{C^\cap} -X_{C_L}-X_{C_R}.
    \end{equation}
    \textbf{(2)} If the contraction curve ends on the loop, like in the case of $\mathcal{C}_2$, then we have that $C^\cap$ and $C^\cup$ coincide with either $C_L$ or $C_R$, for example in  $\mathcal{C}_2$,  $C_R \equiv C^\cap$ and $C_L \equiv C^\cup$. Then in these cases we must a new curve: the complement $C \setminus C^{\cap}$. For example in $\mathcal{C}_2$,  $C \setminus C^{\cap}$ corresponds to the following curve:
    \begin{equation}
    \vcenter{\hbox{\begin{tikzpicture}[line width=0.6,line cap=round,every node/.style={font=\footnotesize}]
		\begin{scope}[scale = 0.5]

			\draw[thick](0,0) circle (0.7);
			\coordinate (2c) at (-3,0);
			\coordinate (2t) at (-4,1);
			\coordinate (2b) at (-4,-1);
			\node (r2) at ($(2c)+(-1,0)$) {$b$}; 
			
			\node at (0,-0.1) {\scriptsize $p$};
			
			\coordinate (4c) at (3,0);
			\coordinate (4t) at (4,1);
			\coordinate (4b) at (4,-1);
			\node (r4) at ($(4c)+(1,0)$) {$d$};

			\coordinate (3l1) at (-3.5,1.5);
			\coordinate (3l2) at (-2.5,0.5);
			\coordinate (3l3) at (-1.1,0.5);

			\coordinate (3r1) at (3.5,1.5);
			\coordinate (3r2) at (2.5,0.5);
			\coordinate (3r3) at (1.1,0.5);
			\coordinate (3aux) at (0,1.25);

			\node (r3) at ($(3r3)!0.5!(3r2)+(0,0.5)$) {$c$};

			\coordinate (1l1) at (-3.5,-1.5);
			\coordinate (1l2) at (-2.5,-0.5);
			\coordinate (1l3) at (-1.1,-0.5);
			\coordinate (1r1) at (3.5,-1.5);
			\coordinate (1r2) at (2.5,-0.5);
			\coordinate (1r3) at (1.1,-0.5);
			\coordinate (1aux) at (0,-1.25);

			\node (r1) at ($(1l3)!0.5!(1l2)+(0,-0.5)$) {$a$};

			\coordinate (2ht) at ($(2c)!0.2!(2t)$);
			\coordinate (2hb) at ($(2c)!0.2!(2b)$);
			\coordinate (4hb) at ($(4c)!0.2!(4b)$);
			\coordinate (4ht) at ($(4c)!0.2!(4t)$);

			\coordinate (phrt) at ($0.5*(0.92388, 0.382683)$);
			\coordinate (phrb) at ($0.5*(0.92388, -0.382683)$);

			\coordinate (phlt) at ($0.5*(-0.92388, 0.382683)$);
			\coordinate (phlb) at ($0.5*(-0.92388, -0.382683)$);

			\coordinate (c1l) at ($(2ht)!0.5!(2hb)$);
			\coordinate (c1r) at ($(phlt)!0.5!(phlb)$);

			\coordinate (c2l) at ($(4ht)!0.5!(4hb)$);
			\coordinate (c2r) at ($(phrt)!0.5!(phrb)$);



			\draw (2c) to (2t);
			\draw (2c) to (2b);

			\draw (4c) to (4t);
			\draw (4c) to (4b);
			\draw (3l1) to (3l2);
			\draw (3l2) to (3l3);
			\draw (3r1) to (3r2);
			\draw (3r2) to (3r3);
			\draw (3r3) to[out=115, in=0] (3aux);
			\draw[dotted] (3aux) to[out=180, in=65] (3l3);

			\draw (1l1) to (1l2);
			\draw (1l2) to (1l3);
			\draw (1r1) to (1r2);
			\draw (1r2) to (1r3);
			\draw[dotted] (1r3) to[out=-115, in=0] (1aux);
			\draw (1aux) to[out=180, in=-65] (1l3);
			\draw[Maroon, thick] (2hb) to[out=135, in=-135] (2ht);
			\draw[Maroon, thick] (4hb) to[out=45, in=-45] (4ht);
			
			\coordinate (mh2) at ($(2hb)!0.5!(2ht)$);
			\coordinate (mh4) at ($(4hb)!0.5!(4ht)$);
			
			\coordinate (auxW) at (180:0.7);
			\coordinate (auxE) at (0:0.7);
			\coordinate (auxN) at (90:0.7);
			\coordinate (auxS) at (-90:0.7);
			
			\draw[Blue, very thick] (mh2) to ($(3l3)!0.5!(1l3)$) to (auxW) to[out=0, in=-90] (auxN);
			


			
		\end{scope}
	\end{tikzpicture}
}}
\rightarrow
	\vcenter{\hbox{\begin{tikzpicture}[line width=0.6,line cap=round,every node/.style={font=\footnotesize}]
		\begin{scope}[scale = 0.5]

			\draw[thick](0,0) circle (0.7);
			\coordinate (2c) at (-3,0);
			\coordinate (2t) at (-4,1);
			\coordinate (2b) at (-4,-1);
			\node (r2) at ($(2c)+(-1,0)$) {$b$}; 
			
			\node at (0,-0.1) {\scriptsize $p$};
			
			\coordinate (4c) at (3,0);
			\coordinate (4t) at (4,1);
			\coordinate (4b) at (4,-1);
			\node (r4) at ($(4c)+(1,0)$) {$d$};

			\coordinate (3l1) at (-3.5,1.5);
			\coordinate (3l2) at (-2.5,0.5);
			\coordinate (3l3) at (-1.1,0.5);

			\coordinate (3r1) at (3.5,1.5);
			\coordinate (3r2) at (2.5,0.5);
			\coordinate (3r3) at (1.1,0.5);
			\coordinate (3aux) at (0,1.25);

			\node (r3) at ($(3r3)!0.5!(3r2)+(0,0.5)$) {$c$};

			\coordinate (1l1) at (-3.5,-1.5);
			\coordinate (1l2) at (-2.5,-0.5);
			\coordinate (1l3) at (-1.1,-0.5);
			\coordinate (1r1) at (3.5,-1.5);
			\coordinate (1r2) at (2.5,-0.5);
			\coordinate (1r3) at (1.1,-0.5);
			\coordinate (1aux) at (0,-1.25);

			\node (r1) at ($(1l3)!0.5!(1l2)+(0,-0.5)$) {$a$};

			\coordinate (2ht) at ($(2c)!0.2!(2t)$);
			\coordinate (2hb) at ($(2c)!0.2!(2b)$);
			\coordinate (4hb) at ($(4c)!0.2!(4b)$);
			\coordinate (4ht) at ($(4c)!0.2!(4t)$);

			\coordinate (phrt) at ($0.5*(0.92388, 0.382683)$);
			\coordinate (phrb) at ($0.5*(0.92388, -0.382683)$);

			\coordinate (phlt) at ($0.5*(-0.92388, 0.382683)$);
			\coordinate (phlb) at ($0.5*(-0.92388, -0.382683)$);

			\coordinate (c1l) at ($(2ht)!0.5!(2hb)$);
			\coordinate (c1r) at ($(phlt)!0.5!(phlb)$);

			\coordinate (c2l) at ($(4ht)!0.5!(4hb)$);
			\coordinate (c2r) at ($(phrt)!0.5!(phrb)$);



			\draw (2c) to (2t);
			\draw (2c) to (2b);

			\draw (4c) to (4t);
			\draw (4c) to (4b);
			\draw (3l1) to (3l2);
			\draw (3l2) to (3l3);
			\draw (3r1) to (3r2);
			\draw (3r2) to (3r3);
			\draw (3r3) to[out=115, in=0] (3aux);
			\draw[dotted] (3aux) to[out=180, in=65] (3l3);

			\draw (1l1) to (1l2);
			\draw (1l2) to (1l3);
			\draw (1r1) to (1r2);
			\draw (1r2) to (1r3);
			\draw[dotted] (1r3) to[out=-115, in=0] (1aux);
			\draw (1aux) to[out=180, in=-65] (1l3);
			\draw[Maroon, thick] (2hb) to[out=135, in=-135] (2ht);
			\draw[Maroon, thick] (4hb) to[out=45, in=-45] (4ht);
			
			\coordinate (mh2) at ($(2hb)!0.5!(2ht)$);
			\coordinate (mh4) at ($(4hb)!0.5!(4ht)$);
			
			\coordinate (auxW) at (180:0.7);
			\coordinate (auxE) at (0:0.7);
			\coordinate (auxN) at (90:0.7);
			\coordinate (auxS) at (-90:0.7);
			
			\draw[Blue,very thick] (mh2) to (auxW);
			\node at ($(3l3)!0.5!(3l2) + (0,0.85)$) {$C\setminus C^\cap$};


			
		\end{scope}
	\end{tikzpicture}
}}
\end{equation}
and so from this type of contractions we get the following combination of $X_\mathcal{C}$:
\begin{equation}
 X_{C^\cup} + X_{C\setminus C^\cap} -X_{C^\cap}.    
\end{equation}
\end{enumerate}

Note that once again, the corresponding minus signs comes exactly from thinking of the relevant subcurve and asking how many extensions we need to produce the full curve, in which case it comes with a $(-1)^{N_e}$. 

Now the main difference comparing with pure gluon LS is that while we can still go back and forth from the monomials to the contraction curves, having a general understanding for the cancellations seems more subtle. This is because the sign of a monomial not only depends on the number of extensions but also on the full pattern of contractions inside the loop -- where the sign depends on the \textit{total} number of intersections. We leave this systematic analysis of the cancellations for fermion LS for future work. Such an analysis would be interesting as it could provide hints into how fermions could be embedded into the surface-integral \footnote{see \cite{Cao:2025lzv,De:2024wsy} for other approaches to include fermions in surface integrals as well as tropical integrals.}, thus bringing this formalism even closer to the real world theory of QCD.

\section{Outlook}
\label{sec:Outlook}

There are several directions for future work that follow naturally from the results we have presented in this paper. An immediate question is whether our graphical understanding of leading singularities can  be leveraged to find an efficient  algorithm for  \textit{analytically} computing them in $D$-dimensions, to be contrasted with the often numerical computation of these objects in practical QCD calculations. 

A more systematic treatment of the story of leading singularities including fermion loops would also clearly be  interesting, most importantly in getting closer to QCD, but also potentially in giving us clues, from the bottom up, for how to incorporate fermions in the surface integral formalism. In another direction, it is interesting to ask how leading singularities/amplitudes with \textit{external} fermion lines can be incorporated in the surface formalism. Recent progress in this front is also presented in \cite{De:2024wsy,Cao:2025lzv}. Nonetheless, just as gluon polarizations are dealt with by thinking about them as being produced/scaffolded by scalars, it is natural to ask whether external fermions with general fermionic wavefunctions can be discovered on cuts of gluon leading singularities.

Finally, it would be interesting to look at the ``theoretical data'' of D-dimensional leading singularities, and see if they have any hidden simplicity hinting at deeper underlying structures, analogous to positive grassmannians and amplituhedra for N=4 SYM. 

Along these lines, it is natural to look at the leading singularities for infinite classes of diagrams, in some simple kinematic limit that lets us compare them on an equal footing, in order to look for natural recursive structures. An example is that of $n$-gon LS for all $n$.  As with any 1-loop LS, we have two different kinematical variables: $X_{i,j}$ and $X_{i,p}$, where the former only involves external momenta while the latter also contains loop momenta. Following this natural separation, we can consider the limit $ X_{i,p} \rightarrow Y$ and $ X_{i,j}\rightarrow X \, $
then we find the following $n$-gon LS up to $n=8$
\begin{equation}
\begin{aligned}
    \text{LS}_{2} &= (2-D)Y^2 \, ,\\
    \text{LS}_{3} &= (2-D)Y^3 + 2(-X^3+3X^2Y)\, ,\\
    \text{LS}_{4} &= (2-D)Y^4- X^4 + 4X^3Y - 2X^2Y^2\, ,\\
     \text{LS}_{5} &= (2-D)Y^5- X^5 + 5X^4Y - 5X^3Y^2\, ,\\
      \text{LS}_{6} &= (2-D)Y^6- X^6 + 6X^5Y - 9X^4Y^2 + 2X^3Y^3\, ,\\  
    \text{LS}_{7} &= (2-D)Y^7- X^7 + 7X^6Y - 14X^5Y^2 + 7X^4Y^3\, ,\\
    \text{LS}_{8} &= (2-D)Y^8- X^8 + 8X^7Y - 20X^6Y^2 + 16X^5Y^3 -2X^4Y^4 \, .
\end{aligned}
\end{equation}
The coefficient of $Y^n$ is always $(2-D)$, which is a simple consequence of the discussion in sec. \ref{sec:CancellationsVRule} and the final result in \eqref{eq:FinalCancLoop}. The remaining polynomial in $X$ has an interesting structure. In particular we see that from the box diagram onwards the coefficient of $X^n$ is always $(-1)$. To analyse the pattern it's useful to recast the LS as 
\begin{equation}
    \text{LS}_n = (2-D)Y^n - \sum_{k=0}^{n/2  } C_k Y^k X^{n-k} \, .
    \label{eq:LSngonXY}
\end{equation}
where the coefficients $C_k$  for $k>4$ follow the recursion relations of the coefficients of the Lucas polynomials \cite{koshy2001fibonacci}\footnote{Note that for the box, the only coefficient that doesn't match the Lucas polynomials is $C_2$, while for the triangle there is only one extra overall factor of $2$.}! Up to $n=8$, the Lucas polynomials, $\text{L}_n$, are given by
\begin{equation}
\begin{aligned}
    \text{L}_{3}(x) &= x^3 - 3x \, , \\
    \text{L}_{4}(x) &= x^4 - 4x^2 + 2 \, ,\\
    \text{L}_{5}(x) &= x^5 - 5x^3 + 5x\, ,
    \end{aligned} \quad \quad \begin{aligned}
    \text{L}_{6}(x) &= x^6 - 6x^4 + 9x^2 - 2\, ,\\
    \text{L}_{7}(x) &= x^7 - 7x^5 + 14x^3 - 7x\, ,\\
    \text{L}_{8}(x) &= x^8 - 8x^6 + 20x^4 - 16x^2 + 2\, ,
\end{aligned}
\end{equation}
and we have checked the exact matching between the $C_k$ and the coefficients of $L_n$ up to $n=13$. As it turns out, this sequence of Lucas polynomials follows a simple recurrence relation given by:
\begin{equation}
    L_n(x) = \begin{cases}
        L_0(x)=2, \, L_1(x)=x,\\
        L_n(x)=x L_{n-1}(x) - L_{n-2}(x).
    \end{cases}
\end{equation}
An explanation for this matching is still lacking, but understanding the simplification of the graphical rule in this $X/Y$ limit would be very interesting, and explain why the residues of the surface integral can be recursively related to each other via the recursion relation above.

\acknowledgments 
We thank Nima Arkani-Hamed, Qu Cao, Song He and Johannes Henn for important discussions. C.F. is supported  by FCT - Fundacao para a Ciencia e Tecnologia, I.P. (2023.01221.BD and DOI  https://doi.org/10.54499/2023.01221.BD). S.C. is supported by the European Union (ERC, UNIVERSE PLUS, 101118787). Views and opinions expressed are however those of the authors only and do not necessarily reflect those of the European Union or the European Research Council Executive Agency. Neither the European Union nor the granting authority can be held responsible for them.

\bibliographystyle{JHEP}\bibliography{Refs}

\end{document}